%% file: DoctoralThesis.tex
\documentclass[12pt,shieldcrest]{ociamthesis} % use older shield crest logo - single-sided (electronic version)
\usepackage{amsmath}
\usepackage{amssymb}
\usepackage{graphicx}
\usepackage{supertabular}
\usepackage{mathrsfs}
\usepackage{subcaption}
\usepackage{graphbox}
\usepackage{color}
\usepackage[table]{xcolor}
\usepackage{hyperref}
\hypersetup{colorlinks, citecolor=black, filecolor=black, linkcolor=black, urlcolor=black}

\newcommand{\ironicon}{\reflectbox?}

\newcommand{\refEq}[1] {(\ref{#1})}

\newcommand{\Sin}[1]{\ensuremath{\sin \left( #1 \right)}}
\newcommand{\Cos}[1]{\ensuremath{\cos \left( #1 \right)}}
\newcommand{\Sec}[1]{\ensuremath{\sec \left( #1 \right)}}

\newcommand{\Tan}[1]{\ensuremath{\tan \left( #1 \right)}}
\newcommand{\ArcTan}[1]{\ensuremath{\text{arctan} \left( #1 \right)}}
\newcommand{\Exp}[1]{\ensuremath{\exp \left( #1 \right)}}
\newcommand{\Ln}[1]{\ensuremath{\ln \left( #1 \right)}}
\newcommand{\tensor}[1]{\ensuremath{\overset{\text{\tiny$\leftrightarrow$}}{#1}}}
\newcommand{\Nabla}{\ensuremath{\vec{\nabla}}}

\newcommand{\Order}[1]{\ensuremath{O \left( #1 \right)}}

\newcommand{\dlpdtheta}{\ensuremath{\frac{\partial l_{p}}{\partial \theta}}}
\newcommand{\dlpdthetaPrime}{\ensuremath{\frac{\partial l_{p}}{\partial \theta'}}}
\newcommand{\romanNum}[1]{\uppercase\expandafter{\romannumeral#1}}

\title{Up-down asymmetric tokamaks}   %note \\[1ex] is a line break in the title

\author{Justin Ball}             %your name
\college{Worcester College}  %your college

\degree{Doctor of Philosophy}     %the degree
\degreedate{Trinity Term 2016}         %the degree date

\begin{document}

%this baselineskip gives sufficient line spacing for an examiner to easily
%markup the thesis with comments
\baselineskip=18pt plus1pt

%set the number of sectioning levels that get number and appear in the contents
\setcounter{secnumdepth}{3}
\setcounter{tocdepth}{2}

\maketitle                  % create a title page from the preamble info
%\include{dedication}        % include a dedication.tex file
\include{abstract}          % include the abstract
\include{acknowlegements}   % include an acknowledgements.tex file

%\begin{romanpages}          % start roman page numbering
\tableofcontents            % generate and include a table of contents
%\listoffigures              % generate and include a list of figures
%\end{romanpages}            % end roman page numbering

%now include the files of latex for each of the chapters etc
%\part{Introduction}
%\label{part:INTRO}
\include{Ch_01_Introduction}

\part{Ideal MHD Equilibrium}
\label{part:MHD}
\include{Ch_02_MHD_GradShafSol}

\include{Ch_03_MHD_ShapingIntuition}

\include{Ch_04_MHD_ShafranovShift}

\include{Ch_05_MHD_LocalEquil}

\part{Turbulent transport}
\label{part:TRANSPORT}
\include{Ch_06_GYRO_Overview}

\include{Ch_07_GYRO_TiltingSymmetry}

\include{Ch_08_GYRO_MomFluxScaling}

\include{Ch_09_GYRO_ShafranovShift}

\include{Ch_10_GYRO_NonMirrorSym}

\include{Ch_11_Conclusions}

%now enable appendix numbering format and include any appendices
\appendix
\part{Appendices}
\include{App_A_MaxShaping}
\include{App_B_MagAxisLoc}

\include{App_C_TurbulentFluxes}
\include{App_D_GeneralCoeff}

\include{App_E_NonmirrorCoeff}
\include{App_F_BetaDependence}

%next line adds the Bibliography to the contents page
\addcontentsline{toc}{chapter}{References}
%uncomment next line to change bibliography name to references
\renewcommand{\bibname}{References}
%\bibliography{refs}        %use a bibtex bibliography file refs.bib
%\bibliographystyle{plain}  %use the plain bibliography style

\bibliographystyle{unsrt}
%\bibliography{/Users/Justin/Documents/Research/Bibliography/references.bib}
\bibliography{references.bib}

\end{document}

%% file: abstract.tex
% !TEX root = /Users/Justin/Documents/Research/Writings/2016DoctoralThesis/DoctoralThesis.tex

\begin{abstractseparate}
Bulk toroidal rotation has proven capable of stabilising both dangerous MHD modes and turbulence. This has allowed existing tokamaks to generate extra fusion power at a fixed size and magnetic field. However, most methods of inducing the plasma to spin do not appear to scale well to larger devices such as ITER or a future power plant. In this thesis, we explore a notable exception: up-down asymmetry in the tokamak magnetic equilibrium. When tokamak flux surfaces are not mirror symmetric about the midplane, turbulence can transport momentum from one surface to the next, creating spontaneous rotation that is ``intrinsic'' to the geometry. We seek to maximise this intrinsic rotation by finding optimal up-down asymmetric flux surface shapes.

First, we use the ideal MHD model to show that low order external shaping (e.g. elongation) is best for creating up-down asymmetric flux surfaces throughout the device. Then, we calculate realistic up-down asymmetric equilibria for input into nonlinear gyrokinetic turbulence analysis. Analytic gyrokinetics shows that, in the limit of fast shaping effects, a poloidal tilt of the flux surface shaping has little effect on turbulent transport. Since up-down symmetric surfaces do not transport momentum, this invariance to tilt implies that devices with mirror symmetry about any line in the poloidal plane will drive minimal rotation. Accordingly, further analytic investigation suggests that non-mirror symmetric flux surfaces with envelopes created by the beating of fast shaping effects may create significantly stronger momentum transport.

Guided by these analytic results, we carry out local nonlinear gyrokinetic simulations of non-mirror symmetric flux surfaces created with the lowest possible shaping effects. First, we consider tilted elliptical flux surfaces with a Shafranov shift and find little increase in the momentum transport when the effect of the pressure profile on the equilibrium is included. We then simulate flux surfaces with independently-tilted elongation and triangularity. These two-mode configurations show a $60\%$ increase over configurations with just elongation or triangularity. A rough analytic estimate indicates that the optimal two-mode configuration can drive rotation with an on-axis Alfv\'{e}n Mach number of $1.5 \%$ in an ITER-like machine.

\end{abstractseparate}

%% file: acknowlegements.tex
% !TEX root = /Users/Justin/Documents/Research/Writings/2016DoctoralThesis/DoctoralThesis.tex

\begin{acknowledgements}

First and foremost, I would like to thank Professor Felix Parra for being a great advisor and a great person. Five years is a long time, yet without his steady guidance it would have been (and felt) so much longer. I am deeply grateful to Felix for his patience and kindness.

On countless occasions during the course of my degree I depended on help from the leaders of our extended research group. Whether it be a practice talk, a last minute recommendation, or a conference bar crawl they were there. Thank you Michael Barnes, Peter Catto, Paul Dellar, Bill Dorland, and Alex Schekochihin!

I would also like to salute my fellow plasma theory students: Ferdinand van Wyk, Alessandro Geraldini, Michael Fox, Michael Hardman, and Greg Colyer. Through countless discussions they have made my lunch, my research, and my life more interesting. I think it is safe to say that we will forever be ``on tour''\ironicon

The biggest surprise of my time at Oxford was how quickly I felt at ease at Worcester College. No matter where I moved, the MCR still felt like home. For this I owe Steve Server, Rachel Kan, Helena Rodriguez, Belinda Lo, Stephanie Tam, Flo Benn, Geoff Pascoe, Sarah Barber, and Katie Light among others. I am especially indebted to Kim Fuellenbach for her company and encouragement during those final long evenings of writing.

Lastly, I thank my family for being a constant and continual source of support throughout my life. Mom, Dad, Elyse, Aunt Jill, Uncle Lex, and Aunt Margie: I believe this thesis is more a result of your efforts than my own.

\end{acknowledgements}

%% file: Ch_01_Introduction.tex
% !TEX root = /Users/Justin/Documents/Research/Writings/2016DoctoralThesis/DoctoralThesis.tex

%===================================================%
%===================================================%
%\chapter{A brief history of energy}
\chapter{Introduction}
\label{ch:historyOfEnergy}
%===================================================%
%===================================================%

Nuclear fusion is a fundamental and universal source of energy. Following the Big Bang nucleosynthesis, the universe was effectively a large cloud of hydrogen with small density fluctuations. In such a cloud there are two dominant sources of free energy: particle rest energy and gravitational potential energy. Through the action of gravity, the density perturbations have been gradually amplified into stars, which possess the conditions necessary to release particle rest energy via the process of nuclear fusion.

While gravity enables stellar fusion, it does not appear to be as attractive of an energy source. The sun, which dominates the energy budget of our solar system, has a capacity to produce $E_{\text{fusion}} \approx M_{\odot} c^{2} \approx 10^{47} J$ of fusion energy, while it only released $E_{\text{gravity}} \approx G M_{\odot}^{2} / R_{\odot} \approx 10^{41} J$ of gravitational potential energy during its entire formation. Here $G$ is the gravitational constant, $M_{\odot}$ is the solar mass, $R_{\odot}$ is the solar radius, and $c$ is the speed of light. On our own planet, the oceans \cite{KanthaOceans2000} alone contain approximately $10^{30} J$ of fusion energy directly accessible through deuterium-deuterium fusion. This is roughly the entire gravitational potential energy in the Earth-moon system. Unfortunately, fully extracting this through tidal power involves the destruction of earth by lunar impact.

\begin{figure}
 \centering
 \includegraphics[width=0.7\textwidth]{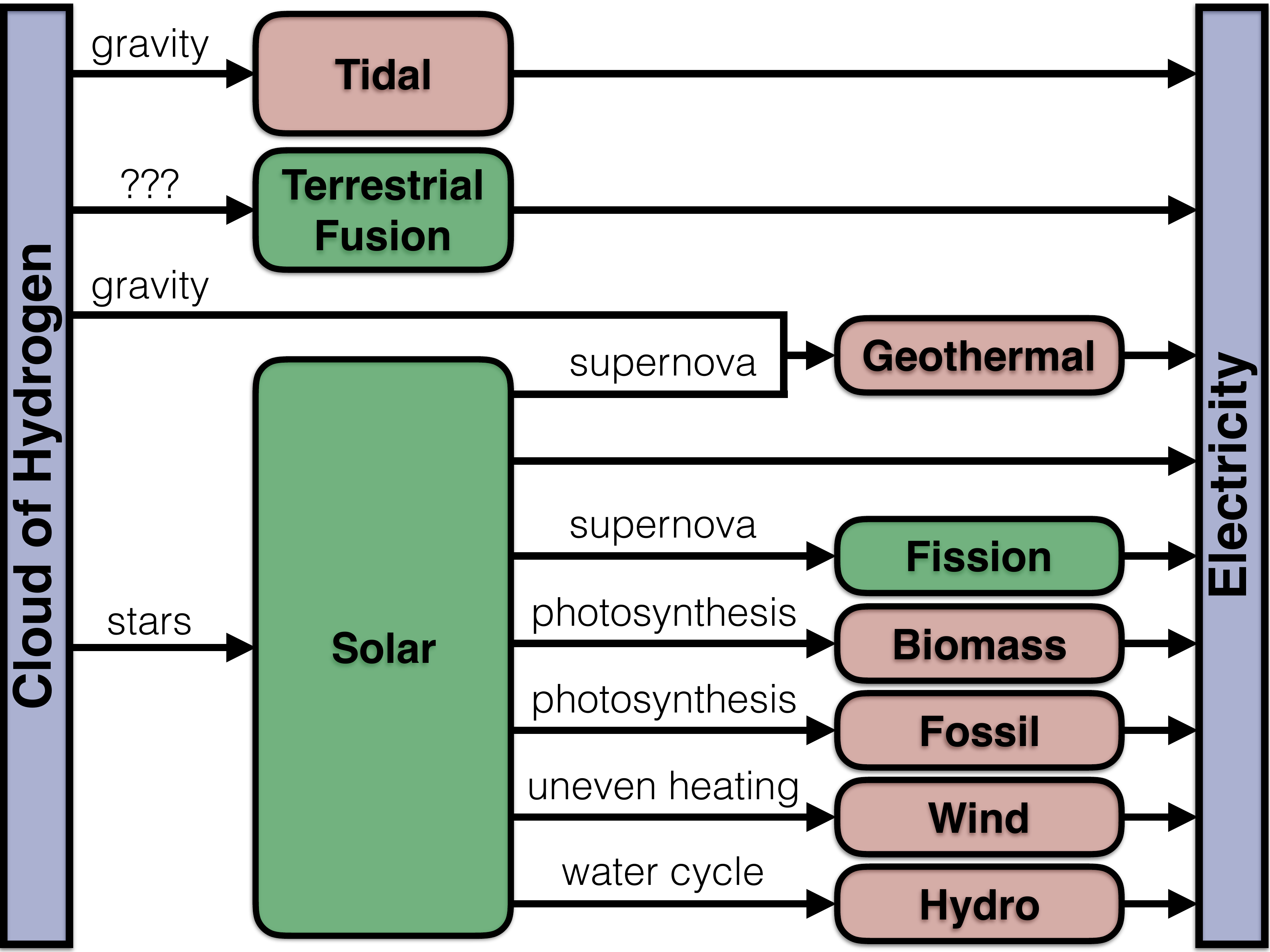}
 \caption{Techniques to generate electricity from an astronomically large cloud of hydrogen.}
 \label{fig:universalElectricityProduction}
\end{figure}

From figure \ref{fig:universalElectricityProduction}, we see that stellar fusion (i.e. solar) is the ultimate drive for nearly all sources of energy on earth. Unfortunately, though the solar energy incident on earth is on average $\sim 10^{5}$ times the current world energy consumption, the local value varies dramatically and unpredictably. Furthermore, the by-products of solar shown in figure \ref{fig:universalElectricityProduction} do not seem promising as we have the intuition that they will contain little energy. This turns out to be true for geothermal, biomass, fossil, wind, and hydro. For example, the total geothermal energy flux arriving at the surface of the earth is only $\sim 0.01\%$ of the solar energy flux and is barely above the current world energy consumption \cite{PollackGeothermalResources1993, IEAworldEnergyStats2015}. However, the nuclear fission of uranium and thorium in breeder reactors is an exception to this intuition, with an energy content that rivals deuterium-deuterium fusion. This is because much of the hydrogen escaped the atmosphere early in the formation of the earth, dramatically increasing the relative abundance of heavy elements compared to most places in the solar system.

Considering the above facts, it appears that here on earth we ultimately have three options:
\begin{itemize}
   \item solar power with energy storage,
   \item nuclear fission using breeder reactors, and/or
   \item terrestrial nuclear fusion.
\end{itemize}
The fact that none of these options are currently competitive with short-term energy solutions motivates this thesis, which will focus exclusively on the last.

%===================================================%
%===================================================%
\section{Terrestrial nuclear fusion}
\label{sec:nuclearFusion}
%===================================================%
%===================================================%

\begin{figure}
 \centering
 (a) \hspace{0.45\textwidth} (b) \hspace{0.45\textwidth}

 \includegraphics[width=0.4\textwidth]{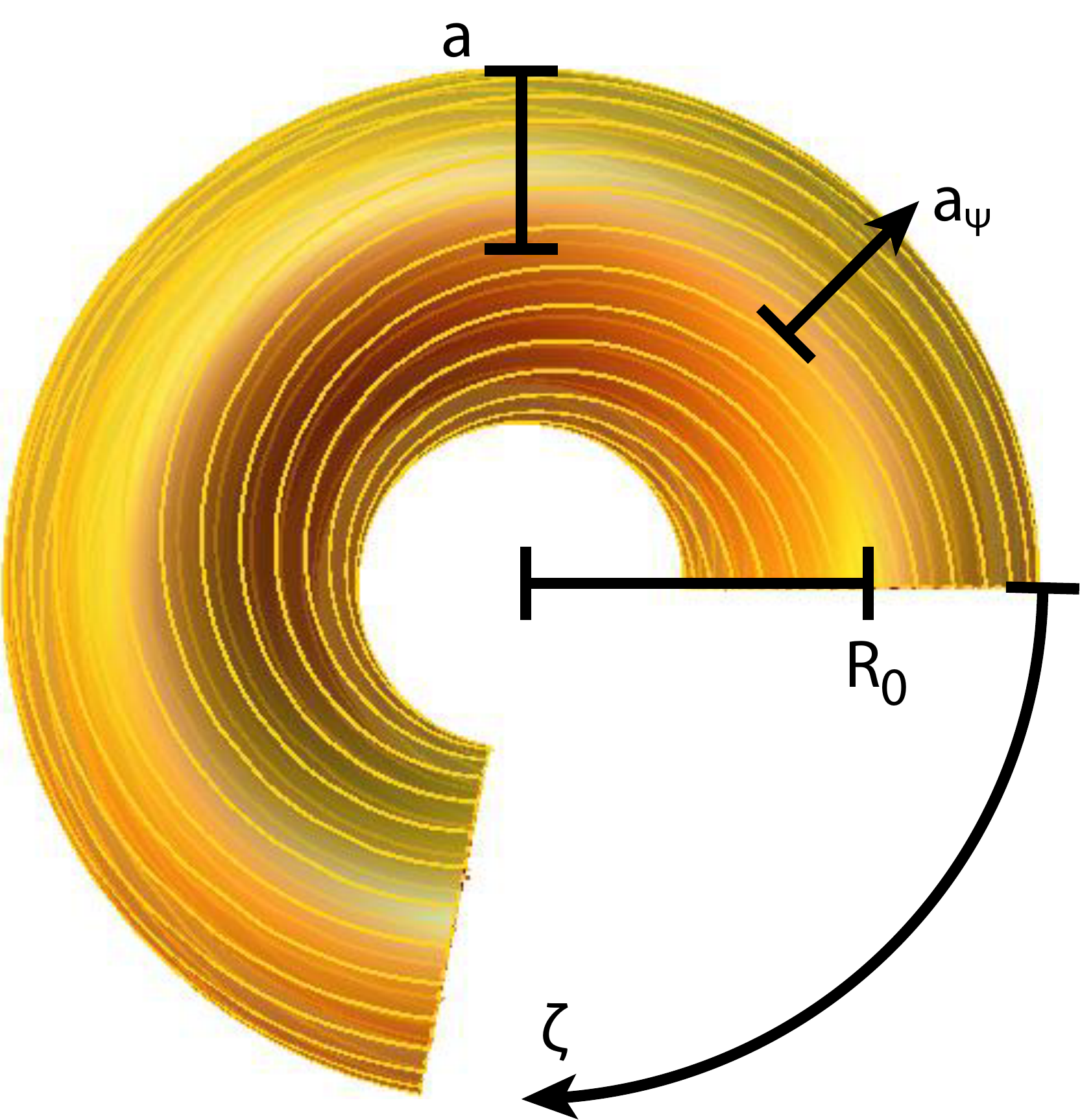}
 \includegraphics[width=0.4\textwidth]{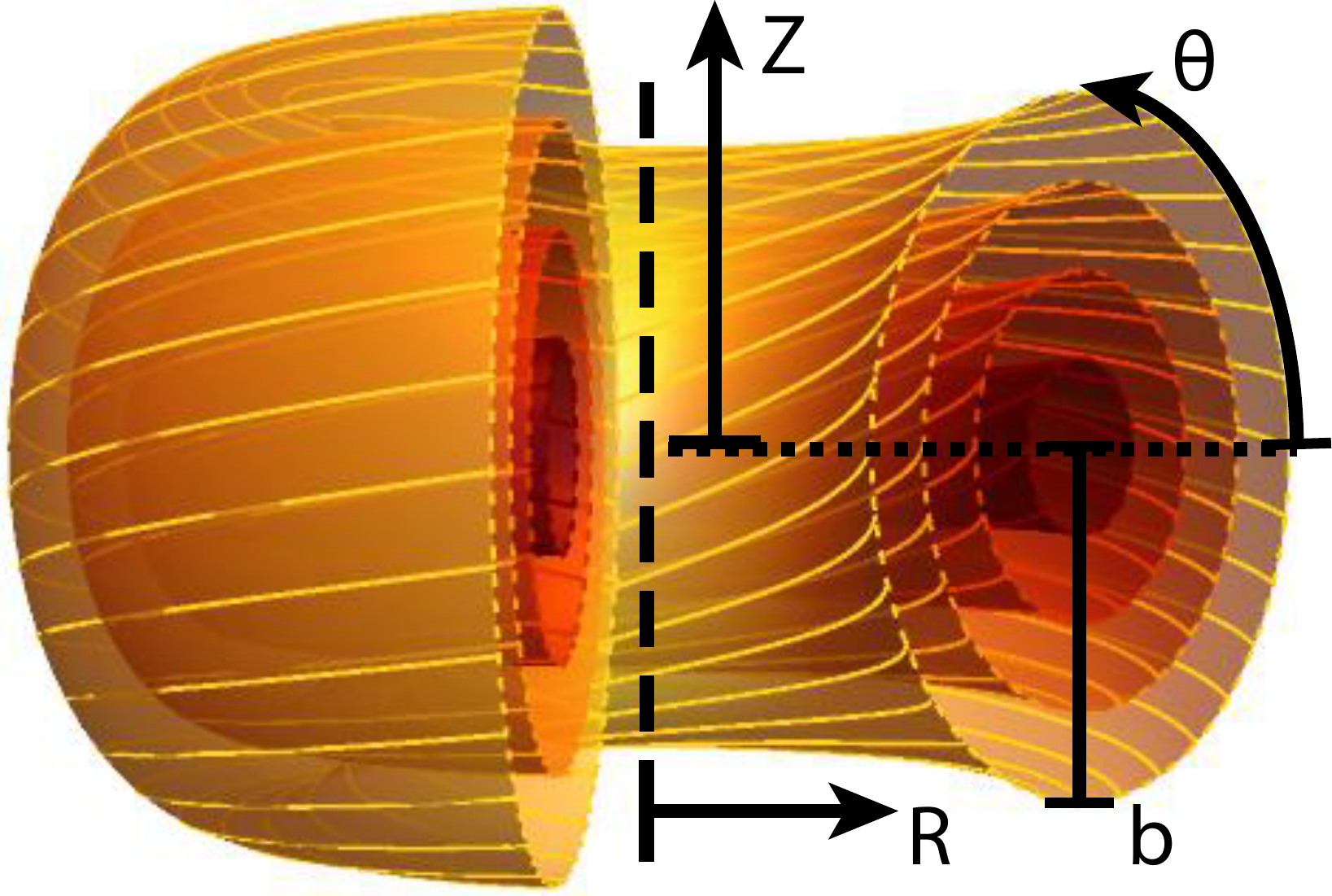}
 \caption{The (a) top and (b) side views of the magnetic field lines that form four elliptical flux surfaces in an example tokamak (with a toroidal section removed for illustrative purposes) showing the axis of symmetry (black, dashed) and the midplane (black, dotted), where $R$ is the major radial coordinate, $Z$ is the axial coordinate, $\zeta$ is the toroidal angle, $a_{\psi}$ is the minor radius flux surface label, $\theta$ is the poloidal angle, $a$ is the device minor radius, $R_{0}$ is the major radial location of the magnetic axis, and $\kappa \equiv b / a$ is the elongation.}
 \label{fig:tokamakGeo}
\end{figure}

Achieving fusion on earth has proven substantially more difficult than originally imagined. No terrestrial fusion device has ever produced more power than it has consumed, a basic requirement for a power plant. The device with the best experimental performance has consistently been the tokamak, a donut-shaped magnetic bottle capable of creating the stellar conditions necessary for fusion. Since the fuel must be astronomically hot, the thermal energy is sufficient to ionise the atoms making plasma, an electrically-charged gas of ions and free electrons. Because the fuel is charged, it is constrained to follow the magnetic field lines in the device (see figure \ref{fig:tokamakGeo}) according to the laws of electromagnetism. Currents in both external magnets and the plasma itself are used to create these magnetic field lines in such a way that they wrap around and close on themselves, forming nested magnetic surfaces known as flux surfaces (see figure \ref{fig:tokamakGeo}).

This would seem to work very well in principle. The energetic charged particles would spiral around the field lines and stream around the device, but never touch a solid surface. Thus, the magnetic field would provide the immense thermal insulation necessary to permit the stellar conditions for fusion to exist only a few metres from the solid material surface of the surrounding vacuum vessel.

However, in practice the enormous temperature gradients give rise to plasma turbulence, which degrades the thermal insulation and determines the performance of the device. If the plasma has stronger turbulence, energy leaks out faster and the plasma must be heated more in order to maintain the same temperature and fusion power. This necessitates more external heating power, which is what causes devices to consume more power than they generate. The fusion power record, achieved in the JET tokamak in 1997, is $16$ MW, $70\%$ of the power needed to heat the device \cite{KeilhackerJETrecord1999}.

Additionally, there is a constraint on how much plasma pressure a given magnetic field can contain. Even though the plasma may be forced to follow magnetic field lines, if the plasma pressure is large enough it can escape confinement by simply dragging the magnetic field with it. This notion is governed by magnetohydrodynamics (MHD) and is formalised through a limit on the plasma $\beta$ called the Troyon limit \cite{TroyonMHDLimit1984},
\begin{align}
   \beta_{N} \equiv \frac{\left( a/\text{m} \right) \left( B_{0}/\text{T} \right)}{\left( I_{p}/\text{MA} \right)} \beta \lesssim 0.03 , \label{eq:TroyonLimit}
\end{align}
where $B_{0}$ is the on-axis magnetic field, $I_{p}$ is the plasma current, $\beta \equiv 2 \mu_{0} p / B_{0}^{2}$ is the plasma beta, $\mu_{0}$ is the vacuum permeability, and $p$ is the plasma pressure. Exceeding this limit typically causes the whole plasma to go unstable, kinking until it makes contact with the vacuum vessel and rapidly cools. The Troyon limit is especially important not just because it constrains the safe operating space, but because it is related to the economics of a power plant. The reactor size and the magnetic field strength are the two most significant factors that determine the capital cost of a device. The plasma pressure is directly related to fusion power density and by that the total amount of power produced. The final quantity appearing is the plasma current, which must be driven externally and often dominates the external power needed to run the device. Hence, the Troyon limit can be thought of as a rough, but direct constraint on the cost of electricity.

%===================================================%
%===================================================%
\section{Toroidal plasma rotation}
\label{sec:rotation}
%===================================================%
%===================================================%

In this context, it is understandable that there has been much work on strategies to exceed the Troyon limit without inducing instability \cite{LiuITERrwmStabilization2004, OkabayashiActiveRWMfeedback2001, GarofaloActiveRWMfeedback2001}. One method, which also has the potential to directly reduce turbulence \cite{RitzRotShearTurbSuppression1990, BurrellShearTurbStabilization1997, BarnesFlowShear2011, HighcockRotationBifurcation2010}, is to use toroidal rotation. When the plasma has an average toroidal flow, interactions with the surrounding vacuum vessel are able to damp bulk plasma instabilities \cite{ReimerdesRWMmachineComp2006}. Experiments have used toroidal rotation to sustain discharges that violate the Troyon limit by a factor of two \cite{GarofaloExpRWMstabilizationD3D2002}. If only for this purpose, it is clear that control of toroidal rotation is beneficial for plasma performance. Unfortunately, the mechanisms that generate toroidal rotation in current experiments do not appear to scale well to future high-performance devices, which will likely be larger and have stronger magnetic fields. One such device that is currently under construction is ITER \cite{AymarITERSummary2001}. Current projections indicate that ITER will not be able to generate sufficiently fast toroidal rotation to allow violation of the Troyon limit. The necessary rotation is difficult to determine, but is estimated \cite{LiuITERrwmStabilization2004} to be in the range of
\begin{align}
   M_{A} \equiv \frac{V_{\zeta}}{v_{\text{Alfv\'{e}n}}} \approx 0.5 \% - 5 \% ,
\end{align}
where $M_{A}$ is the Alfv\'{e}n Mach number of the rotation, $V_{\zeta}$ is the bulk plasma toroidal velocity, and $v_{\text{Alfv\'{e}n}}$ is the Alfv\'{e}n speed. For ITER, one can multiply these values by $10$ to estimate the necessary Mach number, $M_{S} \equiv V_{\zeta} / c_{S} \approx 5 \% - 50 \%$, where $c_{S}$ is the plasma sound speed.

Tokamak plasmas start off at rest, but will start to spin if pushed using external injection of momentum. This is commonly done with beams of neutral particles, which enable current experiments to achieve toroidal rotation with $M_{A} \approx 3 \%$ \cite{GarofaloExpRWMstabilizationD3D2002}. However, since ITER has a much larger plasma, it has significantly more inertia and requires higher velocity neutral beams in order to penetrate to the plasma centre. Since energy is quadratic with velocity and momentum is linear, it can be shown that the ratio of the momentum to energy carried by a neutral beam varies inversely with the beam velocity \cite{ParraMomentumTransitions2011}. Hence, the neutral beams on ITER will be less efficient at driving rotation. Therefore, we should not be surprised that detailed modelling predicts external injection will only be capable of driving rotation with $M_{A} \approx 0.3 \%$ \cite{LiuITERrwmStabilization2004}, significantly less than what is required for violation of the Troyon limit.

%===================================================%
%===================================================%
\section{Up-down asymmetric plasma shaping}
\label{sec:upDownAsym}
%===================================================%
%===================================================%

Alternatively, experiments observe ``intrinsic'' rotation, or rotation spontaneously generated in the absence of external injection. This rotation arises from plasma turbulence moving momentum between flux surfaces and is especially attractive because it does not require any external power. In current experiments the speed of this intrinsic rotation is roughly $M_{A} \approx 1 \%$, but (as we will see in chapter \ref{ch:GYRO_Overview}) it is limited by a poloidal symmetry of tokamak turbulence to be small in $\rho_{\ast} \equiv \rho_{i} / a \ll 1$, the ratio of the ion gyroradius to the tokamak minor radius. Unfortunately, we expect that $\rho_{\ast}$ will get progressively smaller in future devices like ITER or a power plant.

However, there is one mechanism capable of breaking the symmetry of the turbulence to generate lowest order rotation in a stationary plasma: up-down asymmetric plasma shaping. When the tokamak flux surfaces do not have mirror symmetry about the midplane, the momentum transport at the top of the device is no longer guaranteed to cancel the momentum transport at the bottom. Hence large toroidal flows can spontaneously develop. In fact, reference \cite{CamenenPRLExp2010} presents results from the TCV tokamak that have provided the first experimental evidence of intrinsic rotation generated by up-down asymmetry. Consequently, reference \cite{BallMomUpDownAsym2014} performed nonlinear gyrokinetic simulations that are consistent with the TCV results and suggest that up-down asymmetry is a feasible method to generate the current, experimentally-measured rotation levels in reactor-sized devices.

\emph{This thesis will seek the up-down asymmetric flux surface shapes that maximise intrinsic rotation and overall plasma performance. It is separated into two fairly independent lines of inquiry. In part \ref{part:MHD}, we will use the ideal MHD model to calculate practical tokamak equilibria that maximise up-down asymmetric shaping throughout the plasma. In part \ref{part:TRANSPORT}, we will perform nonlinear gyrokinetic analysis of these realistic equilibria to identify the configurations that maximise turbulent momentum transport and minimise turbulent energy transport.}

First, in chapter \ref{ch:MHD_GradShafSol}, we will find solutions for up-down asymmetric MHD equilibria using an expansion in large aspect ratio, given simple radial profiles of the toroidal current and pressure. In chapter \ref{ch:MHD_ShapingIntuition}, we will study how the flux surface shaping in these solutions penetrates from the plasma edge to the magnetic axis in order to identify poloidally-tilted elongation as optimal for maximising up-down asymmetry throughout the device. Next, in chapter \ref{ch:MHD_ShafranovShift}, we will extend our MHD calculation to find the strength and direction of the Shafranov shift in tokamaks with tilted elliptical poloidal cross-sections. In chapter \ref{ch:MHD_LocalEquil}, we will derive local equilibria from the global equilibria of chapters \ref{ch:MHD_GradShafSol} and \ref{ch:MHD_ShafranovShift} to use as input to turbulence simulations.

Chapter \ref{ch:GYRO_Overview} introduces the theoretical model of gyrokinetics, which is thought to govern turbulence in the core of tokamaks. Then the results of references \cite{PeetersMomTransSym2005, ParraUpDownSym2011, SugamaUpDownSym2011} are summarised, which demonstrates a symmetry of the gyrokinetic equation that constrains rotation to be small in up-down symmetric devices. This provides background for chapter \ref{ch:GYRO_TiltingSymmetry}, which presents a new symmetry of the gyrokinetic model. The new symmetry establishes the invariance of turbulent transport to a poloidal tilt of ``fast'' flux surface shaping, where ``fast'' refers to shaping with a small spatial scale. By the up-down symmetry argument, this invariance to poloidal tilt constrains the momentum transport generated by mirror symmetric fast shaping (i.e. has reflectional symmetry about at least one line in the poloidal plane) to be exponentially small in the Fourier mode number of the fast shaping. In chapter \ref{ch:GYRO_MomFluxScaling}, we show that beating fast shaping effects together to produce slowly varying envelopes can generate momentum flux that is only polynomially small in the Fourier mode number of the fast shaping. This, together with an argument showing that mirror symmetric screw pinches have no momentum transport, motivates non-mirror symmetric flux surfaces (i.e. surfaces that do not have mirror symmetry about any line in the poloidal plane) with up-down asymmetric envelopes. Accordingly, chapter \ref{ch:GYRO_ShafranovShift} studies turbulent transport in tilted elliptical flux surfaces that have a Shafranov shift (which breaks the flux surface mirror symmetry) and finds mixed results. As expected, the Shafranov shift can enhance the amount of rotation, but the effect is entirely cancelled when the influence of the pressure gradient on the equilibrium is consistently included. Then, chapter \ref{ch:GYRO_NonMirrorSym} examines non-mirror symmetric configurations created using elongation and triangularity with separate poloidal tilt angles. We identify specific tilt angles that can enhance the momentum transport by $\sim 60\%$ compared to purely elongated configurations and also tend to minimise the energy transport.

Chapter \ref{ch:conclusions} identifies the optimal flux surface geometry for driving intrinsic rotation and uses it to illustrate the most significant results of this thesis.

%% file: Ch_02_MHD_GradShafSol.tex
% !TEX root = /Users/Justin/Documents/Research/Writings/2016DoctoralThesis/DoctoralThesis.tex

\chapter{Global equilibria for arbitrary flux surface shaping}
\label{ch:MHD_GradShafSol}

\begin{quote}
   \emph{Much of this chapter appears in reference \cite{BallShafranovShift2016}.}
\end{quote}

Ideal magnetohydrodynamics (MHD) \cite{FreidbergIdealMHD1987} is a simple, single fluid model that describes the macroscopic behaviour of plasma in a magnetic field. It is valid when the plasma has sufficiently high collisionality, small gyroradius, and small electrical resistivity. Strictly speaking fusion plasmas are not collisional enough for the model to be valid, but for subtle reasons it is empirically accurate for some calculations. In particular, it can be used to calculate the equilibrium magnetic field.

We start by writing the general form for the magnetic field in a tokamak,
\begin{align}
   \vec{B} = I \left( \psi \right) \Nabla \zeta + \Nabla \zeta \times \Nabla \psi , \label{eq:magFieldForm}
\end{align}
where $\psi$ is the poloidal magnetic flux divided by $2 \pi$, $I \left( \psi \right) \equiv R B_{\zeta}$ is the toroidal magnetic field flux function, and $B_{\zeta}$ is the toroidal magnetic field. Noting that $\vec{B} \cdot \Nabla \psi = 0$ we see that the magnetic field lines (and hence the plasma) are confined to nested surfaces of constant $\psi$, which are called flux surfaces. The ideal MHD equilibria of the flux surfaces is governed by the Grad-Shafranov equation \cite{GradGradShafranovEq1958},
\begin{align}
   R^{2} \Nabla \cdot \left( \frac{\Nabla \psi}{R^{2}} \right) = - \mu_{0} R^{2} \left. \frac{\partial p}{\partial \psi} \right|_{R} - I \frac{d I}{d \psi} , \label{eq:gradShafEq}
\end{align}
where the derivative of the pressure is performed holding the major radius constant. We note that with the exception of chapters \ref{ch:GYRO_Overview} and \ref{ch:GYRO_TiltingSymmetry} we will assume the plasma flow is subsonic, meaning that the pressure becomes a flux function and $\left. \partial p / \partial \psi \right|_{R} = d p / d \psi$. Using Ampere's law and \refEq{eq:magFieldForm} we see that the entire right-hand side of the Grad-Shafranov equation is closely related to $j_{\zeta}$, the toroidal current density in the plasma, according to
\begin{align}
   - \mu_{0} R^{2} \frac{d p}{d \psi} - I \frac{d I}{d \psi} = \mu_{0} j_{\zeta} R . \label{eq:toroidalCur}
\end{align}

In order to find flux surface shapes that generate high levels of intrinsic rotation in real experiments, we must first identify practical up-down asymmetric equilibria. There has been significant work on general solutions to the Grad-Shafranov equation \cite{KuiroukidisAnalyticGradShaf2012}, but here we will restrict our attention to several simple, approximate solutions. These solutions will allow us to identify feasible up-down asymmetric geometries as well as determine which features of the equilibria are robust and which are sensitive to the details of the configuration. We will expand the Grad-Shafranov equation in the large aspect ratio limit, i.e. $\epsilon \equiv a / R_{0 b} \ll 1$, where $R_{0 b}$ is the major radial location of the centre of the boundary flux surface. Note that we are expanding in the aspect ratio of the boundary flux surface as it will be more convenient than using the usual aspect ratio, which is based on the major radial location of the magnetic axis (i.e. $R_{0}$). We will also take the typical orderings for a low $\beta$, ohmically heated tokamak \cite{FreidbergIdealMHD1987pg126} of
\begin{align}
   \frac{B_{p}}{B_{0 b}} \sim \epsilon , \hspace{5em}
   \frac{2 \mu_{0} p}{B_{0 b}^{2}} \sim \epsilon^{2} , \label{eq:gradShafOrderings}
\end{align}
where $B_{p} = | \Nabla \psi | / R$ is the poloidal magnetic field and $B_{0 b}$ is the strength of the toroidal magnetic field at the centre of the boundary flux surface. Since we will need to know how the Shafranov shift (i.e. the shift in the magnetic axis due to toroidicity) behaves in up-down asymmetric geometries we must solve the Grad-Shafranov equation both to lowest and next order in $\epsilon$. In order to see the effect of the shapes of the toroidal current and pressure profiles we will look at three simple cases: constant, linear (in poloidal flux) peaked, and linear hollow.

First we must expand \refEq{eq:gradShafEq} in $\epsilon \ll 1$ using $\psi = \psi_{0} + \psi_{1} + \ldots$, $I = I_{0} + I_{1} + I_{2}$, and $p = p_{2}$, where the subscripts indicate the order of the quantity in $\epsilon$.  To $O \left( \epsilon^{-1} B_{0} \right)$ we find that the Grad-Shafranov equation is
\begin{align}
   - I_{0} \frac{d I_{1}}{d \psi_{0}} - I_{1} \frac{d I_{0}}{d \psi_{0}} = 0 . \label{eq:gradShafNegOrder}
\end{align}
Since $I_{0} = R_{0 b} B_{0 b}$ is a constant, this requires that $I_{1}$ also be a constant. We are free to absorb $I_{1}$ into $I_{0}$ and set $I_{1} = 0$. Hence, to $O \left( B_{0} \right)$ the Grad-Shafranov equation is
\begin{align}
   \frac{1}{r} \frac{\partial}{\partial r} \left( r \frac{\partial \psi_{0}}{\partial r} \right) + \frac{1}{r^{2}} \frac{\partial^{2} \psi_{0}}{\partial \theta^{2}} = - \mu_{0} R_{0 b}^{2} \frac{dp_{2}}{d \psi_{0}} - I_{0} \frac{dI_{2}}{d \psi_{0}}  \label{eq:gradShafLowestOrder}
\end{align}
and to $O \left( \epsilon B_{0} \right)$ we find
\begin{align}
   \frac{1}{r} \frac{\partial}{\partial r} \left( r \frac{\partial \psi_{1}}{\partial r} \right) &+ \frac{1}{r^2} \frac{\partial^{2} \psi_{1}}{\partial \theta^{2}} - \psi_{1} \frac{d}{d \psi_{0}} \left( -\mu_{0} R_{0 b}^{2} \frac{d p_{2}}{d \psi_{0}} - I_{0} \frac{d I_{2}}{d \psi_{0}} \right)\label{eq:gradShafNextOrder} \\
   & = - 2 \mu_{0} r R_{0 b} \frac{d p_{2}}{d \psi_{0}} \Cos{\theta} + \frac{\Cos{\theta}}{R_{0 b}} \frac{\partial \psi_{0}}{\partial r} - \frac{\Sin{\theta}}{r R_{0 b}} \frac{\partial \psi_{0}}{\partial \theta} , \nonumber
\end{align}
where $r \equiv \sqrt{\left( R - R_{0 b} \right)^{2} + \left( Z - Z_{0 b} \right)^{2}}$ is the distance from the centre of the boundary flux surface, $\theta \equiv \ArcTan{\left( Z - Z_{0 b} \right) / \left( R - R_{0 b} \right)}$ is the usual cylindrical poloidal angle (see figure \ref{fig:tokamakGeo}), and $Z_{0 b}$ is the axial location of the centre of the boundary flux surface. We note that the $O \left( B_{0} \right)$ Grad-Shafranov equation has cylindrical symmetry (i.e. translational symmetry in $\theta$), unlike the $O \left( \epsilon B_{0} \right)$ equation.

Next, we will parameterize all three current profiles (i.e. constant, peaked, and hollow) by
\begin{align}
   - \mu_{0} R_{0 b}^{2} \frac{dp_{2}}{d \psi_{0}} - I_{0} \frac{dI_{2}}{d \psi_{0}} = \mu_{0} j_{\zeta 0} R_{0 b} = \hat{j}_{0} \left( 1 - \hat{f}_{0} \psi_{0} \right) , \label{eq:currentProfiles}
\end{align}
where $j_{\zeta 0}$ is the lowest order current density in the aspect ratio expansion, $\hat{j}_{0}$ is a positive constant, $\hat{f}_{0} \in \left[ - \psi_{0 b}^{-1}, \psi_{0 b}^{-1} \right]$ determines the slope of the current profile, and $\psi_{0 b}$ is the lowest order value of the poloidal flux on the boundary flux surface (where $\psi$ is taken to vanish at the magnetic axis). The constant current case is achieved by setting $\hat{f}_{0} = 0$, while the hollow current case arises from allowing $\hat{f}_{0}$ to be negative.

Additionally, from \refEq{eq:gradShafNextOrder} we see that it will be necessary to distinguish the contributions to the current from the pressure and magnetic field terms in \refEq{eq:toroidalCur}. Like the toroidal current, we will assume the pressure gradient has the form of
\begin{align}
   - \mu_{0} R_{0 b}^{2} \frac{dp_{2}}{d \psi_{0}} &= \hat{j}_{0 p} \left( 1 - \hat{f}_{0 p} \psi_{0} \right) , \label{eq:pressureProfShape}
\end{align}
where $\hat{j}_{0 p}$ and $\hat{f}_{0 p} \in \left[ - \psi_{0 b}^{-1}, \psi_{0 b}^{-1} \right]$ are constants. By \refEq{eq:currentProfiles}, this pressure profile implies that the toroidal magnetic field flux function term must be
\begin{align}
   - I_{0} \frac{dI_{2}}{d \psi_{0}} &= \hat{j}_{0 I} \left( 1 - \hat{f}_{0 I} \psi_{0} \right) ,
\end{align}
where
\begin{align}
   \hat{j}_{0 I} &\equiv \hat{j}_{0} - \hat{j}_{0 p} \\
   \hat{f}_{0 I} &\equiv \frac{1}{\hat{j}_{0 I}} \left( \hat{j}_{0} \hat{f}_{0} - \hat{j}_{0 p} \hat{f}_{0 p} \right)
\end{align}
are constants.

%===================================================%
%===================================================%
\section{Solutions to the $O ( B_{0} )$ Grad-Shafranov equation}
\label{sec:GradShafSolLowestOrder}
%===================================================%
%===================================================%

In order to solve the $O \left( B_{0} \right)$ Grad-Shafranov equation we will Fourier analyse the magnetic flux in poloidal angle as
\begin{align}
   \psi_{0} \left( r, \theta \right) &= \psi_{0, 0}^{C} \left( r \right) + \sum_{m = 1}^{\infty} \left[ \psi_{0, m}^{C} \left( r \right) \Cos{m \theta} + \psi_{0, m}^{S} \left( r \right) \Sin{m \theta} \right] , \label{eq:gradShafLowestOrderSols}
\end{align}
where $m$ is the poloidal flux surface shaping mode number.
%Note that, by construction, we are setting the $m=1$ coefficient to be zero to emphasise that the Shafranov shift only appears as a next order correction.
Using \refEq{eq:gradShafLowestOrderSols} we can rewrite \refEq{eq:gradShafLowestOrder} as
\begin{align}
   \frac{1}{r} \frac{d}{dr} \left( r \frac{d \psi_{0, m}^{T}}{dr} \right) &+ \left( \hat{f}_{0} \hat{j}_{0} - \frac{m^{2}}{r^{2}} \right) \psi_{0, m}^{T} \left( r \right) = \hat{j}_{0} \delta_{m, 0} , \label{eq:gradShafFourierLowest}
\end{align}
where $\delta_{i, j}$ is the Kronecker delta and $T = C, S$ is a superscript that indicates the sine or cosine mode. The solutions to this equation with zero poloidal flux at the magnetic axis are
\begin{align}
   \psi_{0, 0}^{C} \left( r \right) &= - \frac{1}{\hat{f}_{0}} \left( J_{0} \left( \sqrt{\hat{f}_{0} \hat{j}_{0}} r \right) - 1 \right) \label{eq:gradShafSol0modeFourierLowest} \\
   \psi_{0, m}^{C} \left( r \right) &= C_{0, m} \frac{m! ~ 2^{m}}{\left( \hat{f}_{0} \hat{j}_{0} \right)^{m/2}} J_{m} \left( \sqrt{ \hat{f}_{0} \hat{j}_{0}} r \right) \\
   \psi_{0, m}^{S} \left( r \right) &= S_{0, m} \frac{m! ~ 2^{m}}{\left( \hat{f}_{0} \hat{j}_{0} \right)^{m/2}} J_{m} \left( \sqrt{ \hat{f}_{0} \hat{j}_{0}} r \right) , \label{eq:gradShafSolMmodeFourierLowest}
\end{align}
where $J_{m}$ is the $m^{\text{th}}$ order Bessel function of the first kind. The Fourier coefficients $C_{0, m}$ and $S_{0, m}$ are determined by the boundary conditions at the plasma edge, which is physically controlled by the locations and currents of external plasma shaping coils. Using trigonometric identities, \refEq{eq:gradShafLowestOrderSols} and \refEq{eq:gradShafSol0modeFourierLowest} through \refEq{eq:gradShafSolMmodeFourierLowest} can be rewritten as
\begin{align}
   \psi_{0} \left( r, \theta \right) &= - \frac{1}{\hat{f}_{0}} \left( J_{0} \left( \sqrt{\hat{f}_{0} \hat{j}_{0}} r \right) - 1 \right)  \label{eq:gradShafLowestOrderSolsTilt} \\
   &+ \sum _{m=2}^{\infty} N_{0, m} \frac{m! ~ 2^{m}}{\left( \hat{f}_{0} \hat{j}_{0} \right)^{m/2}} J_{m} \left( \sqrt{\hat{f}_{0} \hat{j}_{0}} r \right) \Cos{m \left( \theta + \theta_{t 0, m} \right)} , \nonumber
\end{align}
where
\begin{align}
   N_{0, m} \equiv \sqrt{ C_{0, m}^{2} + S_{0, m}^{2}} \label{eq:FourierMagDef}
\end{align}
is the magnitude of the Fourier mode and
\begin{align}
   \theta_{t 0, m} \equiv - \frac{1}{m} \ArcTan{ \frac{S_{0, m}}{C_{0, m}}} \label{eq:tiltAngleDef}
\end{align}
is the Fourier mode tilt angle.

Note that for the constant current case  (i.e. $\hat{f}_{0} = 0$), \refEq{eq:gradShafLowestOrderSolsTilt} reduces to
\begin{align}
   \psi_{0} \left( r, \theta \right) &= \frac{\hat{j}_{0}}{4} r^{2} + \sum _{m=2}^{\infty} N_{0, m} r^{m} \Cos{m \left( \theta + \theta_{t 0, m} \right)} . \label{eq:gradShafLowestOrderSolsTiltConst}
\end{align}
To understand the hollow current case, it is helpful to make use of the identity
\begin{align}
   J_{m} \left( i x \right) = i^{m} I_{m} \left( x \right) , \label{eq:modBesselFnFirstKind}
\end{align}
where $I_{m}$ is the $m^{\text{th}}$ order modified Bessel function of the first kind. From this we can demonstrate that \refEq{eq:gradShafLowestOrderSolsTilt} is equivalent to
\begin{align}
   \psi_{0} \left( r, \theta \right) &= \frac{1}{- \hat{f}_{0}} \left( I_{0} \left( \sqrt{- \hat{f}_{0} \hat{j}_{0}} r \right) - 1 \right)  \label{eq:gradShafLowestOrderSolsTiltHollow} \\
   &+ \sum _{m=2}^{\infty} N_{0, m} \frac{m! ~ 2^{m}}{\left( - \hat{f}_{0} \hat{j}_{0} \right)^{m/2}} I_{m} \left( \sqrt{- \hat{f}_{0} \hat{j}_{0}} r \right) \Cos{m \left( \theta + \theta_{t 0, m} \right)} , \nonumber
\end{align}
which can be more easily applied to hollow toroidal current profiles (i.e. $\hat{f}_{0} < 0$).

%===================================================%
%===================================================%
\section{Solutions to the $O ( \epsilon B_{0} )$ Grad-Shafranov equation}
\label{sec:GradShafSolNextOrder}
%===================================================%
%===================================================%

In order to solve the $O \left( \epsilon B_{0} \right)$ equation we again must Fourier analyse the magnetic flux in poloidal angle. The lowest order Fourier-analysed flux is given by \refEq{eq:gradShafLowestOrderSols} and \refEq{eq:gradShafSol0modeFourierLowest} through \refEq{eq:gradShafSolMmodeFourierLowest}. To next order, we can write
\begin{align}
   \psi_{1} \left( r, \theta \right) &= \psi_{1, 0}^{C} \left( r \right) + \sum_{m = 1}^{\infty} \left[ \psi_{1, m}^{C} \left( r \right) \Cos{m \theta} + \psi_{1, m}^{S} \left( r \right) \Sin{m \theta} \right] , \label{eq:gradShafNextOrderSols}
\end{align}
but we still must solve for $\psi_{1, m}^{C} \left( r \right)$ and $\psi_{1, m}^{S} \left( r \right)$ by substituting \refEq{eq:gradShafLowestOrderSols} and \refEq{eq:gradShafNextOrderSols} into \refEq{eq:gradShafNextOrder}. Since $\psi_{1, m}^{C} \left( r \right)$ and $\psi_{1, m}^{S} \left( r \right)$ do not depend on $\theta$, we can take each Fourier component of \refEq{eq:gradShafNextOrder} as a separate equation. This gives
\begin{align}
   \frac{1}{r} \frac{d}{dr} \left( r \frac{d \psi_{1, m}^{T}}{dr} \right) &+ \left( \hat{f}_{0} \hat{j}_{0} - \frac{m^{2}}{r^{2}} \right) \psi_{1, m}^{T} \left( r \right) = \Lambda_{m}^{T} \left( r \right) \label{eq:gradShafFourierNext}
\end{align}
for each Fourier mode $m$, where the inhomogeneous terms are given by $\Lambda_{m}^{T} \left( r \right)$. For $m = 0$ and $T = C$
\begin{align}
\Lambda_{0}^{C} \left( r \right) &\equiv \frac{1}{R_{0}} \left[ \frac{1}{2} \frac{d \psi_{0, 1}^{C}}{dr} + \left( \frac{1}{2 r} - r \hat{f}_{0 p} \hat{j}_{0 p} \right) \psi_{0, 1}^{C} \left( r \right) \right] , \label{eq:inhomoTermsC0}
\end{align}
for $m = 1$ and $T = C$
\begin{align}
\Lambda_{1}^{C} \left( r \right) &\equiv \frac{1}{R_{0}} \left[ \frac{1}{2} \frac{d \psi_{0, 2}^{C}}{dr} + \left( \frac{1}{r} - r \hat{f}_{0 p} \hat{j}_{0 p} \right) \psi_{0, 2}^{C} \left( r \right) \right. \label{eq:inhomoTermsC1} \\
&+ \left. \frac{d \psi_{0, 0}^{C}}{dr} + 2 r \hat{j}_{0 p} \left( 1 - \hat{f}_{0 p} \psi_{0, 0}^{C} \left( r \right) \right) \right] , \nonumber
\end{align}
for $m = 1$ and $T = S$
\begin{align}
\Lambda_{1}^{S} \left( r \right) &\equiv \frac{1}{R_{0}} \left[ \frac{1}{2} \frac{d \psi_{0, 2}^{S}}{dr} + \left( \frac{1}{r} - r \hat{f}_{0 p} \hat{j}_{0 p} \right) \psi_{0, 2}^{S} \left( r \right) \right] , \label{eq:inhomoTermsS1}
\end{align}
and for all other $m$ and $T = C, S$
\begin{align}
\Lambda_{m}^{T} \left( r \right) &\equiv \frac{1}{R_{0}} \left[ \frac{1}{2} \frac{d \psi_{0, m + 1}^{T}}{dr} + \left( \frac{m+1}{2 r} - r \hat{f}_{0 p} \hat{j}_{0 p} \right) \psi_{0, m + 1}^{T} \left( r \right) \right. \label{eq:inhomoTermsTm} \\
&+ \left. \frac{1}{2} \frac{d \psi_{0, m - 1}^{T}}{dr} - \left( \frac{m-1}{2 r} + r \hat{f}_{0 p} \hat{j}_{0 p} \right) \psi_{0, m - 1}^{T} \left( r \right) \right] . \nonumber
\end{align}

Equation \refEq{eq:gradShafFourierNext} can be solved using the method of variation of parameters, yielding
\begin{align}
   \psi_{1, m}^{T} \left( r \right) &= - \frac{\pi}{2} J_{m} \left( \sqrt{\hat{f}_{0} \hat{j}_{0}} r \right) \int_{0}^{r} d r' ~ r' Y_{m} \left( \sqrt{\hat{f}_{0} \hat{j}_{0}} r' \right) \Lambda_{m}^{T} \left( r' \right) \nonumber \\
   &+ \frac{\pi}{2} Y_{m} \left( \sqrt{\hat{f}_{0} \hat{j}_{0}} r \right) \int_{0}^{r} d r' ~ r' J_{m} \left( \sqrt{\hat{f}_{0} \hat{j}_{0}} r' \right) \Lambda_{m}^{T} \left( r' \right) \label{eq:psiFourierNextOrderCoeffs} \\
   &+ T_{1, m} \frac{m! ~ 2^{m}}{\left( \hat{f}_{0} \hat{j}_{0} \right)^{m/2}} J_{m} \left( \sqrt{\hat{f}_{0} \hat{j}_{0}} r \right) , \nonumber
\end{align}
where we have imposed regularity at the origin, $Y_{m}$ is the $m^{\text{th}}$ order Bessel function of the second kind, and $T_{1, m} = C_{1, m}, S_{1, m}$ are Fourier coefficients determined by the boundary conditions at the plasma edge. Combining \refEq{eq:gradShafNextOrderSols}, \refEq{eq:inhomoTermsC0} through \refEq{eq:inhomoTermsTm}, and \refEq{eq:psiFourierNextOrderCoeffs} gives the complete solution to the $O \left( \epsilon B_{0} \right)$ Grad-Shafranov equation for an arbitrary boundary condition. 

To understand the hollow current case (i.e. $\hat{f}_{0} < 0$), we will use \refEq{eq:modBesselFnFirstKind} and the identity
\begin{align}
   Y_{m} \left( i x \right) = i^{m+1} I_{m} \left( x \right) - \frac{2}{\pi} i^{-m} K_{m} \left( x \right) ,
\end{align}
where $K_{m}$ is the $m^{\text{th}}$ order modified Bessel function of the second kind. This enables \refEq{eq:psiFourierNextOrderCoeffs} to be reformulated as
\begin{align}
   \psi_{1, m}^{T} \left( r \right) &= I_{m} \left( \sqrt{- \hat{f}_{0} \hat{j}_{0}} r \right) \int_{0}^{r} d r' ~ r' K_{m} \left( \sqrt{ - \hat{f}_{0} \hat{j}_{0}} r' \right) \Lambda_{m}^{T} \left( r' \right) \nonumber \\
   &- K_{m} \left( \sqrt{- \hat{f}_{0} \hat{j}_{0}} r \right) \int_{0}^{r} d r' ~ r' I_{m} \left( \sqrt{- \hat{f}_{0} \hat{j}_{0}} r' \right) \Lambda_{m}^{T} \left( r' \right) \label{eq:psiFourierNextOrderCoeffsHollow} \\
   &+ T_{1, m} \frac{m! ~ 2^{m}}{\left( - \hat{f}_{0} \hat{j}_{0} \right)^{m/2}} I_{m} \left( \sqrt{- \hat{f}_{0} \hat{j}_{0}} r \right) . \nonumber
\end{align}
For a constant current profile (i.e. $\hat{f}_{0} = 0$), we can take the limit of \refEq{eq:gradShafNextOrderSols}, \refEq{eq:inhomoTermsC0} through \refEq{eq:inhomoTermsTm}, and \refEq{eq:psiFourierNextOrderCoeffs} as $\hat{f}_{0} \hat{j}_{0} \to 0$ to find
\begin{align}
   \psi_{1} \left( r, \theta \right) &= \frac{1}{4 R_{0 b}} \left[ \left( \frac{\hat{j}_{0} + 4 \hat{j}_{0 p}}{4} r^{3} - \frac{\hat{j}_{0} \hat{f}_{0 p} \hat{j}_{0 p}}{12} r^{5} \right) \Cos{\theta} \right. \nonumber \\
   &+ \sum_{m = 2}^{\infty} \left( r^{m+1} - \frac{\hat{f}_{0 p} \hat{j}_{0 p}}{2 \left( m + 1 \right)} r^{m+3} \right) N_{0, m} \Cos{ \left( m - 1 \right) \theta + m \theta_{t 0, m}} \nonumber \\
   &- \left. \sum_{m = 2}^{\infty} \frac{\hat{f}_{0 p} \hat{j}_{0 p}}{m + 2} r^{m+3} N_{0, m} \Cos{\left( m + 1 \right) \theta + m \theta_{t 0, m}} \right] \label{eq:psiNextOrderSolConst} \\
   &+ \sum_{m = 0}^{\infty} r^{m} N_{1, m} \Cos{m \left( \theta + \theta_{t 1, m} \right)} , \nonumber
\end{align}
where $N_{1, m} \equiv \sqrt{ C_{1, m}^{2} + S_{1, m}^{2}}$ is the magnitude of the next order Fourier mode, $\theta_{t 1, m} \equiv - \ArcTan{S_{1, m}/C_{1, m}} / m$ is the next order Fourier mode tilt angle, and we have used \refEq{eq:gradShafLowestOrderSolsTiltConst} with
\begin{align}
%   \lim_{\hat{f}_{0} \to 0} - \frac{1}{\hat{f}_{0}} \left( J_{0} \left( \sqrt{\hat{f}_{0} \hat{j}_{0}} r \right) - 1 \right) &= \frac{\hat{j}_{0}}{4} r^{2} \label{eq:BesselFnExpansionO0} \\
   \lim_{\hat{f}_{0} \hat{j}_{0} \to 0} \frac{m! ~ 2^{m}}{\left( \hat{f}_{0} \hat{j}_{0} \right)^{m/2}} J_{m} \left( \sqrt{\hat{f}_{0} \hat{j}_{0}} r \right) &= r^{m} \label{eq:BesselFnExpansionOn} \\
   \lim_{\hat{f}_{0} \hat{j}_{0} \to 0} Y_{m} \left( \sqrt{\hat{f}_{0} \hat{j}_{0}} r \right) &= - \frac{1}{m \pi} \frac{m! ~ 2^{m}}{\left( \hat{f}_{0} \hat{j}_{0} \right)^{m/2}} r^{-m} \label{eq:modifiedBesselFnExpansionOn}
\end{align}
for $m \neq 0$. The first line of \refEq{eq:psiNextOrderSolConst} contains the direct effect of toroidicity on the equilibrium, i.e. the Shafranov shift. The second and third lines show that a zeroth order shaping mode $m$ splits into two modes, $m-1$ and $m+1$, at first order. The last line contains the homogeneous solution, which enables an arbitrary boundary condition to be satisfied.

%% file: Ch_03_MHD_ShapingIntuition.tex
% !TEX root = /Users/Justin/Documents/Research/Writings/2016DoctoralThesis/DoctoralThesis.tex

\chapter{Radial penetration of flux surface shaping}
\label{ch:MHD_ShapingIntuition}

\begin{quote}
   \emph{Much of this chapter appears in reference \cite{BallShapingPenetration2015}.}
\end{quote}

This chapter uses a series of independent arguments to show that tokamaks with lower order shaping modes and a more hollow current profile will better allow shaping to penetrate to the magnetic axis. This provides intuition for existing analytic \cite{LaoGradShafExpansion1981, CerfonSolovevGradShafSols2010, AtanasiuGradShafSols2004} and numerical \cite{RomanelliShapingPenetration2000, LaoShapeAndCurrent1985} results concerning how flux surface shaping penetrates in the ideal MHD model.

Here we will use the large aspect ratio solutions found in chapter \ref{ch:MHD_GradShafSol} to investigate the effects of both free parameters in the lowest order Grad-Shafranov equation: the boundary condition and the toroidal current profile (see \refEq{eq:gradShafEq} and \refEq{eq:toroidalCur}). Although the motivation is to create up-down asymmetric flux surfaces near the magnetic axis, the main results of this chapter also apply to the penetration of traditional up-down symmetric plasma shaping. Additionally, the following derivations are appropriate to treat the Shafranov shift, but it will not be investigated specifically. This is because it is formally small in aspect ratio and, in isolation, does not create up-down asymmetry. As we will explore in chapter \ref{ch:MHD_ShafranovShift}, the Shafranov shift becomes up-down asymmetric when the flux surfaces already have an up-down asymmetric shape. Hence it can enhance existing up-down asymmetry, but cannot create asymmetry by itself.

The traditional argument concerning shaping penetration \cite{LaoGradShafExpansion1981, RodriguesMHDupDownAsym2014, BizarroUpDownAsymGradShafEq2014} uses a Taylor expansion of the poloidal flux about the magnetic axis to find
\begin{align}
   \psi \left( R, Z \right) &\approx \frac{1}{2} \left. \frac{\partial^{2} \psi}{\partial R^{2}} \right|_{R_{0}, Z_{0}} \left( R - R_{0} \right)^{2} + \left. \frac{\partial^{2} \psi}{\partial R \partial Z} \right|_{R_{0}, Z_{0}} \left( R - R_{0} \right) \left( Z - Z_{0} \right) \label{eq:TaylorExpansion} \\
   &+ \frac{1}{2} \left. \frac{\partial^{2} \psi}{\partial Z^{2}} \right|_{R_{0}, Z_{0}} \left( Z - Z_{0} \right)^{2} , \nonumber
\end{align}
where $Z_{0}$ is the axial location of the magnetic axis. Here we have imposed that at the magnetic axis the poloidal flux vanishes and is at a minimum. This implies that the constant and linear terms in the Taylor expansion are zero. Hence, no matter what external fields shape the plasma, close enough to the magnetic axis the flux surface ellipticity will dominate over higher order shaping effects. This argument fails if all the second order Taylor coefficients are zero. However, very close to the magnetic axis the plasma current can be assumed to be constant (since the slope of the current must be zero on axis), so \refEq{eq:gradShafLowestOrderSolsTiltConst} must be a valid equilibrium in the region. Therefore, we see that, in order for the second order Taylor coefficients to vanish, the on-axis toroidal current density must be zero. This prevents closed, nested flux surfaces \cite{RodriguesGradShafVanishingCurrent2013}. Thus, the case in which the second order Taylor coefficients vanish is uninteresting.

While the argument based on the Taylor expansion around the magnetic axis is compelling, it says nothing about how shaping behaves away from the magnetic axis or how triangularity penetrates in the absence of elongation. A more sophisticated version of this argument is presented in references \cite{BallMomUpDownAsym2014, RodriguesMHDupDownAsym2014}, which includes effects from having a linear toroidal current profile.

In section \ref{sec:calculation}, we show that the shaping of a given flux surface depends on the magnitude of the poloidal variation of the poloidal magnetic field on the flux surface. Then, in section \ref{sec:fluxSurfaceShapeEffect}, we use this dependence to study why different flux surface shapes penetrate better than others. In section \ref{sec:effectOfCurrentProfile}, we explore a limit of the Grad-Shafranov equation that separates the effects of magnetic pressure and tension. In this limit we clearly see how the current profile affects shaping penetration.

%===================================================%
%===================================================%
\section{Quantifying shaping penetration}
\label{sec:calculation}
%===================================================%
%===================================================%

First, we will define the parameter
\begin{align}
   \Delta \left( a_{\psi} \right) \equiv \frac{b_{\psi} \left( a_{\psi} \right)}{a_{\psi}} , \label{eq:deltaDef}
\end{align}
where $a_{\psi}$ is the minimum distance of the flux surface from the magnetic axis and $b_{\psi}$ is the maximum distance of the flux surface from the magnetic axis. For circular flux surfaces without a Shafranov shift $\Delta = 1$. Since the definitions of $a_{\psi}$ and $b_{\psi}$ are based on the magnetic axis, $\Delta \neq 1$ for circular flux surfaces with a Shafranov shift. We note that $\Delta$ reduces to the typical definition of elongation (usually denoted by $\kappa$) when the flux surfaces are purely elliptical without a Shafranov shift.

Taking a derivative of \refEq{eq:deltaDef} we find the change in $\Delta$ across a flux surface is given by
\begin{align}
   \frac{d \Delta}{da_{\psi}} = \frac{1}{a_{\psi}} \frac{d b_{\psi}}{d a_{\psi}} - \frac{b_{\psi} \left( a_{\psi} \right)}{a_{\psi}^{2}} . \label{eq:kappaDeriv}
\end{align}
The derivative $d b_{\psi} / d a_{\psi}$ can be calculated from the poloidal magnetic flux,
\begin{align}
   \psi = \frac{1}{2 \pi} \oint_{-\pi}^{\pi} d \zeta \int_{0}^{r} d r' R B_{p} . \label{eq:poloidalFluxDef}
\end{align}
We note that \refEq{eq:poloidalFluxDef} is only valid along the integration path connecting the radial minimum on each flux surface, $a_{\psi}$, and the path connecting the radial maximum on each flux surface, $b_{\psi}$. This is because, at the flux surface radial extrema, the poloidal field is necessarily perpendicular to the usual cylindrical radial direction. Using implicit differentiation and evaluating on both of these integration paths, \refEq{eq:poloidalFluxDef} gives
\begin{align}
   \frac{d a_{\psi}}{d \psi} = \frac{1}{\left. R B_{p} \right|_{a}} \label{eq:minorRadiusFluxDeriv} \\
   \frac{d b_{\psi}}{d \psi} = \frac{1}{\left. R B_{p} \right|_{b}} . \label{eq:majorRadiusFluxDeriv}
\end{align}
Here $|_{a}$ and $|_{b}$ indicate the quantity should be evaluated at the poloidal locations of the minimum and maximum radial positions on a given flux surface. Therefore, we find that \refEq{eq:kappaDeriv} becomes
\begin{align}
   \frac{a_{\psi}}{\Delta} \frac{d \Delta}{da_{\psi}} = \frac{1}{\Delta} \frac{ \left. R B_{p} \right|_{a}}{\left. R B_{p} \right|_{b}}  - 1 , \label{eq:kappaDerivFields}
\end{align}
which is only a consequence of geometry and the definition of the poloidal flux. In current experiments \cite{LaoShapeAndCurrent1985,HofmannTCVRecordElong2002,BrixFluxSurfShapes2008,LaoDIIIDfluxSurfShapes2005} this quantity is generally between $0$ and $0.3$, but, as additional shaping is generally advantageous, the goal would be to make it as negative as possible. We will use \refEq{eq:kappaDerivFields} to understand why different flux surface shapes (elongated, triangular, etc.) penetrate better from the edge to the core and how the toroidal current profile affects this penetration.

%===================================================%
%===================================================%
\section{Effect of flux surface shape}
\label{sec:fluxSurfaceShapeEffect}
%===================================================%
%===================================================%

\begin{figure}
 \hspace{0.052\textwidth} (a) \hspace{0.248\textwidth} (b) \hspace{0.248\textwidth} (c) \hspace{0.1\textwidth}

 \centering
 \includegraphics[height=0.32\textwidth]{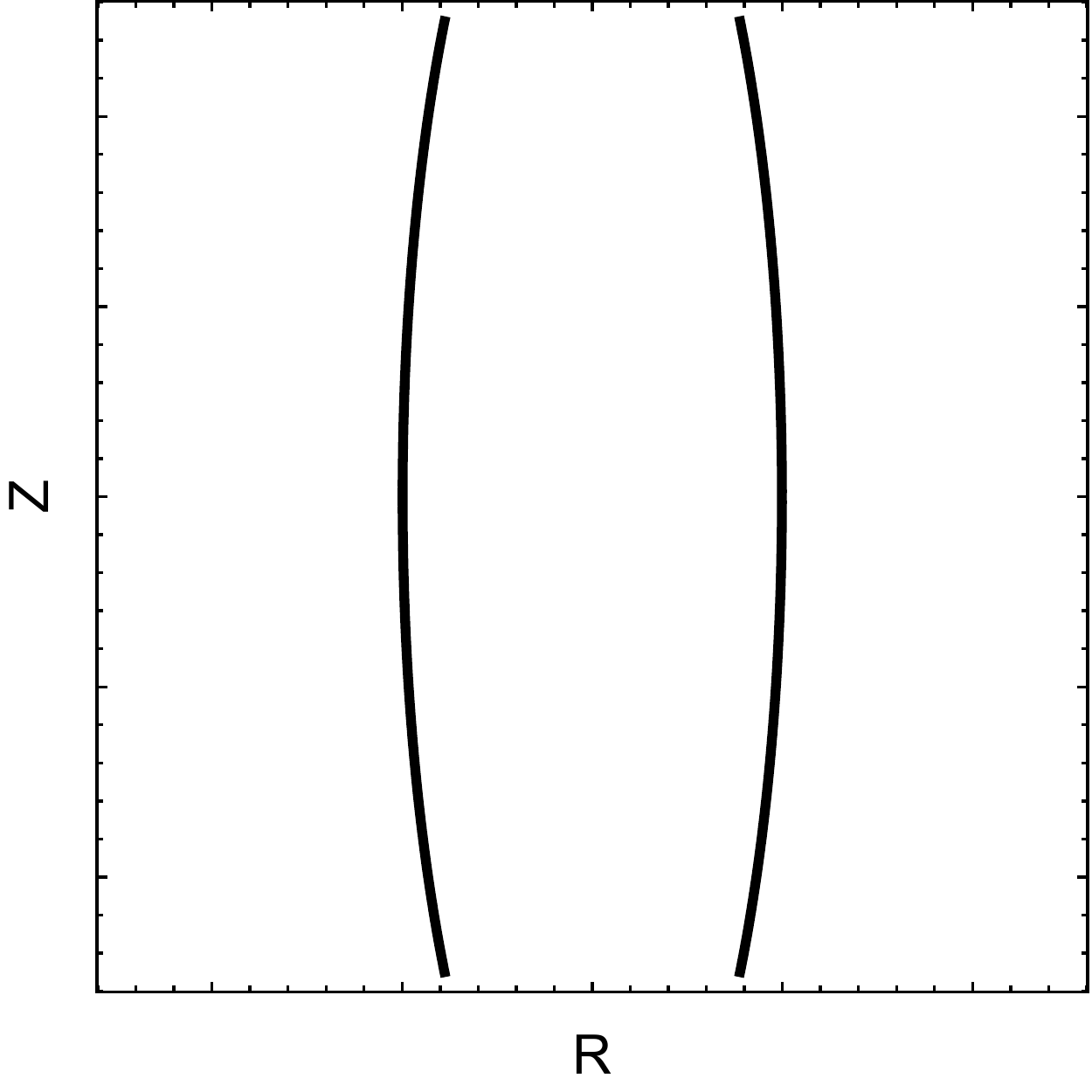}
 \includegraphics[height=0.32\textwidth]{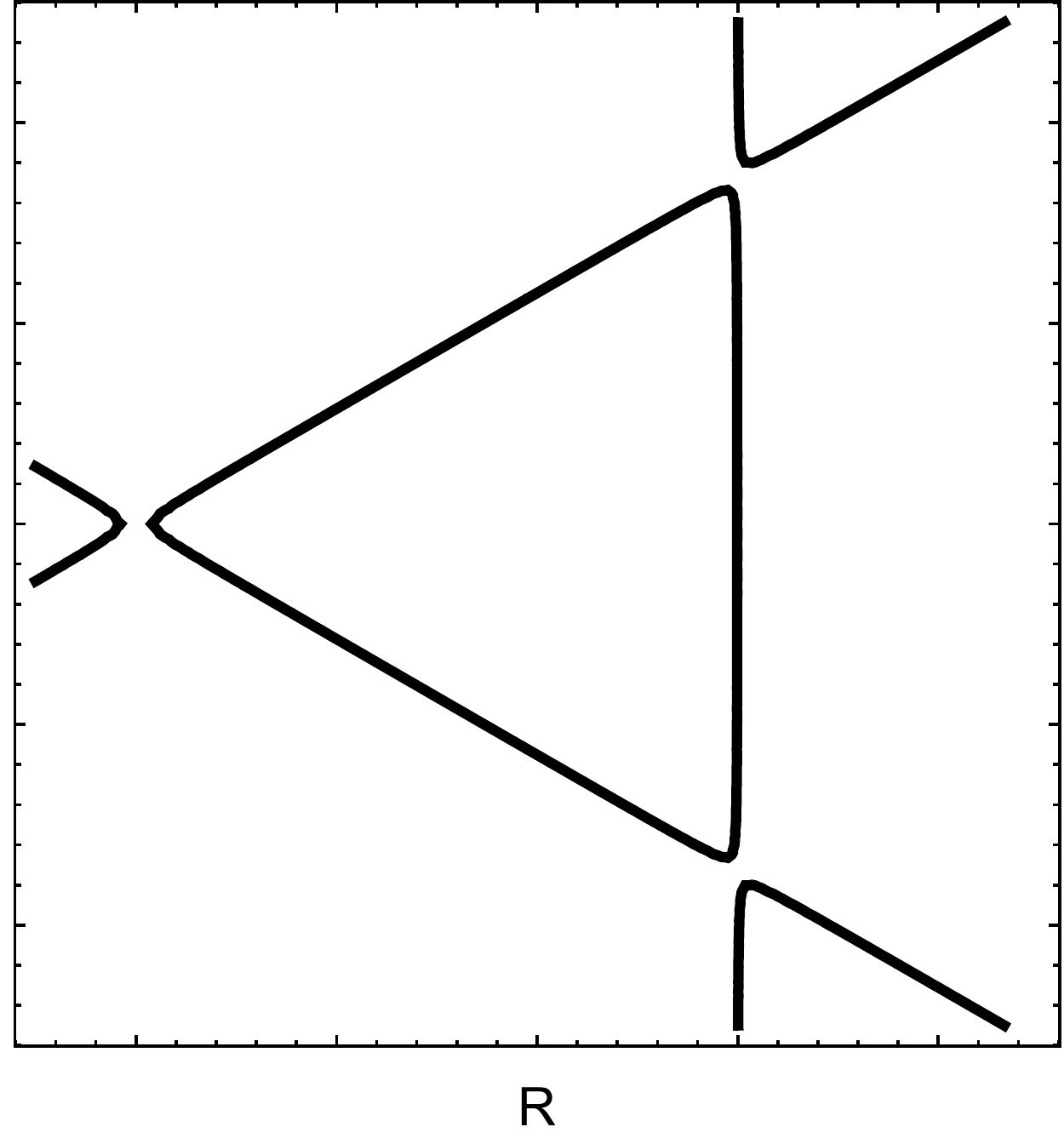}
 \includegraphics[height=0.32\textwidth]{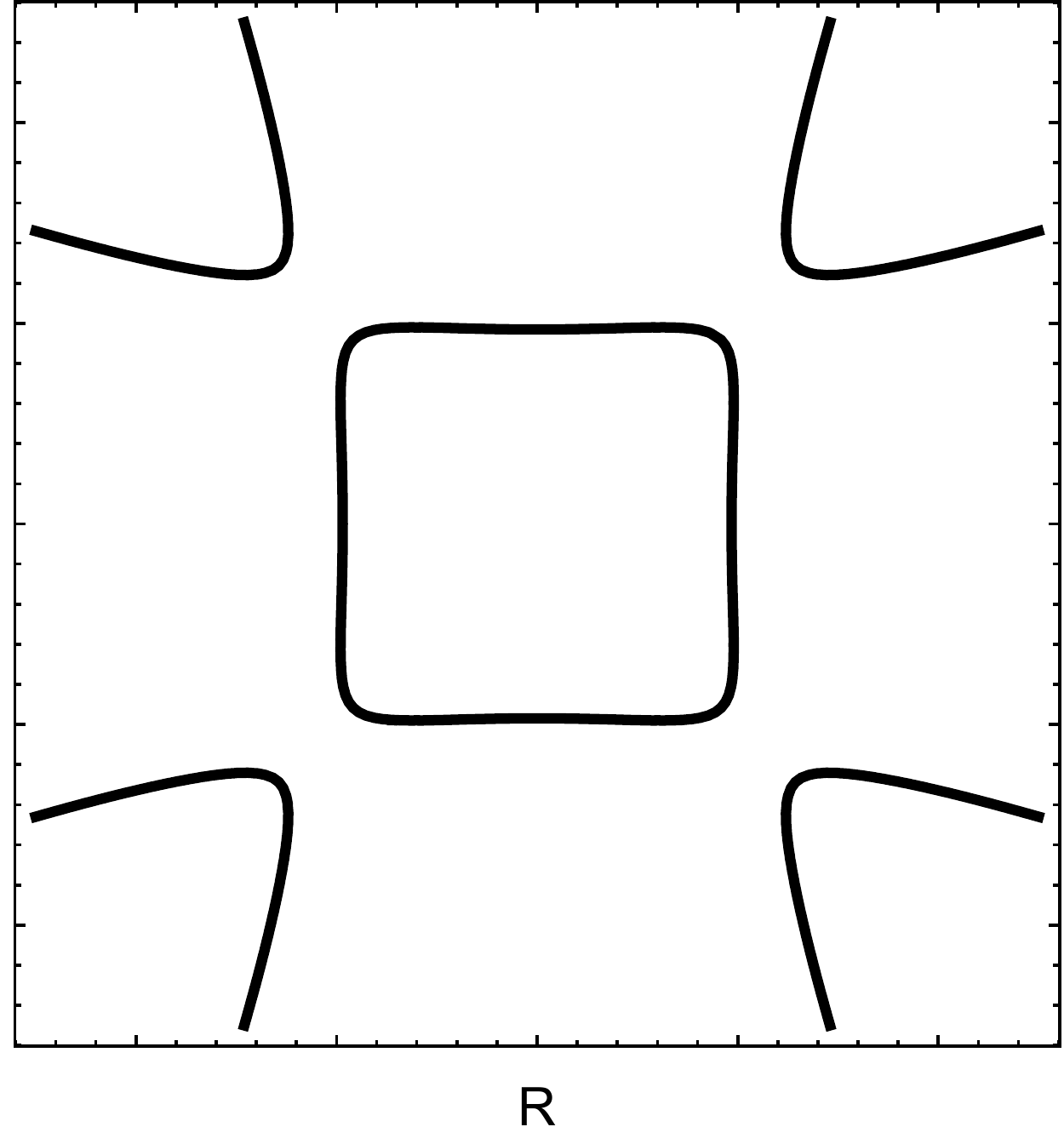}
 \caption{The (a) $m=2$, (b) $m=3$, and (c) $m=4$ strongly shaped flux surface shapes.}
 \label{fig:equilibriaFluxShapes}
\end{figure}

In this section we will compare different flux surface shapes and show that lower order shaping effects penetrate from the plasma boundary to the magnetic axis more effectively. First, we must determine which shapes to consider and argue that comparisons between them are fair. We will use large aspect ratio equilibria produced with a constant toroidal current profile because it is a reasonable approximation of experimental profiles and the solutions are simple cylindrical harmonics given by \refEq{eq:gradShafLowestOrderSolsTiltConst}. From these equilibria we will investigate each cylindrical harmonic shaping effect in isolation by creating strongly shaped flux surfaces, specifically those that approach having magnetic field nulls (see figure \ref{fig:equilibriaFluxShapes}). These configurations are created by including only a single shaping mode $m$ in \refEq{eq:gradShafLowestOrderSolsTiltConst} with the maximum possible value of $\Delta$ as calculated in appendix \ref{app:maxShaping}. This $\Delta$ is given by the numerical solution of \refEq{eq:sepShapingCond} and can be converted to the Fourier shaping coefficient using
\begin{align}
	N_{0, m} = \frac{\hat{j}_{0}}{4} \frac{\Delta^{2} - 1}{\Delta^{m} + 1} a_{\psi}^{2-m} . \label{eq:deltaToFourier}
\end{align}
Equation \refEq{eq:deltaToFourier} is a consequence of the definitions of $a_{\psi}$ as well as $\Delta$ and can be derived from $\psi_{0} \left( a_{\psi}, - \theta_{t m} \right) = \psi_{0} \left( \Delta a_{\psi}, \left( \pi / m \right) - \theta_{t m} \right)$ and \refEq{eq:gradShafLowestOrderSolsTiltConst}. We also need to solve for the relationship between the poloidal flux and the minor radius, which can be found to be
\begin{align}
   \psi_{0} \left( a_{\psi} \right) = \frac{\hat{j}_{0}}{4} a_{\psi}^{2} + N_{0, m} a_{\psi}^{m} \label{eq:gradShafranovSolPsi_a}
\end{align}
from \refEq{eq:gradShafLowestOrderSolsTiltConst} and the definition of $a_{\psi}$. These configurations will be analytically tractable and exaggerate the effects we mean to investigate. It should be noted that we expect flux surfaces with higher order shaping to be more difficult to create experimentally. This is because they have more magnetic field nulls, so they require more poloidal shaping magnets and more total external current to create.

From Ampere's law we find that $\left. R B_{p} \right|_{a} \approx \left(S_{p} / L_{p} \right) \mu_{0} j_{\zeta} R$ is a finite quantity, where $S_{p}$ is the poloidal area enclosed by the flux surface and $L_{p}$ is the poloidal perimeter. Additionally, we note that $\left. R B_{p} \right|_{b}$ approaches zero because we have chosen configurations that nearly have magnetic nulls. This reveals that, as the flux surface shaping is increased, the ratio of poloidal fields in \refEq{eq:kappaDerivFields} diverges to positive infinity. This implies that $d \Delta / d a_{\psi}$ is positive and large, i.e. it will be impossible to maintain strong shaping from the boundary to the magnetic axis. While this is true for nearly all configurations, there is one caveat: when the shaping parameter $\Delta$ also diverges to infinity. Then, $d \Delta / d a_{\psi}$ can be finite and negative. This makes the $m=2$ cylindrical harmonic shaping effect special because flux surfaces with arbitrarily large elongation are possible. Additionally, the $m=1$ mode is an exception as it is impossible to create magnetic field nulls with a pure Shafranov shift. However, all pure shaping effects above $m=2$ cannot make flux surfaces that are both closed and have arbitrarily large shaping.

Lastly, we note that \refEq{eq:deltaToFourier} directly determines how different flux surface shaping effects penetrate radially (given a constant current profile). In general, solving \refEq{eq:deltaToFourier} for $\Delta \left( a_{\psi} \right)$ cannot be done analytically, but after expanding to lowest order in $\Delta - 1 \ll 1$ we find that
\begin{align}
	\Delta - 1 = \left( \Delta_{b} - 1 \right) \rho^{m-2} , \label{eq:radialPenetration}
\end{align}
where $\Delta_{b}$ is the shaping parameter of the outermost flux surface, $\rho \equiv a_{\psi} / a$ is the usual normalised minor radial coordinate, and $a$ is the tokamak minor radius (i.e. $a_{\psi}$ of the outermost flux surface). From this we see that, to lowest order in aspect ratio, a constant current profile does not alter the externally applied elongation \cite{BallMomUpDownAsym2014, RodriguesMHDupDownAsym2014, ChristiansenCurrentDist1982} (meaning that $d \Delta / da_{\psi} = 0$). Furthermore, we see that all higher order shaping effects have exponentially poor radial penetration in $m$. Therefore, elongation will penetrate throughout the plasma better than all higher order shaping modes.

%===================================================%
%===================================================%
\section{Effect of the toroidal current profile}
\label{sec:effectOfCurrentProfile}
%===================================================%
%===================================================%

As we compare configurations with different toroidal current profiles, we will choose to keep the external flux surface shape fixed. Therefore, from \refEq{eq:kappaDerivFields} we conclude that changing the current profile, while maintaining a constant boundary flux surface shape, only affects the shaping penetration by altering $\left. R B_{p} \right|_{a} / \left. R B_{p} \right|_{b}$.

In order to calculate the ratio of the poloidal fields we will start with the toroidal component of Ampere's law,
\begin{align}
   \left( \vec{\nabla} \times \vec{B} \right) \cdot \hat{e}_{\zeta} = \mu_{0} j_{\zeta} .
\end{align}
Noting that $\vec{B} = I \vec{\nabla} \zeta + \vec{B}_{p}$, we see that
\begin{align}
  \left( \vec{\nabla} \times \vec{B}_{p} \right) \cdot R \vec{\nabla} \zeta = \mu_{0} j_{\zeta} .
\end{align}
Since $\vec{B}_{p} = \vec{\nabla} \zeta \times \vec{\nabla} \psi$, we know that $\vec{\nabla} \zeta = \vec{\nabla} \psi \times \vec{B}_{p} / \left| \vec{\nabla} \psi \right|^{2}$. Making this substitution and using a number of vector identities on the quantity $\vec{B}_{p} \times \left( \vec{\nabla} \times \vec{B}_{p} \right)$ we find that
\begin{align}
   R \frac{\vec{\nabla} \psi}{\left| \vec{\nabla} \psi \right|^{2}} \cdot \left( \vec{\nabla} \vec{B}_{p} \right) \cdot \vec{B}_{p} - \frac{R B_{p}^{2}}{\left| \vec{\nabla} \psi \right|^{2}} \hat{b}_{p} \cdot \left( \vec{\nabla} \hat{b}_{p} \right) \cdot \vec{\nabla} \psi = \mu_{0} j_{\zeta} ,
\end{align}
where $\hat{b}_{p} \equiv \vec{B}_{p} / B_{p}$ is the poloidal field unit vector. Using the definition of the poloidal field curvature,
\begin{align}
   \kappa_{p} \equiv - \left( \hat{b}_{p} \cdot \vec{\nabla} \hat{b}_{p} \right) \cdot \frac{\Nabla \psi}{\left| \Nabla \psi \right|} \label{eq:polCurvDef} ,
\end{align}
together with $\vec{\nabla} \psi = R B_{p} \hat{e}_{\psi}$ gives
\begin{align}
   \frac{R}{2} \frac{\vec{\nabla} \psi}{\left| \vec{\nabla} \psi \right|^{2}} \cdot \vec{\nabla} \left( B_{p}^{2} \right) + B_{p} \kappa_{p} = \mu_{0} j_{\zeta} , \label{eq:gradShafranovSimple}
\end{align}
a rearranged form of \refEq{eq:gradShafEq}, the Grad-Shafranov equation. We choose this form because it clearly separates the effects of poloidal magnetic pressure in the first term and field line tension in the second, while the right hand side is constant on a flux surface to lowest order in aspect ratio. Equation \refEq{eq:gradShafranovSimple} is a different way to express the conclusion reached in reference \cite{ChristiansenCurrentDist1982}: in non-circular flux surfaces, the current profile determines the gradient of the shaping. We can determine the poloidal magnetic field from the current profile using \refEq{eq:gradShafranovSimple}, which can then be related to the gradient of the shaping through \refEq{eq:kappaDerivFields}.

\begin{figure}
 \hspace{0.04\textwidth} (a) \hspace{0.255\textwidth} (b) \hspace{0.255\textwidth} (c) \hspace{0.1\textwidth}
 \begin{center}
  \includegraphics[width=0.3\textwidth]{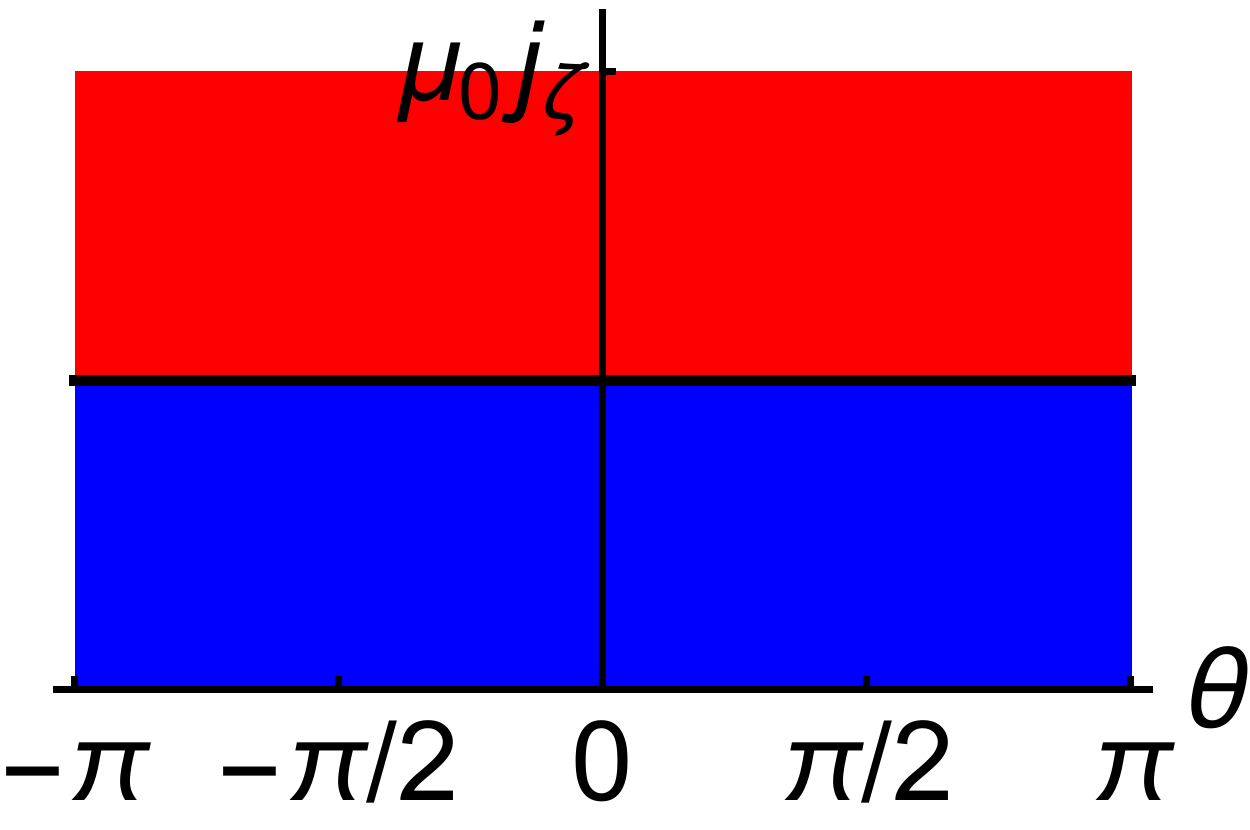}
  \includegraphics[width=0.3\textwidth]{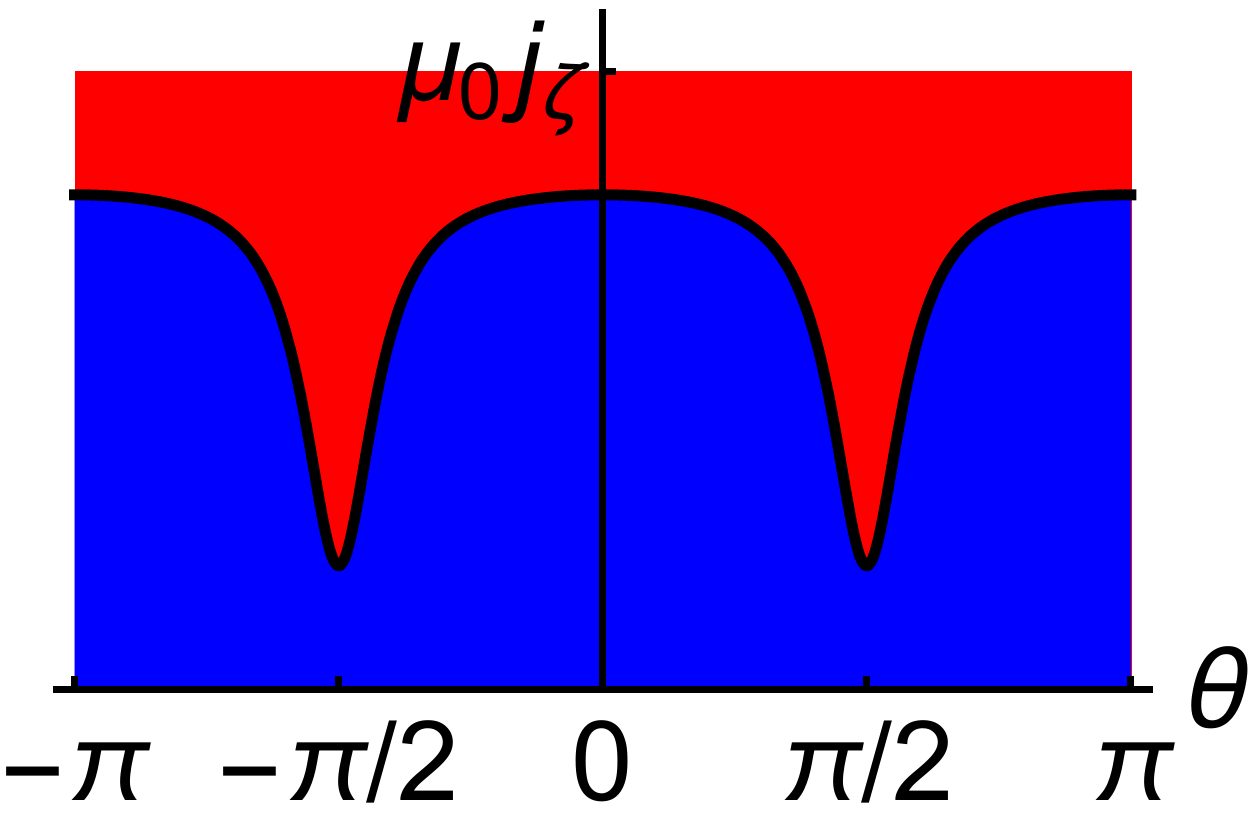}
  \includegraphics[width=0.3\textwidth]{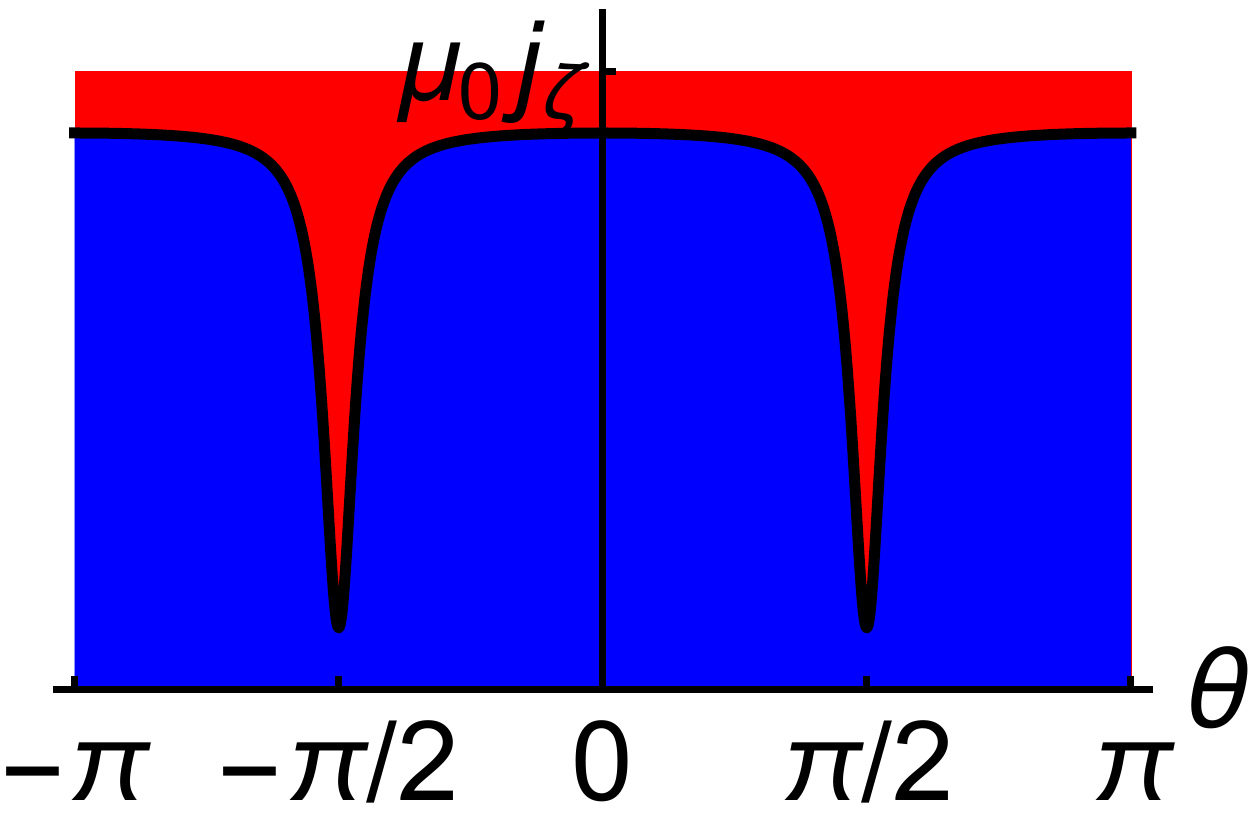}
 \end{center}
 \caption{A stacked area graph showing, to lowest order in aspect ratio, the contributions of the magnetic pressure (blue, below the curve) and tension (red, above the curve) terms from \refEq{eq:gradShafranovSimple} on an elongated flux surface with (a) $\Delta = 1$, (b) $\Delta = 2$, and (c) $\Delta = 3$.}
 \label{fig:pressureTensionBalance}
\end{figure}

We apply \refEq{eq:gradShafranovSimple} to strongly shaped flux surfaces, which causes the first and second terms to vary dramatically with the poloidal location. We will assume that, at the poloidal location of the minimum radial position, the field lines become straight and the curvature term vanishes. Additionally, since the poloidal derivative necessarily vanishes at this location, the gradient can be converted according to the chain rule as
\begin{align}
   \left. \vec{\nabla} \left( B_{p}^{2} \right) \right|_{a} = \left. \vec{\nabla} \psi \right|_{a} \frac{d a_{\psi}}{d \psi} \frac{d}{d a_{\psi}}  \left( \left. B_{p}^{2} \right|_{a} \right) .
\end{align}
Then \refEq{eq:minorRadiusFluxDeriv} and \refEq{eq:gradShafranovSimple} can be used to find
\begin{align}
   \left. B_{p} \right|_{a} = \mu_{0} \int_{0}^{a_{\psi}} d a_{\psi}' \left. j_{\zeta} \right|_{a} \left( a_{\psi}' \right) . \label{eq:gradShafranovPressure}
\end{align}
Furthermore, we assume that, at the poloidal location of the maximum radial position, the magnetic pressure term is small, giving
\begin{align}
   \left. B_{p} \right|_{b} = \left. \frac{\mu_{0} j_{\zeta}}{\kappa_{p}} \right|_{b} \label{eq:gradShafranovTension}
\end{align}
from \refEq{eq:gradShafranovSimple}. The integral in \refEq{eq:gradShafranovPressure} assumes that the separation between magnetic pressure and tension must be valid over the entire radial profile, not just on the flux surface of interest. If the flux surfaces are circular over a substantial region near the axis, \refEq{eq:gradShafranovPressure} is no longer accurate. For the $m=2$ mode with a constant current profile, \refEq{eq:gradShafranovPressure} and \refEq{eq:gradShafranovTension} are exact in the limits of $\Delta \rightarrow \infty$ and $\epsilon \rightarrow 0$ (see figure \ref{fig:pressureTensionBalance}). This is because, in these conditions, the flux surface exactly maintains its shape as it penetrates the plasma \cite{BallMomUpDownAsym2014, RodriguesMHDupDownAsym2014, ChristiansenCurrentDist1982}. One can use an exact solution (given by \refEq{eq:gradShafLowestOrderSolsTilt}) to estimate that \refEq{eq:gradShafranovPressure} and \refEq{eq:gradShafranovTension} are only accurate to about 20\% for a linear peaked current profile with $\hat{f}_{0} \psi_{0 b} = 0.2$ and an elongation of $\Delta = 2$. These equations are not exact for other shaping modes, but we will keep the derivation completely general because approximate results may still be useful and other exact limits may exist for different current profiles.

Substituting \refEq{eq:gradShafranovPressure} and \refEq{eq:gradShafranovTension} into \refEq{eq:kappaDerivFields} we find that
\begin{align}
   \frac{a_{\psi}}{\Delta} \frac{d \Delta}{da_{\psi}} = \frac{\left. \kappa_{p} \right|_{b}}{\Delta} \frac{\left. R \right|_{a} \int_{0}^{a_{\psi}} d a' \left. j_{\zeta}\right|_{a} \left( a_{\psi}' \right)}{\left. R \right|_{b} \left. j_{\zeta} \right|_{b}}  - 1 . \label{eq:kappaDerivCurrents}
\end{align}
Since we are considering a fixed flux surface shape, we can solve for the required current profile properties to locally permit the shape to penetrate (i.e. $d \Delta / d a_{\psi} = 0$) and find
\begin{align}
   \frac{\left. \kappa_{p} \right|_{b}}{\Delta} = \frac{ \left. R \right|_{b} \left. j_{\zeta c} \right|_{b}}{\left. R \right|_{a} \int_{0}^{a} d a_{\psi}' \left. j_{\zeta c} \right|_{a} \left( a_{\psi}' \right)} . \label{eq:geometricQuantites}
\end{align}
Here $j_{\zeta c}$ is any toroidal current density profile that ensures $d \Delta / d a_{\psi} = 0$ locally. We are guaranteed that a solution to \refEq{eq:geometricQuantites} exists for every boundary flux surface shape because, by different choices of $j_{\zeta c}$, we can make the right-hand side span the full range of $\left[ 0, \infty \right)$. Furthermore, due to the integral, this requirement can be satisfied by many different profiles.

Solving for this constant shape penetration case is useful because we are comparing configurations holding the flux surface shape constant, so both $\left. \kappa_{p} \right|_{b}$ and $\Delta$ will stay fixed. Substituting \refEq{eq:geometricQuantites} into \refEq{eq:kappaDerivCurrents}, we find that
\begin{align}
   \frac{a_{\psi}}{\Delta} \frac{d \Delta}{da_{\psi}} = \frac{\left. j_{\zeta c} \right|_{b}}{\left. j_{\zeta} \right|_{b}} \frac{ \int_{0}^{a_{\psi}} d a_{\psi}' \left. j_{\zeta} \right|_{a} \left( a_{\psi}' \right)}{\int_{0}^{a_{\psi}} d a_{\psi}' \left. j_{\zeta c} \right|_{a} \left( a_{\psi}' \right)}  - 1 . \label{eq:verificationEq}
\end{align}
By normalising this equation, we see that the total plasma current can be scaled without changing the flux surface shapes (by scaling the external currents accordingly). In other words, we can multiply $j_{\zeta c}$ or $j_{\zeta}$ by any numerical factor without changing any flux surface shapes. Equation \refEq{eq:verificationEq} is a differential equation for $\Delta \left( a_{\psi} \right)$, which can be solved giving
\begin{align}
   \frac{\Delta \left( a_{\psi} \right)}{\Delta_{b}} = \text{exp} &\left( -\int_{a_{\psi}}^{a} d a_{\psi}' \frac{1}{a_{\psi}'} \left( \frac{\left. j_{\zeta c} \right|_{b} \left( a_{\psi}' \right)}{\left. j_{\zeta} \right|_{b} \left( a_{\psi}' \right)} \frac{\int_{0}^{a_{\psi}'} d a_{\psi}'' \left. j_{\zeta} \right|_{a} \left( a_{\psi}'' \right)}{\int_{0}^{a_{\psi}'} d a_{\psi}'' \left. j_{\zeta c} \right|_{a} \left( a_{\psi}'' \right)} - 1 \right) \right) . \label{eq:exactShapingProfile}
\end{align}

\begin{figure}
 \centering
 \includegraphics[width=0.5\textwidth]{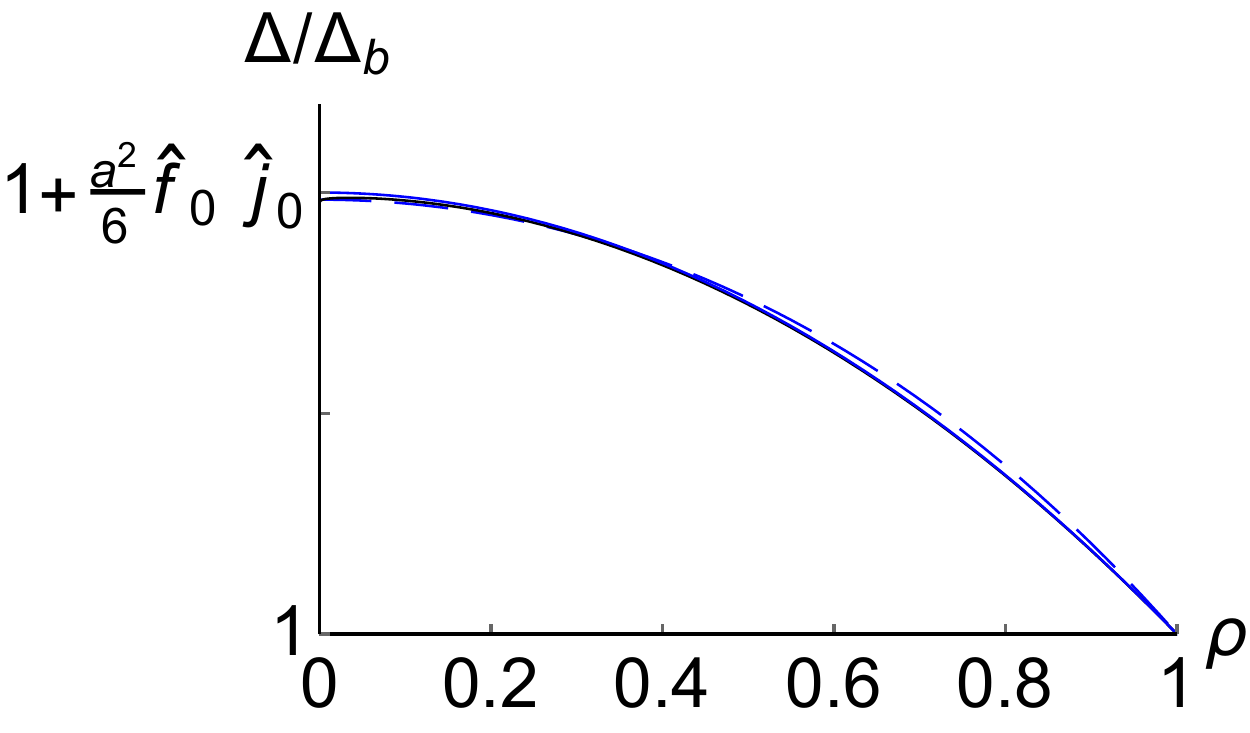}
 \caption{The exact radial shaping profile (black, solid) along with \refEq{eq:exactShapingProfile} (blue, dashed) and \refEq{eq:quadraticShapingProfile} (blue, solid), which are nearly indistinguishable, for elongated flux surfaces in the limit that $\hat{f}_{0} \ll 1$, $\epsilon \rightarrow 0$, and $\Delta_{b} \gg 1$.}
 \label{fig:limitVerification}
\end{figure}

This equation gives the radial profile of the flux surface shaping, but it is only exact when the separation of the two terms in \refEq{eq:gradShafranovSimple} is valid over the entire radial profile. For example, elongated flux surfaces with a linear current profile defined by \refEq{eq:currentProfiles} have an exact solution in the limits that $\hat{f}_{0} \ll 1$, $\epsilon \rightarrow 0$, and $\Delta_{b} \gg 1$. Using these limits, we can simplify \refEq{eq:exactShapingProfile} to
\begin{align}
   \frac{\Delta \left( \rho \right)}{\Delta_{b}} = 1 + \frac{a^{2}}{6} \hat{f}_{0} \hat{j}_{0} \left( 1 - \rho^{2} \right) . \label{eq:quadraticShapingProfile}
\end{align}
Figure \ref{fig:limitVerification} shows good agreement between this simple quadratic profile, \refEq{eq:exactShapingProfile}, and the exact numerical solution calculated from \refEq{eq:gradShafLowestOrderSolsTilt}.

\begin{figure}
	\begin{center}
		(a) \hspace{0.36\textwidth} (b) \hspace{0.39\textwidth}
		
		\includegraphics[height=0.23\textwidth]{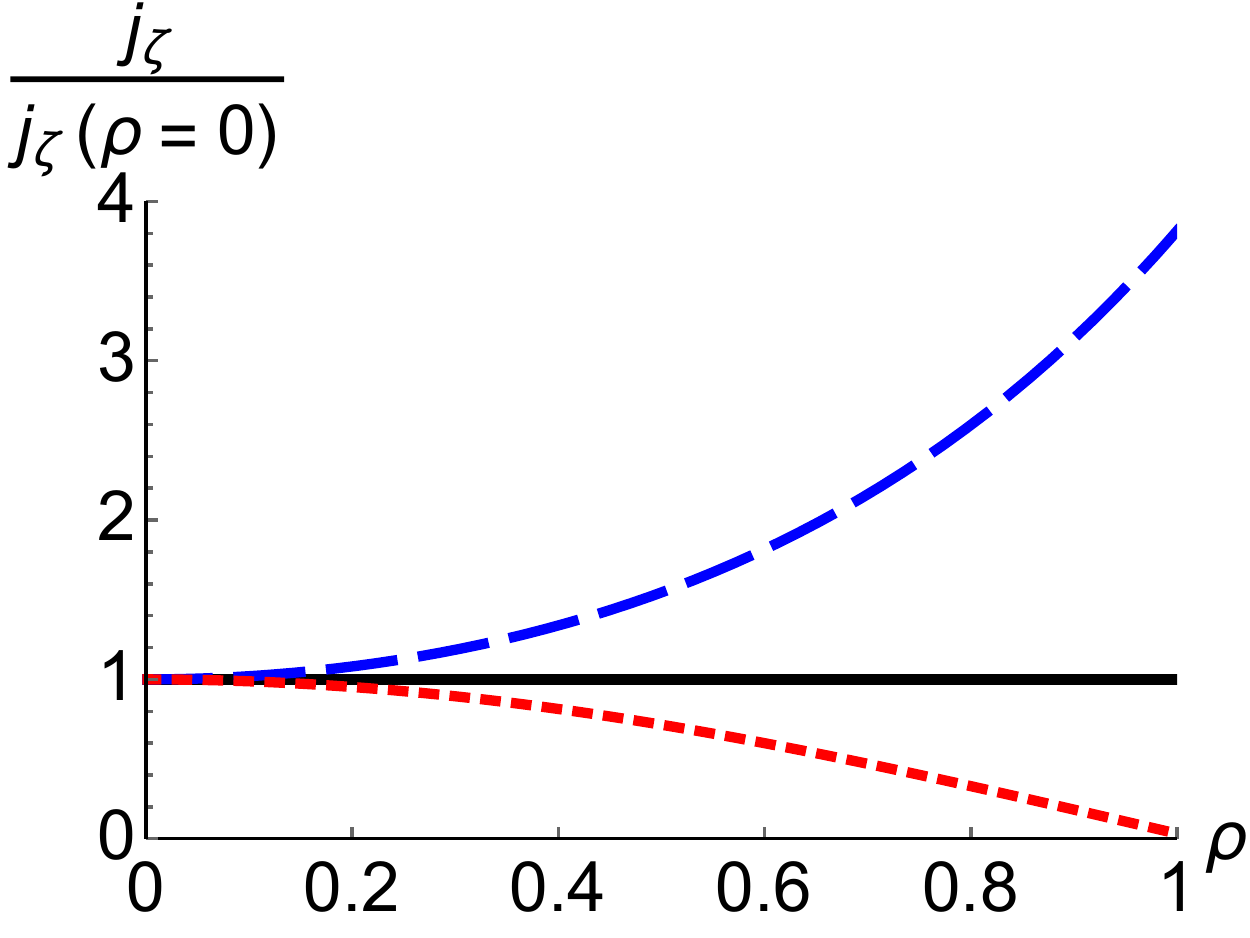}
		\hspace{0.1\textwidth}
		\includegraphics[height=0.23\textwidth]{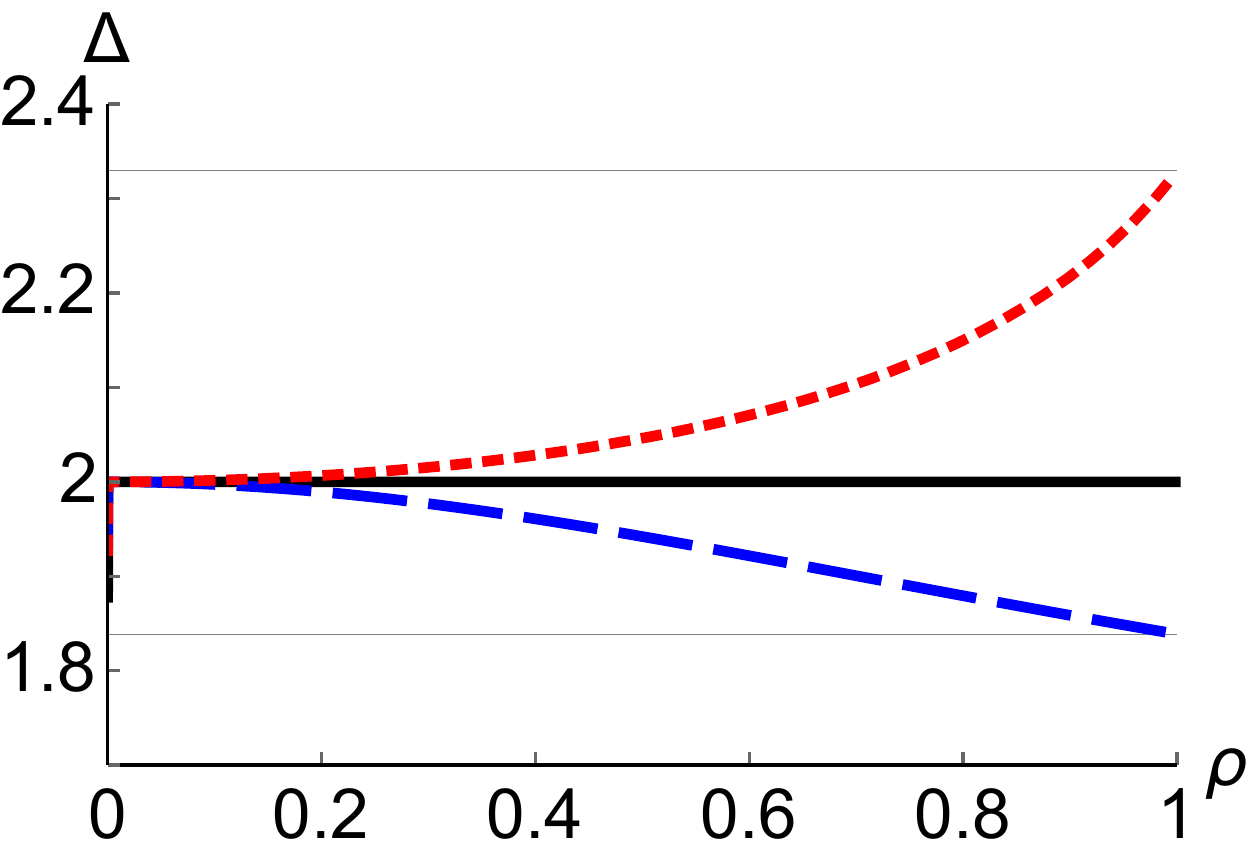}
		
		(c) \hspace{0.3\textwidth}
		
		\includegraphics[height=0.32\textwidth]{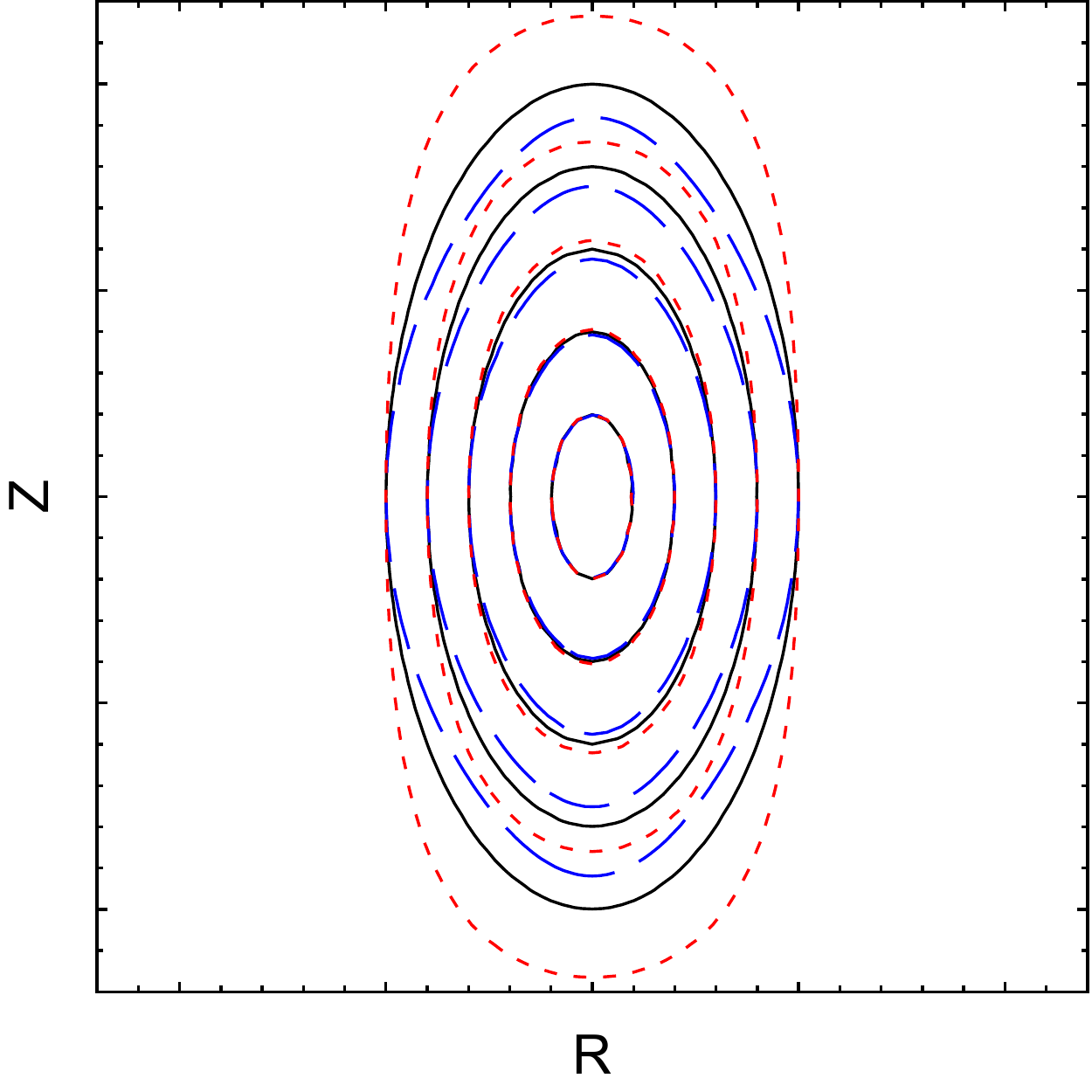}
	\end{center}
	
	\begin{center}
		(d) \hspace{0.36\textwidth} (e) \hspace{0.39\textwidth}
		
		\includegraphics[height=0.23\textwidth]{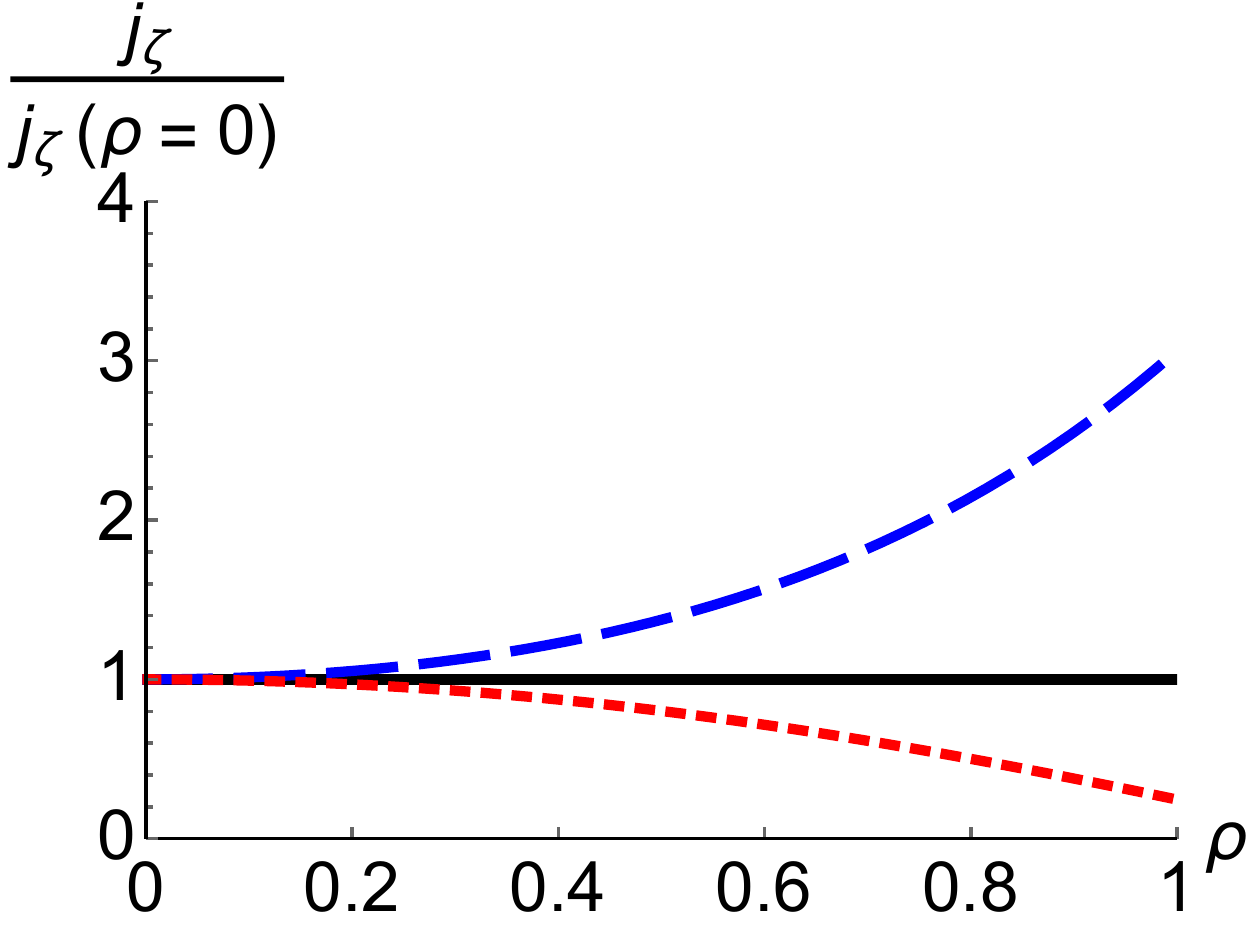}
		\hspace{0.1\textwidth}
		\includegraphics[height=0.23\textwidth]{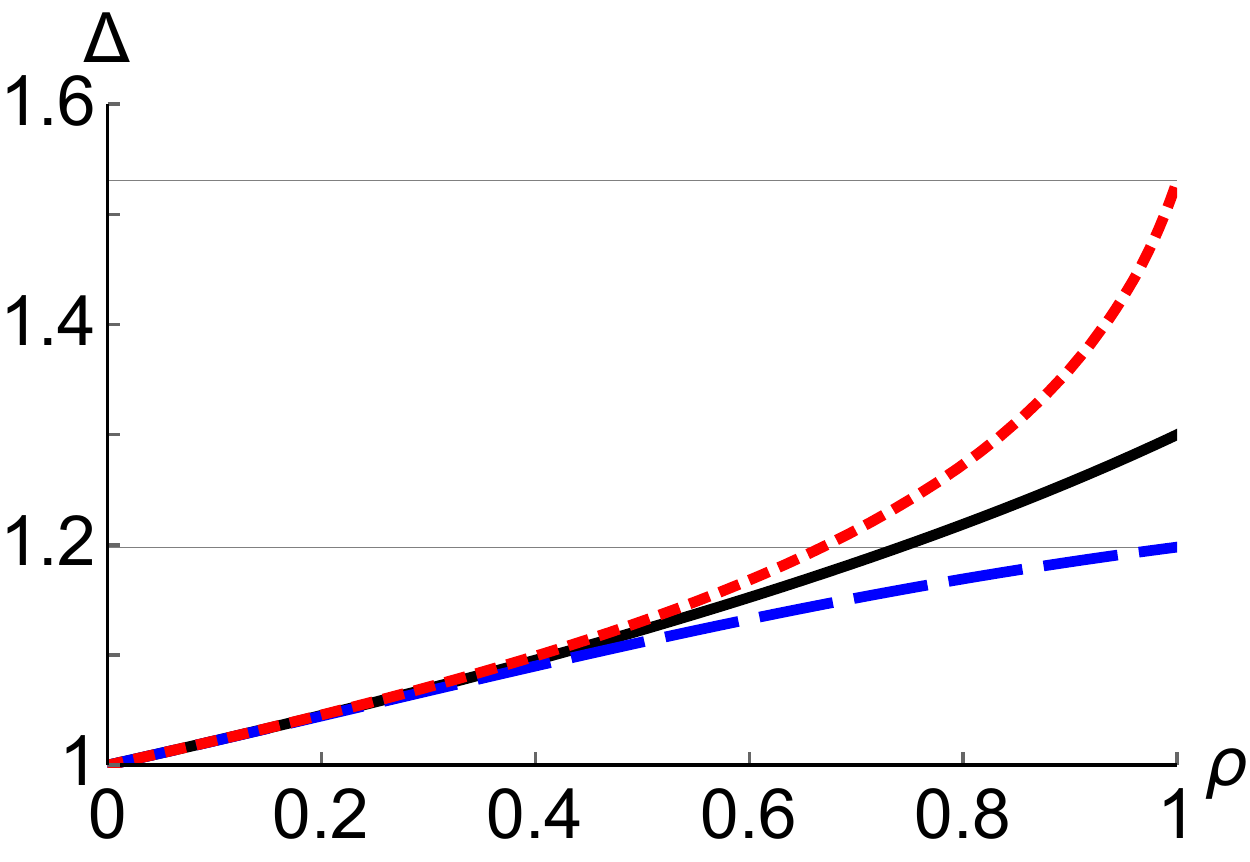}
		
		(f) \hspace{0.3\textwidth}
		
		\includegraphics[height=0.32\textwidth]{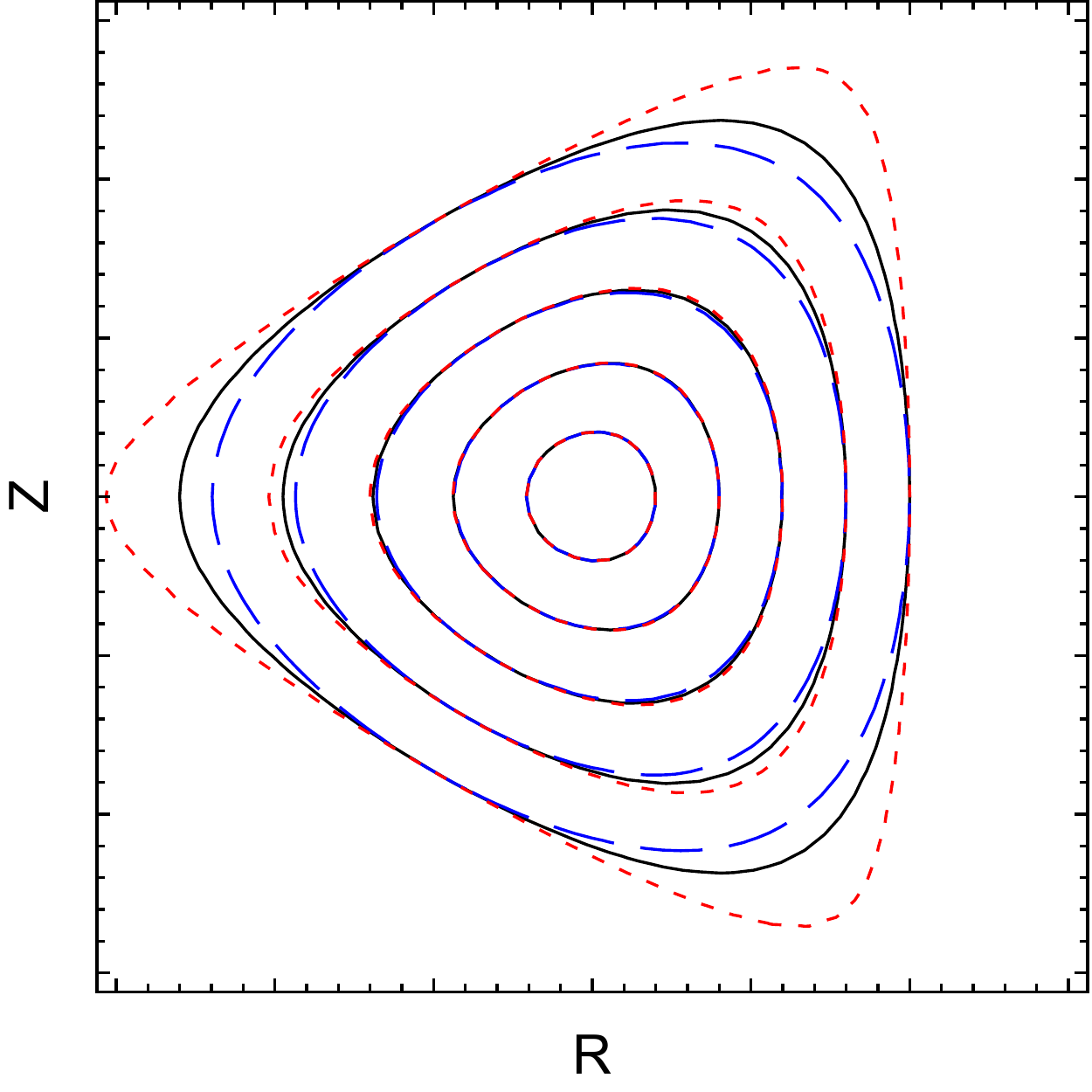}
	\end{center}	
	\caption{The (a,d) normalised radial current profile, (b,e) flux surface shapes, and (c,f) shaping profile for solutions to the Grad-Shafranov equation to lowest order in aspect ratio with constant (black, solid), hollow (blue, dashed), and peaked (red, dotted) toroidal current profiles with (a,b,c) elongated or (d,e,f) triangular boundary conditions.}
	\label{fig:exShaping}
\end{figure}

Since $j_{\zeta c}$ can be scaled arbitrarily, \refEq{eq:verificationEq} can be further simplified by choosing $\left. j_{\zeta c} \right|_{b}$ to be $\left. j_{\zeta} \right|_{b}$, the toroidal current on the flux surface of interest, giving
\begin{align}
   \frac{\rho}{\Delta} \frac{d \Delta}{d \rho} = \frac{\int_{0}^{\rho} d \rho' \left. j_{\zeta} \right|_{a} \left( \rho' \right)}{\int_{0}^{\rho} d \rho' \left. j_{\zeta c} \right|_{a} \left( \rho' \right)}  - 1 \label{eq:kappaDerivFinal}
\end{align}
at a specific radial location. This shows that the shaping penetration only depends on the amount of toroidal current within the flux surface compared with the constant shape penetration case. Profiles that are more hollow will help shaping penetrate into the plasma. What happens is, as the on-axis current is lowered, the shaping and $\left. R B_{p} \right|_{b}$ stay constant (maintained by the external magnets), while $\left. R B_{p} \right|_{a}$ decreases because of the drop in the total plasma current. From \refEq{eq:kappaDerivFields} we see that a change in the ratio of these magnetic fields allows the shaping to penetrate radially. Analogously, peaked current profiles will tend to limit the shaping to the edge. In figure \ref{fig:exShaping}, we plot \refEq{eq:gradShafLowestOrderSolsTilt} for different boundary conditions and values of $\hat{f}_{0}$. From figure \ref{fig:exShaping}(a,b,c), we see that achieving an on-axis elongation of 2 with a peaked current profile requires a 25\% greater edge elongation than it would with a hollow profile. Figure \ref{fig:exShaping}(d,e,f) shows that triangular flux surface shaping is only large near the boundary, as would be expected from the arguments in both the introduction to this chapter and section \ref{sec:fluxSurfaceShapeEffect}. However, we still observe that the shaping penetrates more effectively with a hollow current profile, relative to a peaked profile. This, along with \refEq{eq:kappaDerivFinal}, suggests that the beneficial effect of hollow current profiles for shaping penetration is general to all flux surface shapes (see references \cite{BallMomUpDownAsym2014, RodriguesMHDupDownAsym2014} for a different approach to the same problem). Numerical evidence of this using EFIT equilibrium reconstruction on simulated experimental data can be seen in figure 5(b) of reference \cite{LaoShapeAndCurrent1985}.

%% file: Ch_04_MHD_ShafranovShift.tex
% !TEX root = /Users/Justin/Documents/Research/Writings/2016DoctoralThesis/DoctoralThesis.tex

\chapter{Global equilibria with a Shafranov shift and tilted elliptical boundary}
\label{ch:MHD_ShafranovShift}

\begin{quote}
   \emph{Much of this chapter appears in reference \cite{BallShafranovShift2016}.}
\end{quote}

In order to model a realistic Shafranov shift we must know how it depends on the free parameters that appear in the next order (in large aspect ratio) Grad-Shafranov equation: the boundary flux surface, the current profile, and the pressure profile. We will restrict our investigation to using a tilted elliptical boundary because the MHD analysis in chapter \ref{ch:MHD_ShapingIntuition} suggests that low modes penetrate most effectively. We will explicitly calculate how the Shafranov shift depends on the tilt angle of the elliptical boundary flux surface (parameterized by $\theta_{\kappa b}$).
% for a tokamak with constant current and pressure gradient profiles (i.e. $\hat{f}_{0} = 0$ and $\hat{f}_{0 p} = 0$). Additionally, 
We will argue that the Shafranov shift is insensitive to the shape of the current and pressure profiles (parameterized by $\hat{f}_{0}$ and $\hat{f}_{0 p}$ respectively) when the geometry, plasma current, and average $d p / d \psi$ is kept fixed. Doing so makes the gyrokinetic simulations presented in chapter \ref{ch:GYRO_ShafranovShift} more widely applicable, as they use equilibria derived assuming constant current and pressure gradient profiles. In order to accomplish this, we require a general solution for the magnitude and direction of the Shafranov shift in tokamaks with a tilted elliptical boundary as well as linear current and pressure gradient profiles.

\begin{figure}
	\centering
	\includegraphics[width=0.4\textwidth]{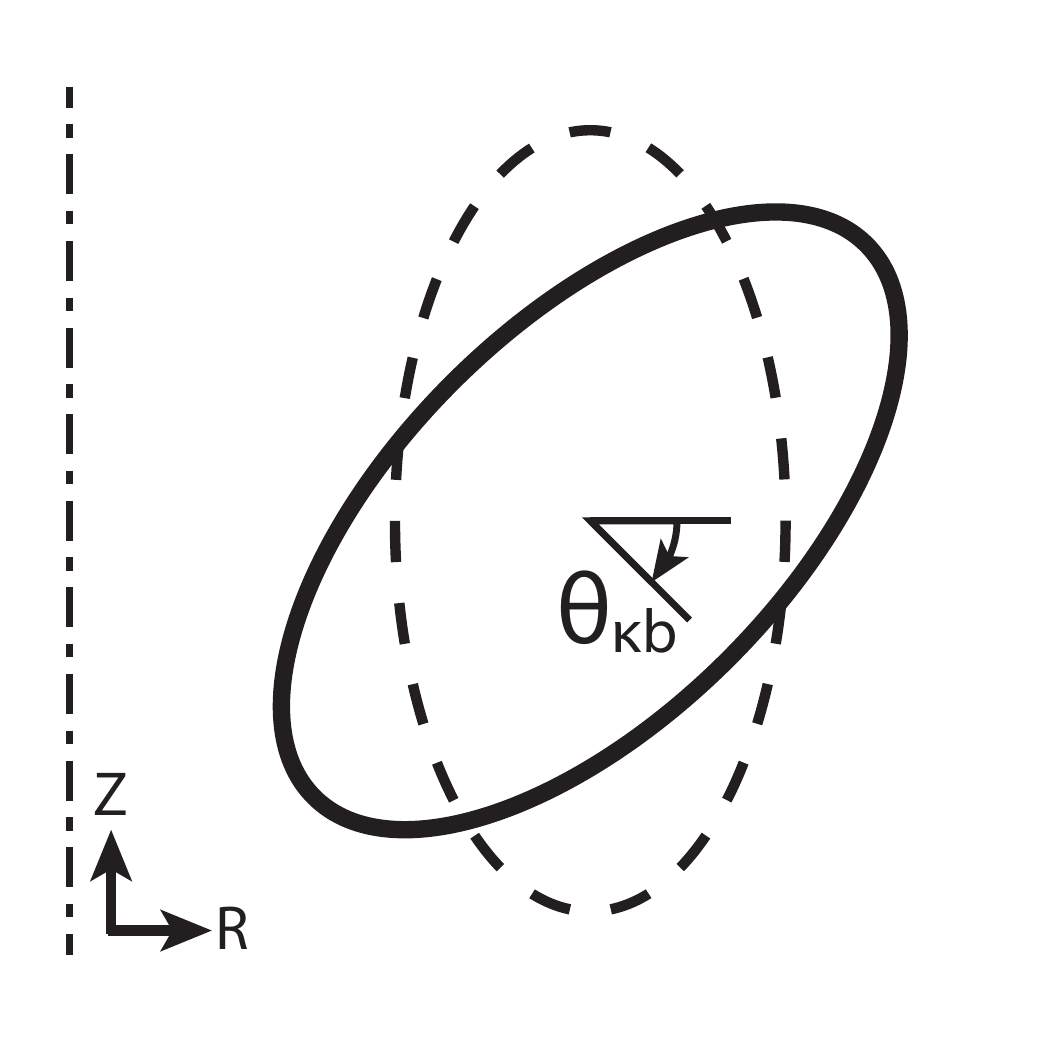}
	\caption{The boundary flux surface specified by \refEq{eq:boundarySurf} for two values of the boundary elongation tilt angle, $\theta_{\kappa b} = \pi / 4$ (solid) and $\theta_{\kappa b} = 0$ (dotted), where the axis of axisymmetry is indicated by a dash-dotted line.}
	\label{fig:thetaKappaDef}
\end{figure}

Together \refEq{eq:gradShafLowestOrderSolsTilt}, \refEq{eq:gradShafNextOrderSols}, \refEq{eq:inhomoTermsC0} through \refEq{eq:inhomoTermsTm}, and \refEq{eq:psiFourierNextOrderCoeffs} give this general solution to $O \left( \epsilon B_{0} \right)$, which is sufficient to capture the behaviour of the Shafranov shift. However, we still must determine the Fourier coefficients $N_{0, m}$, $\theta_{t 0, m}$, $C_{1, m}$, and $S_{1, m}$ in order to create a tilted elliptical boundary flux surface. To do so we require the poloidal flux to be constant on the boundary, parameterized in polar form by
\begin{align}
	r_{b} \left( \theta \right) = \frac{\sqrt{2} \kappa_{b} a}{\sqrt{\kappa_{b}^{2} + 1 + \left( \kappa_{b}^{2} - 1 \right) \Cos{2 \left( \theta + \theta_{\kappa b} \right)}}} , \label{eq:boundarySurf}
\end{align}
where $\kappa_{b}$ is the elongation of the boundary flux surface, $\theta_{\kappa b}$ is the boundary tilt angle, and $a$ is the tokamak minor radius (i.e. the minor radial position of the boundary flux surface at $\theta = - \theta_{\kappa b}$).

%===================================================%
%===================================================%
\section{Solution to the $O ( B_{0} )$ Grad-Shafranov equation for a tilted elliptical boundary condition}
\label{sec:numCoeffCalcLowest}
%===================================================%
%===================================================%

To calculate $N_{0, m}$ and $\theta_{t 0, m}$ we substitute \refEq{eq:boundarySurf} into \refEq{eq:gradShafLowestOrderSolsTilt} to give
\begin{align}
   \psi_{0} \left( r_{b} \left( \theta \right), \theta \right) = \psi_{0 b} , \label{eq:gradShafBoundaryCondLowestOrder}
\end{align}
where $\psi_{0 b}$ is the value of the poloidal flux on the plasma boundary. Since $\psi_{0 b}$ is a constant we know that $\psi_{0} \left( r_{b} \left( \theta \right), \theta \right)$ does not depend on $\theta$. In theory, ensuring that this is true for all values of $\theta$ determines all of the lowest order Fourier coefficients. However, the exact solution for these coefficients is not analytic, so we will resort to a numerical solution. Before we do so we will note that, because the lowest order Grad-Shafranov equation has cylindrical symmetry, the only angle intrinsic to the problem is $\theta_{\kappa b}$, which is introduced by the boundary condition. This implies that
\begin{align}
   \theta_{t 0, m} = \theta_{\kappa b} \label{eq:tiltAngleSol}
\end{align}
for all $m$, which suggests that it will be useful to define a new poloidal angle
\begin{align}
   \theta_{s} \equiv \theta + \theta_{\kappa b} . \label{eq:shiftThetaDef}
\end{align}
Furthermore, since an ellipse has mirror symmetry about exactly two axes, we know that $N_{0, m} = 0$ for odd $m$.

To determine $N_{0, m}$ for even $m$ we will take the Fourier series of $\psi_{0} \left( r_{b} \left( \theta_{s} \right), \theta_{s} \right) - \psi_{0 b}$. Truncating the series at a large mode number $m_{\text{max}}$ gives a long series of cosine terms. Requiring that the coefficient of each term must individually vanish gives a numerical approximation for all $N_{0, m}$ with $m \leq m_{\text{max}}$. In the limit that $m_{\text{max}} \to \infty$ this approximation approaches the exact solution, though in practice $m_{\text{max}} \approx 10$ was found to achieve sufficient precision for our purposes. This was determined by visually assessing how well the solution matched the boundary condition at the plasma edge.

%===================================================%
%===================================================%
\section{Solution to the $O ( \epsilon B_{0} )$ Grad-Shafranov equation for a tilted elliptical boundary condition}
\label{sec:numCoeffCalcNext}
%===================================================%
%===================================================%

To next order we must determine $C_{1, m}$ and $S_{1, m}$ such that
\begin{align}
   \psi_{1} \left( r_{b} \left( \theta \right), \theta \right) = \psi_{1 b} \label{eq:gradShafBoundaryCondNextOrder}
\end{align}
is true, where $\psi_{1 b}$ is the next order value of the poloidal flux on the boundary flux surface. This is done in a similar manner to the lowest order equations, except the Grad-Shafranov equation no longer has cylindrical symmetry and we must evaluate the integral in \refEq{eq:psiFourierNextOrderCoeffs}. The lack of symmetry means that we do not automatically know the tilt angle of the modes. However, since $\psi_{0}$ only has even Fourier mode numbers, it can be shown that \refEq{eq:gradShafNextOrder} only has odd Fourier modes. Hence, $C_{1, m} = S_{1, m} = 0$ for even $m$. 

To calculate $C_{1, m}$ and $S_{1, m}$ for odd $m$ we take $\psi_{1} \left( r, \theta \right)$ from \refEq{eq:gradShafNextOrderSols} and Taylor expand in $\hat{f}_{0} \hat{j}_{0} a^{2} \ll 1$ to $O \left( \left( \hat{f}_{0} \hat{j}_{0} a^{2} \right)^{f_{\text{max}}} \right)$. This allows us to analytically calculate the integrals appearing in \refEq{eq:psiFourierNextOrderCoeffs} because the Bessel functions become summations of polynomials. We can now substitute \refEq{eq:boundarySurf} and find the Fourier series of $\psi_{1} \left( r_{b} \left( \theta \right), \theta \right) - \psi_{1 b}$ to mode number $m_{\text{max}}$. Again, we require that all of the Fourier coefficients must individually vanish, which produces a numerical approximation for each $C_{1, m}$ and $S_{1, m}$ with $m \leq m_{\text{max}}$. A value of $f_{\text{max}} \approx 10$ was found to give a sufficiently accurate solution.

For a hollow current profile, we repeat the entire above process except for using \refEq{eq:gradShafLowestOrderSolsTiltHollow} instead of \refEq{eq:gradShafLowestOrderSolsTilt} and \refEq{eq:psiFourierNextOrderCoeffsHollow} instead of \refEq{eq:psiFourierNextOrderCoeffs}. While the above process also works for the case of a constant toroidal current profile, it has an analytic solution, which we derive in appendix \ref{app:exactMagAxisLoc}.

In order to understand the effect of changing the current and pressure profiles in a single experimental device, we will choose to keep the major radial location of the centre of the boundary flux surface ($R_{0 b}$), the minor radius ($a$), the edge elongation ($\kappa_{b}$), the total plasma current ($I_{p}$), and an estimate of the average pressure gradient ($p_{\text{axis}}/\psi_{0 b}$, i.e. the on-axis pressure divided by the lowest order edge poloidal flux) fixed. In order to keep these parameters fixed as we change the current and pressure profiles we must calculate how they enter into both $\hat{j}_{0}$ and $\hat{j}_{0 p}$. Calculating $\hat{j}_{0 p}$ is straightforward, as we can directly integrate \refEq{eq:pressureProfShape} over poloidal flux to find
\begin{align}
   \hat{j}_{0 p} = \mu_{0} R_{0 b}^{2} \frac{p_{\text{axis}}}{\psi_{0 b}} \left( 1 - \frac{\hat{f}_{0 p} \psi_{0 b}}{2} \right)^{-1} . \label{eq:jNpEstimate}
\end{align}
To calculate $\hat{j}_{0}$ we start with the definition of the plasma current,
\begin{align}
I_{p} \equiv \int d S j_{\zeta} = \int_{0}^{2 \pi} d \theta_{s} \int_{0}^{r_{b} \left( \theta_{s} \right)} d r j_{\zeta} r , \label{eq:totalPlasmaCurrent}
\end{align}
where $S$ is the poloidal cross-sectional surface. Since we are only searching for a simple estimate, we will use \refEq{eq:currentProfiles} to rewrite \refEq{eq:totalPlasmaCurrent} as
\begin{align}
I_{p} = \int_{0}^{2 \pi} d \theta_{s} \int_{0}^{r_{b} \left( \theta_{s} \right)} d r \frac{\hat{j}_{0}}{\mu_{0} R_{0 b}} \left( 1 - \hat{f}_{0} \psi_{0} \right) r ,
\end{align}
which is accurate to lowest order in aspect ratio. Substituting the boundary shape (i.e. \refEq{eq:boundarySurf}) and the constant current solution for $\psi_{0} \left( r, \theta_{s} \right)$ (i.e. \refEq{eq:gradShafLowestOrderSolsTiltConst}, \refEq{eq:tiltAngleSol}, \refEq{eq:boundaryFluxConst}, and \refEq{eq:fourierShapingConst}) allows us to directly take the integral to find
\begin{align}
\hat{j}_{0} = \mu_{0} \frac{I_{p}}{\pi a^{2} \kappa_{b}} R_{0 b} \left( 1 - \frac{\hat{f}_{0} \psi_{0 b}}{2} \right)^{-1} + \Order{\hat{f}_{0}^{2} \hat{j}_{0}^{2} a^{4}} . \label{eq:jNestimate}
\end{align}
The $\Order{\hat{f}_{0}^{2} \hat{j}_{0}^{2} a^{4}}$ error arises from the fact that we used the constant current solution for $\psi_{0} \left( r, \theta_{s} \right)$, which is only accurate to lowest order in $\hat{f}_{0} \hat{j}_{0} a^{2} \ll 1$. This means that as we change $\hat{f}_{0 p}$ and $\hat{f}_{0}$ we must change $\hat{j}_{0 p}$ and $\hat{j}_{0}$ according to \refEq{eq:jNpEstimate} and \refEq{eq:jNestimate} respectively.

\begin{figure}
 \centering
 \includegraphics[width=0.6\textwidth]{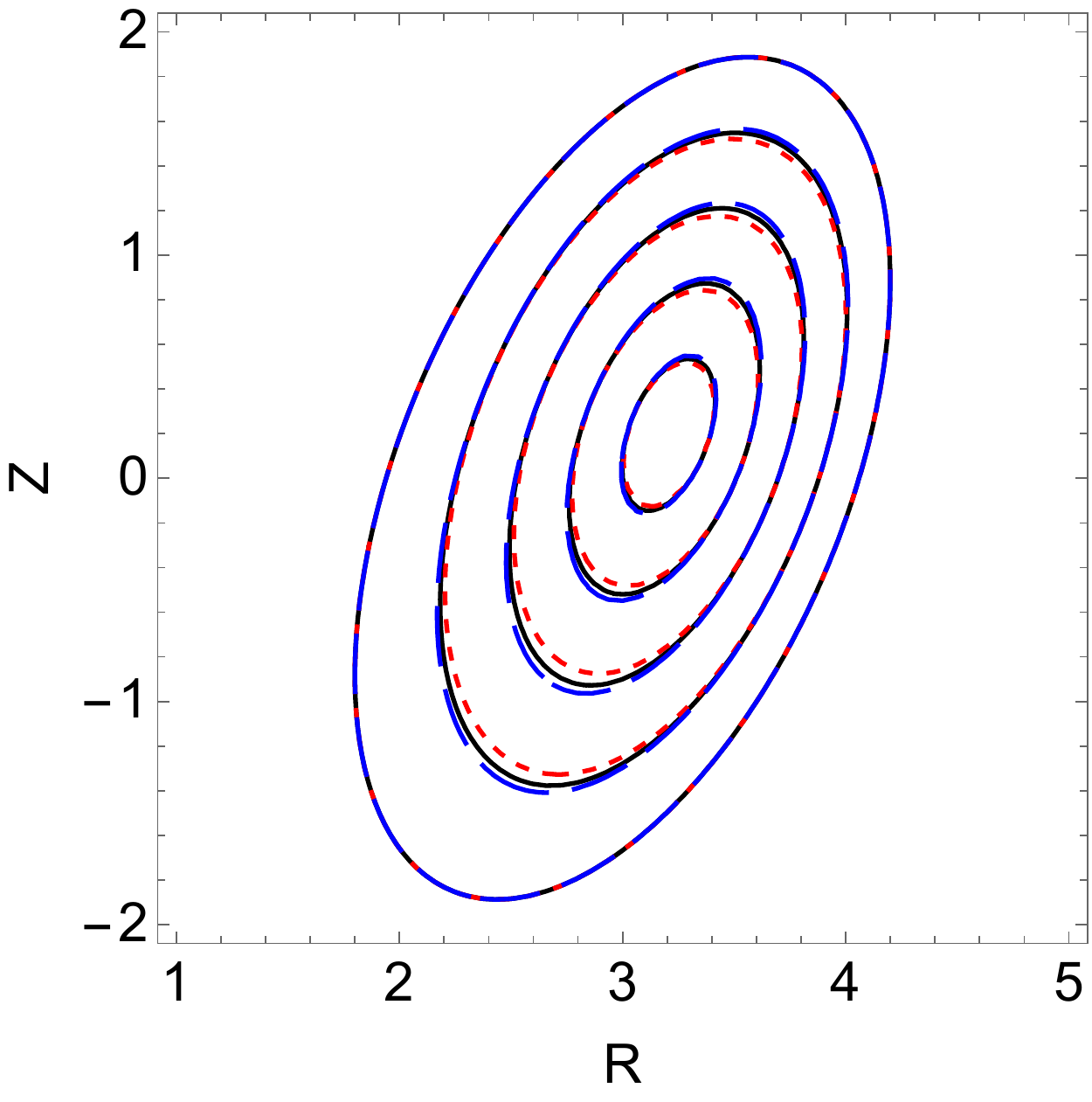}
 \caption{Calculated flux surfaces for $\hat{f}_{0} \psi_{0 b} = \hat{f}_{0 p} \psi_{0 b} = 0$ (black, solid), $\hat{f}_{0} \psi_{0 b} = \hat{f}_{0 p} \psi_{0 b} = 0.4$ (red, dotted), and $\hat{f}_{0} \psi_{0 b} = \hat{f}_{0 p} \psi_{0 b} = - 0.4$ (blue, dashed).}
 \label{fig:gradShafCalcSol}
\end{figure}

In figure \ref{fig:gradShafCalcSol} we plot the calculated flux surfaces resulting from three different current profiles, setting $\hat{f}_{0 p} = \hat{f}_{0}$. We use inputs of
\begin{align}
	R_{0 b} = 3 , \hspace{5em}
	a = 1 , \hspace{5em}
	\kappa_{b} = 2 , \label{eq:ShafShiftInputs}
\end{align}
(we have normalised all lengths to the minor radius), and
\begin{align}
\frac{\hat{j}_{0 p}}{\hat{j}_{0}} \approx \frac{\pi a^{2} \kappa_{b} R_{0 b}}{I_{p}} \frac{p_{\text{axis}}}{\psi_{0 b}} \approx 0.7 \label{eq:jNconstantsRatio}
\end{align}
(from projections for ITER \cite{AymarITERSummary2001}). Additionally, we choose to plot the case of $\theta_{\kappa b} = \pi / 8$ because nonlinear gyrokinetic simulations have shown this value to be optimal for generating rotation (see figure \ref{fig:momHeatFluxRatio} and reference \cite{BallMomUpDownAsym2014}). Note that the $\psi_{0 b}$ appearing in \refEq{eq:jNconstantsRatio} is part of $p_{\text{axis}} / \psi_{0 b}$, so it is fixed for all three profiles and can be calculated for a constant current profile from \refEq{eq:boundaryFluxConst}. In figure \ref{fig:gradShafCalcSol} we see that the current profile has an effect on the penetration of elongation from the boundary to the magnetic axis. This indicates that hollower current profiles better support elongation throughout the plasma, which is consistent with the results of chapter \ref{ch:MHD_ShapingIntuition} as well as previous work \cite{BallMomUpDownAsym2014, RodriguesMHDupDownAsym2014}. However, given these parameters, the Shafranov shift is not visibly altered, even with the significant changes to the current profile.

\begin{figure}
 \begin{center}
   (a) \hspace{0.36\textwidth}
  
  \includegraphics[height=0.5\textwidth]{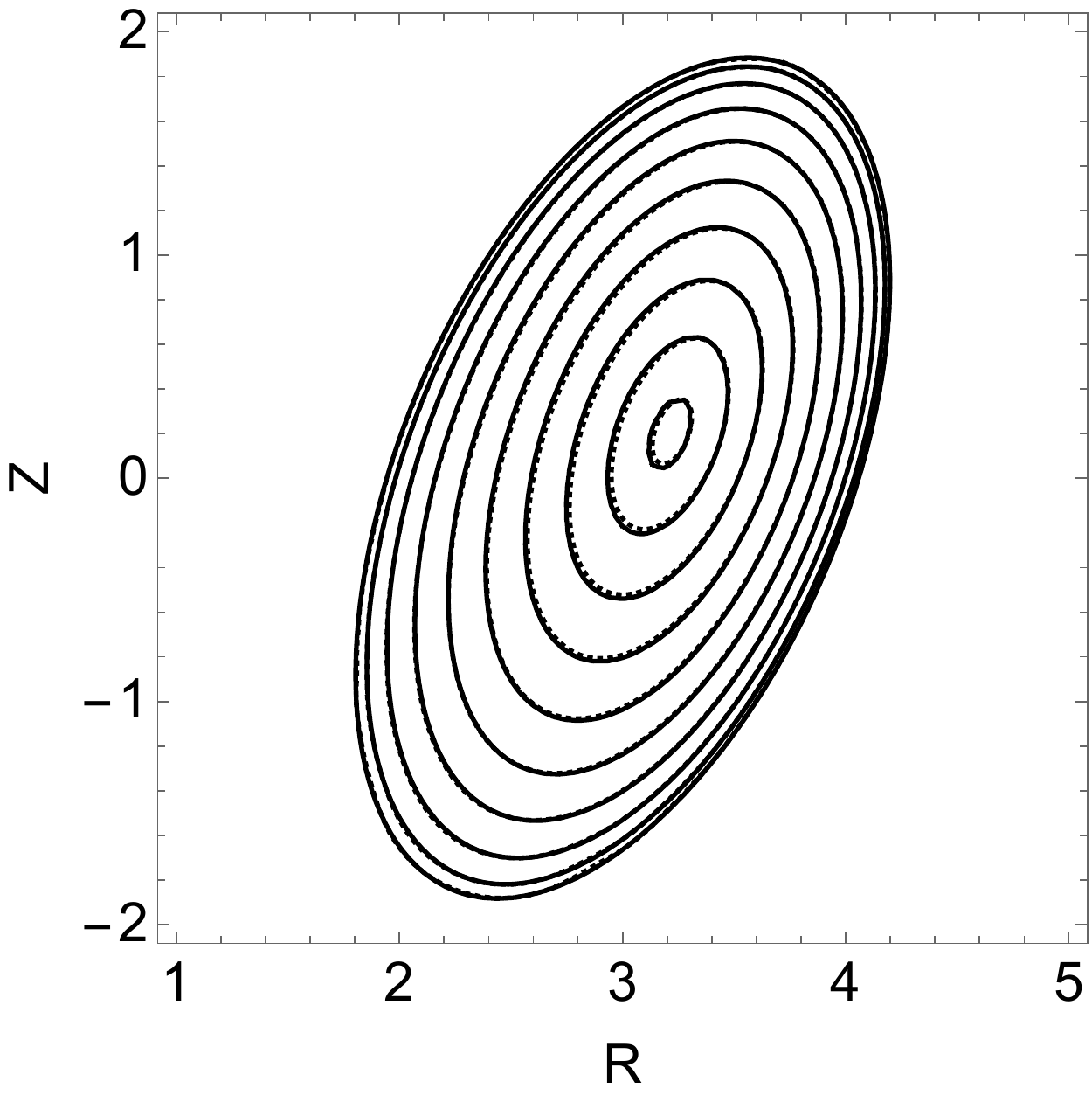}
  
  (b) \hspace{0.41\textwidth} (c) \hspace{0.38\textwidth}
  
  \includegraphics[height=0.5\textwidth]{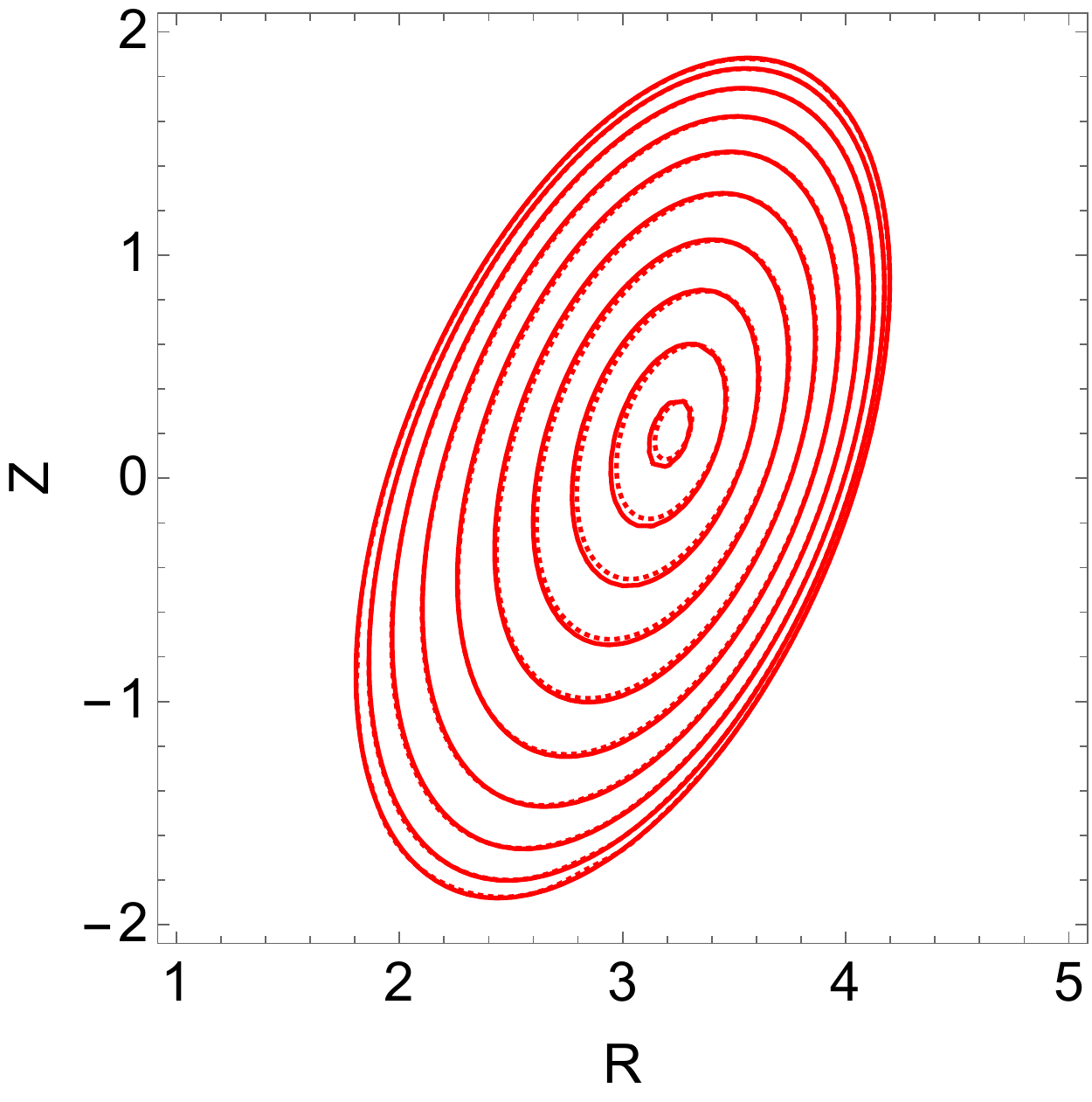}
  \includegraphics[height=0.5\textwidth]{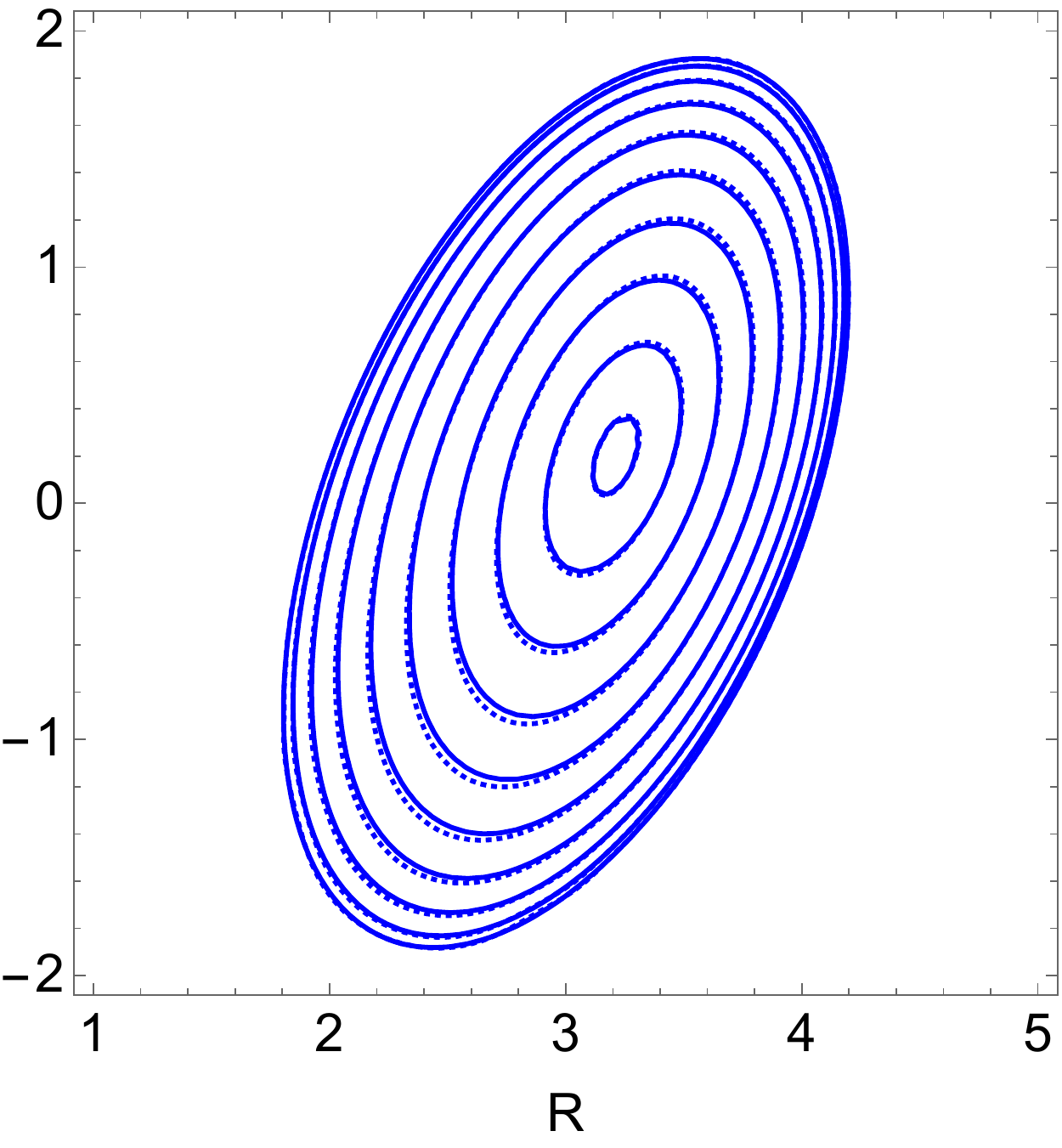}
 \end{center}
 \caption{Flux surfaces calculated by both ECOM (dotted) and analytically (solid) for (a) $\hat{f}_{0} \psi_{0 b} = \hat{f}_{0 p} \psi_{0 b} = 0$ (black), (b) $\hat{f}_{0} \psi_{0 b} = \hat{f}_{0 p} \psi_{0 b} = 0.4$ (red), and (c) $\hat{f}_{0} \psi_{0 b} = \hat{f}_{0 p} \psi_{0 b} = - 0.4$ (blue).}
 \label{fig:gradShafSolComp}
\end{figure}

In order to verify our calculation, we compared our results with the ECOM code \cite{LeeECOM2015}, a fixed boundary equilibrium solver capable of modelling up-down asymmetric configurations. In figure \ref{fig:gradShafSolComp} we see a direct graphical comparison between ECOM and the results of our calculation that were shown in figure \ref{fig:gradShafCalcSol}. The two sets of results agree well, especially for the constant and hollow current profile cases. We believe that the most significant source of error is finite aspect ratio effects in our analytic calculation, which arise from the assumption that $\epsilon = 1 / 3 \ll 1$. Hence, formally we would only expect the analytic calculation to be accurate to about $\epsilon^{2} \sim 10\%$.

\begin{figure}
 \centering
 \includegraphics[width=0.45\textwidth]{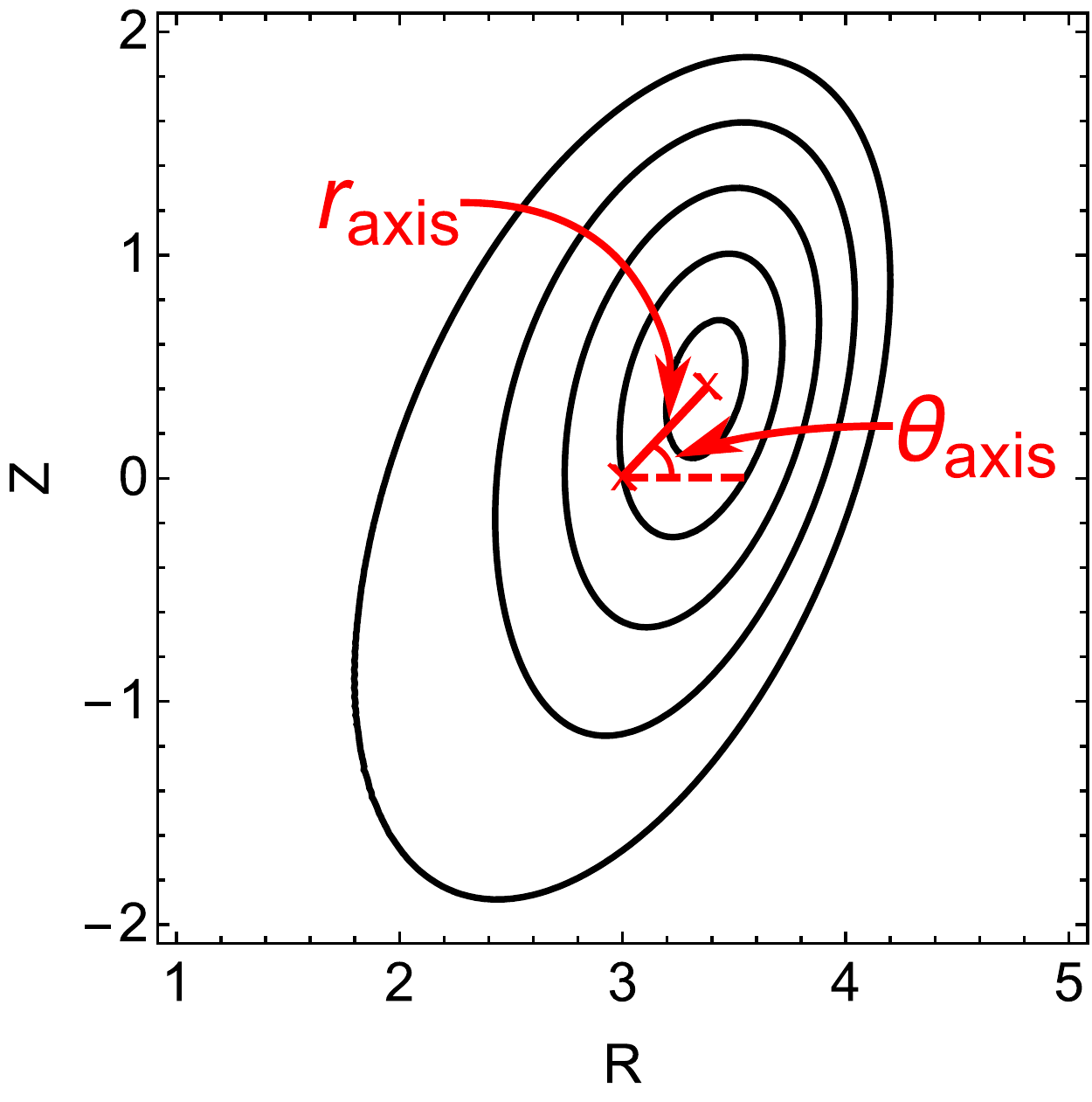}
 \caption{Example flux surfaces showing the geometric meaning of the parameters $r_{\text{axis}}$ and $\theta_{\text{axis}}$, the minor radial and poloidal locations of the magnetic axis respectively.}
 \label{fig:geoMagAxisLoc}
\end{figure}

%===================================================%
%===================================================%
\section{Location of the magnetic axis}
\label{sec:ECOMcomparison}
%===================================================%
%===================================================%

We can obtain the Shafranov shift from our calculation by numerically solving the equation
\begin{align}
   \left. \Nabla \left( \psi_{0} \left( r, \theta \right) + \psi_{1} \left( r, \theta \right) \right) \right|_{r = r_{\text{axis}}, \theta = \theta_{\text{axis}}} = 0 \label{eq:magAxisCondition}
\end{align}
using \refEq{eq:gradShafLowestOrderSolsTilt}, \refEq{eq:gradShafNextOrderSols}, \refEq{eq:inhomoTermsC0} through \refEq{eq:inhomoTermsTm}, \refEq{eq:psiFourierNextOrderCoeffs}, and \refEq{eq:tiltAngleSol} as well as our numerical solutions for $N_{0, m}$, $C_{1, m}$, and $S_{1, m}$. Here $r_{\text{axis}}$ and $\theta_{\text{axis}}$ are the minor radial and poloidal location of the magnetic axis respectively, as indicated in figure \ref{fig:geoMagAxisLoc}. For the special case of a tilted elliptical boundary with a constant toroidal current profile (i.e. $\hat{f}_{0} = 0$) we can exactly solve \refEq{eq:magAxisCondition} as shown in appendix \ref{app:exactMagAxisLoc}. Equations \refEq{eq:magAxisPoloidalLocExact} and \refEq{eq:magAxisRadialLocExactCondition} give the exact location of the magnetic axis when considering the poloidal flux to lowest order and next order in $\epsilon \ll 1$.

\begin{figure}
 \hspace{0.04\textwidth} (a) \hspace{0.4\textwidth} (b) \hspace{0.25\textwidth}
 \begin{center}
  \includegraphics[width=0.45\textwidth]{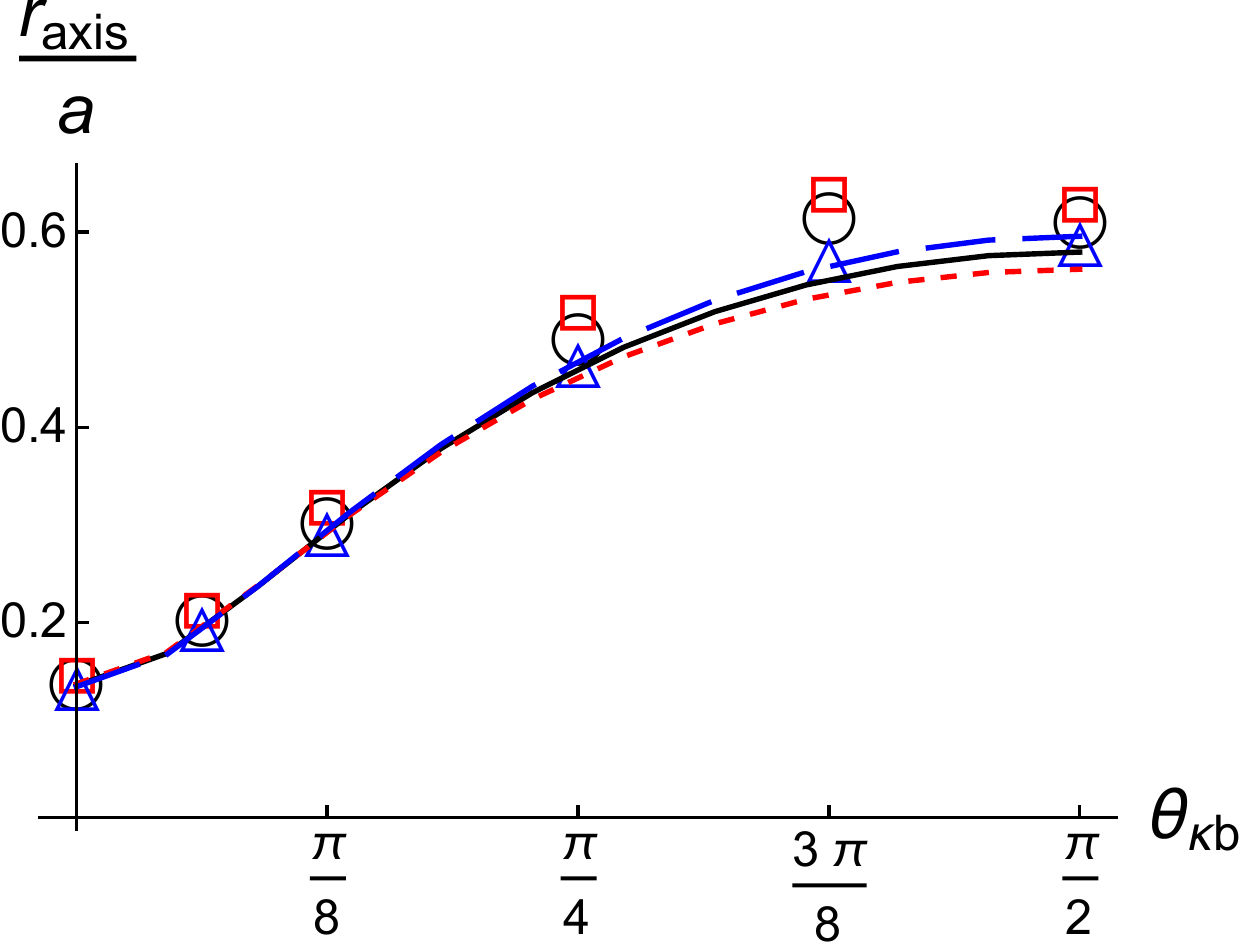}
  \includegraphics[width=0.45\textwidth]{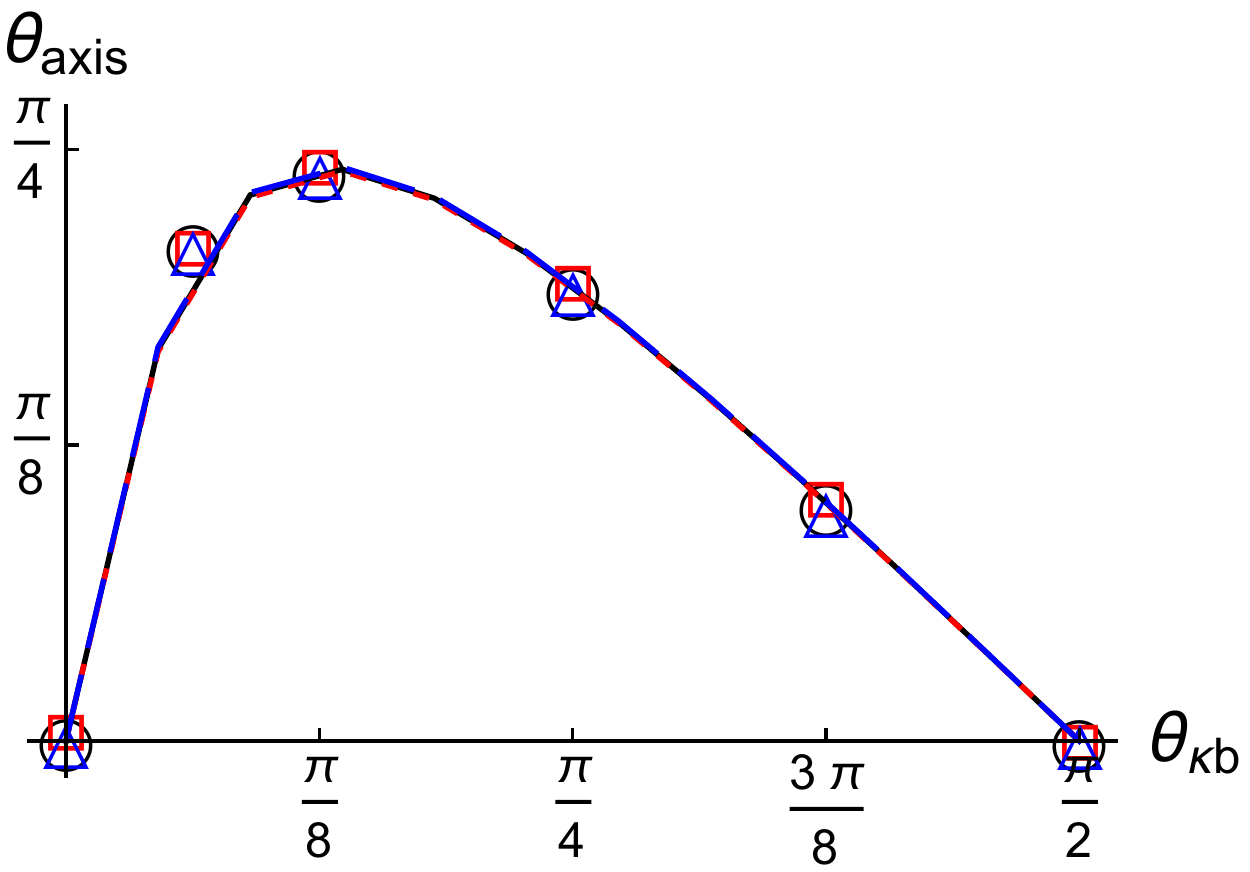}
 \end{center}
 \caption{The (a) minor radial and (b) poloidal location of the magnetic axis for constant ($\hat{f}_{0} \psi_{0 b} = \hat{f}_{0 p} \psi_{0 b} = 0$) (black, solid, circles), linear peaked ($\hat{f}_{0} \psi_{0 b} = \hat{f}_{0 p} \psi_{0 b} = 0.4$) (red, dotted, squares), and linear hollow ($\hat{f}_{0} \psi_{0 b} = \hat{f}_{0 p} \psi_{0 b} = - 0.4$) (blue, dashed, triangles) current/pressure gradient profiles, calculated analytically (lines) and by ECOM (points).}
 \label{fig:ShafShiftLoc}
\end{figure}

In figure \ref{fig:ShafShiftLoc} we show the location of the magnetic axis for different boundary tilt angles as we vary the shape of the current/pressure profile (by changing $\hat{f}_{0}$ and keeping $\hat{f}_{0 p} = \hat{f}_{0}$). In this scan we hold the geometry, $I_{p}$, and $p_{\text{axis}} / \psi_{0 b}$ fixed at the values determined by \refEq{eq:ShafShiftInputs} and \refEq{eq:jNconstantsRatio}. For the most part, we see reasonable quantitative agreement between our theoretical results and ECOM. However, the trend of $r_{\text{axis}}$ with $\hat{f}_{0} \psi_{0 b}$ at large tilt angles is inconsistent between the two calculations. This appears to be a breakdown in our inverse aspect ratio expansion as the analytical calculation and ECOM become consistent at smaller tilt angles (where the effective aspect ratio is larger) and if the aspect ratio is directly increased. An important property of figure \ref{fig:ShafShiftLoc}, which is supported by both the analytic and ECOM calculations, is the insensitivity of the Shafranov shift to significant changes in the shape of the current profile. Both the magnitude and the direction of the Shafranov shift change very little between the different current profiles. This is especially true in the domain of $\theta_{\kappa b} \in \left[ 0, \pi / 4 \right]$, which is the range of tilt angles that seem most promising for implementing in an experiment \cite{CamenenPRLExp2010, BallMomUpDownAsym2014}. This means that, even though we will only run gyrokinetic simulations of equilibria with constant current and pressure gradient profiles, we expect the Shafranov shift to have a similar effect in equilibria with other profiles. 

Conversely, we see that the boundary elongation tilt angle has a large effect, not just on the direction of the Shafranov shift, but also its magnitude. This is intuitive because we know that, for an ellipse with $\kappa = 2$, the midplane chord length is twice as long in the $\theta_{\kappa b} = \pi / 2$ geometry as it is in the $\theta_{\kappa b} = 0$ geometry. Lastly, we see that the direction of the Shafranov shift varies considerably, but it is purely outwards for the $0$ and $\pi / 2$ tilt angles as expected. We note that (except at $\theta_{\kappa b} = 0$ and $\theta_{\kappa b} = \pi / 2$) it does not align with the lines of symmetry of the ellipse, so it breaks the mirror symmetry of the configuration.

\begin{figure}
 \hspace{0.04\textwidth} (a) \hspace{0.4\textwidth} (b) \hspace{0.25\textwidth}
 \begin{center}
  \includegraphics[width=0.45\textwidth]{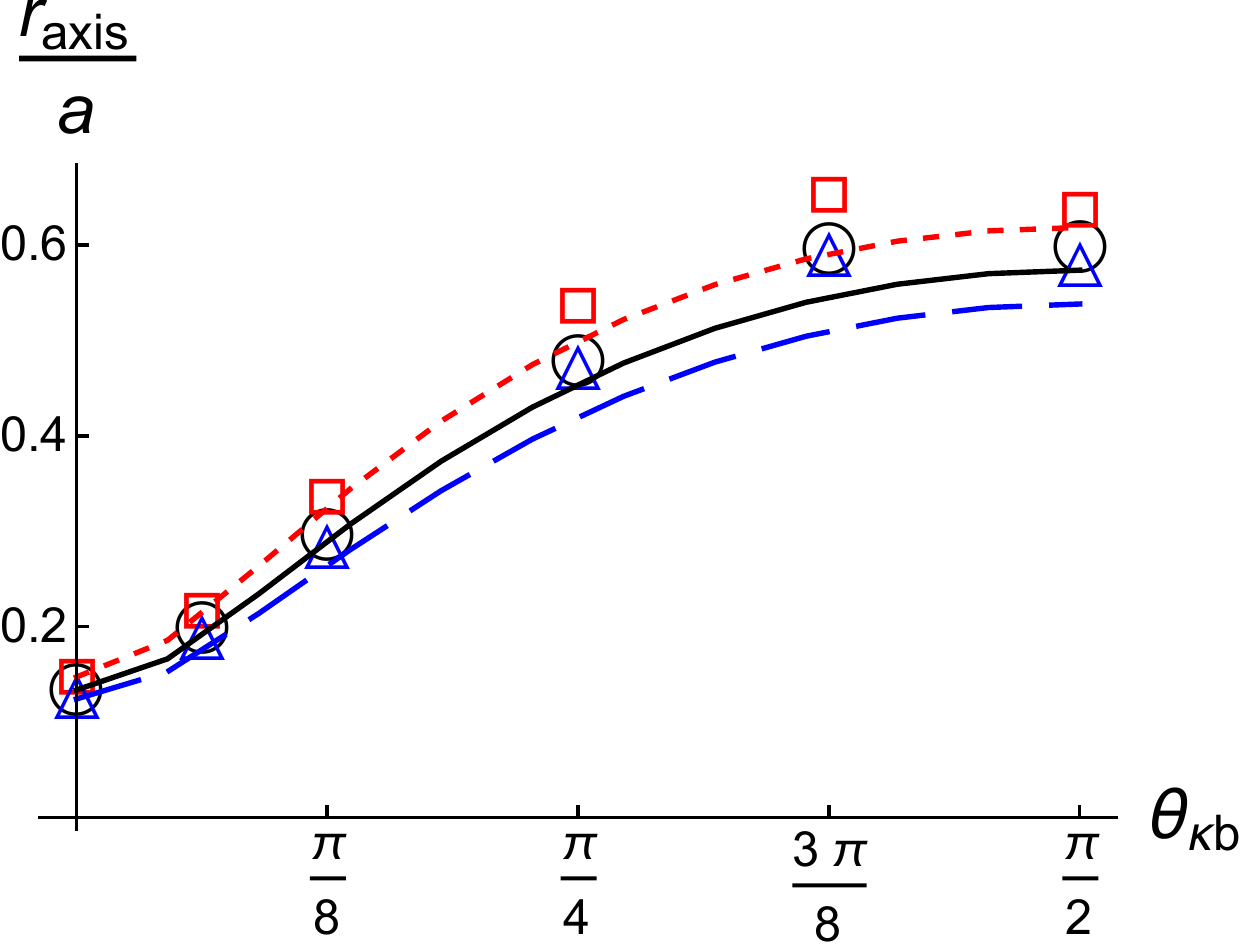}
  \includegraphics[width=0.45\textwidth]{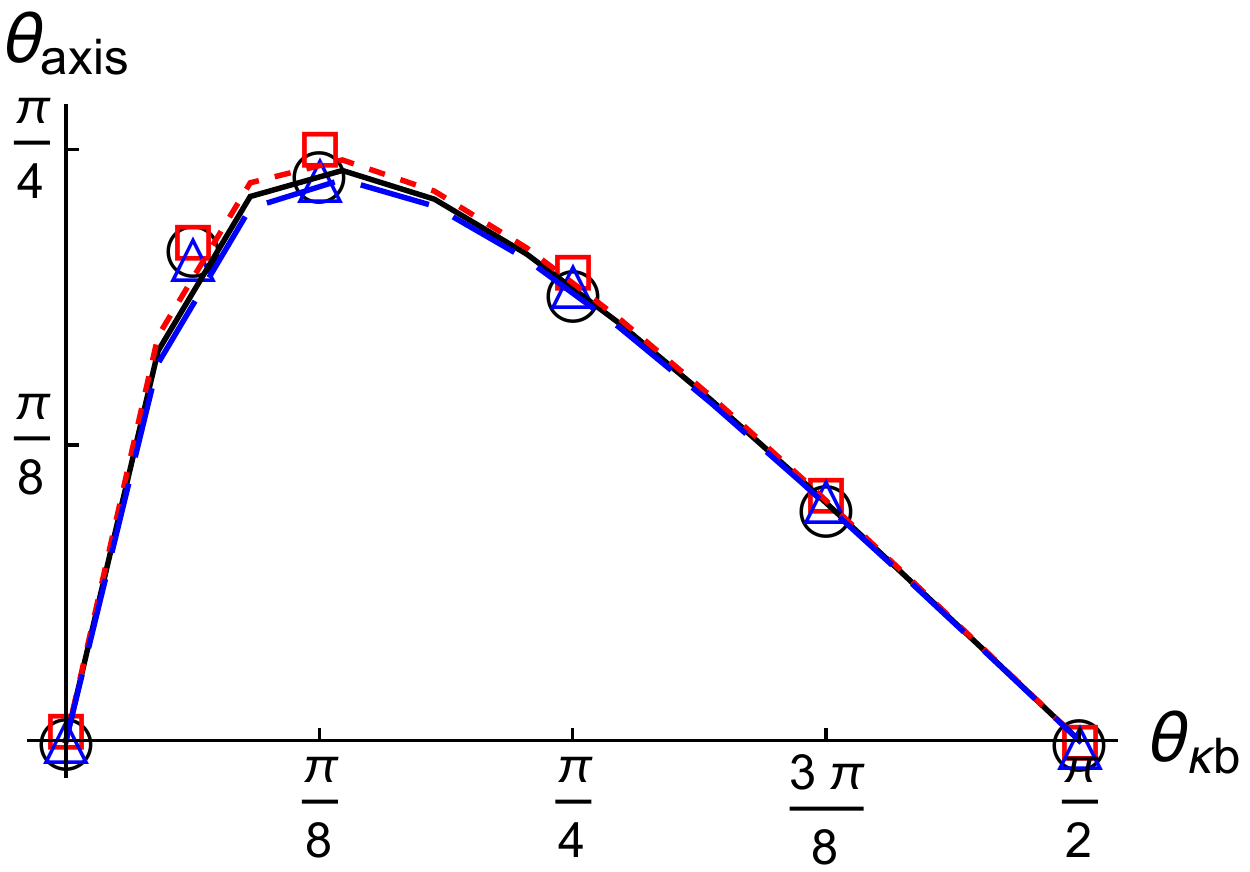}
 \end{center}
 \caption{The (a) minor radial and (b) poloidal location of the magnetic axis for constant ($\hat{f}_{0 p} \psi_{0 b} = 0$) (black, solid, circles), linear peaked ($\hat{f}_{0 p} \psi_{0 b} = 0.4$) (red, dotted, squares), and linear hollow ($\hat{f}_{0 p} \psi_{0 b} = - 0.4$) (blue, dashed, triangles) pressure gradient profiles, calculated analytically (lines) and by ECOM (points) for a constant current profile.}
 \label{fig:ShafShiftLocPressureProf}
\end{figure}

In figure \ref{fig:ShafShiftLocPressureProf} we show the location of the magnetic axis as we vary the shape of the pressure profile (by changing $\hat{f}_{0 p}$) with a constant current profile (i.e. $\hat{f}_{0} = 0$), while holding the geometry, $I_{p}$, and $p_{\text{axis}} / \psi_{0 b}$ fixed. We see good quantitative agreement between the results of appendix \ref{app:exactMagAxisLoc} and ECOM. Additionally, it appears that varying the profile of the pressure gradient while holding its average fixed has little effect on the Shafranov shift. We note that, in general, varying the pressure gradient has a large effect on the magnitude of the Shafranov shift, but not when $I_{p}$ and $p_{\text{axis}} / \psi_{0 b}$ are held constant. This is important as it justifies using our MHD results for the Shafranov shift with a constant $d p / d \psi$ profile as input for gyrokinetic simulations that are based on ITER, which we will take to have a constant $d p / d r_{\psi}$ profile \cite{AymarITERSummary2001}. Even though this is formally inconsistent, our analysis suggests the Shafranov shift in a configuration with constant $d p / d \psi$ will be a reasonable estimate of the Shafranov shift in a configuration with constant $d p / d r_{\psi}$ (as long as the geometry, $I_{p}$, and $p_{\text{axis}} / \psi_{0 b}$ are the same).

%% file: Ch_05_MHD_LocalEquil.tex
% !TEX root = /Users/Justin/Documents/Research/Writings/2016DoctoralThesis/DoctoralThesis.tex

\chapter{Derivation of local Miller equilibria}
\label{ch:MHD_LocalEquil}

\begin{quote}
   \emph{Much of this chapter appears in references \cite{BallShafranovShift2016} and \cite{BallMomFluxScaling2016}.}
\end{quote}

In this chapter we will take the simple, but physical large aspect ratio global MHD equilibria found in chapters \ref{ch:MHD_GradShafSol} and \ref{ch:MHD_ShafranovShift} and derive the corresponding Miller local equilibria \cite{MillerGeometry1998} for use in gyrokinetic simulations. The Miller local equilibrium model includes the shape of the flux surface of interest as an input and also requires the radial derivative of the flux surface shape in order to calculate the local poloidal field. We will derive two distinct local equilibrium specifications from the lowest order constant current global equilibrium. The first, the ``Expanded'' specification, is a simple Fourier series useful for analytic calculations. The second, the ``Exact'' specification, is more appropriate for creating realistic flux surface shapes for use in gyrokinetic simulations. These specifications can produce arbitrary flux surface shaping, which is specified by an infinite series of modes with independent tilt angles $\theta_{t m}$. Then, we will calculate the local Shafranov shift from the constant current global equilibrium assuming a constant $d p / d \psi$ profile and a tilted elliptical boundary flux surface.

%===================================================%
%===================================================%
\section{Specification of arbitrary flux surface shaping}
\label{sec:lowestOrderMillerEquil}
%===================================================%
%===================================================%

To calculate the lowest order Miller local equilibrium, we will start with
\begin{align}
   \psi_{0} \left( r, \theta \right) &= \frac{\hat{j}_{0}}{4} r^{2} + N_{0, m} r^{m} \Cos{m \left( \theta + \theta_{t 0, m} \right)} , \label{eq:gradShafConstLowestOrderSolsTiltSingleMode}
\end{align}
which is just the constant current global equilibrium (i.e. \refEq{eq:gradShafLowestOrderSolsTiltConst}) when considering only one shaping mode $m$. We will define a new parameter
\begin{align}
\Delta_{m} \left( a_{\psi, m} \right) \equiv \frac{b_{\psi, m} \left( a_{\psi, m} \right)}{a_{\psi, m}} , \label{eq:deltamDef}
\end{align}
which is similar to \refEq{eq:deltaDef}, but quantifies the magnitude of flux surface shaping from each poloidal shaping effect $m$ in isolation. Here $a_{\psi, m}$ is the minimum distance of the flux surface from the magnetic axis if all other shaping modes are ignored in \refEq{eq:gradShafLowestOrderSols}. Similarly, $b_{\psi, m}$ is the maximum distance of the flux surface from the magnetic axis if all other shaping modes are ignored. For circular flux surfaces without a Shafranov shift $\Delta_{m} = 1$ for all $m$. Since the definitions of $a_{\psi, m}$ and $b_{\psi, m}$ are based on the magnetic axis, $\Delta_{1} \neq 1$ for circular flux surfaces with a Shafranov shift. We note that $\Delta_{2}$ is the typical definition of the elongation usually denoted by $\kappa$. The parameter $\Delta_{m}$ can be related to the Fourier coefficients used in chapter \ref{ch:MHD_GradShafSol} by substituting \refEq{eq:gradShafConstLowestOrderSolsTiltSingleMode} into $\psi_{0} \left( a_{\psi, m}, - \theta_{t 0, m} \right) = \psi_{0} \left( \Delta_{m} a_{\psi, m}, \left( \pi / m \right) - \theta_{t 0, m} \right)$. For a constant current profile this gives the relation
\begin{align}
N_{0, m} = \frac{\hat{j}_{0}}{4} \frac{\Delta_{m}^{2} - 1}{\Delta_{m}^{m} + 1} a_{\psi, m}^{2-m} , \label{eq:deltaToFourierSingleMode}
\end{align}
which is analogous to \refEq{eq:deltaToFourier}. On a given flux surface, we can use \refEq{eq:gradShafranovSolPsi_a}, \refEq{eq:gradShafConstLowestOrderSolsTiltSingleMode}, and \refEq{eq:deltaToFourierSingleMode} to find
\begin{align}
	\left( \frac{r}{a_{\psi, m}} \right)^{2} + \frac{\Delta_{m}^{2} - 1}{\Delta_{m}^{m} + 1} \left( \frac{r}{a_{\psi, m}} \right)^{m} \Cos{m \left( \theta + \theta_{t m} \right)} = \frac{\Delta_{m}^{m} + \Delta_{m}^{2}}{\Delta_{m}^{m} + 1} , \label{eq:gradShafranovSolr}
\end{align}
where we have let $\theta_{t 0, m} = \theta_{t m}$ for notational simplicity. We would like to exactly solve this equation for $r$ to get a polar expression for each flux surface shape, but it is not analytic in general. Our method of dealing with this will distinguish two different geometry specifications.

%===================================================%
%===================================================%
\subsection{Expanded flux surface specification}
\label{subsec:expandedSpec}
%===================================================%
%===================================================%

To derive the ``Expanded'' flux surface specification we will expand \refEq{eq:gradShafranovSolr} in $\Delta_{m} - 1 \ll 1$ (i.e. weak shaping) and assume the flux surface is circular to lowest order to find the solution of
\begin{align}
	r \left( a_{\psi, m}, \theta \right) &= a_{\psi, m} \left( 1 + \frac{\Delta_{m} - 1}{2} \left( 1 - \Cos{m \left( \theta + \theta_{t m} \right)} \right) \right) . \label{eq:radialExpandedSingleModeFluxSurfGeo}
\end{align}
However, in this work we will want to study geometries with shaping from more than one mode number. In order to parameterize these configurations we will simply superimpose the different effects, in keeping with \refEq{eq:radialExpandedSingleModeFluxSurfGeo}, as
\begin{align}
	r \left( r_{\psi}, \theta \right) &= r_{\psi} \left( 1 - \sum_{m} \frac{\Delta_{m} - 1}{2} \Cos{m \left( \theta + \theta_{t m} \right)} \right) , \label{eq:radialExpandedFluxSurfGeo}
\end{align}
where we have defined a new flux surface label
\begin{align}
	r_{\psi} &\equiv a_{\psi} \left( 1 + \sum_{m} \frac{\Delta_{m} - 1}{2} \right) \label{eq:rPsiDef}
\end{align}
to make the specification as simple as possible. Then, by choosing $r_{\psi} = r_{\psi 0}$ we pick a particular flux surface of interest with the shape
\begin{align}
	r_{0} \left( \theta \right) &= r_{\psi 0} \left( 1 - \sum_{m} \frac{\Delta_{m} - 1}{2} \Cos{m \left( \theta + \theta_{t m} \right)} \right) \label{eq:fluxSurfofInterestShapeWeak}
\end{align}
to be the centre of the local equilibrium. Note that we are free to prescribe this shape however we wish as external coils can be used to arbitrarily shape any single flux surface in the global MHD equilibrium.

In fact, the change in the flux surface shape with minor radius is the important quantity determined by the global equilibrium. To calculate it we will directly differentiate \refEq{eq:radialExpandedFluxSurfGeo} to find
\begin{align}
	\left. \frac{\partial r}{\partial r_{\psi}} \right|_{r_{\psi 0}, \theta} &= 1 - \sum_{m} \bigg[ \left( \frac{\Delta_{m} - 1}{2} + \frac{r_{\psi 0}}{2} \frac{d \Delta_{m}}{d r_{\psi}} \right) \Cos{m \left( \theta + \theta_{t m} \right)} \label{eq:fluxSurfofInterestDerivWeak} \\
	&- m \left( \Delta_{m} - 1 \right) \frac{r_{\psi 0}}{2} \frac{d \theta_{t m}}{d r_{\psi}} \Sin{m \left( \theta + \theta_{t m} \right)} \bigg] \nonumber
\end{align}
to lowest order in $\Delta_{m} - 1 \ll 1$, where all quantities are evaluated on the flux surface of interest. The values of $d \Delta_{m} / d r_{\psi}$ and $d \theta_{t m} / d r_{\psi}$ are unknown, but can be calculated from the constant current global equilibrium.

To estimate $d \Delta_{m} / d r_{\psi}$ from the global equilibrium, we will also expand \refEq{eq:deltaToFourierSingleMode} to lowest order in the weak shaping limit to get
\begin{align}
	N_{0, m} = \frac{\hat{j}_{0}}{4} \left( \Delta_{m} - 1 \right) r_{\psi}^{2 - m} . \label{eq:T0mAllModes}
\end{align}
Remembering that $N_{0, m}$ and $\hat{j}_{0}$ are constants of the equilibrium, we can differentiate this implicitly to find
\begin{align}
	\frac{d \Delta_{m}}{d r_{\psi}} &= \left( m - 2 \right) \frac{\Delta_{m} - 1}{r_{\psi 0}} \label{eq:shapingRadialDeriv_rPsi}
\end{align}
to lowest order in $\Delta_{m} - 1 \ll 1$ on the flux surface of interest. Lastly, since $\theta_{t m} = \theta_{t 0, m}$ is a constant defined by \refEq{eq:tiltAngleDef} we know that
\begin{align}
	\frac{d \theta_{t m}}{d r_{\psi}} = 0 . \label{eq:tiltAngleDeriv}
\end{align}

%===================================================%
%===================================================%
\subsection{Exact flux surface specification}
\label{subsec:exactSpec}
%===================================================%
%===================================================%

While the Expanded flux surface specification is simple, the $m = 2$ mode does not exactly correspond to elongation and single-mode flux surfaces become unrealistic at large shaping. This motivates the ``Exact'' flux surface specification, which is found by exactly solving \refEq{eq:gradShafranovSolr} for the $m=2$ case to get
\begin{align}
   r \left( a_{\psi, 2}, \theta \right) = \frac{a_{\psi, 2} \Delta_{2}}{\sqrt{1 + \left( \Delta_{2}^{2} - 1 \right) \cos^{2} \left( \theta + \theta_{t 2} \right)}} . \label{eq:elongExactShape}
\end{align}
This is the polar equation for an ellipse and matches \refEq{eq:boundarySurf}. From this we will extrapolate a simple generalisation for arbitrary $m$ of
\begin{align}
   r \left( a_{\psi, m}, \theta \right) = \frac{a_{\psi, m} \Delta_{m}}{\sqrt{1 + \left( \Delta_{m}^{2} - 1 \right) \cos^{2} \left( m \left( \theta + \theta_{t m} \right) / 2 \right)}} . \label{eq:exactFluxSurfGeo}
\end{align}
This particular generalisation is acceptable because it is consistent with \refEq{eq:gradShafranovSolr} and \refEq{eq:radialExpandedSingleModeFluxSurfGeo} to $\Order{\Delta_{m} - 1}$ in the weak shaping limit  (i.e. the error is $\Order{\left( \Delta_{m} - 1 \right)^{2}}$) and satisfies
\begin{align}
	r \left( a_{\psi, m}, - \theta_{t m} \right) &= a_{\psi, m} \\
	\frac{r \left( a_{\psi, m}, \pi/m - \theta_{t m} \right)}{r \left( a_{\psi, m}, - \theta_{t m} \right)} &= \Delta_{m} .
\end{align}

Repeating the method used for the Expanded specification, we will superimpose the different shaping effects from \refEq{eq:exactFluxSurfGeo} to find
\begin{align}
r \left( a_{\psi}, \theta \right) &= a_{\psi} \left[ 1 + \sum_{m} \left( \frac{\Delta_{m}}{\sqrt{1 + \left( \Delta_{m}^{2} - 1 \right) \cos^{2} \left( m \left( \theta + \theta_{t m} \right) / 2 \right)}} - 1 \right) \right] . \label{eq:fluxSurfShapeExact}
\end{align}
The plus and minus $1$ terms were added to ensure that $r \left( a_{\psi}, \theta \right) = a_{\psi}$ when $\Delta_{m} = 1$ for all $m$. Note that strictly speaking in writing \refEq{eq:fluxSurfShapeExact} (and \refEq{eq:rPsiDef}) we have somewhat modified our definition of the flux surface label $a_{\psi}$. In section \ref{sec:calculation} we had defined it as the minimum radial location on a given flux surface, but from \refEq{eq:fluxSurfShapeExact} we see that this is not true for geometries with multiple shaping modes that are not aligned. Instead it is defined by \refEq{eq:fluxSurfShapeExact} from the definitions of $r$ and $\theta$. Evaluating \refEq{eq:fluxSurfShapeExact} at $a_{\psi} = a_{\psi 0}$ we find
\begin{align}
	r_{0} \left( \theta \right) &= a_{\psi 0} \left[ 1 + \sum_{m} \left( \frac{\Delta_{m}}{\sqrt{1 + \left( \Delta_{m}^{2} - 1 \right) \cos^{2} \left( m \left( \theta + \theta_{t m} \right) / 2 \right)}} - 1 \right) \right] , \label{eq:fluxSurfofInterestShapeExact}
\end{align}
the shape of the flux surface of interest.

Directly differentiating \refEq{eq:fluxSurfShapeExact} gives
\begin{align}
   \left. \frac{\partial r}{\partial a_{\psi}} \right|_{a_{\psi 0}, \theta} &= 1 + \sum_{m} \left[ - 1 + \frac{\Delta_{m}}{\sqrt{1 + \left( \Delta_{m}^{2} - 1 \right) \cos^{2} \left( m \left( \theta + \theta_{t m} \right) / 2 \right)}} \right. \nonumber \\
   &\times \left( 1 + \frac{a_{\psi 0}}{\Delta_{m}} \frac{d \Delta_{m}}{d a_{\psi}} - \frac{a_{\psi 0} \cos^{2} \left( m \left( \theta + \theta_{t m} \right) / 2 \right)}{1 + \left( \Delta_{m}^{2} - 1 \right) \cos^{2} \left( m \left( \theta + \theta_{t m} \right) / 2 \right)} \right. \label{eq:fluxSurfofInterestDerivExact} \\
   &\times \left. \left. \left( \Delta_{m} \frac{d \Delta_{m}}{d a_{\psi}} - m \frac{\Delta_{m}^{2} - 1}{2} \frac{d \theta_{t m}}{d a_{\psi}} \Tan{m \left( \theta + \theta_{t m} \right) / 2} \right) \right) \right] , \nonumber
\end{align}
where we have evaluated all quantities on the flux surface of interest. We can estimate that $d \Delta_{m} / d a_{\psi} = \left( m - 2 \right) \left( \Delta_{m} - 1 \right) / a_{\psi 0}$ and $d \theta_{t m} / d a_{\psi} = 0$ to lowest order in weak shaping from \refEq{eq:rPsiDef}, \refEq{eq:shapingRadialDeriv_rPsi}, \refEq{eq:tiltAngleDeriv}, and the chain rule.

%===================================================%
%===================================================%
\section{Shafranov shift in tilted elliptical tokamaks}
\label{sec:nextOrderMillerEquil}
%===================================================%
%===================================================%

\begin{figure}
 \centering
 \includegraphics[width=0.45\textwidth]{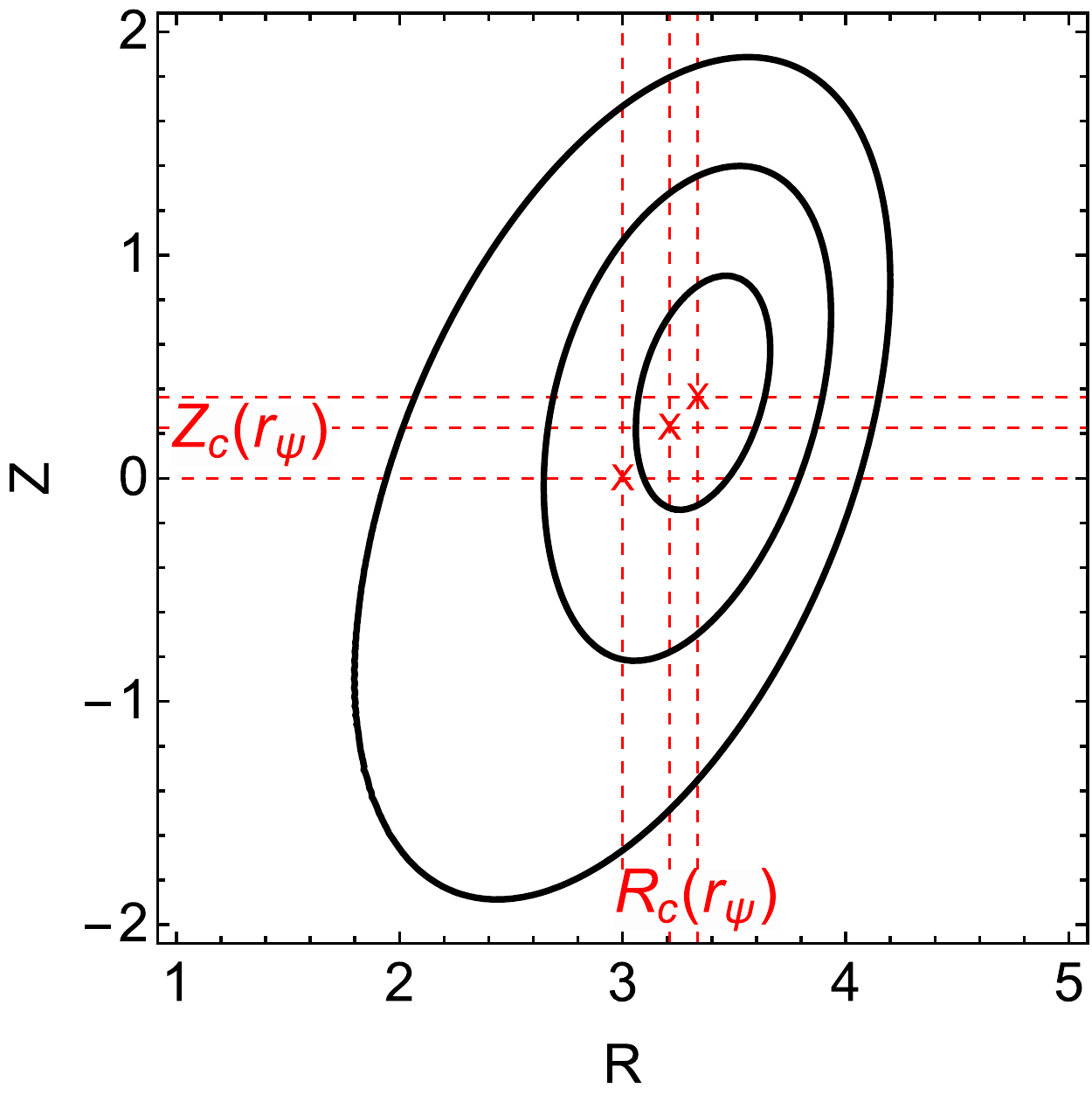}
 \caption{Example flux surfaces showing the geometric meaning of the parameters $R_{c} \left( a_{\psi} \right)$ and $Z_{c} \left( a_{\psi} \right)$, the major radial and axial locations of the centre of each flux surface respectively.}
 \label{fig:geoFluxSurfCenter}
\end{figure}

\begin{figure}
 \centering
 \includegraphics[width=0.5\textwidth]{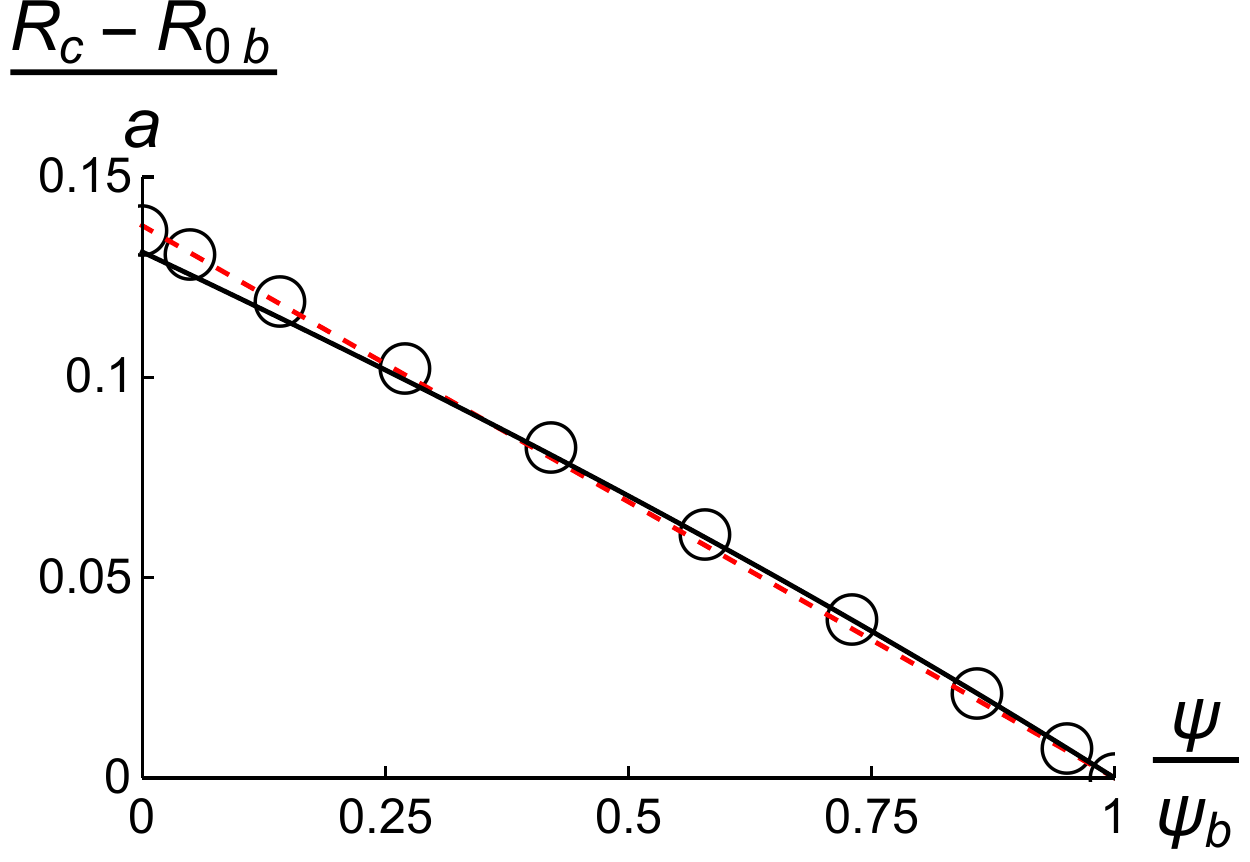}
 \caption{The shift in the centre of flux surfaces (relative to the centre of the boundary flux surface $R_{0 b}$) as a function of normalised poloidal flux for a constant current profile and $\theta_{\kappa b} = 0$, according to ECOM (black, circles) and our analytic calculation (black, solid) with a linear best fit (red, dotted).}
 \label{fig:shiftWithRadius}
\end{figure}

The Miller geometry specification captures the Shafranov shift through local values of $d R_{c} / d a_{\psi}$ and $d Z_{c} / d a_{\psi}$, where $R_{c} \left( a_{\psi} \right)$ and $Z_{c} \left( a_{\psi} \right)$ indicate the location of the centre of each flux surface as shown in figure \ref{fig:geoFluxSurfCenter}. In order to model a realistic geometry, we will include its effect in the Exact specification by calculating local values of $d R_{c} / d a_{\psi}$ and $d Z_{c} / d a_{\psi}$ for arbitrary tilt angle from our global MHD results. Specifically, we will use the dependence of the global Shafranov shift on tilt angle calculated for constant current and $d p / d \psi$ profiles (i.e. the solid black line shown in figure \ref{fig:ShafShiftLoc}).

First we will assume that $d R_{c} / d \psi$ and $d Z_{c} / d \psi$ are constant from the boundary flux surface to the magnetic axis. In figure \ref{fig:shiftWithRadius}, we plot our analytic solution (using the coefficients calculated in appendix \ref{app:exactMagAxisLoc}) and ECOM results to show that this assumption is satisfied for the case of a vertically-elongated boundary. Additionally, using \refEq{eq:gradShafLowestOrderSolsTiltConst} and \refEq{eq:fourierShapingConst} we see that
\begin{align}
   \psi \propto a_{\psi}^{2}
\end{align}
for a constant current profile and an exactly elliptical boundary. Therefore, using that $\psi = \psi_{b}$ at $a_{\psi} = a$, one can calculate the constant of proportionality to show that
\begin{align}
   \frac{d \psi}{d a_{\psi}} = 2 \frac{\psi_{b}}{a} \rho ,
\end{align}
where $\rho \equiv a_{\psi} / a$ is the usual normalised minor radial flux surface label. Hence, the local Shafranov shift can be written as
\begin{align}
   \left. \frac{d R_{c}}{d a_{\psi}} \right|_{a_{\psi 0}} &= \left. \frac{d \psi}{d a_{\psi}} \right|_{a_{\psi 0}} \frac{d R_{c}}{d \psi} = \left( 2 \frac{\psi_{b}}{a} \rho_{0} \right) \frac{R_{0 b} - R_{0}}{\psi_{b} - 0} = - 2 \rho_{0} \frac{r_{\text{axis}}}{a} \Cos{\theta_{\text{axis}}} \label{eq:radialShafShiftLocal} \\
   \left. \frac{d Z_{c}}{d a_{\psi}} \right|_{a_{\psi 0}} &= \left. \frac{d \psi}{d a_{\psi}} \right|_{a_{\psi 0}} \frac{d Z_{c}}{d \psi} = \left( 2 \frac{\psi_{b}}{a} \rho_{0} \right) \frac{Z_{0 b} - Z_{0}}{\psi_{b} - 0} = - 2 \rho_{0} \frac{r_{\text{axis}}}{a} \Sin{\theta_{\text{axis}}} ,  \label{eq:axialShafShiftLocal}
\end{align}
where $\rho_{0} \equiv a_{\psi 0} / a$, $a_{\psi 0}$ is the value of $a_{\psi}$ on the flux surface of interest and the coordinate system is defined such that the boundary flux surface is centred at $\left( R = R_{0 b}, Z = Z_{0 b} \right)$.

%===================================================%
%===================================================%
\section{Summary}
\label{sec:summaryMillerEquil}
%===================================================%
%===================================================%

In summary, we have defined two different sets of expressions for the flux surface shape and its derivative. The first, which we call the ``Expanded'' parameterization, is a simple Fourier parameterization given by \refEq{eq:fluxSurfofInterestShapeWeak}, \refEq{eq:fluxSurfofInterestDerivWeak}, and
\begin{align}
	r \left( r_{\psi}, \theta \right) &= r_{0} \left( \theta \right) + \left. \frac{\partial r}{\partial r_{\psi}} \right|_{r_{\psi 0}, \theta} \left( r_{\psi} - r_{\psi 0} \right) \label{eq:gradShafranovLocalEqWeak} \\
	R \left( r_{\psi}, \theta \right) &= R_{c 0} + r \left( r_{\psi}, \theta \right) \Cos{\theta} \label{eq:geoMajorRadiusWeak} \\
	Z \left( r_{\psi}, \theta \right) &= Z_{c 0} + r \left( r_{\psi}, \theta \right) \Sin{\theta} , \label{eq:geoAxialWeak}
\end{align}
where the flux surface of interest is centred at $\left( R = R_{c 0}, Z = Z_{c 0} \right)$. This shaping parameterization will be useful for theoretical scaling calculations (see chapters \ref{ch:GYRO_TiltingSymmetry} and \ref{ch:GYRO_MomFluxScaling}).

The second set, which we call the ``Exact'' parameterization, is consistent with the Expanded parameterization to next order in the weak shaping expansion. It is given by \refEq{eq:fluxSurfofInterestShapeExact}, \refEq{eq:fluxSurfofInterestDerivExact}, and
\begin{align}
	r \left( a_{\psi}, \theta \right) &= r_{0} \left( \theta \right) + \left. \frac{\partial r}{\partial a_{\psi}} \right|_{a_{\psi 0}, \theta} \left( a_{\psi} - a_{\psi 0} \right) \label{eq:gradShafranovLocalEqExact} \\
	R_{c} \left( a_{\psi} \right) &= R_{c 0} + \left. \frac{d R_{c}}{d a_{\psi}} \right|_{a_{\psi 0}} \left( a_{\psi} - a_{\psi 0} \right) \\
	Z_{c} \left( a_{\psi} \right) &= Z_{c 0} + \left. \frac{d Z_{c}}{d a_{\psi}} \right|_{a_{\psi 0}} \left( a_{\psi} - a_{\psi 0} \right) \\
	R \left( a_{\psi}, \theta \right) &= R_{c} \left( a_{\psi} \right) + r \left( a_{\psi}, \theta \right) \Cos{\theta} \label{eq:geoMajorRadiusExact} \\
	Z \left( a_{\psi}, \theta \right) &= Z_{c} \left( a_{\psi} \right) + r \left( a_{\psi}, \theta \right) \Sin{\theta} . \label{eq:geoAxialExact}
\end{align}
In this parameterization the $m=2$ mode exactly corresponds to elliptical flux surfaces, which will be useful for realistic numerical simulations (see chapters \ref{ch:GYRO_ShafranovShift} and \ref{ch:GYRO_NonMirrorSym}). We note that to treat the Shafranov shift we calculate $\left. d R_{c} / d a_{\psi} \right|_{a_{\psi 0}}$ and $\left. d Z_{c} / d a_{\psi} \right|_{a_{\psi 0}}$ for an ITER-like pressure profile using \refEq{eq:radialShafShiftLocal} and \refEq{eq:axialShafShiftLocal} as well as the constant current results shown in figure \ref{fig:ShafShiftLoc} (and given in appendix \ref{app:exactMagAxisLoc}).

Both of these prescriptions require either $d \Delta_{m} / d a_{\psi}$ or $d \Delta_{m} / d r_{\psi}$, which were estimated from the global equilibrium.

%% file: Ch_06_GYRO_Overview.tex
% !TEX root = /Users/Justin/Documents/Research/Writings/2016DoctoralThesis/DoctoralThesis.tex

\chapter{Overview of gyrokinetics}
\label{ch:GYRO_Overview}

\begin{quote}
   \emph{Much of this chapter appears in reference \cite{BallMirrorSymArg2016}.}
\end{quote}

Gyrokinetics has many variations \cite{LeeGenFreqGyro1983, LeeParticleSimGyro1983, DubinHamiltonianGyro1983, HahmGyrokinetics1988, SugamaGyroTransport1996, SugamaHighFlowGyro1998, BrizardGyroFoundations2007, ParraGyrokineticLimitations2008, ParraLagrangianGyro2011, AbelGyrokineticsDeriv2012}. It is based on the expansion of the Fokker-Planck and Maxwell's equations in $\rho_{\ast} \equiv \rho_{i} / a \ll 1$, where $\rho_{i}$ is the ion gyroradius and $a$ is the tokamak minor radius. This model investigates plasma behaviour with timescales much slower than the ion gyrofrequency $\Omega_{i}$ and the electron gyrofrequency $\Omega_{e}$ (i.e. $\omega \ll \Omega_{i} \ll \Omega_{e}$), but retains the finite size of the gyroradius by assuming that the perpendicular wavenumber of the turbulence is comparable to the ion gyroradius (i.e. $k_{\perp} \rho_{i} \sim 1$ where $k_{\perp}$ is the characteristic wavenumber of the turbulence perpendicular to the magnetic field). In this limit, the six dimensions of velocity space reduce to five because the particle gyrophase can be ignored. As such, gyrokinetics evolves rings of charge as they generate and respond to electric and magnetic fields. In this thesis we will use $\delta f$ gyrokinetics, which assumes that the turbulence arises from perturbations to the distribution function that are small compared to the background (i.e. $f_{s 1} \ll f_{s 0}$, where $f_{s 0}$ is the background distribution function for species $s$ and $f_{s 1}$ is the lowest order perturbation). These particular choices have been shown experimentally to be appropriate for modelling core turbulence \cite{McKeeTurbulenceScale2001}. Furthermore, we will assume the plasma is sufficiently collisional so that the background distribution function is Maxwellian, 
\begin{align}
   f_{s 0} = F_{M s} \equiv n_{s} \left( \frac{m_{s}}{2 \pi T_{s}} \right)^{3/2} \text{exp} \left( - \frac{m_{s} w^{2}}{2 T_{s}} \right) . \label{eq:maxwellianDef}
\end{align}
Here $n_{s}$ is the density of species $s$, $m_{s}$ is the particle mass, $T_{s}$ is the temperature, $\vec{w} \equiv \vec{v} - R \Omega_{\zeta} \hat{e}_{\zeta}$ is the velocity shifted into the rotating frame, and $\Omega_{\zeta} \left( \psi \right) = V_{\zeta} / R$ is the toroidal rotation frequency. We note that $\Omega_{\zeta}$ is a flux function and $R \Omega_{\zeta} \sim v_{th i}$ in the high flow regime, where $v_{th i} \equiv \sqrt{2 T_{i} / m_{i}}$ is the ion thermal speed. To lowest order in $\rho_{\ast} \ll 1$, it can be shown that all species rotate at $\Omega_{\zeta} = - d \Phi_{-1} / d \psi$, where $\Phi_{-1} \sim \rho_{\ast}^{-1} T_{e} / e$ is the lowest order electrostatic potential and a flux function \cite{HintonNeoclassicalRotation1985, ConnorMomTransport1987, CattoRotation1987} while $e$ is the proton electric charge. Though $T_{s}$ and $\Omega_{\zeta}$ are flux functions, the centrifugal force can cause the density to vary on a flux surface according to \cite{MaschkeEquilWithRotation1980}
\begin{align}
   n_{s} \left( \psi, \theta \right) = \eta_{s} \left( \psi \right) \Exp{ \frac{m_{s} R^{2} \Omega_{\zeta}^{2}}{2 T_{s}} - \frac{Z_{s} e \Phi_{0}}{T_{s}}} , \label{eq:pseudoDensity}
\end{align}
where $\eta_{s} \left( \psi \right)$ is the pseudo-density flux function, $Z_{s}$ is the electric charge number, and $\Phi_{0}$ is the next order electrostatic potential. We can find $\Phi_{0}$ by imposing quasineutrality,
\begin{align}
   \sum_{s} Z_{s} e n_{s} = \sum_{s} Z_{s} e \eta_{s} \left( \psi \right) \Exp{ \frac{m_{s} R^{2} \Omega_{\zeta}^{2}}{2 T_{s}} - \frac{Z_{s} e \Phi_{0}}{T_{s}}} = 0 . \label{eq:backgroundQuasi}
\end{align}

From the assumption that $k_{\perp} \rho_{i} \sim 1$ (remembering our expansion in $\rho_{i} / a \ll 1$), we know that the background plasma quantities vary little on the scale of the turbulence in the directions perpendicular to the background magnetic field. Neglecting this small variation is called the local approximation and it motivates periodic boundary conditions in the perpendicular directions. Ballooning coordinates \cite{BeerBallooingCoordinates1995} are generally used in local gyrokinetics to model turbulence in a flux tube, a long narrow domain that follows a single field line. These boundary conditions allow us to Fourier analyse in the poloidal flux $\psi$ (which parameterizes the radial direction) and in
\begin{align}
   \alpha \equiv \zeta - I \left( \psi \right) \left. \int_{\theta_{\alpha} \left( \psi \right)}^{\theta} \right|_{\psi} d \theta' \left( R^{2} \vec{B} \cdot \vec{\nabla} \theta' \right)^{-1} - \Omega_{\zeta} t \label{eq:alphaDef}
\end{align}
(which parameterizes the direction perpendicular to the field lines, but within the flux surface). Note the free parameter $\theta_{\alpha} \left( \psi \right)$, which determines the field line selected by $\alpha = 0$ on each flux surface and will be important in chapter \ref{ch:GYRO_TiltingSymmetry}.

The high-flow, Fourier analysed gyrokinetic equation can be written as \cite{ParraUpDownSym2011}
\begin{align}
   \frac{\partial h_{s}}{\partial t} &+ w_{||} \hat{b} \cdot \vec{\nabla} \theta \left. \frac{\partial h_{s}}{\partial \theta} \right|_{w_{||}, \mu} + i \left( k_{\psi} v_{d s \psi} + k_{\alpha} v_{d s \alpha} \right) h_{s} + a_{s ||} \left. \frac{\partial h_{s}}{\partial w_{||}} \right|_{\theta, \mu} - \sum_{s'} \langle C_{ss'}^{\left( l \right)} \rangle_{\varphi} \nonumber \\
&+ \left\{ \langle \chi \rangle_{\varphi}, h_{s} \right\} = \frac{Z_{s} e F_{M s}}{T_{s}} \frac{\partial \langle \chi \rangle_{\varphi}}{\partial t} - v_{\chi s \psi} F_{M s} \left[ \frac{1}{n_{s}} \left. \frac{\partial n_{s}}{\partial \psi} \right|_{\theta} \right. \label{eq:gyrokineticEq} \\
 &+ \left. \frac{m_{s} I w_{||}}{B T_{s}} \frac{d \Omega_{\zeta}}{d \psi} + \frac{Z_{s} e}{T_{s}} \left. \frac{\partial \Phi_{0}}{\partial \psi} \right|_{\theta} - \frac{m_{s} R \Omega_{\zeta}^{2}}{T_{s}} \left. \frac{\partial R}{\partial \psi} \right|_{\theta} + \left( \frac{m_{s} w^{2}}{2 T_{s}} - \frac{3}{2} \right) \frac{1}{T_{s}} \frac{d T_{s}}{d \psi} \right] , \nonumber
\end{align}
where the coordinates are $t$ (the time), $\theta$ (the poloidal angle), $k_{\psi}$ (the radial wavenumber), $k_{\alpha}$ (the poloidal wavenumber), $w_{||}$ (the parallel velocity in the rotating frame), $\mu \equiv m_{s} w_{\perp}^{2} / 2 B$ (the magnetic moment), and we have already eliminated $\varphi$ (the gyrophase) by gyroaveraging. The unknowns are
\begin{align}
   h_{s} \equiv \left\langle \left\langle \left( f_{s 1} + \frac{Z_{s} e \phi}{T_{s}} F_{M s} \right) \Exp{-i k_{\psi} \psi - i k_{\alpha} \alpha} \right\rangle_{\Delta \psi} \right\rangle_{\Delta \alpha} \label{eq:distFnFourierAnalysis}
\end{align}
(the Fourier-analysed nonadiabatic portion of the distribution function) and the fields contained in
\begin{align}
   \left\langle \chi \right\rangle_{\varphi} \equiv J_{0} \left( k_{\perp} \rho_{s} \right) \left( \phi - w_{||} A_{||} \right) + \frac{1}{\Omega_{s}} \frac{\mu B}{m_{s}} \frac{2 J_{1} \left( k_{\perp} \rho_{s} \right)}{k_{\perp} \rho_{s}} B_{||}
\end{align}
(the Fourier analysed gyroaveraged generalised potential). We note that $\left\langle \ldots \right\rangle_{\Delta \psi} \equiv \Delta \psi^{-1} \int_{\Delta \psi} d \psi \left( \ldots \right)$ is a coarse-grain average over the radial distance $\Delta \psi$ (which is larger then the scale of the turbulence, but smaller than the scale of the device), $\left\langle \ldots \right\rangle_{\Delta \alpha} \equiv \Delta \alpha^{-1} \int_{\Delta \alpha} d \alpha \left( \ldots \right)$ is a coarse-grain average over the poloidal distance $\Delta \alpha$ (which is larger then the scale of the turbulence, but smaller than the scale of the device), $\left\langle \ldots \right\rangle_{\varphi}$ is the gyroaverage at fixed guiding centre, $J_{n} \left( \ldots \right)$ is the $n$th order Bessel function of the first kind, $\phi$ is the Fourier analysed perturbed electrostatic potential, $A_{||}$ is the Fourier analysed perturbed magnetic vector potential, $B_{||}$ is the component of the Fourier analysed perturbed magnetic field parallel to the background magnetic field,
\begin{align}
   k_{\perp} = \sqrt{k_{\psi}^{2} \left| \vec{\nabla} \psi \right|^{2} + 2 k_{\psi} k_{\alpha} \vec{\nabla} \psi \cdot \vec{\nabla} \alpha + k_{\alpha}^{2} \left| \vec{\nabla} \alpha \right|^{2}} \label{eq:kperpDef}
\end{align}
is the perpendicular wavevector, $\rho_{s} \equiv \sqrt{2 \mu B / m_{s}} / \Omega_{s}$ is the gyroradius, $\Omega_{s} \equiv Z_{s} e B / m_{s}$ is the gyrofrequency, and $Z_{s}$ is the species charge number.

The drift coefficients are given by
\begin{align}
   v_{d s \psi} &\equiv \vec{v}_{d s} \cdot \vec{\nabla} \psi \label{eq:radialGCdriftvelocity} \\
   &= \left( - \frac{I}{B} \frac{\partial \Phi_{0}}{\partial \theta} - \frac{I \left( m_{s} w_{||}^{2} + \mu B \right)}{m_{s} \Omega_{s} B} \frac{\partial B}{\partial \theta} + \frac{2 B R \Omega_{\zeta} w_{||}}{\Omega_{s}} \frac{\partial R}{\partial \theta} + \frac{I R \Omega_{\zeta}^{2}}{\Omega_{s}} \frac{\partial R}{\partial \theta} \right) \hat{b} \cdot \vec{\nabla} \theta \nonumber
\end{align}
and
\begin{align}
   v_{d s \alpha} &\equiv \vec{v}_{d s} \cdot \vec{\nabla} \alpha = - \frac{\partial \Phi_{0}}{\partial \psi} + \frac{\partial \Phi_{0}}{\partial \theta} \frac{\hat{b} \cdot \left( \vec{\nabla} \theta \times \vec{\nabla} \alpha \right)}{B} \nonumber \\
   &- \frac{m_{s} w_{||}^{2} + \mu B}{m_{s} \Omega_{s}} \left( \frac{\partial B}{\partial \psi} - \frac{\partial B}{\partial \theta} \frac{\hat{b} \cdot \left( \vec{\nabla} \theta \times \vec{\nabla} \alpha \right)}{B} \right) - \frac{\mu_{0} w_{||}^{2}}{B \Omega_{s}} \left. \frac{\partial p}{\partial \psi} \right|_{R} \label{eq:alphaGCdriftvelocity} \\
   &+ \frac{2 \Omega_{\zeta} w_{||}}{\Omega_{s}} \hat{e}_{\zeta} \cdot \left( \vec{\nabla} \alpha \times \vec{\nabla} R \right) + \frac{m_{s} R \Omega_{\zeta}^{2}}{Z_{s} e} \left( \frac{\partial R}{\partial \psi} - \frac{\partial R}{\partial \theta} \frac{\hat{b} \cdot \left( \vec{\nabla} \theta \times \vec{\nabla} \alpha \right)}{B} \right) , \nonumber
\end{align}
where $p \equiv \sum_{s} n_{s} T_{s}$ is the plasma pressure and
\begin{align}
   \left. \frac{\partial p}{\partial \psi} \right|_{R} = \left. \frac{\partial p}{\partial \psi} \right|_{\theta} - \sum_{s} n_{s} m_{s} R \Omega_{\zeta}^{2} \left. \frac{\partial R}{\partial \psi} \right|_{\theta} . \label{eq:pressureGradDeriv}
\end{align}
The parallel acceleration is given by
\begin{align}
   a_{s ||} = \left( - \frac{\mu}{m_{s}} \frac{\partial B}{\partial \theta} - \frac{Z_{s} e}{m_{s}} \frac{\partial \Phi_{0}}{\partial \theta} + R \Omega_{\zeta}^{2} \frac{\partial R}{\partial \theta} \right) \hat{b} \cdot \vec{\nabla} \theta , \label{eq:parallelAcceleration}
\end{align}
$C_{s s'}^{\left( l \right)}$ is the linearized collision operator, the nonlinear term is
\begin{align}
  \left\{ \langle \chi \rangle_{\varphi}, h_{s} \right\} = \sum_{k_{\psi}', k_{\alpha}'} \left( k_{\psi}' k_{\alpha} - k_{\psi} k_{\alpha}' \right) \left\langle \chi \right\rangle_{\varphi} \left( k_{\psi}', k_{\alpha}' \right) h_{s} \left( k_{\psi} - k_{\psi}', k_{\alpha} - k_{\alpha}' \right) , \label{eq:nonlinearTerm}
\end{align}
and
\begin{align}
   v_{\chi s \psi} \equiv i k_{\alpha} \left\langle \chi \right\rangle_{\varphi} .
\end{align}

In order to solve for $\phi$, $A_{||}$, and $B_{||}$ we also need the Fourier analysed quasineutrality equation \cite{ParraUpDownSym2011}
\begin{align}
      \phi = 2 \pi \left( \sum_{s} \frac{Z_{s}^{2} e^{2} n_{s}}{T_{s}} \right)^{-1} \sum_{s} \frac{Z_{s} e B}{m_{s}} \int dw_{||} \int d \mu J_{0} \left( k_{\perp} \rho_{s} \right) h_{s} , \label{eq:quasineut}
\end{align}
parallel current equation \cite{ParraUpDownSym2011}
\begin{align}
      A_{||} = \frac{2 \pi \mu_{0}}{k_{\perp}^{2}} \sum_{s} \frac{Z_{s} e B}{m_{s}} \int dw_{||}  \int d \mu J_{0} \left( k_{\perp} \rho_{s} \right) w_{||} h_{s} , \label{eq:parCur}
\end{align}
and perpendicular current equation \cite{ParraUpDownSym2011}
\begin{align}
      B_{||} = -2 \pi \mu_{0} \sum_{s} \frac{B}{m_{s}} \int dw_{||} \int d \mu \frac{2 J_{1} \left( k_{\perp} \rho_{s} \right)}{k_{\perp} \rho_{s}} \mu h_{s} . \label{eq:perpCur}
\end{align}
Equations \refEq{eq:gyrokineticEq}, \refEq{eq:quasineut}, \refEq{eq:parCur}, and \refEq{eq:perpCur} comprise the nonlinear electromagnetic gyrokinetic model, in the presence of rotation, which we will use in chapter \ref{ch:GYRO_TiltingSymmetry}. These equations simplify considerably when the plasma is assumed to be electrostatic (i.e. $A_{||} = B_{||} = 0$) and stationary (i.e. $\Omega_{\zeta} = 0$) as is done in chapters \ref{ch:GYRO_MomFluxScaling}, \ref{ch:GYRO_ShafranovShift}, and \ref{ch:GYRO_NonMirrorSym}.

Solving the gyrokinetic model for $h_{s}$, $\phi$, $A_{||}$, and $B_{||}$ allows us to calculate the turbulent radial fluxes of particles, momentum, and energy as well as the turbulent energy exchange between species. These are the only turbulent quantities needed to evolve the transport equations for particles, momentum, and energy \cite{ParraUpDownSym2011, SugamaHighFlowGyro1998, AbelGyrokineticsDeriv2012}. The full expressions are written in appendix \ref{app:fluxes}. Here we give only the electrostatic contribution to the particle flux
\begin{align}
   \Gamma_{s}^{\phi} &\equiv - \left\langle R \left\langle \left\langle \int d^{3} w \underline{h}_{s} \hat{e}_{\zeta} \cdot \delta \vec{E} \right\rangle_{\Delta \psi} \right\rangle_{\Delta t} \right\rangle_{\psi} \label{eq:partFluxDef} \\
  &= \frac{4 \pi^{2} i}{m_{s} V'} \left\langle \sum_{k_{\psi}, k_{\alpha}} k_{\alpha} \oint d \theta J B \phi \left( k_{\psi}, k_{\alpha} \right) \int dw_{||} d \mu ~ h_{s} \left( - k_{\psi}, - k_{\alpha} \right) J_{0} \left( k_{\perp} \rho_{s} \right) \right\rangle_{\Delta t} , \label{eq:partFlux}
\end{align}
the momentum flux
\begin{align}
   \Pi_{\zeta s}^{\phi} &\equiv - \left\langle R \left\langle \left\langle \int d^{3} w \underline{h}_{s} m_{s} R \left( \vec{w} \cdot \hat{e}_{\zeta} + R \Omega_{\zeta} \right) \hat{e}_{\zeta} \cdot \delta \vec{E} \right\rangle_{\Delta \psi} \right\rangle_{\Delta t} \right\rangle_{\psi} \label{eq:momFluxDef} \\
  &= \frac{4 \pi^{2} i}{V'} \left\langle \sum_{k_{\psi}, k_{\alpha}} k_{\alpha} \oint d \theta J B \phi \left( k_{\psi}, k_{\alpha} \right) \int dw_{||} d \mu ~ h_{s} \left( - k_{\psi}, - k_{\alpha} \right) \right. \label{eq:momFlux} \\
   &\times \left. \left[ \left( \frac{I}{B} w_{||} + R^{2} \Omega_{\zeta} \right) J_{0} \left( k_{\perp} \rho_{s} \right) + \frac{i}{\Omega_{s}} \frac{k^{\psi}}{B} \frac{\mu B}{m_{s}} \frac{2 J_{1} \left( k_{\perp} \rho_{s} \right)}{k_{\perp} \rho_{s}} \right] \right\rangle_{\Delta t} , \nonumber
\end{align}
the energy flux
\begin{align}
   Q_{s}^{\phi} &\equiv - \left\langle R \left\langle \left\langle \int d^{3} w \underline{h}_{s} \left( \frac{m_{s}}{2} w^{2} + Z_{s} e \Phi_{0} - \frac{m_{s}}{2} R^{2} \Omega_{\zeta}^{2} \right) \hat{e}_{\zeta} \cdot \delta \vec{E} \right\rangle_{\Delta \psi} \right\rangle_{\Delta t} \right\rangle_{\psi} \label{eq:heatFluxDef} \\
  &= \frac{4 \pi^{2} i}{V'} \left\langle \sum_{k_{\psi}, k_{\alpha}} k_{\alpha} \oint d \theta J B \phi \left( k_{\psi}, k_{\alpha} \right) \int dw_{||} d \mu ~ h_{s} \left( - k_{\psi}, - k_{\alpha} \right) \right. \label{eq:heatFlux} \\
   &\times \left. \left( \frac{w^{2}}{2} + \frac{Z_{s} e \Phi_{0}}{m_{s}} - \frac{1}{2} R^{2} \Omega_{\zeta}^{2} \right) J_{0} \left( k_{\perp} \rho_{s} \right) \right\rangle_{\Delta t} , \nonumber
\end{align}
and the turbulent energy exchange between species
\begin{align}
   P_{Q s}^{\phi} &\equiv \left\langle \left\langle \left\langle \int d^{3} w Z_{s} e \underline{h}_{s} \frac{\partial \underline{\phi}}{\partial t} \right\rangle_{\Delta \psi} \right\rangle_{\Delta t} \right\rangle_{\psi} \label{eq:heatingDef} \\
   &= \frac{4 \pi^{2}}{V'} \left\langle \sum_{k_{\psi}, k_{\alpha}} \oint d \theta J \Omega_{s} \frac{\partial}{\partial t} \left( \phi \left( k_{\psi}, k_{\alpha} \right) \right) \int dw_{||} d \mu ~ h_{s} \left( - k_{\psi}, - k_{\alpha} \right) J_{0} \left( k_{\perp} \rho_{s} \right) \right\rangle_{\Delta t} . \label{eq:heating}
\end{align}
Here $\underline{h}_{s} \equiv f_{s 1} + Z_{s} e \underline{\phi} F_{M s} / T_{s}$ is the nonadiabatic portion of the distribution function, $\underline{\left( \ldots \right)}$ indicates the quantity has not been Fourier analysed, $\delta \vec{E} = - \vec{\nabla}_{\perp} \underline{\phi}$ is the turbulent electric field, $\left\langle \ldots \right\rangle_{\psi} \equiv \left( 2 \pi / V' \right) \oint_{0}^{2 \pi} d \theta J \left( \ldots \right)$ is the flux surface average, $\left\langle \ldots \right\rangle_{\Delta t} \equiv \Delta t^{-1} \int_{\Delta t} dt \left( \ldots \right)$ is a coarse-grain average over a time $\Delta t$ (which is longer than the turbulent decorrelation time, but shorter than the transport time), $V' \equiv 2 \pi \oint d\theta J$, $J \equiv \left| \vec{B} \cdot \vec{\nabla} \theta \right|^{-1}$ is the Jacobian, and $k^{\psi} \equiv \vec{k}_{\perp} \cdot \vec{\nabla} \psi = k_{\psi} \left| \vec{\nabla} \psi \right|^{2} + k_{\alpha} \vec{\nabla} \psi \cdot \vec{\nabla} \alpha$.

%===================================================%
%===================================================%
\section{Estimating intrinsic momentum transport}
\label{sec:momTransportEst}
%===================================================%
%===================================================%

Equations \refEq{eq:momFluxDef} and \refEq{eq:heatFluxDef} allow us to calculate the local momentum flux and energy flux respectively. The energy flux signifies how much power must be injected to maintain the temperature gradient specified in \refEq{eq:gyrokineticEq}. A lower energy flux is desirable as it means that less external heating power is needed to maintain a fixed temperature profile. Similarly, the momentum flux tells how much external momentum must be injected to maintain the specified rotation shear (at a given value of rotation). However, in this work we will use GS2 \cite{DorlandETGturb2000}, a local $\delta f$ gyrokinetic code, to self-consistently calculate the nonlinear turbulent fluxes of momentum and energy generated at zero rotation and rotation shear. From this information we can estimate the intrinsic ability of a given geometry to drive rotation by following the analysis of reference \cite{BallMomUpDownAsym2014}.

First, we will Taylor expand the momentum flux around zero rotation and rotation shear to get the usual momentum transport equation \cite{ParraIntrinsicRotReview2012},
\begin{align}
   \left\langle \Pi_{\zeta i} \right\rangle_{t} - P_{\Pi i} n_{i} m_{i} R_{c}^{2} \Omega_{\zeta} - D_{\Pi i} n_{i} m_{i} R_{c}^{2} \frac{d \Omega_{\zeta}}{d a_{\psi}} \approx 0 ,
\end{align}
where $\left\langle \Pi_{\zeta i} \right\rangle_{t}$ is the time-averaged intrinsic ion momentum flux (i.e. the momentum flux calculated for $\Omega_{\zeta} = d \Omega_{\zeta} / d a_{\psi} = 0$), $P_{\Pi i}$ is the momentum pinch, $D_{\Pi i}$ is the momentum diffusivity (i.e. the kinematic viscosity), and $R_{c}$ is the major radial location of the centre of a given flux surface. By neglecting the momentum pinch we find
\begin{align}
\left\langle \Pi_{\zeta i} \right\rangle_{t} \approx D_{\Pi i} n_{i} m_{i} R_{c}^{2} \frac{d \Omega_{\zeta}}{d a_{\psi}} , \label{eq:momFluxEstimate}
\end{align}
a balance between rotation diffusion and the intrinsic momentum flux. Doing so is conservative as the momentum pinch can only ever enhance the level of rotation, maybe by as much as a factor of three \cite{PeetersMomPinch2007}. We will also write the energy flux as a diffusive term \cite{FreidbergFusionEnergy2007pg452} according to
\begin{align}
\left\langle Q_{i} \right\rangle_{t} \approx - D_{Q i} n_{i} \frac{d T_{i}}{d a_{\psi}} , \label{eq:heatFluxEstimate}
\end{align}
where $\left\langle Q_{i} \right\rangle_{t}$ is the time-averaged energy flux. Combining these two equations through the turbulent ion Prandtl number $Pr_{i} \equiv D_{\Pi i} / D_{Q i}$ gives
\begin{align}
\frac{1}{v_{th i}} \frac{d \left( R_{c} \Omega_{\zeta} \right)}{d a_{\psi}} \approx \frac{-1}{2 Pr_{i}} \left( \frac{v_{th i}}{R_{c}} \frac{\left\langle \Pi_{\zeta i} \right\rangle_{t}}{\left\langle Q_{i} \right\rangle_{t}} \right) \frac{d}{d a_{\psi}} \Ln{T_{i}} , \label{eq:rotationGradEstSimple}
\end{align}
where we used that $T_{i} = m_{i} v_{th i}^{2} / 2$. Doing this is useful because the Prandtl number is expected to be both $O \left( 1 \right)$ and unaffected by changes in tokamak parameters. We now substitute the Alfv\'{e}n Mach number, $M_{A} \equiv \left| R_{c} \Omega_{\zeta} \right| \sqrt{\mu_{0} n_{i} m_{i}} / B_{0}$, as it is the relevant quantity for stabilising MHD modes. Assuming that $n_{e} = n_{i}$, $T_{e} = T_{i}$, and $\eta_{i} \equiv \left( d \Ln{T_{i}} / d a_{\psi} \right) / \left( d \Ln{n_{i}} / d a_{\psi} \right) \gg 1$ allows us to use \refEq{eq:rotationGradEstSimple} to estimate the Alfv\'{e}n Mach number profile as
\begin{align}
M_{A} \left( \rho \right) \approx \left| \int_{1}^{\rho} d \rho' \frac{1}{2 \sqrt{2} Pr_{i} \left( \rho' \right)} \left( \frac{v_{th i} \left( \rho' \right)}{R_{c} \left( \rho' \right)} \frac{\left\langle \Pi_{\zeta i} \left( \rho' \right) \right\rangle_{t}}{\left\langle Q_{i} \left( \rho' \right) \right\rangle_{t}} \right) \frac{\beta' \left( \rho' \right)}{\sqrt{\beta \left( \rho' \right)}} \right| . \label{eq:rotationGradEst}
\end{align}
Notice that this expression is in terms of $\left( v_{th i} / R_{c} \right) \left\langle \Pi_{\zeta i} \right\rangle_{t} / \left\langle Q_{i} \right\rangle_{t}$, a normalised parameter that indicates how strongly a given geometry drives rotation from the turbulent fluxes of momentum and energy. The remainder of this thesis will be focused on finding the geometries that maximise this momentum transport figure of merit as well as minimise the energy flux.

First, in this chapter we will briefly outline the argument for why the intrinsic rotation in up-down symmetric tokamaks must be small in $\rho_{\ast} \ll 1$. Then, in chapter \ref{ch:GYRO_TiltingSymmetry} we will present a similar argument demonstrating that the momentum flux from fast mirror symmetric flux surface shaping (i.e. shaping with poloidal variation on a small spatial scale) must be exponentially small in the Fourier mode numbers of the fast shaping. This motivates the calculation in chapter \ref{ch:GYRO_MomFluxScaling}, which shows that flux surfaces with slowly varying envelopes created by the beating of fast shaping can generate momentum flux that is only polynomially small. We also argue that a mirror symmetric tokamak has no momentum transport in the screw pinch limit. Accordingly, in chapters \ref{ch:GYRO_ShafranovShift} and \ref{ch:GYRO_NonMirrorSym} we search for the optimal configurations in the space of non-mirror symmetric geometries created by the beating of low order shaping modes. Chapter \ref{ch:GYRO_ShafranovShift} reveals that introducing the Shafranov shift into tilted elliptical flux surfaces breaks mirror symmetry and enhances the rotation. Unfortunately, this enhancement is entirely cancelled by including the effect of the pressure profile on the equilibrium, which is needed to be consistent. However, in chapter \ref{ch:GYRO_NonMirrorSym} we use independently tilted elongation and triangularity to directly break mirror symmetry and significantly increase the momentum transport. Lastly, in chapter \ref{ch:conclusions} we present the optimal geometry, as indicated by the analysis of this thesis, and comment on some overarching conclusions.

%===================================================%
%===================================================%
\section{Momentum flux from up-down symmetric flux surface shaping}
\label{sec:upDownSymArg}
%===================================================%
%===================================================%

In this thesis we are concerned with the effect of geometry on the turbulent fluxes. All of the information concerning the tokamak geometry enters the gyrokinetic model via ten geometric coefficients: $B$, $\hat{b} \cdot \vec{\nabla} \theta$, $v_{d s \psi}$, $v_{d s \alpha}$, $a_{s ||}$, $\left| \vec{\nabla} \psi \right|^{2}$, $\vec{\nabla} \psi \cdot \vec{\nabla} \alpha$, $\left| \vec{\nabla} \alpha \right|^{2}$, $R$, and $\left. \partial R / \partial \psi \right|_{\theta}$. We note that, when the plasma is stationary (i.e. $\Omega_{\zeta} = 0$), only eight coefficients appear as the terms containing $R$ and $\left. \partial R / \partial \psi \right|_{\theta}$ vanish. In appendix \ref{app:genGeoCoeff} we show the full, explicit calculation of these geometric coefficients in the context of the Miller local equilibrium model introduced in chapter \ref{ch:MHD_LocalEquil}.

As shown by references \cite{PeetersMomTransSym2005, ParraUpDownSym2011, SugamaUpDownSym2011}, in an up-down symmetric tokamak all of the geometric coefficients have a well defined parity, which has important consequences for the overall symmetry properties of the gyrokinetic model. In an up-down symmetric tokamak, the coefficients $v_{d s \psi}$, $a_{s ||}$, and $\Nabla \psi \cdot \Nabla \alpha$ are necessarily odd in $\theta$, while $\hat{b} \cdot \Nabla \theta$, $B$, $v_{d s \alpha}$, $\left| \Nabla \psi \right|^{2}$, $\left| \Nabla \alpha \right|^{2}$, $R$, and $\left. \partial R / \partial \psi \right|_{\theta}$ are even. This means that, when the rotation and rotation shear are zero, the equations become invariant to the $\left( k_{\psi}, k_{\alpha}, \theta, w_{||}, \mu, t \right) \rightarrow \left( - k_{\psi}, k_{\alpha}, - \theta, - w_{||}, \mu, t \right)$ coordinate system transformation, which is not true in up-down asymmetric devices. This symmetry means that, given any solution $h_{s} \left( k_{\psi}, k_{\alpha}, \theta, w_{||}, \mu, t \right)$, we can construct a second solution $- h_{s} \left( - k_{\psi}, k_{\alpha}, - \theta, - w_{||}, \mu, t \right)$ that will also satisfy the gyrokinetic equations. From \refEq{eq:momFlux} (or \refEq{eq:totalMomFlux}, \refEq{eq:speciesMomFlux}, and \refEq{eq:elecMagMomFlux}) we see that this second solution will have a momentum flux that cancels that of the first. These two solutions are each valid for different initial conditions, but since the turbulence is presumed to be chaotic, both solutions will arise within a turbulent decorrelation time (statistically speaking). This demonstrates that, in the gyrokinetic limit, the time-averaged momentum flux must be zero in a stationary, up-down symmetric tokamak.

%% file: Ch_07_GYRO_TiltingSymmetry.tex
% !TEX root = /Users/Justin/Documents/Research/Writings/2016DoctoralThesis/DoctoralThesis.tex

\chapter{Mirror symmetry: Scaling of momentum flux with shaping mode number}
\label{ch:GYRO_TiltingSymmetry}

\begin{quote}
   \emph{Much of this chapter appears in reference \cite{BallMirrorSymArg2016}.}
\end{quote}

\begin{figure}
 \hspace{0.04\textwidth} (a) \hspace{0.4\textwidth} (b) \hspace{0.25\textwidth}
 \begin{center}
  \includegraphics[width=0.4\textwidth]{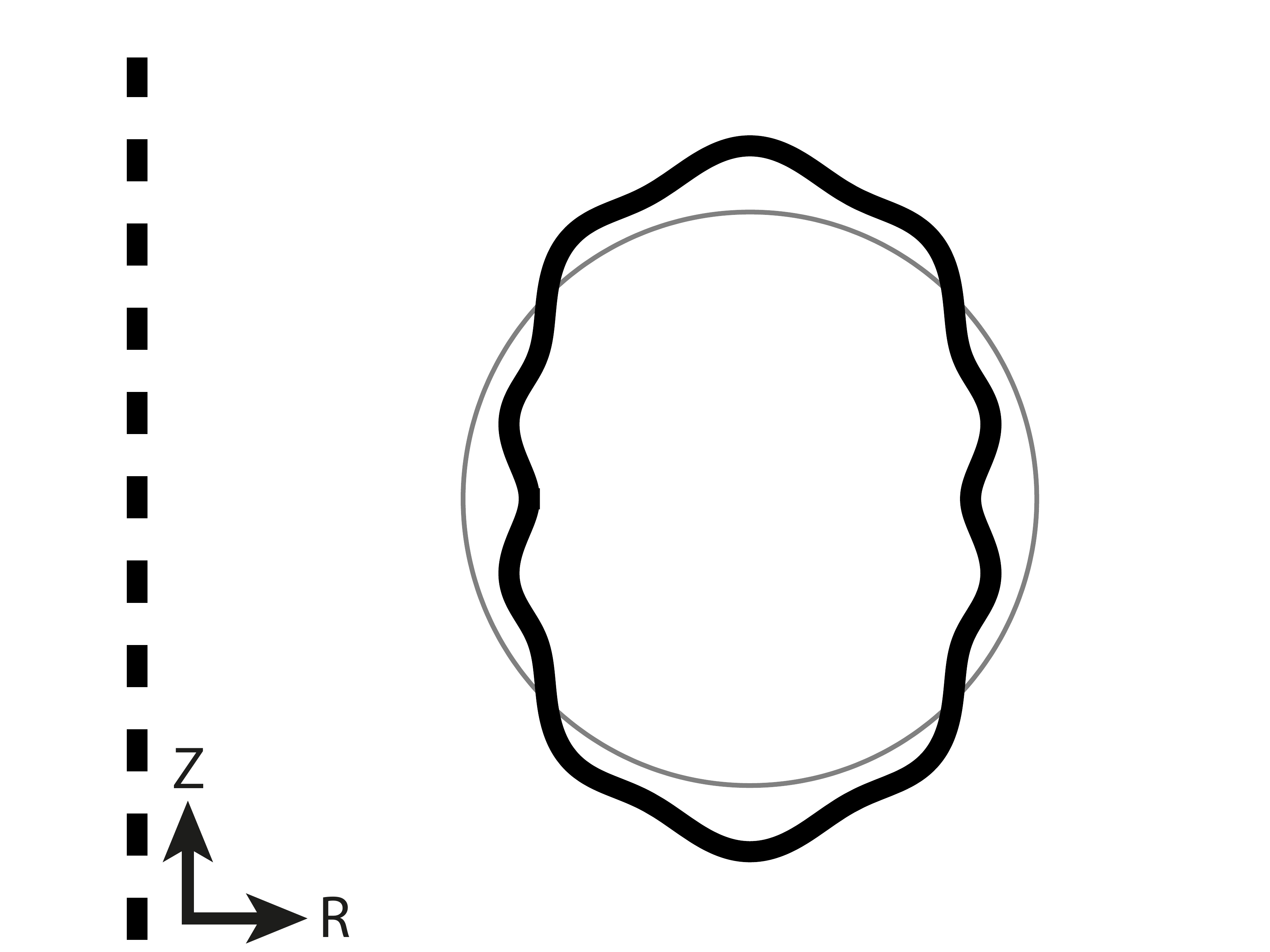}
  \includegraphics[width=0.4\textwidth]{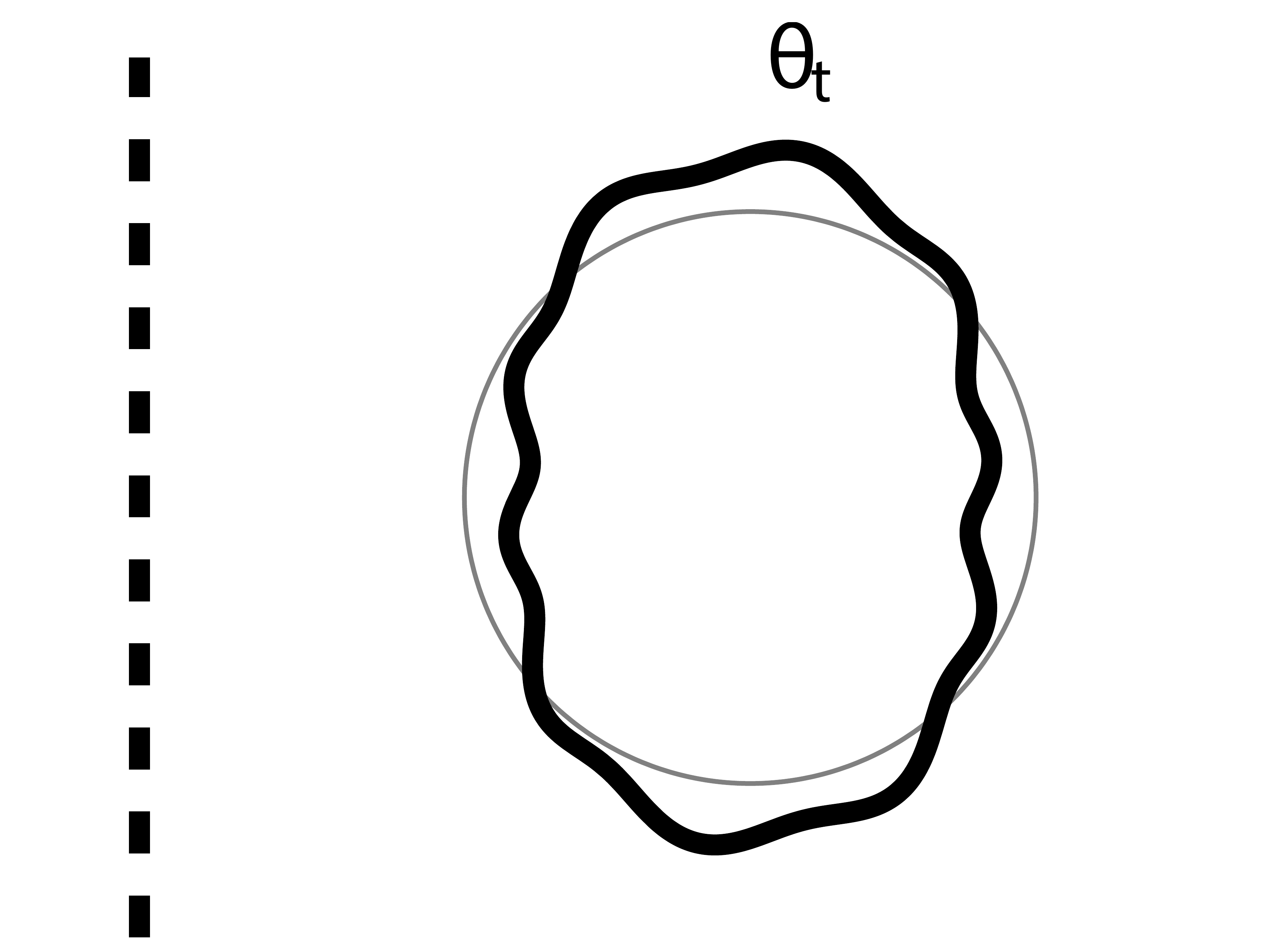}
 \end{center}
 \caption{Example poloidal cross-section of a tokamak (a) without any tilt and (b) with the fast flux surface shaping effects tilted by an angle $\theta_{t}$, noting that slow shaping effects like the elongation and envelopes (not shown) are not rotated. Circular flux surfaces are shown in grey for comparison and the axis of axisymmetry is indicated by a dashed line.}
 \label{fig:tiltedGeometry}
\end{figure}

In this chapter, we demonstrate a new symmetry of the local, high-flow, electromagnetic $\delta f$ gyrokinetic equations. This symmetry means that poloidally rotating all ``fast'' flux surface shaping (i.e. poloidal variation with a spatial scale much smaller then the connection length) by a single tilt angle (as shown in figure \ref{fig:tiltedGeometry}) has little effect on the transport properties of a tokamak. More broadly, it indicates that turbulence is insensitive to the interactions between flux surface variation on different poloidal scales.
	
To establish this tilting symmetry we expand the high-flow gyrokinetic equations in the limit of large flux surface shaping Fourier mode number. Here we distinguish between fast flux surface shaping and shaping with a large mode number because different large mode number shaping effects can beat together to create rapid variation with an envelope that varies on the slow connection length scale. We will see that gyrokinetics is symmetric to a tilt in the rapid variation, but not a tilt in the slowly varying envelope. This is intuitive as we expect turbulent eddies to extend along the field line and average over rapid poloidal variation, but still respond to large-scale flux surface shaping. Therefore, we would expect the effect of tilting flux surface shaping should diminish as poloidal flux surface variation becomes faster. However, what is surprising is that this symmetry proves that the effect diminishes exponentially, rather than polynomially. Hence we find that tilting fast flux surface shaping to create up-down asymmetry has an exponentially small effect on the turbulent momentum flux.

In section \ref{sec:upDownSymArg} we presented an argument showing that up-down symmetric devices generate no momentum transport in the usual lowest order gyrokinetics. This argument, together with the tilting symmetry presented in this chapter, will show that flux surfaces with mirror symmetry across some line in the poloidal plane can only generate exponentially small momentum transport, in the limit of fast shaping effects. This is because mirror symmetric flux surfaces can be transformed into up-down symmetric flux surfaces by poloidally tilting of all the shaping effects by a single global tilt angle. Consequently, this establishes a distinction between devices with mirror symmetric flux surfaces and devices without mirror symmetry, which may have important consequences for flux surface shaping of any mode number. Additionally, the exponential scaling suggests that generating rotation using up-down asymmetric triangularity or squareness will be significantly less effective than up-down asymmetric elongation, which is consistent with previous work \cite{BallMomUpDownAsym2014}. The tilting symmetry also indicates that the geometry used in the TCV up-down asymmetry experiments \cite{CamenenPRLExp2010} is close to the optimal mirror symmetric shape for generating large rotation \cite{BallMomUpDownAsym2014}, but this has not been tested experimentally. Regardless, a significant enhancement over the TCV results may still be found in the space of non-mirror symmetric shapes.

While the practical implications of the tilting symmetry are most relevant to momentum transport, it also applies to energy and particle transport. For example, references \cite{WeisenShapeOnConfinement1997, CamenenTriConfinement2007} look at the effect of elongation and triangularity on the energy confinement time in TCV. From the scaling presented in this chapter, we would expect that tilting triangularity or higher order shaping would have a smaller effect on energy confinement compared with tilting elongation. This can have significance for purely up-down symmetric configurations. For example, horizontal elongation can be thought of as vertical elongation with a $90^{\circ}$ tilt just as changing the sign of triangularity is equivalent to tilting the triangularity by $180^{\circ}$. Therefore, we would expect switching from vertical to horizontal elongation would have a larger effect on the energy confinement time than changing the sign of the triangularity.

Section \ref{sec:analyticArg} of this chapter contains the analytic analysis demonstrating the poloidal tilting symmetry of fast flux surface shaping, while section \ref{sec:numResultsSym} presents the results of nonlinear local gyrokinetic simulations. These simulations are aimed at providing numerical verification of the analytic work.

%===================================================%
%===================================================%
\section{Poloidal tilting symmetry of high order flux surface shaping}
\label{sec:analyticArg}
%===================================================%
%===================================================%

In this section we will show the tilting symmetry of fast flux surface shaping in the nonlinear local $\delta f$ gyrokinetic model. First, we will start with results from appendix \ref{app:genGeoCoeff}, which gives a detailed calculation of the geometric coefficients from the Miller local equilibrium specification. Then, we will use the Expanded local flux surface specification derived in chapter \ref{ch:MHD_LocalEquil} to prescribe arbitrarily-shaped flux surfaces using Fourier analysis. This reveals how arbitrary high order shaping enters into the gyrokinetic equations. Finally we will expand the gyrokinetic equations in the limit of large flux surface shaping Fourier mode number and show that tilting fast shaping does not affect particle, momentum, or energy transport.

%===================================================%
\subsection{Geometric coefficients}
\label{subsec:millerGeo}
%===================================================%

To specify the background tokamak equilibrium for our local gyrokinetic model we will use the generalisation of the Miller local equilibrium model derived in chapter \ref{ch:MHD_LocalEquil}. The Miller prescription approximates the equilibrium around a single flux surface of interest when given: $R_{c 0}$ (the major radial location of the centre of the flux surface of interest), $r_{0} \left( \theta \right)$ (the shape of the flux surface of interest), $\left. \partial r / \partial r_{\psi} \right|_{r_{\psi 0}, \theta}$ (how the shape of the flux surface of interest changes with minor radius), and several additional scalar quantities.  We note that a local equilibrium is exactly mirror symmetric if $r_{0} \left( \theta \right) = r_{0} \left( - \theta + \theta_{0} \right)$ and $\left. \partial r / \partial r_{\psi} \right|_{\psi_{0}, \theta} = \left. \partial r / \partial r_{\psi} \right|_{\psi_{0}, -\theta + \theta_{0}}$ for some $\theta_{0}$, otherwise it is non-mirror symmetric. Similarly, if $r_{0} \left( \theta \right) = r_{0} \left( - \theta \right)$ and $\left. \partial r / \partial r_{\psi} \right|_{\psi_{0}, \theta} = \left. \partial r / \partial r_{\psi} \right|_{\psi_{0}, -\theta}$, then the local equilibrium is exactly up-down symmetric (as well as mirror symmetric), otherwise it is up-down asymmetric.  We will completely specify the geometry of the equilibrium using the Expanded prescription, given by \refEq{eq:fluxSurfofInterestShapeWeak}, \refEq{eq:fluxSurfofInterestDerivWeak}, and \refEq{eq:gradShafranovLocalEqWeak} through \refEq{eq:geoAxialWeak} while assuming $Z_{c 0} = 0$ without loss of generality.

The four scalar quantities needed for the Miller equilibrium are commonly taken to be $I \equiv R B_{\zeta}$ (the toroidal field flux function),
\begin{align}
   q \equiv \frac{I}{2 \pi} \left. \oint_{0}^{2 \pi} \right|_{\psi} d \theta \left( R^{2} \vec{B}_{p} \cdot \vec{\nabla} \theta \right)^{-1} \label{eq:safetyFactorDef}
\end{align}
(the safety factor), $dq/dr_{\psi}$ (the magnetic shear), and $d p / d r_{\psi}$ (the pressure gradient) of the flux surface of interest. However, when the plasma is rotating quickly the pressure ceases to be a flux function because the density varies poloidally. Therefore, we replace $d p / d r_{\psi}$ with $\left. \partial p / \partial r_{\psi} \right|_{R}$, which requires four additional, species-dependent parameters: $\eta_{s} T_{s}$ (the pseudo-pressure), $d \left( \eta_{s} T_{s} \right) / d r_{\psi}$ (the derivative of the pseudo-pressure), $m_{s} \Omega_{\zeta}^{2} / 2 T_{s}$ (a rotational frequency parameter), and $d \left( m_{s} \Omega_{\zeta}^{2} / 2 T_{s} \right) / d r_{\psi}$ (the derivative of the rotational frequency parameter). These four species-dependent parameters specify the pressure (and its radial gradient) on the flux surface of interest by using \refEq{eq:pseudoDensity}. 

The functions $r_{0} \left( \theta \right)$ and $\left. \partial r / \partial r_{\psi} \right|_{r_{\psi 0}, \theta}$ allow us to calculate poloidal derivatives of any order as well as the first order radial derivatives: $\left. \partial R / \partial r_{\psi} \right|_{\theta}$ and $\left. \partial Z / \partial r_{\psi} \right|_{\theta}$ (but not higher order radial derivatives). As shown in appendix \ref{app:genGeoCoeff} this is enough information to calculate seven of the ten geometric coefficients, specifically $\hat{b} \cdot \Nabla \theta$, $B$, $v_{d s \psi}$, $a_{s ||}$, $\left| \Nabla \psi \right|^{2}$, $R$, and $\left. \partial R / \partial \psi \right|_{\theta}$. However, calculating $\Nabla \alpha$, which appears in the other three coefficients, is more complex as it involves second order radial derivatives of the flux surface shape and an integral. In the Miller local equilibrium, second order radial derivatives must be calculated by ensuring that the Grad-Shafranov equation is satisfied. Referring to \refEq{eq:IalphaDef} and \refEq{eq:gradAlphaFinal}, we see that doing so produces
\begin{align}
\vec{\nabla} \alpha &= - \left( I \left. \int_{\theta_{\alpha}}^{\theta} \right|_{\psi} d \theta' \left\{ \frac{1}{R^{2} B_{p}} \left. \frac{\partial l_{p}}{\partial \theta'} \right|_{\psi} \left[ \frac{1}{I} \frac{d I}{d \psi} + \frac{I}{R^{2} B_{p}^{2}} \frac{d I}{d \psi} + \frac{\mu_{0}}{B_{p}^{2}} \left. \frac{\partial p}{\partial \psi} \right|_{R} \right. \right. \right. \nonumber \\
&- \left. \left. \frac{2}{R^{2} B_{p}} \left( \dlpdthetaPrime \right)^{-1} \frac{\partial Z}{\partial \theta'} + \frac{2 \kappa_{p}}{R B_{p}} \right] \right\} - \left[ \frac{I}{R^{4} B_{p}^{3}} \left. \frac{\partial l_{p}}{\partial \theta'} \right|_{\psi} \Nabla \psi \cdot \Nabla \theta' \right]_{\theta' = \theta_{\alpha}}^{\theta' = \theta} \label{eq:gradAlphaMiller} \\
&- \left. \left[ \frac{I}{R^{2} B_{p}} \left. \frac{\partial l_{p}}{\partial \theta} \right|_{\psi} \right]_{\theta = \theta_{\alpha}} \frac{d \theta_{\alpha}}{d \psi} + \frac{d \Omega_{\zeta}}{d \psi} t \right) \vec{\nabla} \psi - \frac{I}{R^{2} B_{p}} \left. \frac{\partial l_{p}}{\partial \theta} \right|_{\psi} \vec{\nabla} \theta + \vec{\nabla} \zeta , \nonumber
\end{align}
where
\begin{align}
\kappa_{p} = \left( \dlpdtheta \right)^{-3} \left( \frac{\partial R}{\partial \theta} \frac{\partial^{2} Z}{\partial \theta^{2}} - \frac{\partial^{2} R}{\partial \theta^{2}} \frac{\partial Z}{\partial \theta}  \right) \label{eq:poloidalCurv}
\end{align}
is the poloidal magnetic field curvature (defined by \refEq{eq:polCurvDef} such that the inwards normal direction is positive), $l_{p}$ is the poloidal arc length defined such that
\begin{align}
\left. \frac{\partial l_{p}}{\partial \theta} \right|_{\psi} &= \sqrt{\left. \frac{\partial \vec{r}}{\partial \theta} \right|_{\psi} \cdot \left. \frac{\partial \vec{r}}{\partial \theta} \right|_{\psi}} = \sqrt{\left( \left. \frac{\partial R}{\partial \theta} \right|_{\psi} \right)^{2} + \left( \left. \frac{\partial Z}{\partial \theta} \right|_{\psi} \right)^{2}} , \label{eq:dLpdtheta}
\end{align}
and all quantities are evaluated on the flux surface of interest. Note that $d I / d \psi$ can be found from $dq/dr_{\psi}$ using \refEq{eq:dIdpsiFullForm}.

%===================================================%
\subsection{Asymptotic expansion}
\label{subsec:asymptoticExpansion}
%===================================================%

Now we will investigate the effect of high order flux surface shaping, first on the geometric coefficients and then on the fluxes of particles, momentum, and energy. We can always Fourier analyse the flux surface shape and its derivative (without loss of generality) to write \refEq{eq:fluxSurfofInterestShapeWeak} and \refEq{eq:fluxSurfofInterestDerivWeak}. Next, using trigonometric identities we can convert \refEq{eq:fluxSurfofInterestShapeWeak} and \refEq{eq:fluxSurfofInterestDerivWeak} to
\begin{align}
r_{0} \left( \theta, z \right) &= r_{\psi 0} \Bigg( 1 - \sum_{l = 0}^{\infty} \sum_{k = k_{\text{min}}}^{k_{\text{max}}} \frac{\Delta_{k + l m_{c}} - 1}{2} \big[ \Cos{l \left( z + m_{c} \theta_{t m} \right)} \Cos{k \left( \theta + \theta_{t m} \right)} \nonumber \\
   -& \Sin{l \left( z + m_{c} \theta_{t m} \right)} \Sin{k \left( \theta + \theta_{t m} \right)} \big] \Bigg) \label{eq:fluxSurfaceSpecScaleSep} \\
\left. \frac{\partial r}{\partial r_{\psi}} \right|_{r_{\psi 0}, \theta, z} &= 1 - \sum_{l = 0}^{\infty} \sum_{k = k_{\text{min}}}^{k_{\text{max}}} \bigg[ \left( \frac{\Delta_{k + l m_{c}} - 1}{2} + \frac{r_{\psi 0}}{2} \frac{d \Delta_{k + l m_{c}}}{d r_{\psi}} \right)  \label{eq:fluxSurfaceChangeSpecScaleSep} \\
   \times \big[& \Cos{l \left( z + m_{c} \theta_{t m} \right)} \Cos{k \left( \theta + \theta_{t m} \right)} - \Sin{l \left( z + m_{c} \theta_{t m} \right)} \Sin{k \left( \theta + \theta_{t m} \right)} \big] \nonumber \\
   - ( & k + l m_{c} ) \left( \Delta_{k + l m_{c}} - 1 \right) \frac{r_{\psi 0}}{2} \frac{d \theta_{t \left(k + l m_{c} \right)}}{d r_{\psi}} \nonumber \\
   \times \big[& \Sin{l \left( z + m_{c} \theta_{t m} \right)} \Cos{k \left( \theta + \theta_{t m} \right)} + \Cos{l \left( z + m_{c} \theta_{t m} \right)} \Sin{k \left( \theta + \theta_{t m} \right)} \big] \bigg] \nonumber
\end{align}
respectively, where $m_{c}$ is the characteristic Fourier mode number of the fast shaping,
\begin{align}
z \equiv m_{c} \theta \label{eq:zDef}
\end{align}
is a fast spatial scale poloidal coordinate, $k \equiv m - l m_{c}$, and $m_{c}$ is in the range $\left[ k_{\text{min}}, k_{\text{max}} \right]$. Additionally, we require $k_{\text{max}} - k_{\text{min}} = m_{c} - 1$ to ensure that the summation is over all possible modes. This is the form of a two dimensional Fourier decomposition in the slow spatial scale coordinate $\theta$ and the fast spatial scale coordinate $z$. The definition of $l$ contains the physics of the scale separation as it divides the poloidal variation into fast and slow components. Later, using numerical results we will motivate $l \equiv \left\lfloor \left( m + 2 \right) / m_{c} \right\rfloor$, where $\left\lfloor x \right\rfloor$ is the floor function that gives the integer value $n$ such that $n \leq x < n + 1$ for any real number $x$. This definition of $l$ means that variation at least as rapid as the $m_{c} - 2$ Fourier mode is considered fast, while any lower modes are considered slow. However, analytically we will expand in $m_{c} \gg 1$ to investigate the effect of high order flux surface shaping on a traditionally shaped equilibrium. This means that the precise definition of $l$ does not matter because the scales are sufficiently distinct by construction of the expansion.

The separation of scales in \refEq{eq:fluxSurfaceSpecScaleSep} and \refEq{eq:fluxSurfaceChangeSpecScaleSep}, e.g. $r_{0} \left( \theta, z \right)$, means that
\begin{align}
  \left. \frac{\partial}{\partial \theta} \right|_{w_{||}, \mu} = \left. \frac{\partial}{\partial \theta} \right|_{z, w_{||}, \mu} + m_{c} \left. \frac{\partial}{\partial z} \right|_{\theta, w_{||}, \mu} . \label{eq:poloidalDerivTransform}
\end{align}
Additionally, because we are only interested in bulk behaviour, we will eventually average quantities in $z$ using
\begin{align}
\overline{\left( \ldots \right)} \equiv \frac{1}{2 \pi} \left. \oint_{-\pi}^{\pi} \right|_{\theta} dz \left( \ldots \right) . \label{eq:zAvg}
\end{align}

%===================================================%
\subsection{Gyrokinetic tilting symmetry}
\label{subsec:gyroSym}
%===================================================%

This section contains an analytic calculation that demonstrates a symmetry of the gyrokinetic model, when expanding in $m_{c} \gg 1$. Since turbulent eddies are generally quite extended along the field line, we expect them to effectively average over the small scale magnetic variations created by fast flux surface shaping. Therefore, we would anticipate that tilting such shaping should have a minimal effect on the turbulence. However, the unexpected result of this calculation is that tilting fast flux surface shaping has an exponentially small effect on the turbulent fluxes in $m_{c} \gg 1$, rather than a polynomial effect. This argument only relies on $m_{c} \gg 1$ and does \textit{not} presume that the flux surface shaping is weak.

We will start with a completely general local equilibrium, with flux surfaces specified by $r_{0} \left( \theta, z \left( \theta \right) \right)$ and $\left. \partial r / \partial r_{\psi} \right|_{r_{\psi 0}, \theta, z \left( \theta \right)}$ (see \refEq{eq:fluxSurfaceSpecScaleSep} and \refEq{eq:fluxSurfaceChangeSpecScaleSep}). Using this specification we will compare two different geometries that are identical except for the form of $z \left( \theta \right)$. In the untilted case $z \left( \theta \right) = z_{u} \left( \theta \right) \equiv m_{c} \theta$, while in the tilted case $z \left( \theta \right) = z_{t} \left( \theta \right) \equiv m_{c} \left( \theta + \theta_{t} \right)$. We see that the tilted case translates all the fast poloidal variation by a single global tilt angle. Note that this is different than simply translating all of the shaping effects with large Fourier mode numbers. Due to the form of \refEq{eq:fluxSurfaceSpecScaleSep}, the fast variation is tilted, while any slowly varying envelopes that might have been created by high mode number shaping effects are not. Introducing the tilt into the form of $z \left( \theta \right)$ alters the equilibrium and in principle changes the transport properties, but we will show its effect is exponentially small when expanding in $m_{c} \gg 1$.

Although we just presented two specific examples of $z \left( \theta \right)$, we are free to calculate the geometric coefficients for a completely general $z \left( \theta \right)$. From the form of the ten geometric coefficients (see appendix \ref{app:genGeoCoeff}) we see that $z \left( \theta \right) $ only enters as $z$, derivatives of $z$, and in the integral over poloidal angle contained in $\vec{\nabla} \alpha$ (see \refEq{eq:gradAlphaMiller}). This means that we can indicate the poloidal dependence of any geometric coefficient, $Q_{\text{geo}} \in \left\{ B, \hat{b} \cdot \vec{\nabla} \theta, v_{d s \psi}, v_{d s \alpha}, a_{s ||}, \left| \vec{\nabla} \psi \right|^{2}, \vec{\nabla} \psi \cdot \vec{\nabla} \alpha, \left| \vec{\nabla} \alpha \right|^{2}, R, \left. \partial R / \partial \psi \right|_{\theta} \right\}$, by writing it as
\begin{align}
   Q_{\text{geo}} \left( \theta, z, \frac{\partial z}{\partial \theta}, \frac{\partial^{2} z}{\partial \theta^{2}}, \left. \int_{\theta_{\alpha}}^{\theta} \right|_{\psi} d \theta' F_{\alpha} \left( \theta', z \left( \theta' \right), \frac{\partial z}{\partial \theta'}, \frac{\partial^{2} z}{\partial \theta'^{2}} \right) - \left[ \frac{1}{R^{2} B_{p}^{2}} \left. \frac{\partial l_{p}}{\partial \theta} \right|_{\psi} \right]_{\theta = \theta_{\alpha}} \frac{d \theta_{\alpha}}{d \psi} \right) , \label{eq:Qgeo}
\end{align}
where
\begin{align}
   F_{\alpha} \left( \theta, z \left( \theta \right), \frac{\partial z}{\partial \theta}, \frac{\partial^{2} z}{\partial \theta^{2}} \right) &\equiv \frac{1}{R^{2} B_{p}} \left. \frac{\partial l_{p}}{\partial \theta} \right|_{\psi} \Bigg[ \frac{1}{I} \frac{d I}{d \psi} - \frac{1}{B_{p}} \left. \frac{\partial B_{p}}{\partial \psi} \right|_{\theta} - \frac{2}{R} \left. \frac{\partial R}{\partial \psi} \right|_{\theta} \nonumber \\
    &+ \left( \left. \frac{\partial l_{p}}{\partial \theta} \right|_{\psi} \right)^{-1} \left. \frac{\partial}{\partial \psi} \right|_{\theta} \left( \left. \frac{\partial l_{p}}{\partial \theta} \right|_{\psi} \right) \Bigg] \label{eq:FalphaDef}
\end{align}
is a periodic function of both $\theta$ and $z$.

Now we will compare the untilted equilibrium ($z \left( \theta \right) = z_{u} \left( \theta \right) \equiv m_{c} \theta$) and the equilibrium with tilted fast shaping effects ($z \left( \theta \right) = z_{t} \left( \theta \right) \equiv m_{c} \left( \theta + \theta_{t} \right)$). Since the only difference between the two cases is contained in the form of $z \left( \theta \right)$, we only need to look for differences in the arguments of \refEq{eq:Qgeo}. We immediately see that $\partial z_{u} / \partial \theta = \partial z_{t} / \partial \theta = m_{c}$ and all higher order poloidal derivatives are zero. Hence, we can eliminate the derivatives to write the geometric coefficients as
\begin{align}
   Q_{\text{geo}} \left( \theta, z, G_{\alpha} \left( \theta, z \left( \theta \right) \right) + H_{\alpha} \right) \label{eq:QgeoSimple}
\end{align}
for both cases, where we choose to define
\begin{align}
   G_{\alpha} \left( \theta, z \left( \theta \right) \right) \equiv& \left. \int_{\theta_{\alpha}}^{\theta} \right|_{\psi} d \theta' F_{\alpha} \left( \theta', z \left( \theta' \right) \right) \label{eq:GalphaThetaDef} \\
   H_{\alpha} \equiv& - \left[ \frac{1}{R^{2} B_{p}^{2}} \left. \frac{\partial l_{p}}{\partial \theta} \right|_{\psi} \right]_{\theta = \theta_{\alpha}} \frac{d \theta_{\alpha}}{d \psi} . \label{eq:HalphaThetaDef}
\end{align}

As we will now show, we can also eliminate the integral $\left. \int_{\theta_{\alpha}}^{\theta} \right|_{\psi} d \theta' F_{\alpha} \left( \theta', z \left( \theta' \right) \right)$, in addition to the derivatives. To do so, we start by defining the operator
\begin{align}
   \Lambda \left[ g \right] \left( \theta, z \right) &\equiv \left. \int_{z_{0}}^{z} \right|_{\theta} d z' \Big( g \left( \theta, z' \right) - \overline{g \left( \theta, z \right)} \Big) , \label{eq:Idef}
\end{align}
where the integral over $z$ is done holding $\theta$ constant, $g \left( \theta, z \right)$ is a yet unspecified function that is periodic in both $\theta$ and $z$, and $z_{0}$ is chosen such that
\begin{align}
  \overline{\Lambda \left[ g \right] \left( \theta, z \right)} = 0 \label{eq:z0cond}
\end{align}
(which can always be found when $g$ is periodic in $z$). Taking the total derivative in $\theta$ we find
\begin{align}
   \frac{d}{d \theta} \Lambda \left[ g \right] \left( \theta, z \left( \theta \right) \right) &= \left. \frac{\partial}{\partial \theta} \right|_{z} \Lambda \left[ g \right] + m_{c} \left. \frac{\partial}{\partial z} \right|_{\theta} \Lambda \left[ g \right]
\end{align}
from \refEq{eq:poloidalDerivTransform}, where we have taken $d z / d \theta = m_{c}$. Substituting in \refEq{eq:Idef} and rearranging gives
\begin{align}
   g \left( \theta, z \right) - \overline{g \left( \theta, z \right)} &= \frac{1}{m_{c}} \frac{d}{d \theta} \Lambda \left[ g \right] - \frac{1}{m_{c}} \left. \frac{\partial}{\partial \theta} \right|_{z} \Lambda \left[ g \right] . \label{eq:gDiff}
\end{align}
Integrating \refEq{eq:gDiff} with respect to $\theta$ and then using
\begin{align}
   \left. \frac{\partial}{\partial \theta} \right|_{z} \Lambda \left[ g \right] = \Lambda \left[ \left. \frac{\partial g}{\partial \theta} \right|_{z} \right] \label{eq:dLambdadThetaCond}
\end{align}
gives
\begin{align}
   \left. \int_{\theta_{\alpha}}^{\theta} \right|_{\psi} d \theta' \Big( g \left( \theta', z \left( \theta' \right) \right) - \overline{g \left( \theta', z \right)} \Big) &= \frac{1}{m_{c}} \Big( \Lambda \left[ g \right] \left( \theta, z \right) - \Lambda \left[ g \right] \left( \theta_{\alpha}, z \left( \theta_{\alpha} \right) \right) \Big) \label{eq:recursionRel} \\
   &- \frac{1}{m_{c}} \left. \int_{\theta_{\alpha}}^{\theta} \right|_{\psi} d \theta' \Lambda \left[ \left. \frac{\partial g}{\partial \theta'} \right|_{z} \right] . \nonumber
\end{align}
Now since $g \left( \theta, z \right)$ is an unspecified periodic function, we can always make the substitution $g \left( \theta, z \right) \rightarrow \Lambda \left[ \left. \partial g/ \partial \theta \right|_{z} \right]$ because $\Lambda \left[ \left. \partial g/ \partial \theta \right|_{z} \right]$ is also a periodic function. Doing so in \refEq{eq:recursionRel} gives
\begin{align}
   \left. \int_{\theta_{\alpha}}^{\theta} \right|_{\psi} d \theta' \Lambda \left[ \left. \frac{\partial g}{\partial \theta'} \right|_{z} \right] &= \frac{1}{m_{c}} \left( \Lambda^{2} \left[ \left. \frac{\partial g}{\partial \theta} \right|_{z} \right] \left( \theta, z \left( \theta \right) \right) - \Lambda^{2} \left[ \left. \frac{\partial g}{\partial \theta} \right|_{z} \right]  \left( \theta_{\alpha}, z \left( \theta_{\alpha} \right) \right) \right) \label{eq:recursionRelSecondIteration} \\
   &- \frac{1}{m_{c}} \left. \int_{\theta_{\alpha}}^{\theta} \right|_{\psi} d \theta' \Lambda^{2} \left[ \left. \frac{\partial^2 g}{\partial \theta'^2} \right|_{z} \right] . \nonumber
\end{align}
The second term on the left-hand side vanished because the definition of $z_{0}$ requires \refEq{eq:z0cond}, so we know $\left. \partial \overline{\Lambda \left[ g \right]} / \partial \theta \right|_{z} = \overline{\Lambda \left[ \left. \partial g/ \partial \theta \right|_{z} \right]} = 0$ from \refEq{eq:dLambdadThetaCond}. Here $\Lambda^{i} \left[ \ldots \right]$ indicates that the operator $\Lambda \left[ \ldots \right]$ is applied $i$ times. Substituting \refEq{eq:recursionRelSecondIteration} into the last term of \refEq{eq:recursionRel}, we see that \refEq{eq:recursionRel} is a recursion relation that can be put in the form of an infinite series,
\begin{align}
   \left. \int_{\theta_{\alpha}}^{\theta} \right|_{\psi} d \theta' \Big( g \left( \theta', z \left( \theta' \right) \right) - \overline{g \left( \theta', z \right)} \Big) &= \sum_{p = 1}^{\infty} \frac{\left( - 1 \right)^{p-1}}{m_{c}^{p}} \left( \Lambda^{p} \left[ \left. \frac{\partial^{p-1} g}{\partial \theta^{p-1}} \right|_{z} \right] \left( \theta, z \left( \theta \right) \right) \right. \\
   &- \left. \Lambda^{p} \left[ \left. \frac{\partial^{p-1} g}{\partial \theta^{p-1}} \right|_{z} \right]  \left( \theta_{\alpha}, z \left( \theta_{\alpha} \right) \right) \right) . \nonumber
\end{align}
Finally choosing the form $g \left( \theta, z \right) = F_{\alpha} \left( \theta, z \right)$ and rearranging we can calculate the integral appearing in the geometric coefficients (see \refEq{eq:GalphaThetaDef}) to be
\begin{align}
   G_{\alpha} \left( \theta, z \left( \theta \right) \right) &= \left.\int_{\theta_{\alpha}}^{\theta} \right|_{\psi} d \theta' \overline{F_{\alpha} \left( \theta', z \right)} + \sum_{p = 1}^{\infty} \frac{\left( - 1 \right)^{p-1}}{m_{c}^{p}} \left( \Lambda^{p} \left[ \left. \frac{\partial^{p-1} F_{\alpha}}{\partial \theta^{p-1}} \right|_{z} \right] \left( \theta, z \left( \theta \right) \right) \right. \nonumber \\
   &- \left. \Lambda^{p} \left[ \left. \frac{\partial^{p-1} F_{\alpha}}{\partial \theta^{p-1}} \right|_{z} \right] \left( \theta_{\alpha}, z \left( \theta_{\alpha} \right) \right) \right) . \label{eq:generalAlphaInt}
\end{align}

For the untilted case we can set $\theta_{\alpha} = d \theta_{\alpha} / d \psi = 0$ and use \refEq{eq:generalAlphaInt} to prove that the quantity appearing in the geometric coefficients (see \refEq{eq:QgeoSimple}) is
%\begin{align}
%   G_{\alpha} \left( \theta, z_{u} \left( \theta \right) \right) &+ H_{\alpha} = \left. \int_{0}^{\theta} \right|_{\psi} d \theta' F_{\alpha} \left( \theta', z_{u} \left( \theta' \right) \right) .
%\end{align}
%Hence substituting \refEq{eq:generalAlphaInt} (remembering that $\theta_{\alpha} = 0$) gives
\begin{align}
   G_{\alpha} \left( \theta, z_{u} \left( \theta \right) \right) &+ H_{\alpha} = \left. \int_{0}^{\theta} \right|_{\psi} d \theta' \overline{F_{\alpha} \left( \theta', z \right)} \label{eq:untiltAlphaInt} \\
   &+ \sum_{p = 1}^{\infty} \frac{\left( - 1 \right)^{p-1}}{m_{c}^{p}} \left( \Lambda^{p} \left[ \left. \frac{\partial^{p-1} F_{\alpha}}{\partial \theta^{p-1}} \right|_{z} \right] \left( \theta, z_{u} \left( \theta \right) \right) - \Lambda^{p} \left[ \left. \frac{\partial^{p-1} F_{\alpha}}{\partial \theta^{p-1}} \right|_{z} \right] \left( 0, 0 \right) \right) . \nonumber
\end{align}
In the tilted case ($z = z_{t} = m_{c} \left( \theta + \theta_{t} \right)$) we can carefully choose
\begin{align}
   \frac{d \theta_{\alpha}}{d \psi} &= \left[ \frac{1}{R^{2} B_{p}^{2}} \left. \frac{\partial l_{p}}{\partial \theta} \right|_{\psi} \right]_{\theta = \theta_{\alpha}}^{-1} \left\{ \left. \int_{\theta_{\alpha}}^{0} \right|_{\psi} d \theta' \overline{F_{\alpha} \left( \theta', z \right)} \right. \label{eq:thetaAlphaDerivTilted} \\
   &+ \left. \sum_{p = 1}^{\infty} \frac{\left( - 1 \right)^{p-1}}{m_{c}^{p}} \left( \Lambda^{p} \left[ \left. \frac{\partial^{p-1} F_{\alpha}}{\partial \theta^{p-1}} \right|_{z} \right] \left( 0, 0 \right) - \Lambda^{p} \left[ \left. \frac{\partial^{p-1} F_{\alpha}}{\partial \theta^{p-1}} \right|_{z} \right] \left( \theta_{\alpha}, z_{t} \left( \theta_{\alpha} \right) \right) \right) \right\} \nonumber
\end{align}
to substitute into \refEq{eq:HalphaThetaDef}, giving 
\begin{align}
   G_{\alpha} \left( \theta, z_{t} \left( \theta \right) \right) &+ H_{\alpha} = \left. \int_{0}^{\theta} \right|_{\psi} d \theta' \overline{F_{\alpha} \left( \theta', z \right)} \label{eq:tiltAlphaInt} \\
   &+ \sum_{p = 1}^{\infty} \frac{\left( - 1 \right)^{p-1}}{m_{c}^{p}} \left( \Lambda^{p} \left[ \left. \frac{\partial^{p-1} F_{\alpha}}{\partial \theta^{p-1}} \right|_{z} \right] \left( \theta, z_{t} \left( \theta \right) \right) - \Lambda^{p} \left[ \left. \frac{\partial^{p-1} F_{\alpha}}{\partial \theta^{p-1}} \right|_{z} \right] \left( 0, 0 \right) \right) . \nonumber
\end{align}
This exactly matches \refEq{eq:untiltAlphaInt} (except for replacing $z_{u}$ with $z_{t}$) and means the entire effect of the tilt can be contained in the functional form of $z$. To make things as simple as possible we also choose
\begin{align}
   \theta_{\alpha} = 0 \label{eq:thetaAlphaTilted}
\end{align}
for the tilted case.

The choice of $d \theta_{\alpha} / d \psi$ in \refEq{eq:thetaAlphaDerivTilted} means that the geometric coefficients for both the untilted and tilted cases can be written in the form
\begin{align}
   Q_{\text{geo}} \left( \theta, z \right) , \label{eq:QgeoFinal}
\end{align}
where $z = z_{u}$ for the untilted case and $z = z_{t}$ for the tilted case. Therefore, we know that
\begin{align}
   Q_{\text{geo}}^{t} \left( \theta, z_{u} \right) = Q_{\text{geo}}^{u} \left( \theta, z_{u} + m_{c} \theta_{t} \right)
\end{align}
for each of the geometric coefficients, where the superscript $u$ indicates the quantity in the untilted configuration and the superscript $t$ indicates the tilted configuration.

Since all of the geometric coefficients depend separately on both $\theta$ and $z$ we know that $h_{s}$, $\phi$, $A_{||}$, and $B_{||}$ must also. Apart from the geometric coefficients, the only way the poloidal coordinate enters the gyrokinetic equations is through the poloidal derivative in the streaming term (i.e. the second term of \refEq{eq:gyrokineticEq}). However, \refEq{eq:poloidalDerivTransform} is appropriate for both $z = z_{u}$ and $z = z_{t}$. Hence, using any solution to the gyrokinetic equation for the untilted case, $\left\{ h_{s}^{u} \left( \theta, z_{u} \right), \phi^{u} \left( \theta, z_{u} \right), A_{||}^{u} \left( \theta, z_{u} \right), B_{||}^{u} \left( \theta, z_{u} \right) \right\}$, we can construct a solution for the tilted case,
\begin{align}
  &\left\{ h_{s}^{t} \left( \theta, z_{u} \right), \phi^{t} \left( \theta, z_{u} \right), A_{||}^{t} \left( \theta, z_{u} \right), B_{||}^{t} \left( \theta, z_{u} \right) \right\} \label{eq:symSolToAsymSol} \\ 
  &\hspace{3em} = \left\{ h_{s}^{u} \left( \theta, z_{u} + m_{c} \theta_{t} \right), \phi^{u} \left( \theta, z_{u} + m_{c} \theta_{t} \right), A_{||}^{u} \left( \theta, z_{u} + m_{c} \theta_{t} \right), B_{||}^{u} \left( \theta, z_{u} + m_{c} \theta_{t} \right) \right\} , \nonumber
\end{align}
given our choices for the free parameter $\theta_{\alpha} \left( \psi \right)$ in the definition of $\alpha$ (i.e. \refEq{eq:thetaAlphaDerivTilted} and \refEq{eq:thetaAlphaTilted}). Because the average over $z$ (see \refEq{eq:zAvg}) can always be shifted by $m_{c} \theta_{t}$ without affecting the result these two solution sets give the same large scale turbulent fluxes and turbulent energy exchange between species, e.g. in the electrostatic limit they are
\begin{align}
  \Gamma_{s}^{\phi} &= \frac{4 \pi^{2} i}{m_{s} V'} \left\langle \sum_{k_{\psi}, k_{\alpha}} k_{\alpha} \oint d \theta \overline{J B \phi \left( k_{\psi}, k_{\alpha} \right) \int dw_{||} d \mu ~ h_{s} \left( - k_{\psi}, - k_{\alpha} \right) J_{0} \left( k_{\perp} \rho_{s} \right)} \right\rangle_{\Delta t} \label{eq:partFluxAvg} \\
  \Pi_{\zeta s}^{\phi} &= \frac{4 \pi^{2} i}{V'} \left\langle \sum_{k_{\psi}, k_{\alpha}} k_{\alpha} \oint d \theta \overline{J B \phi \left( k_{\psi}, k_{\alpha} \right) \int dw_{||} d \mu ~ h_{s} \left( - k_{\psi}, - k_{\alpha} \right)} \right. \label{eq:momFluxAvg} \\
   & \left. \overline{\times \left[ \left( \frac{I}{B} w_{||} + R^{2} \Omega_{\zeta} \right) J_{0} \left( k_{\perp} \rho_{s} \right) + \frac{i}{\Omega_{s}} \frac{k^{\psi}}{B} \frac{\mu B}{m_{s}} \frac{2 J_{1} \left( k_{\perp} \rho_{s} \right)}{k_{\perp} \rho_{s}} \right]} \right\rangle_{\Delta t} \nonumber \\
  Q_{s}^{\phi} &= \frac{4 \pi^{2} i}{V'} \left\langle \sum_{k_{\psi}, k_{\alpha}} k_{\alpha} \oint d \theta \overline{J B \phi \left( k_{\psi}, k_{\alpha} \right) \int dw_{||} d \mu ~ h_{s} \left( - k_{\psi}, - k_{\alpha} \right)} \right. \label{eq:heatFluxAvg} \\
   & \left. \overline{\times \left( \frac{w^{2}}{2} + \frac{Z_{s} e \Phi_{0}}{m_{s}} - \frac{1}{2} R^{2} \Omega_{\zeta}^{2} \right) J_{0} \left( k_{\perp} \rho_{s} \right)} \right\rangle_{\Delta t} \nonumber \\
   P_{Q s}^{\phi} &= \frac{4 \pi^{2}}{V'} \left\langle \sum_{k_{\psi}, k_{\alpha}} \oint d \theta \overline{J \Omega_{s} \frac{\partial}{\partial t} \left( \phi \left( k_{\psi}, k_{\alpha} \right) \right) \int dw_{||} d \mu ~ h_{s} \left( - k_{\psi}, - k_{\alpha} \right) J_{0} \left( k_{\perp} \rho_{s} \right)} \right\rangle_{\Delta t} . \label{eq:heatingAvg}
\end{align}
Looking at the full electromagnetic fluxes and the turbulent energy exchange between species (see appendix \ref{app:fluxes}) we see that they also remain unchanged by the tilt.

Since we relied on expanding in $m_{c} \gg 1$ to separate scales in \refEq{eq:partFluxAvg} through \refEq{eq:heatingAvg}, this argument can only give the fluxes as an expansion in powers of $1 / m_{c}$, not the unexpanded quantity. We already know that, since the two configuration are not \textit{exactly} identical, they will in general produce different fluxes. However, the above argument proves the two configurations must have the same fluxes to all orders in $1 / m_{c}$. This demonstrates that, while the fluxes from the two configurations can be different, the difference does not scale polynomially and so cannot scale more strongly than $\sim \Exp{- \beta m_{c}^{\gamma}}$, where $\beta$ and $\gamma$ are both positive and do not depend on $m_{c}$.

%===================================================%
\subsection{Accuracy of the local equilibrium approximation}
\label{subsec:localEqApprox}
%===================================================%

We finish with an important remark concerning our use of an approximate local MHD equilibrium, as opposed to the full global MHD equilibria. Although there was no problem in the Miller local equilibrium, it may not be possible to exactly tilt the fast flux surface shaping poloidally in a real global equilibrium. We can always Fourier analyse a flux surface shape and its radial derivative (see \refEq{eq:fluxSurfofInterestShapeWeak} and \refEq{eq:fluxSurfofInterestDerivWeak}). We can also use the external shaping coils to arbitrarily tilt the fast shaping of the flux surface of interest. However, the way that the radial derivative of the flux surface shape changes with tilt is set by the global MHD equilibrium and is not under our control (as it is in the Miller local equilibrium approximation). The global equilibrium in a screw pinch has cylindrical symmetry, but in a tokamak toroidal effects may preclude tilting the radial derivative of the flux surface shape in exactly the same manner we tilted the flux surface shape itself.

This means that, strictly speaking, when we introduce $z_{t} \left( \theta \right) = m_{c} \left( \theta + \theta_{t} \right)$ into the derivative of the flux surface shape we may no longer be modelling a physically possible tokamak. However, we can construct a proof by induction to show that the effect of toroidicity on mode $m_{c}$ must be exponentially small in $m_{c} \gg 1$. We start by rearranging the Grad-Shafranov equation (i.e. \refEq{eq:gradShafEq}) as
\begin{align}
   R \Nabla \cdot \left( \frac{\Nabla \psi}{R} \right) + I \frac{d I}{d \psi} = \frac{1}{R} \vec{\nabla} R \cdot \vec{\nabla} \psi - \mu_{0} R^{2} \left. \frac{\partial p}{\partial \psi} \right|_{R} . \label{eq:gradShafranovMod}
\end{align}
The left side of this equation is completely cylindrically symmetric, while the right side contains all of the toroidal effects. To $\Order{B_{0}}$ in the large aspect ratio limit the Grad-Shafranov equation does not include toroidicity (see \refEq{eq:gradShafLowestOrder}). This is the base case of the inductive argument and demonstrates that $\psi_{0}$ has tilting symmetry, i.e. $\psi_{0}^{t} \left( \theta, z \right) = \psi_{0}^{u} \left( \theta, z + m_{c} \theta_{t} \right)$. To $\Order{\epsilon^{i} B_{0}}$ the Grad-Shafranov equation only depends on $R$, derivatives of $R$, and $\psi_{j} \left( \theta, z \right)$ (where $i$ and $j$ are integers and $0 \leq j < i$). During the proof of the gyrokinetic tilting symmetry (see section \ref{subsec:gyroSym}), we showed that $R$ and the derivatives of $R$ have the appropriate tilting symmetry. Furthermore, by the complete induction hypothesis we know that $\psi_{j} \left( \theta, z \right)$ is tilting symmetric for all $0 \leq j < i$. Therefore, $\psi$ follows the same tilting symmetry as gyrokinetics to all orders. Hence, tilting the radial derivative of the flux surface shape in the same manner as the flux surface shape produces a local Miller equilibrium that only differs from the actual equilibrium by an exponentially small error in $m_{c} \gg 1$.

As an illustration of this, reference \cite{HakkarainenEquilibrium1990} uses an equilibrium that is circular to lowest order in aspect ratio to show that at $\Order{B_{0}}$ the Grad-Shafranov equation is entirely cylindrically symmetric, at $\Order{\epsilon B_{0}}$ toroidicity introduces a natural shift (i.e. the Shafranov shift), at $\Order{\epsilon^{2} B_{0}}$ toroidicity introduces a natural elongation, and at $\Order{\epsilon^{3} B_{0}}$ toroidicity introduces a natural triangularity. This indicates that, in a global equilibrium, toroidicity introduces an $O \left( \epsilon^{m} \right)$ modification to the $m^{\text{th}}$ Fourier mode of a flux surface. Therefore, the error introduced into the geometric coefficients by ignoring this effect in the local equilibrium approximation is $O \left( \epsilon^{m_{\text{min}}} \right)$, where $m_{\text{min}}$ is the smallest mode number that is tilted. This error is exponentially small in $m_{c} \gg 1$, hence it does not change our result that tilting the equilibrium has an exponentially small effect on the turbulent fluxes.

%===================================================%
\section{Numerical results}
\label{sec:numResultsSym}
%===================================================%

In this section we will give numerical results to test the analytic conclusions of the previous section. We use GS2 to calculate the nonlinear turbulent fluxes of momentum and energy generated by a given geometry. We investigate the influence of the shape of the flux surface of interest by scanning $m_{c}$, the characteristic mode number of the fast poloidal shaping. The geometry is specified using the generalisation of the Miller local equilibrium model presented in chapter \ref{ch:MHD_LocalEquil}. The flux surface shapes (shown in figure \ref{fig:simGeoMirror}) are parameterized by \refEq{eq:fluxSurfaceSpecScaleSep} through \refEq{eq:zDef} as well as \refEq{eq:rPsiDef}, with only one high order mode, $m = m_{c}$ (which corresponds to $l = 1$ and $k = 0$). We will choose
\begin{align}
 \Delta_{m} - 1 = \frac{3}{2} m_{c}^{-2} \label{eq:reasonableShaping}
\end{align}
because it has a physical scaling (see appendix \ref{app:maxShaping}) and seems reasonably similar to the typical magnitude of flux surface shaping effects in experiments. For example, regular polygons have $\Delta_{m} - 1 = \Sec{\pi / m} - 1 \sim m^{-2}$, so we see that exceeding this scaling necessarily leads to flux surfaces with convex regions. With the exception of ``bean-shaped'' tokamaks \cite{GrimmBeanShape1985}, practically all proposed experimental configurations have purely concave flux surfaces, so we know they respect this scaling. We will also take $d \Delta_{m} / d r_{\psi} = \left( m_{c} - 2 \right) \left( \Delta_{m} - 1 \right) / r_{\psi 0}$ (appropriate for a constant current profile according to \refEq{eq:shapingRadialDeriv_rPsi}) and $d \theta_{t m} / d r_{\psi} = 0$ (appropriate for a constant current profile according to \refEq{eq:tiltAngleDeriv}) in the scan to give the neighbouring flux surfaces a reasonable shape. Up-down asymmetric geometries are created by fixing the tilt angle at $\theta_{t m} = \pi / \left( 2 m_{c} \right)$, the angle halfway between neighbouring up-down symmetric configurations (at $\theta_{t m} = 0$ and $\theta_{t m} = \pi / m_{c}$), while the up-down symmetric configurations have $\theta_{t m} = 0$.

Except for the shape, all parameters of the flux surface of interest are fixed at Cyclone base case values \cite{DimitsCycloneBaseCase2000} of
\begin{align}
 \rho_{0} &\equiv a_{\psi 0} / a = 0.54, \hspace{3em}
 R_{c 0} / a = 3, \hspace{5.35em}
 q = 1.4,  \label{eq:cycloneBaseCase} \\
 \hat{s} &= 0.8, \hspace{7.25em}
 a / L_{T s} = 2.3, \hspace{2.5em}
 a / L_{n s} = 0.733 \nonumber
\end{align}
for the minor radius, major radius, safety factor, magnetic shear, temperature gradient, and density gradient respectively. All simulations are electrostatic and collisionless. The fluxes calculated by GS2 are normalised to their gyroBohm values of
\begin{align}
 \Pi_{gB} &\equiv \rho_{\ast}^{2} n_{i} a m_{i} v_{th i}^{2} \label{eq:momFluxGyroBohm} \\
 Q_{gB} &\equiv \rho_{\ast}^{2} n_{i} T_{i} v_{th i} , \label{eq:energyFluxGyroBohm}
\end{align}
where $n_{i}$ is the ion density, $m_{i}$ is the ion mass, $T_{i}$ is the local ion temperature, and $v_{th i} \equiv \sqrt{2 T_{i} / m_{i}}$ is the local ion thermal speed.

We will compare the numerical scans in $m_{c}$ (shown in figure \ref{fig:simGeoMirror}) to the analytic theory in two different manners. From \refEq{eq:symSolToAsymSol} we expect that, using the poloidal distribution of any flux from a given geometry, it should be possible to predict the flux from any geometry that is identical except for a poloidal tilt of the fast variation. First, we will directly investigate this by comparing the poloidal dependence of the fluxes of particles, momentum, and energy in just such geometries.  Then, we will show that the change in the total fluxes due to tilting fast shaping disappears in the limit of $m_{c} \gg 1$.

%===================================================%
\subsection{Poloidal structure of fluxes}
\label{subsec:polStructComp}
%===================================================%

\begin{figure}
 \centering

 \includegraphics[width=0.18\textwidth]{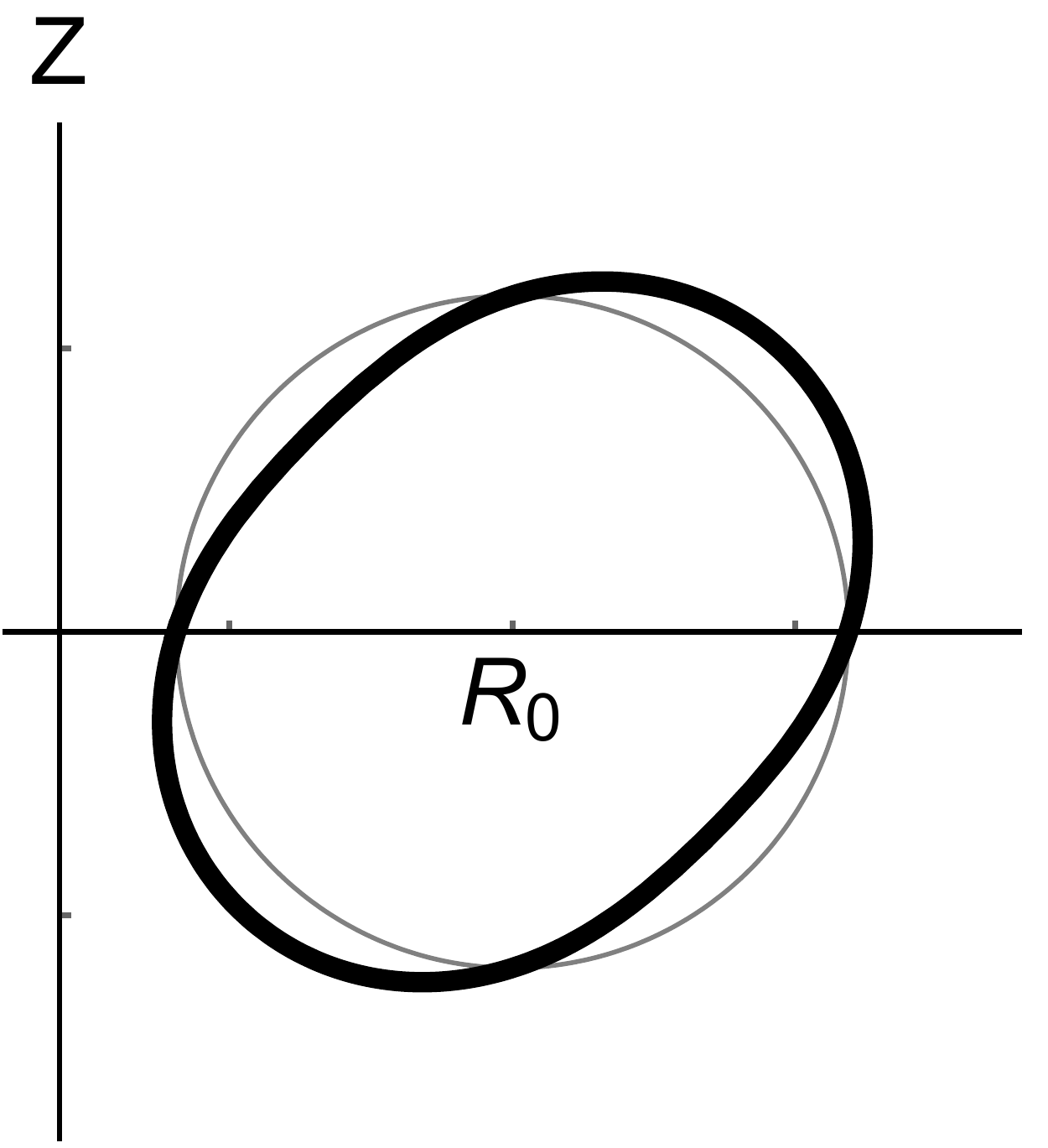}
 \includegraphics[width=0.18\textwidth]{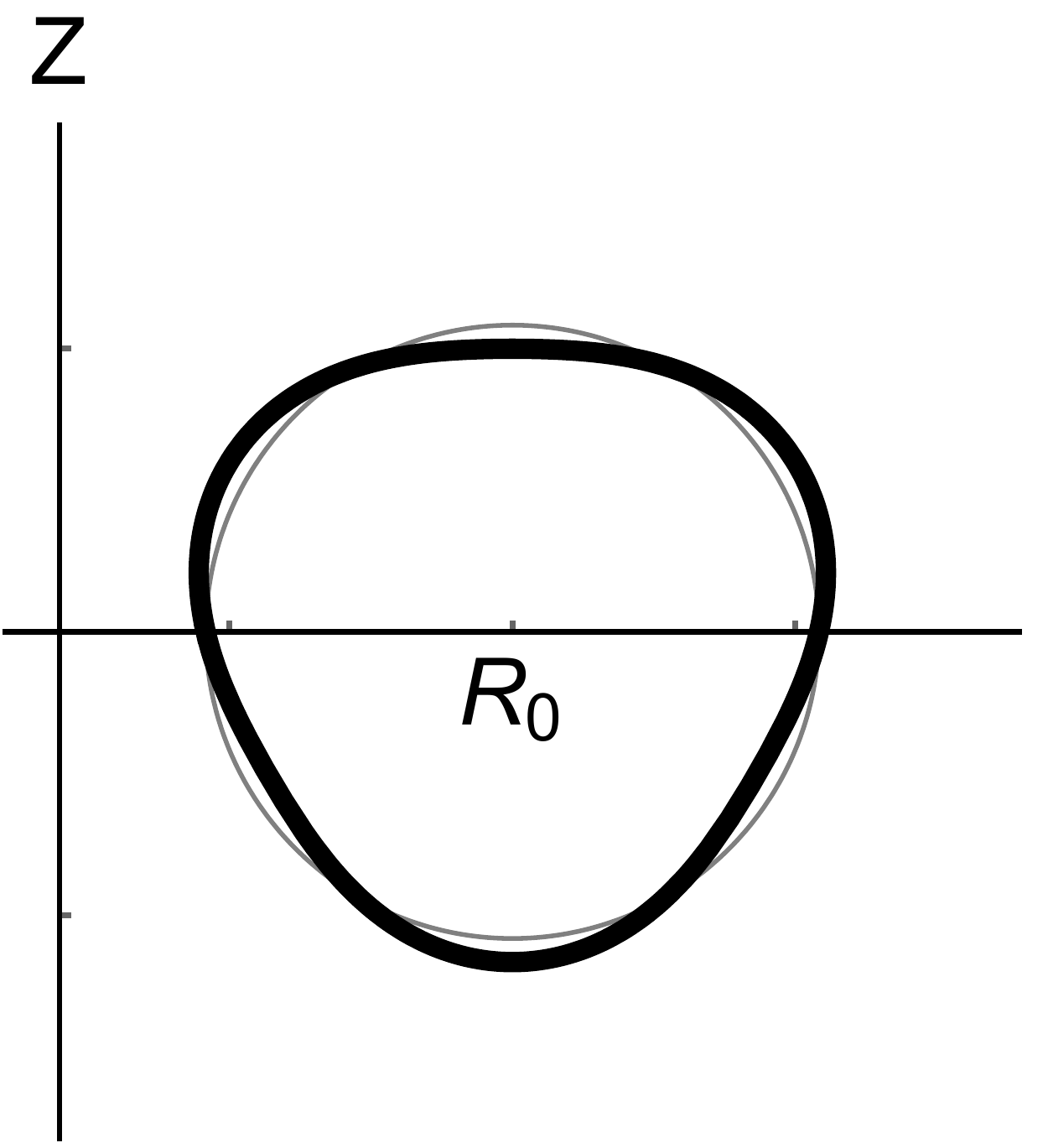}
 \includegraphics[width=0.18\textwidth]{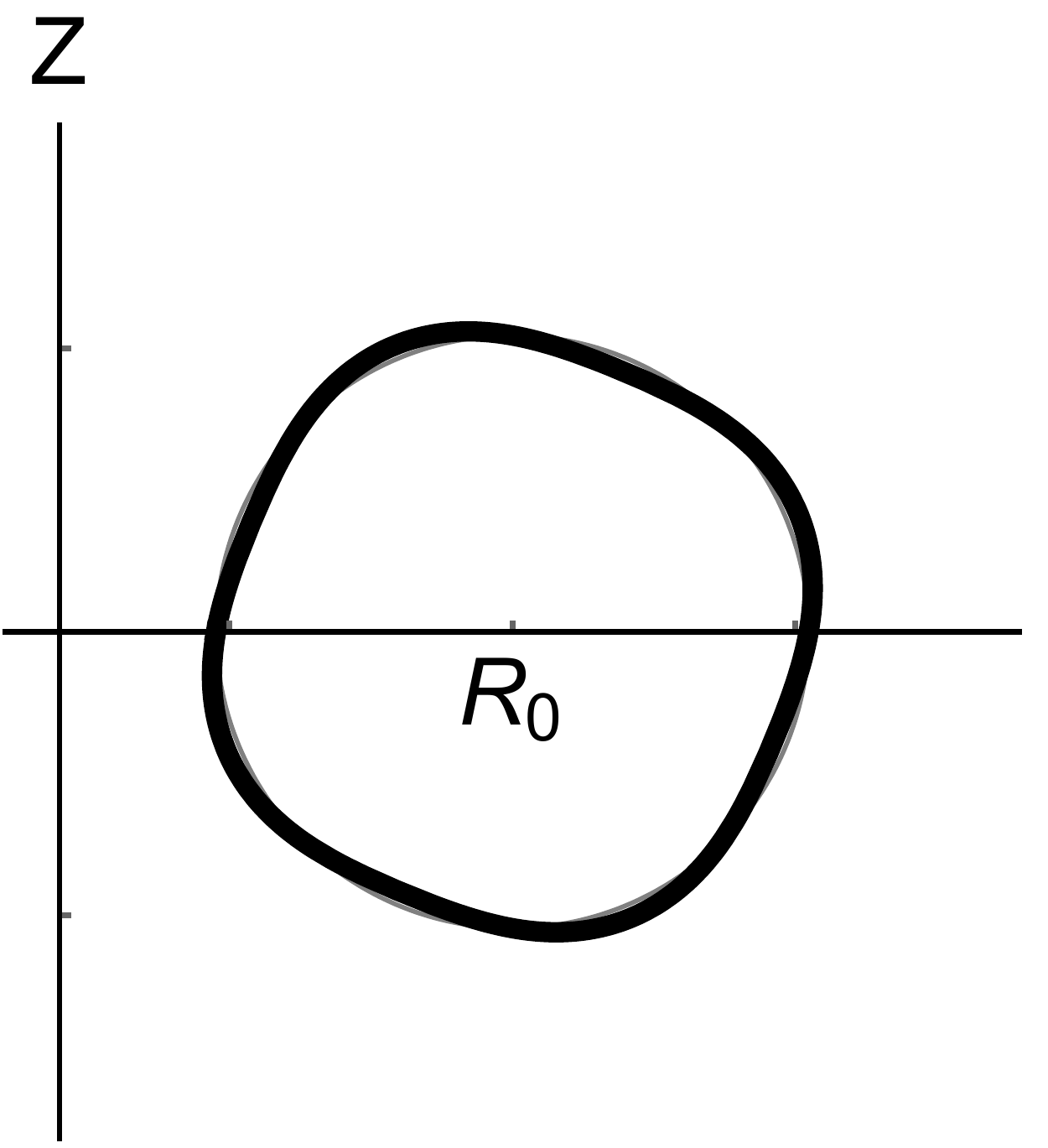}
 \includegraphics[width=0.18\textwidth]{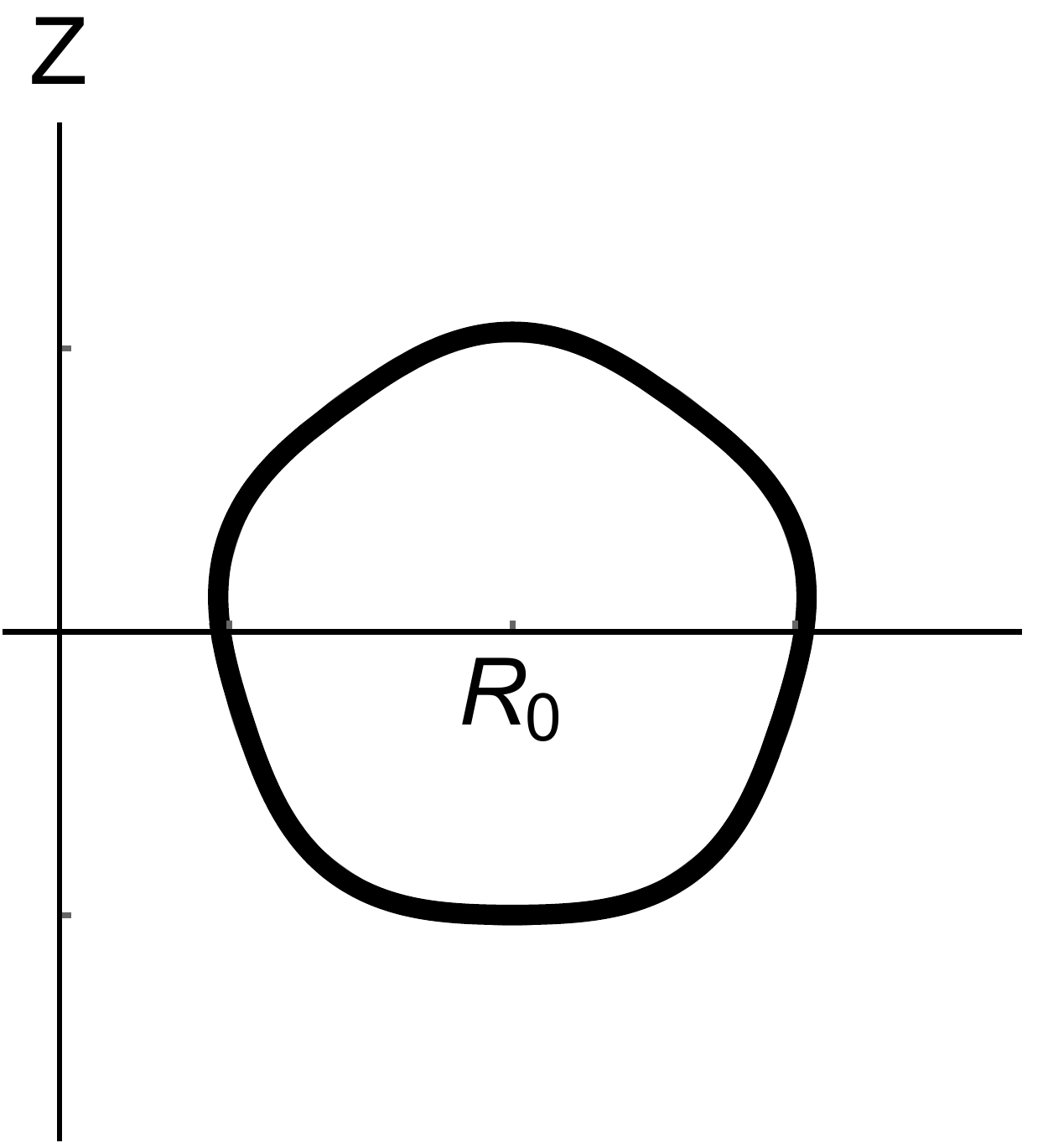}
 \includegraphics[width=0.18\textwidth]{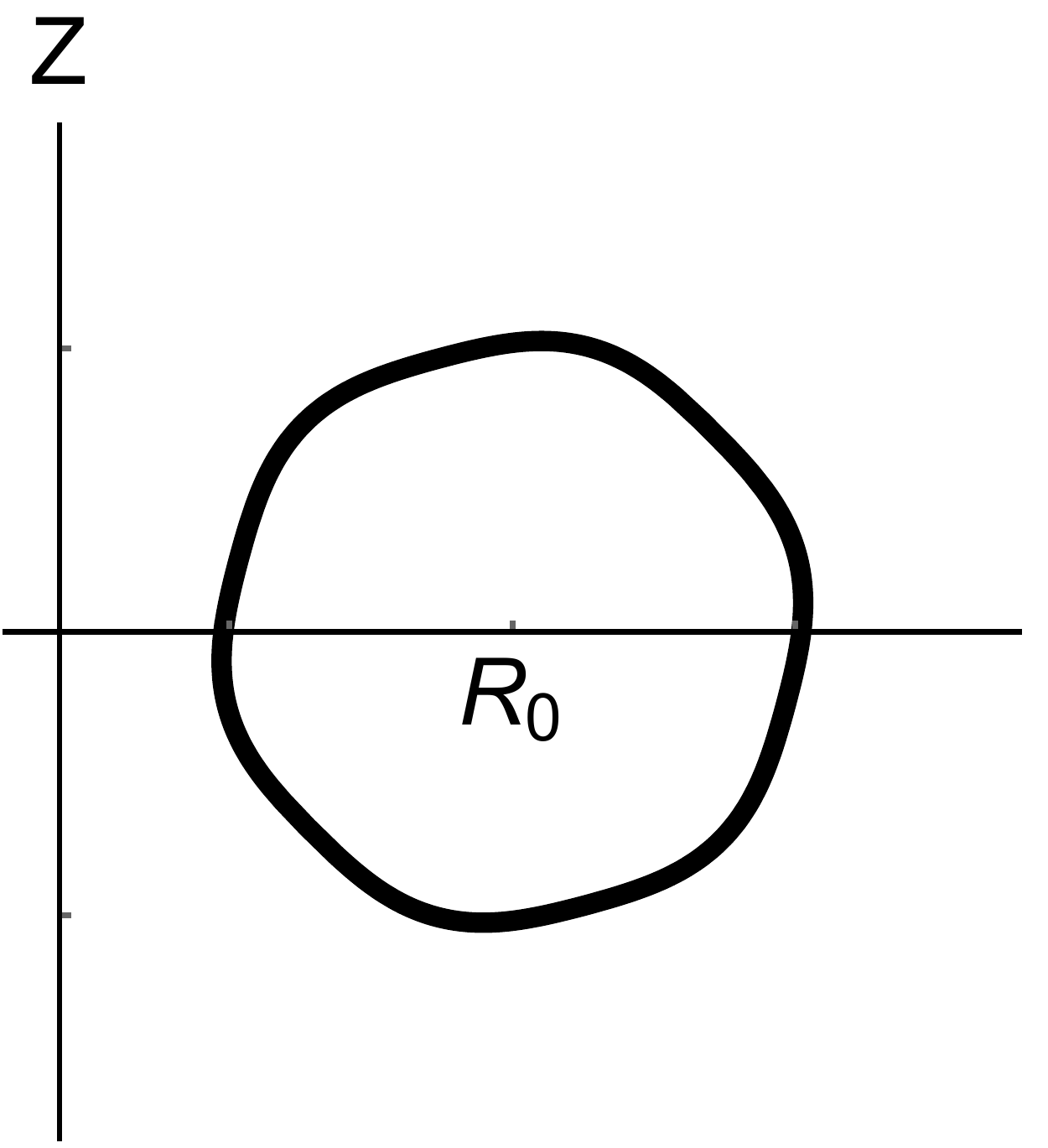}
 \includegraphics[width=0.0111\textwidth]{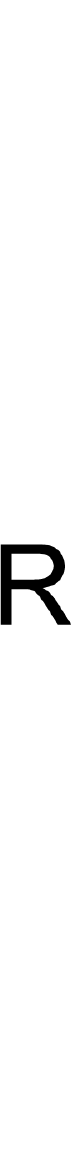}

 \includegraphics[width=0.18\textwidth]{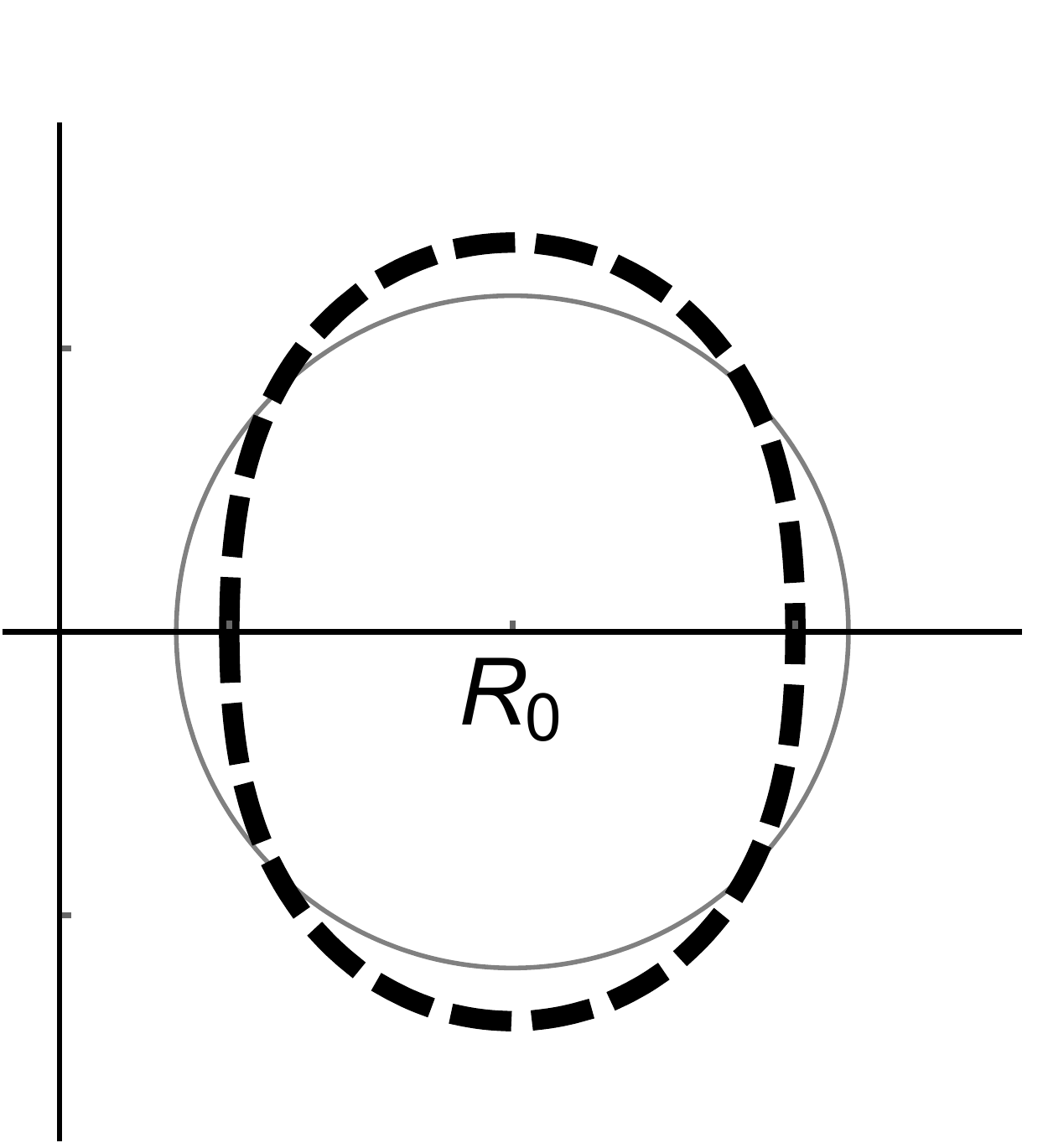}
 \includegraphics[width=0.18\textwidth]{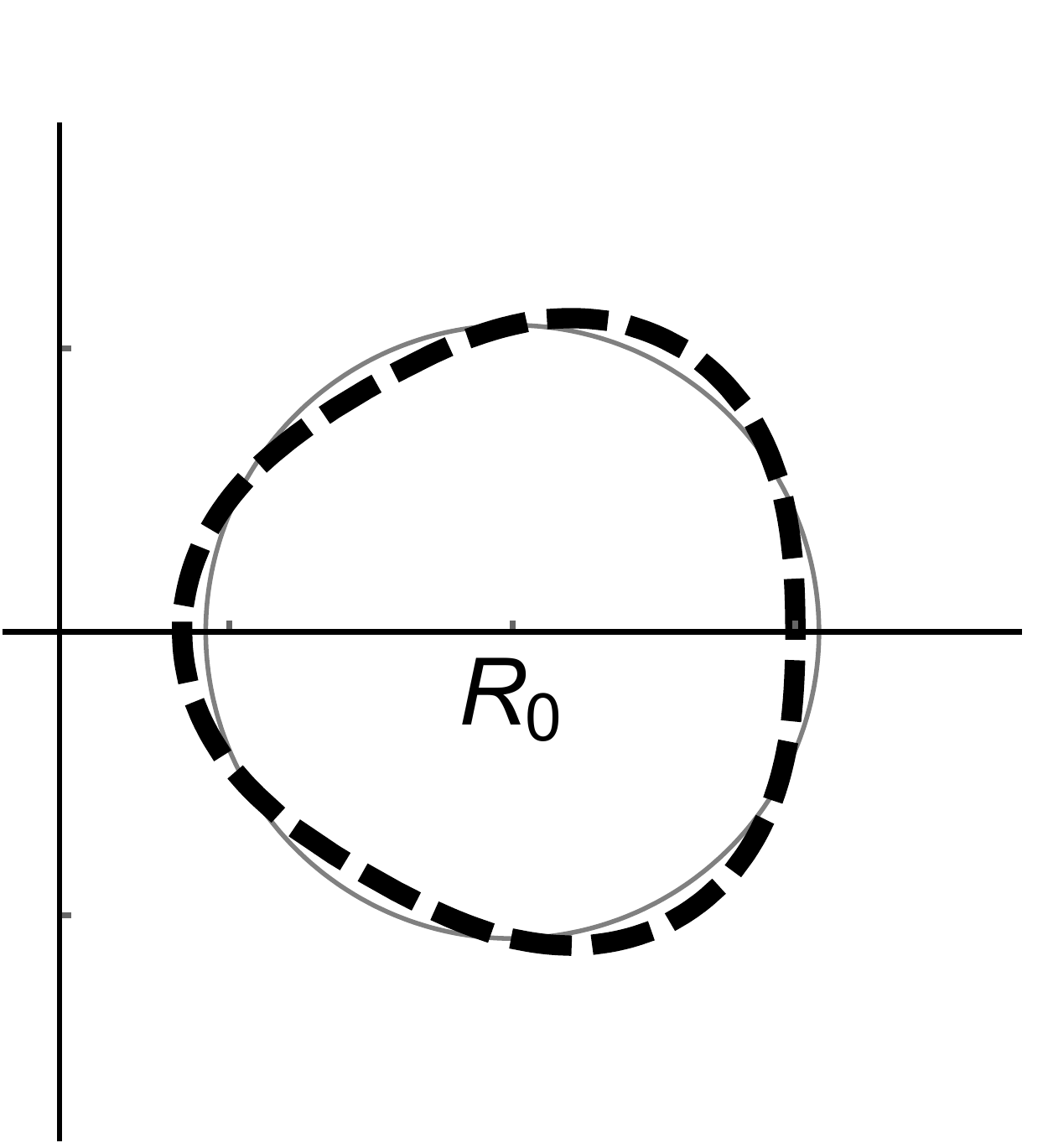}
 \includegraphics[width=0.18\textwidth]{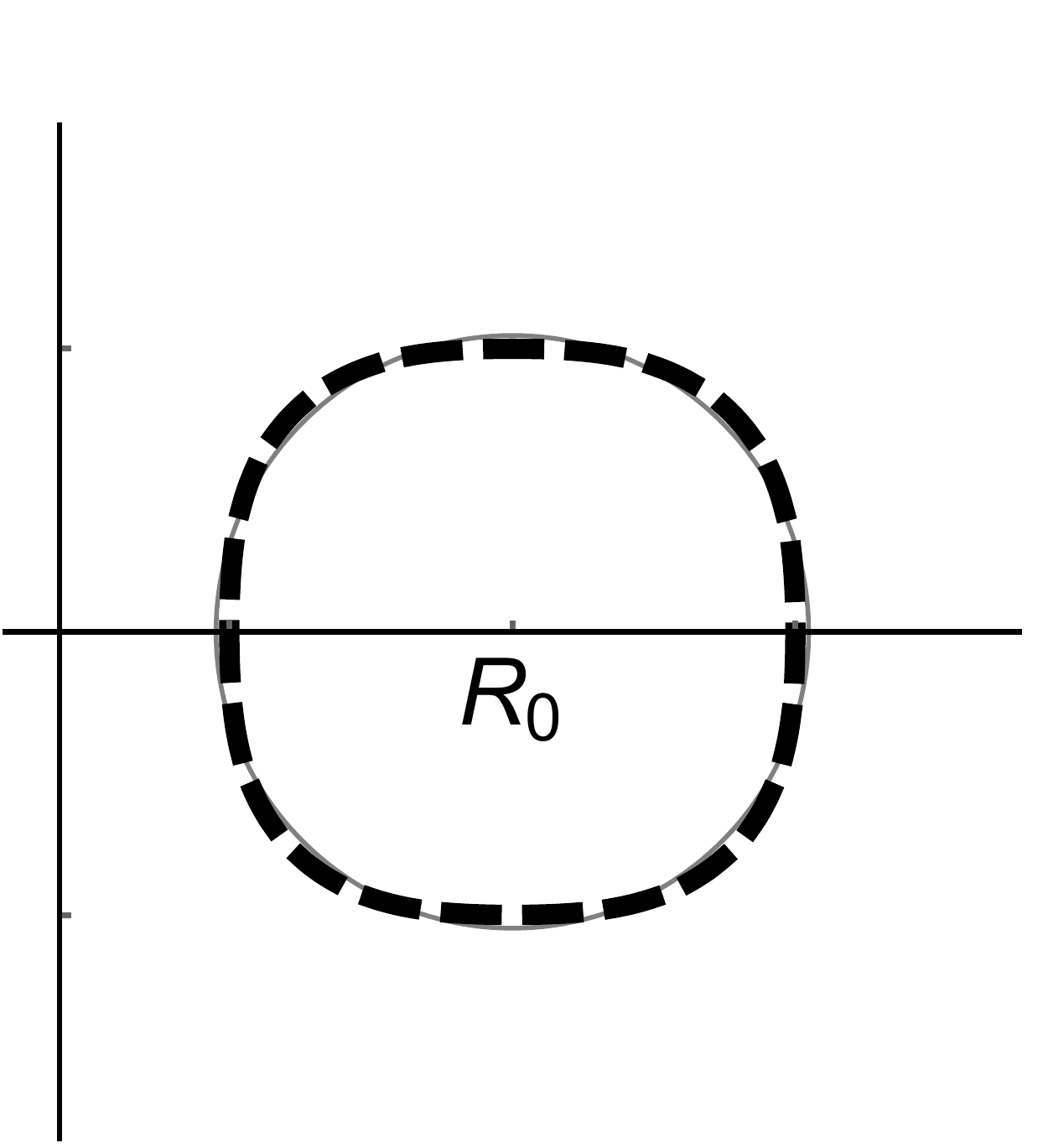}
 \includegraphics[width=0.18\textwidth]{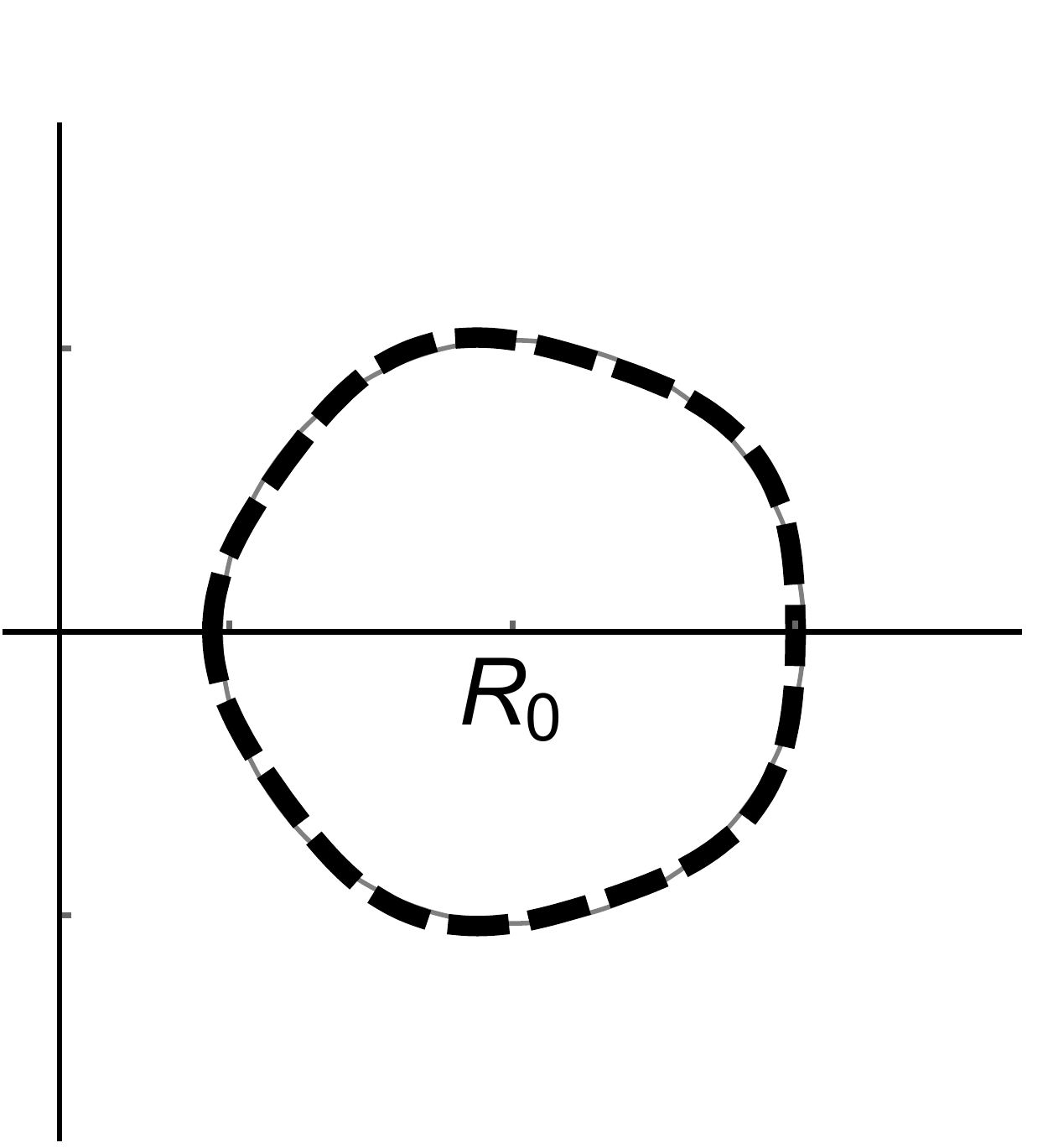}
 \includegraphics[width=0.18\textwidth]{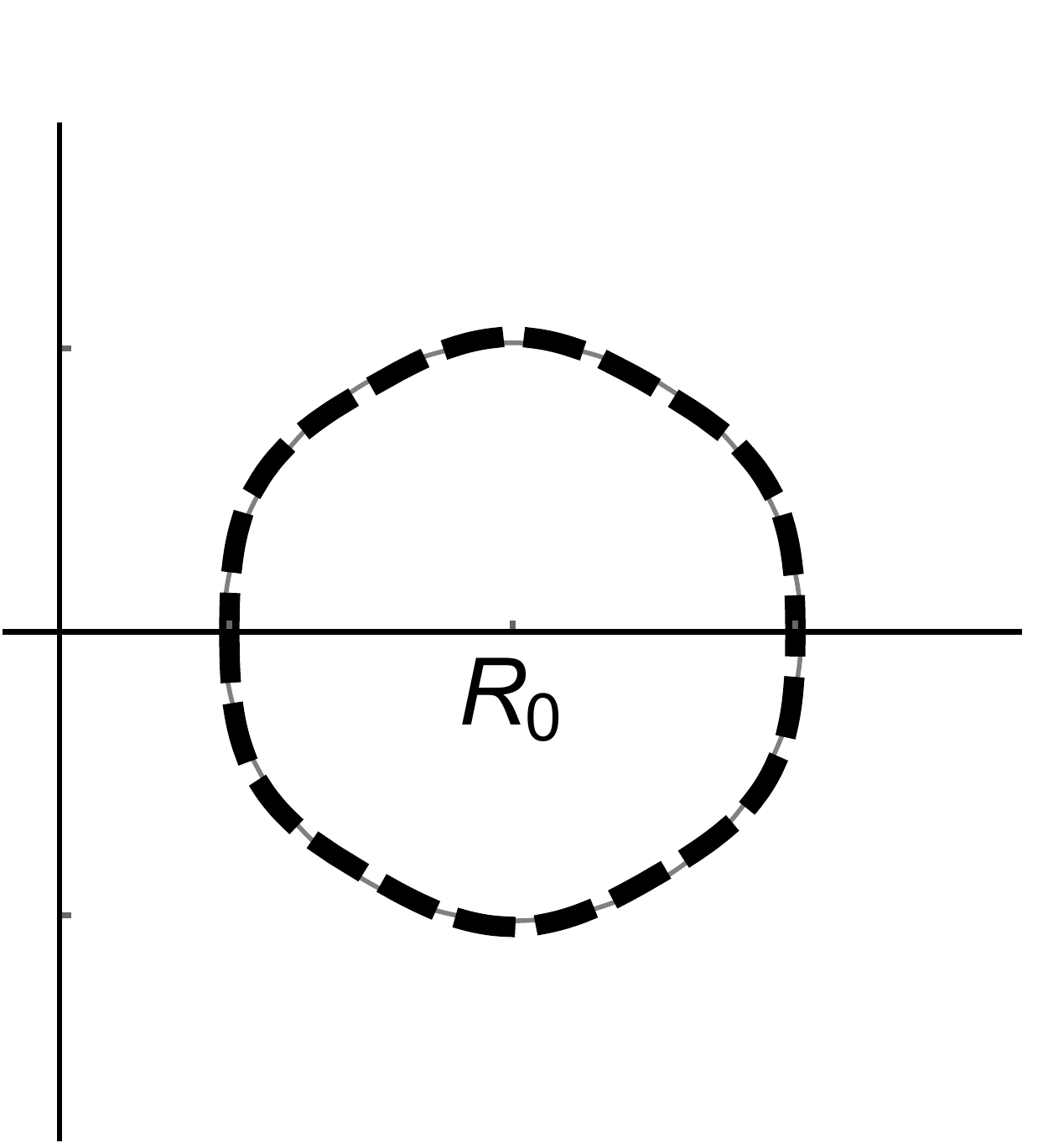}
 \includegraphics[width=0.0111\textwidth]{figs/Ch_07/xAxisLabelR.pdf}

 \caption{The $m_{c}=2$ through $m_{c}=6$ flux surface geometries in the tilted (solid) and up-down symmetric (dashed) configurations, with circular flux surfaces shown for comparison (grey).}
 \label{fig:simGeoMirror}
\end{figure}

In section \ref{subsec:gyroSym}, we presented an analytic argument showing that (when expanding in $m_{c} \gg 1$) the solution to the gyrokinetic equation for a given geometry can be used to generate the solution to any geometry that is identical, except for a global tilt of the fast poloidal variation. This relationship, given by \refEq{eq:symSolToAsymSol}, makes predictions about the poloidal distributions of fluxes. In appendix \ref{app:fluxes} we define the poloidal distribution of the different quantities that appear in the gyrokinetic model. In the electrostatic limit they are given by
\begin{align}
  \gamma_{s}^{\phi} &\equiv - R \left\langle \left\langle \int d^{3} w \underline{h}_{s} \hat{e}_{\zeta} \cdot \delta \vec{E} \right\rangle_{\Delta \psi} \right\rangle_{\Delta t} \label{eq:polDistPartFlux} \\
  \pi_{\zeta s}^{\phi} &\equiv - R \left\langle \left\langle \int d^{3} w \underline{h}_{s} m_{s} R \left( \vec{w} \cdot \hat{e}_{\zeta} + R \Omega_{\zeta} \right) \hat{e}_{\zeta} \cdot \delta \vec{E} \right\rangle_{\Delta \psi} \right\rangle_{\Delta t} \label{eq:polDistMomFlux} \\
  q_{s}^{\phi} &\equiv - R \left\langle \left\langle \int d^{3} w \underline{h}_{s} \left( \frac{m_{s}}{2} w^{2} + Z_{s} e \Phi_{0} - \frac{m_{s}}{2} R^{2} \Omega_{\zeta}^{2} \right) \hat{e}_{\zeta} \cdot \delta \vec{E} \right\rangle_{\Delta \psi} \right\rangle_{\Delta t} \label{eq:polDistHeatFlux} \\
  p_{Q s}^{\phi} &\equiv \left\langle \left\langle \int d^{3} w Z_{s} e \underline{h}_{s} \frac{\partial \underline{\phi}}{\partial t} \right\rangle_{\Delta \psi} \right\rangle_{\Delta t} , \label{eq:polDistHeating}
\end{align}
which are just \refEq{eq:partFluxDef}, \refEq{eq:momFluxDef}, \refEq{eq:heatFluxDef}, and \refEq{eq:heatingDef} without the flux surface average (e.g. $Q_{s} = \left\langle q_{s} \right\rangle_{\psi}$). Specifically, using \refEq{eq:symSolToAsymSol} the analytic theory predicts that we should find
\begin{align}
  \gamma_{s}^{t} \left( \theta, z \right) &= \gamma_{s}^{u} \left( \theta, z + m_{c} \theta_{t} \right) \label{eq:symSolToAsymSolPart} \\
  \pi_{\zeta s}^{t} \left( \theta, z \right) &= \pi_{\zeta s}^{u} \left( \theta, z + m_{c} \theta_{t} \right) \label{eq:symSolToAsymSolMom} \\
  q_{s}^{t} \left( \theta, z \right) &= q_{s}^{u} \left( \theta, z + m_{c} \theta_{t} \right) \label{eq:symSolToAsymSolHeat} \\
  p_{Q s}^{t} \left( \theta, z \right) &= p_{Q s}^{u} \left( \theta, z + m_{c} \theta_{t} \right) , \label{eq:symSolToAsymSolEnergyExchange}
\end{align}
where the superscript $u$ indicates the geometry is up-down symmetric (i.e. untilted), $t$ indicates the geometry is tilted, and for our chosen geometries $\theta_{t} = \theta_{t m}$. By simulating several up-down symmetric geometries (see the bottom row of figure \ref{fig:simGeoMirror}) and their corresponding tilted geometries (see the top row of figure \ref{fig:simGeoMirror}) we can numerically verify \refEq{eq:symSolToAsymSolPart} through \refEq{eq:symSolToAsymSolEnergyExchange}. We will focus on the ion momentum flux because the symmetry has particularly profound consequences for it, but the analysis in this section can be applied to any of the fluxes.

We should note that GS2 automatically takes $\theta_{\alpha} \left( \psi \right) = 0$ in its definition of $\alpha$ (see \refEq{eq:alphaDef}), so we have to be careful about making numerical predictions from our analytic results. In general, converting between our definition of $\alpha$ and the GS2 definition, $\alpha_{\text{GS2}}$, involves accounting for a shift in $\alpha$ and $\vec{\nabla} \alpha$ of
\begin{align}
   \delta \left( \alpha \right) &= - I \left. \int_{\theta_{\alpha}}^{0} \right|_{\psi} d \theta' \left( R^{2} \vec{B} \cdot \vec{\nabla} \theta' \right)^{-1} \label{eq:GS2alphaShift} \\
   \delta \left( \vec{\nabla} \alpha \right) &= - I \left( \left. \int_{\theta_{\alpha}}^{0} \right|_{\psi} d \theta' F_{\alpha} \left( \theta' \right) - \left[ \frac{1}{R^{2} B_{p}} \left. \frac{\partial l_{p}}{\partial \theta} \right|_{\psi} \right]_{\theta = \theta_{\alpha}} \frac{d \theta_{\alpha}}{d \psi} \right) \vec{\nabla} \psi \label{eq:GS2gradAlphaShift}
\end{align}
respectively. However, given our specific choices in \refEq{eq:thetaAlphaDerivTilted} and \refEq{eq:thetaAlphaTilted} we see that
\begin{align}
   \delta \left( \alpha \right) &= 0 \\
   \delta \left( \vec{\nabla} \alpha \right) &= I \sum_{p = 1}^{\infty} \frac{\left( - 1 \right)^{p-1}}{m_{c}^{p}} \left( \Lambda^{p} \left[ \left. \frac{\partial^{p-1} F_{\alpha}}{\partial \theta^{p-1}} \right|_{z} \right] \left( 0, 0 \right) \right. \\
   &- \left. \Lambda^{p} \left[ \left. \frac{\partial^{p-1} F_{\alpha}}{\partial \theta^{p-1}} \right|_{z} \right] \left( 0, m_{c} \theta_{t} \right) \right) \vec{\nabla} \psi . \nonumber
\end{align}
The only effect of the shift in $\alpha$ is to introduce a phase factor of $\Exp{- i k_{\alpha} \delta \left( \alpha \right)}$ in the Fourier analysed turbulent quantities $h_{s}$, $\phi$, $A_{||}$, and $B_{||}$ (e.g. \refEq{eq:distFnFourierAnalysis}). The shift in $\vec{\nabla} \alpha$ only enters the gyrokinetic model through the quantity
\begin{align}
   \vec{k}_{\perp} &= k_{\psi} \vec{\nabla} \psi + k_{\alpha} \vec{\nabla} \alpha = \left[ k_{\psi} + k_{\alpha} \frac{\partial \vec{r}}{\partial \psi} \cdot \delta \left( \vec{\nabla} \alpha \right) \right] \vec{\nabla} \psi + k_{\alpha} \vec{\nabla} \alpha_{\text{GS2}} . \label{eq:wavenumberShift}
\end{align}
Fortunately, neither of these changes has an effect on \refEq{eq:symSolToAsymSolPart} through \refEq{eq:symSolToAsymSolEnergyExchange}. The phase factor cancels because all transport is driven by the beating of two turbulent quantities (see appendix \ref{app:fluxes}): one with the complex conjugate taken, the other without. As seen in \refEq{eq:wavenumberShift}, the tilt of $\vec{\nabla} \alpha$ can be taken into account by shifting $k_{\psi}$. Since the fluxes we are looking at involve the sum over all of wavenumber space, shifting flux from one wavenumber to another does not alter the total value.

\begin{figure}
	\hspace{0.04\textwidth} (a) \hspace{0.4\textwidth} (b) \hspace{0.25\textwidth}
	\begin{center}
		\includegraphics[width=0.45\textwidth]{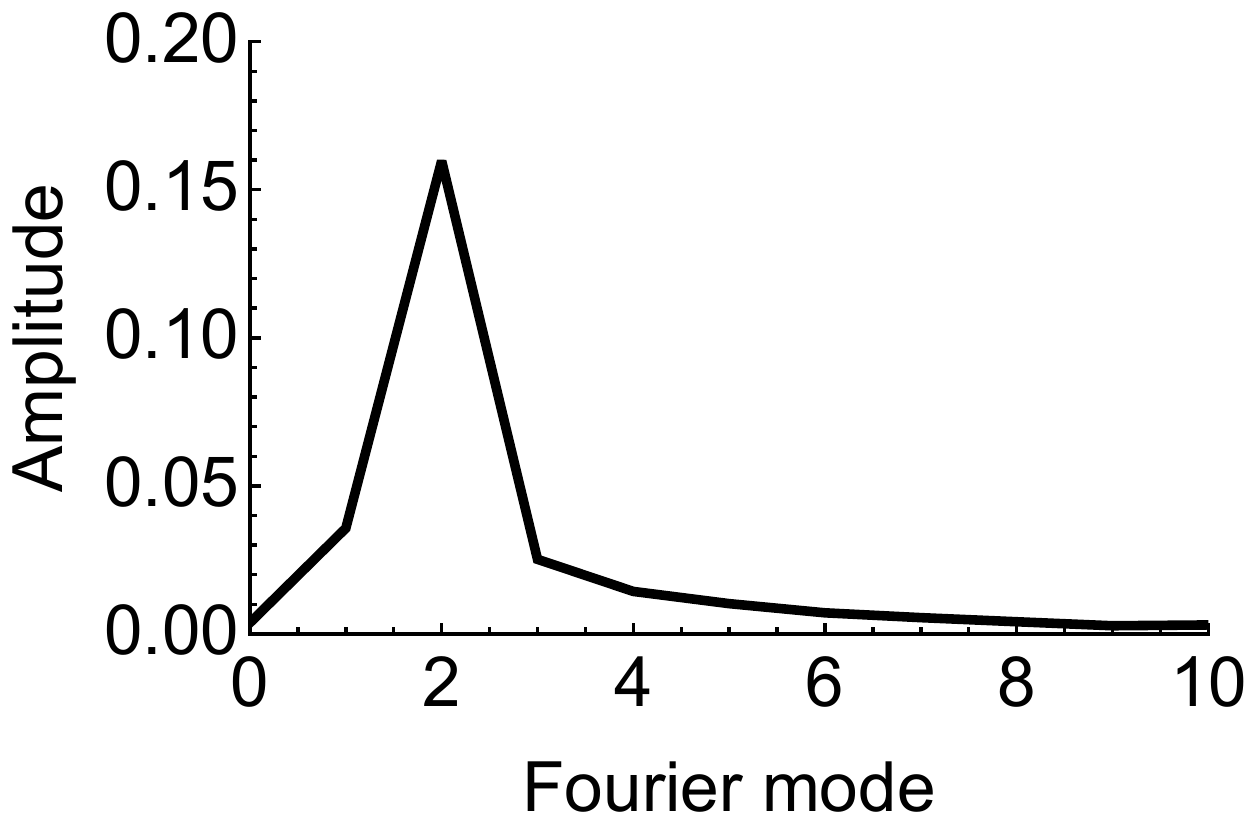}
		\includegraphics[width=0.45\textwidth]{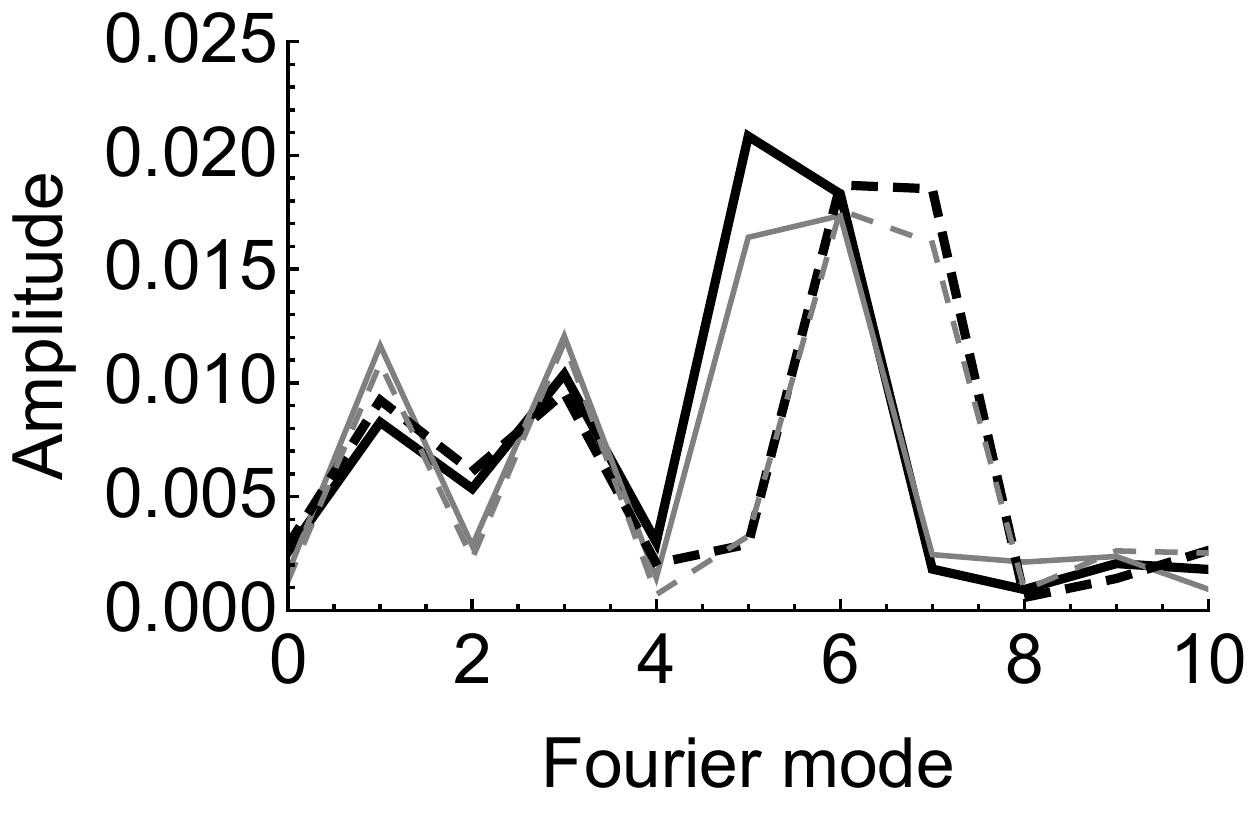}
	\end{center}
	\caption{(a) The Fourier spectrum of the poloidal distribution of the ion momentum flux generated in circular flux surfaces. \\ (b) The Fourier spectrum of the poloidal distribution of ion momentum flux after subtracting the flux generated by circular flux surfaces (shown in (a)) for up-down symmetric (grey) and tilted (black) configurations in the $m_{c} = 7$ (solid) and $m_{c} = 8$ (dashed) geometries.}
	\label{fig:FourierSpectrum}
\end{figure}

Because GS2 is not constructed to separate the two spatial scales represented by $\theta$ and $z$, our simulations give $\pi_{\zeta s} \left( \theta \right) = \pi_{\zeta s} \left( \theta, z \left( \theta \right) \right)$ rather than $\pi_{\zeta s} \left( \theta, z \right)$. Therefore, we have to take the data produced by GS2, separate the dependences on the fast and slow poloidal coordinate, and tilt only the fast spatial variation. Repeating the analysis from section \ref{subsec:asymptoticExpansion}, we start by Fourier analysing the poloidal distribution of momentum flux from GS2,
\begin{align}
   \pi_{\zeta s}^{u} \left( \theta \right) = \sum_{m = 1}^{\infty} P_{m} \Sin{m \theta} , \label{eq:GS2outputDist}
\end{align}
in the untilted case. Figure \ref{fig:FourierSpectrum}(a) shows a typical Fourier spectrum for the momentum flux calculated by GS2. Since the untilted case is up-down symmetric, we know from section \ref{sec:upDownSymArg} that the momentum flux distribution must be odd in $\theta$, so we have neglected the even terms in \refEq{eq:GS2outputDist}. These even terms must be retained when considering the particle or energy fluxes. As in section \ref{subsec:asymptoticExpansion}, we want to transform \refEq{eq:GS2outputDist} into the form of a two dimensional Fourier series in the two separate spatial scales, e.g.
\begin{align}
   \pi_{\zeta s}^{u} \left( \theta, z \right) = \sum_{l = 0}^{\infty} \sum_{k = k_{\text{min}}}^{k_{\text{max}}} P_{k + l m_{c}} \Big( \Sin{l  z} \Cos{k \theta} + \Cos{l  z} \Sin{k \theta} \Big) . \label{eq:2dFourierSeries}
\end{align}
Using some trigonometric identities and \refEq{eq:zDef} it can be shown that if we choose to define $k$ as
\begin{align}
   k &\equiv m - l m_{c} , \label{eq:kDef}
\end{align}
then we can transform \refEq{eq:GS2outputDist} into \refEq{eq:2dFourierSeries} as long as $k_{max} - k_{min} = m_{c} - 1$.

The definition of $l$ contains the physics of the scale separation and consequently will strongly affect how well we match GS2 results. The definition of $l$ controls which Fourier harmonics (enumerated by $m$) are mapped to $l = 0$ (and remain untilted), as opposed to $l = 1$ (which are tilted by $m_{c} \theta_{t}$), $l = 2$ (which are tilted by $2 m_{c} \theta_{t}$), etc. Intuitively we expect modes with $m \approx 1$ should remain untilted (i.e. map to $l = 0$), modes with $m \approx m_{c}$ should map to $l = 1$, and modes with $m \approx 2 m_{c}$ should map to $l = 2$. This general intuition motivates some sort of rounding to integers. The specific form of
\begin{align}
   l &\equiv \left\lfloor \frac{m + 2}{m_{c}} \right\rfloor \label{eq:lDef}
\end{align}
(where $\left\lfloor x \right\rfloor$ is the floor function) was chosen in accordance with figure \ref{fig:FourierSpectrum}(b). We see that, as the shaping effect mode number $m_{c}$ is increased, the $m_{c} - 2$ and $m_{c} - 1$ Fourier terms of the momentum flux track with it, while all lower modes stay roughly constant. Unsurprisingly, this definition of $l$ was also found to produce the best agreement between theory and GS2 data. Our choice for $k$ and $l$ requires that $k_{\text{min}} = - 2$ and $k_{\text{max}} = m_{c} - 3$ in order to include all modes in the summation, meaning \refEq{eq:2dFourierSeries} becomes
\begin{align}
   \pi_{\zeta s}^{u} \left( \theta, z \right) = \sum_{l = 0}^{\infty} \sum_{k = - 2}^{m_{c} - 3} P_{k + l m_{c}} \Big( \Sin{l  z} \Cos{k \theta} + \Cos{l  z} \Sin{k \theta} \Big) .
\end{align}
Now we can use \refEq{eq:symSolToAsymSolMom} to construct
\begin{align}
   \pi_{\zeta s}^{t} \left( \theta, z \right) = \sum_{l = 0}^{\infty} \sum_{k = - 2}^{m_{c} - 3} P_{k + l m_{c}} \Big( \Sin{l  z + l m_{c} \theta_{t}} \Cos{k \theta} + \Cos{l  z + l m_{c} \theta_{t}} \Sin{k \theta} \Big) , \label{eq:tiltedMomPrediction}
\end{align}
a prediction for the distribution of momentum flux in the tilted geometry.

\begin{figure}
 \hspace{0.04\textwidth} (a) \hspace{0.4\textwidth} (b) \hspace{0.25\textwidth}
 \begin{center}
  \includegraphics[width=0.45\textwidth]{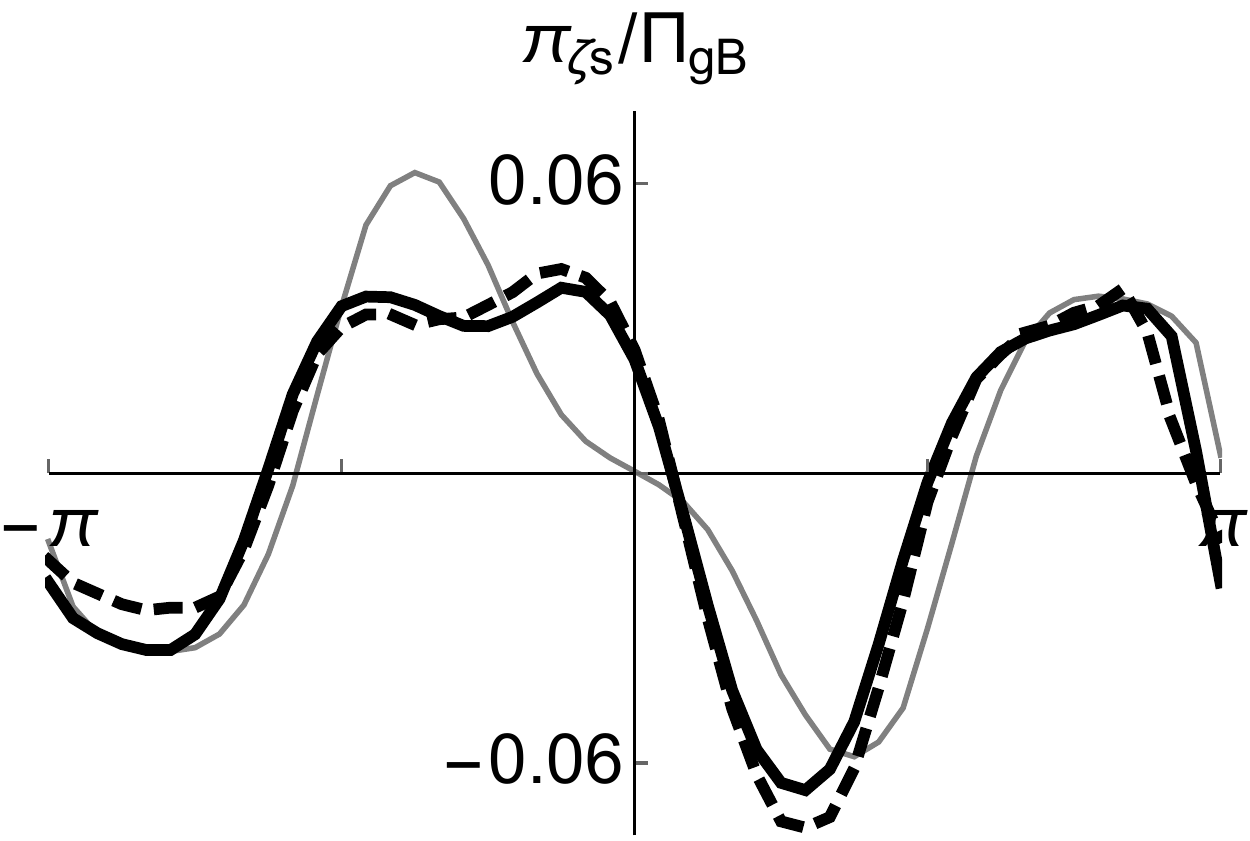}
  \includegraphics[width=0.45\textwidth]{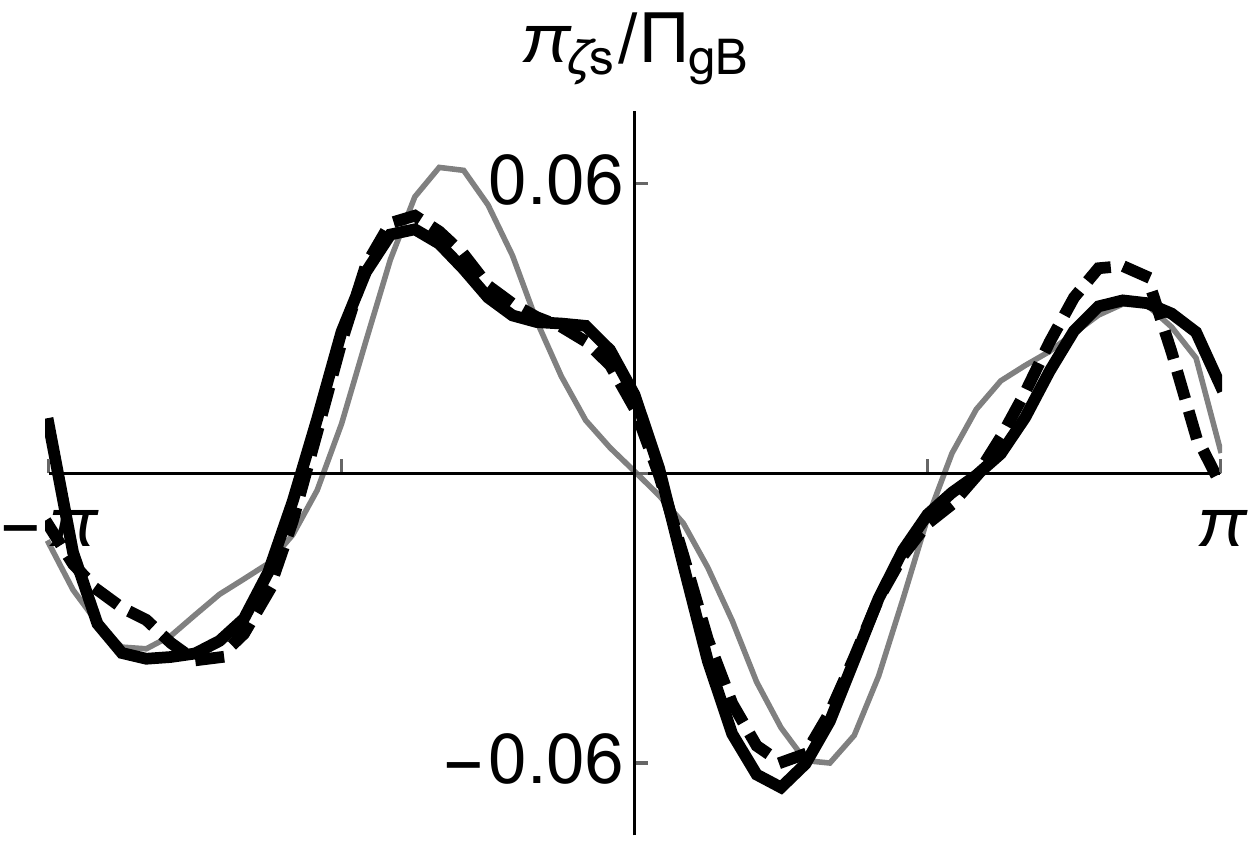}
  \raisebox{6.4\height}{$\theta$}
 \end{center}
 
 \hspace{0.04\textwidth} (c) \hspace{0.4\textwidth} (d) \hspace{0.25\textwidth}
 \begin{center}
  \includegraphics[width=0.45\textwidth]{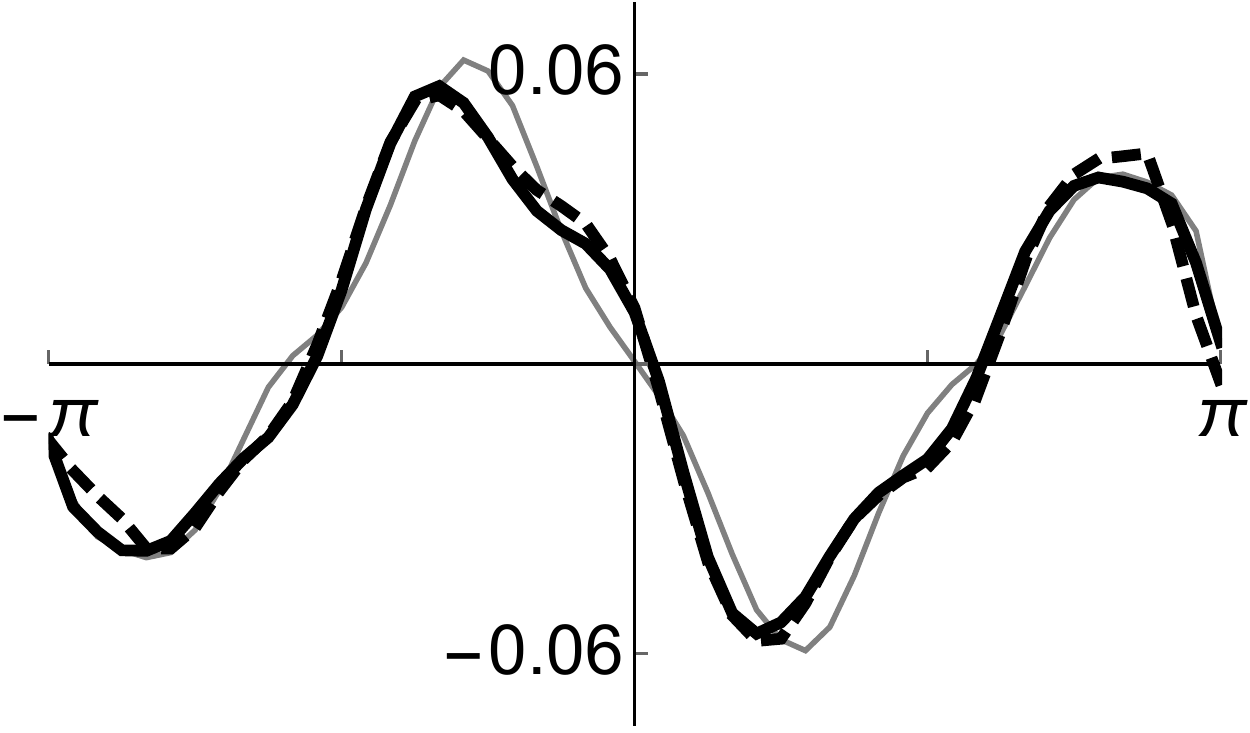}
  \includegraphics[width=0.45\textwidth]{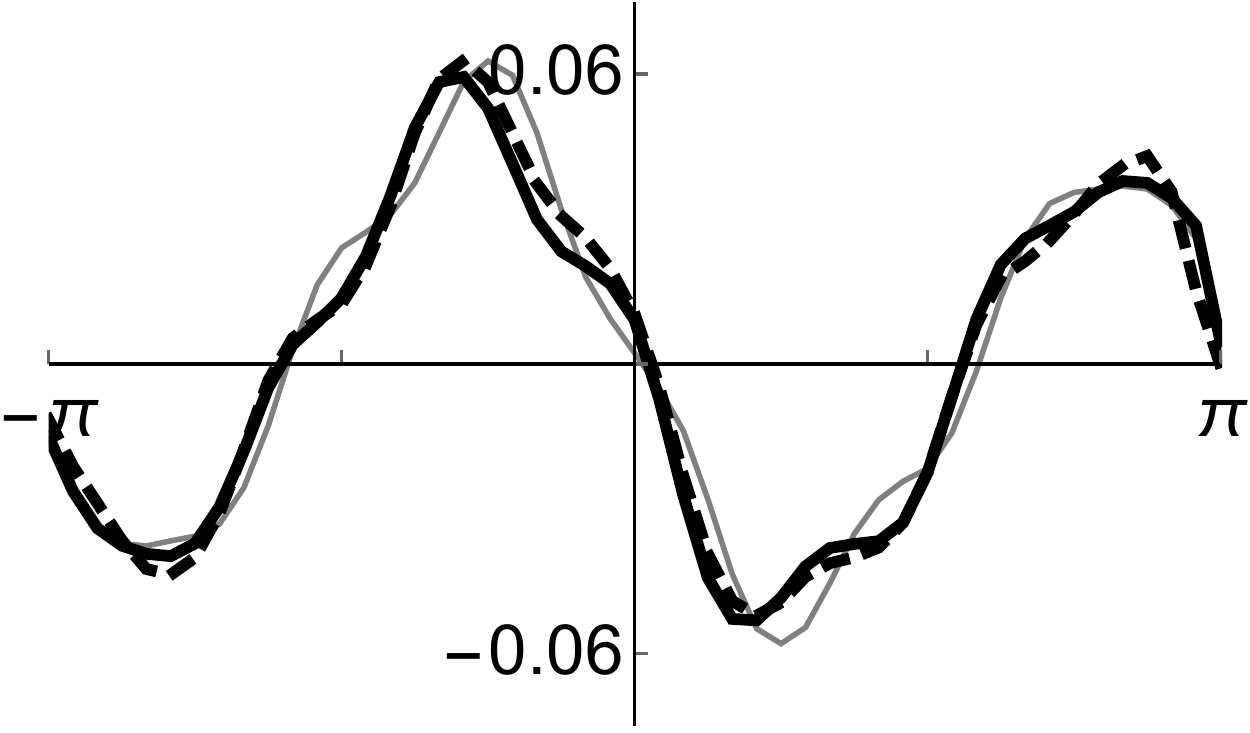}
  \raisebox{6.4\height}{$\theta$}
  \end{center}

 \caption{The full poloidal distribution of the ion momentum flux (see \refEq{eq:polDistMomFlux}) for the tilted geometry (black, thick), up-down symmetric geometry transformed according to the analytic argument (dashed, thick), and up-down symmetric geometry without any transformation (grey, thin), using (a) $m_{c} = 5$, (b) $m_{c} = 6$, (c) $m_{c} = 7$, and (d) $m_{c} = 8$ geometries (see figure \ref{fig:simGeoMirror}). The momentum flux is normalised to the gyroBohm value.}
 \label{fig:fullMomProfiles}
\end{figure}

\begin{figure}
 \begin{center}
  \includegraphics[width=0.55\textwidth]{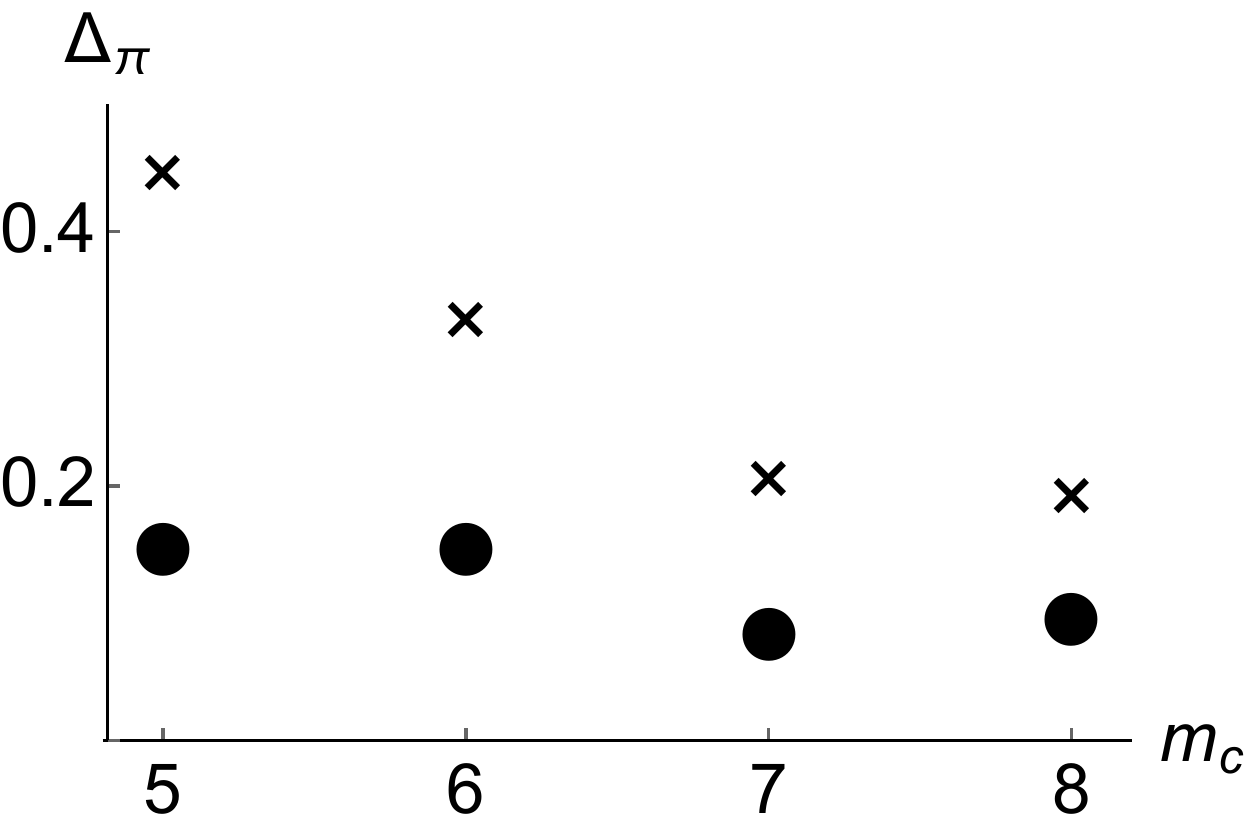}
 \end{center}
 \caption{The fractional error (i.e. \refEq{eq:momDistFracError}) between the poloidal distribution of the momentum flux in the tilted geometry and the distribution predicted from the untilted geometry (circles), with the fractional error between the tilted and untilted (without any adjustment) shown as a control (crosses).}
 \label{fig:momDistFracError}
\end{figure}

Fundamentally, in this comparison we are testing the truth of \refEq{eq:symSolToAsymSol} and \refEq{eq:symSolToAsymSolMom}, which we used in deriving \refEq{eq:tiltedMomPrediction}. In figure \ref{fig:fullMomProfiles}, we use the numerical results from the untilted configuration and \refEq{eq:tiltedMomPrediction} to generate what we expect the momentum flux to be in the corresponding tilted configuration. Visually we see good agreement between the analytic expectation and the actual GS2 results. In figure \ref{fig:momDistFracError} we quantify the agreement by calculating the fractional error according to
\begin{align}
   \Delta_{\pi} &\equiv \frac{\oint_{- \pi}^{\pi} d \theta \left| \pi_{\zeta s}^{\text{act}} \left( \theta \right) - \pi_{\zeta s}^{\text{calc}} \left( \theta \right) \right|}{\oint_{- \pi}^{\pi} d \theta \left| \pi_{\zeta s}^{\text{act}} \left( \theta \right) \right|} , \label{eq:momDistFracError}
\end{align}
where $\pi_{\zeta s}^{\text{act}} \left( \theta \right)$ is the GS2 momentum flux distribution from the tilted geometry (the thick black lines in figure \ref{fig:fullMomProfiles}) and $\pi_{\zeta s}^{\text{calc}} \left( \theta \right)$ is either the predicted distribution calculated analytically from the GS2 untilted result (the dashed black lines in figure \ref{fig:fullMomProfiles}) or the raw GS2 untilted result (the thin grey lines in figure \ref{fig:fullMomProfiles}) to serve as a control. As is also apparent from figure \ref{fig:fullMomProfiles}, when we look at geometries with larger values of $m_{c}$ we find better agreement between the titled geometry and the analytic prediction from the up-down symmetric geometry. The agreement breaks down significantly below $m_{c} = 5$ and we have enough information to understand why. Extrapolating from figure \ref{fig:FourierSpectrum}(b), we would expect $m_{c} = 4$ shaping to be problematic because it generates an ion momentum flux distribution with a strong $m=2$ Fourier mode. This would be indistinguishable from the dominant $m=2$ Fourier mode, which is from toroidicity (see figure \ref{fig:FourierSpectrum}(a)). Since we cannot separate these two contributions (the one from the $m_{c} = 4$ shaping and the one from toroidicity), it is not possible to translate the contribution from the $m_{c} = 4$ shaping as is appropriate.

These numerical results verify \refEq{eq:symSolToAsymSol} and the analytic derivation of section \ref{subsec:gyroSym}. Additionally, though not shown here, the poloidal distributions of particle, momentum, and energy flux (for both ions and electrons) all agree with theory in a similar manner to what is seen in figure \ref{fig:fullMomProfiles}.

%===================================================%
\subsection{Change in total fluxes with tilt}
\label{subsec:totalFluxComp}
%===================================================%

\begin{figure}
 \begin{center}
  \includegraphics[width=0.6\textwidth]{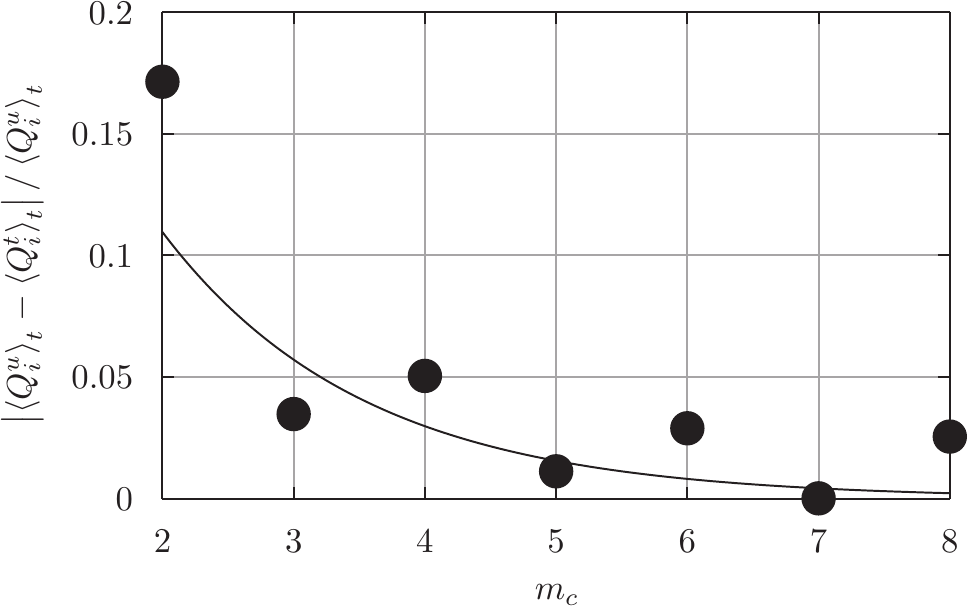}
 \end{center}
 \caption{The fractional difference in the ion energy flux between up-down symmetric and tilted geometries (see figure \ref{fig:simGeoMirror}) as a function of $m_{c}$ (points), with an exponential fit of the form $K \Exp{- \beta m_{c}}$ (line).}
 \label{fig:heatFluxChange}
\end{figure}

From section \ref{subsec:gyroSym} we expect the change in the turbulent fluxes (i.e. \refEq{eq:partFluxAvg} through \refEq{eq:heatFluxAvg}) due to the tilt of fast shaping effects to be exponentially small in $m_{c} \gg 1$. In figure \ref{fig:heatFluxChange}, we show the fractional difference between the ion energy flux from an up-down symmetric configuration and the corresponding tilted configuration, for the geometries of figure \ref{fig:simGeoMirror}. We see that the difference is consistent with an exponential as expected. It is most pronounced for the $m_{c} = 2$ case and rapidly diminishes at higher $m_{c}$.

%===================================================%
\section{Consequences for momentum flux in mirror symmetric tokamaks}
\label{subsec:mirrorSymShaping}
%===================================================%

The gyrokinetic symmetry presented in this chapter demonstrates that a poloidal translation of all fast poloidal variation (i.e. that of order $m_{c}$) by a single tilt angle has an exponentially small effect in $m_{c} \gg 1$ on the turbulent transport. Additionally, we know from the argument reviewed in section \ref{sec:upDownSymArg} that up-down symmetric flux surfaces generate no momentum flux in the gyrokinetic model. Since, by definition, mirror symmetric tokamaks must have mirror symmetry about some line in the poloidal plane, we can rotate all of the shaping effects by a single tilt angle until the line of mirror symmetry is coincident with the midplane. Hence, because all mirror symmetric flux surfaces can be generated by tilting up-down symmetric surfaces, we know that the momentum flux in mirror symmetric tokamaks cannot scale more strongly than $\Pi_{\zeta s} \sim \Exp{- \beta m_{c}^{\gamma}}$, where $\beta$ and $\gamma$ are both positive and do not depend on $m_{c}$. This exponential scaling is true for all flux surfaces that have mirror symmetry about any line in the poloidal plane, not just those with mirror symmetry about the midplane (i.e. up-down symmetry). This argument only relies on the conditions needed for the symmetry, namely $m_{c} \gg 1$. It does \textit{not} presume that the flux surface shaping is weak.

\begin{figure}
    \hspace{0.04\textwidth} (a) \hspace{0.255\textwidth} (b) \hspace{0.255\textwidth} (c) \hspace{0.1\textwidth}
	\begin{center}
		\includegraphics[width=0.3\textwidth]{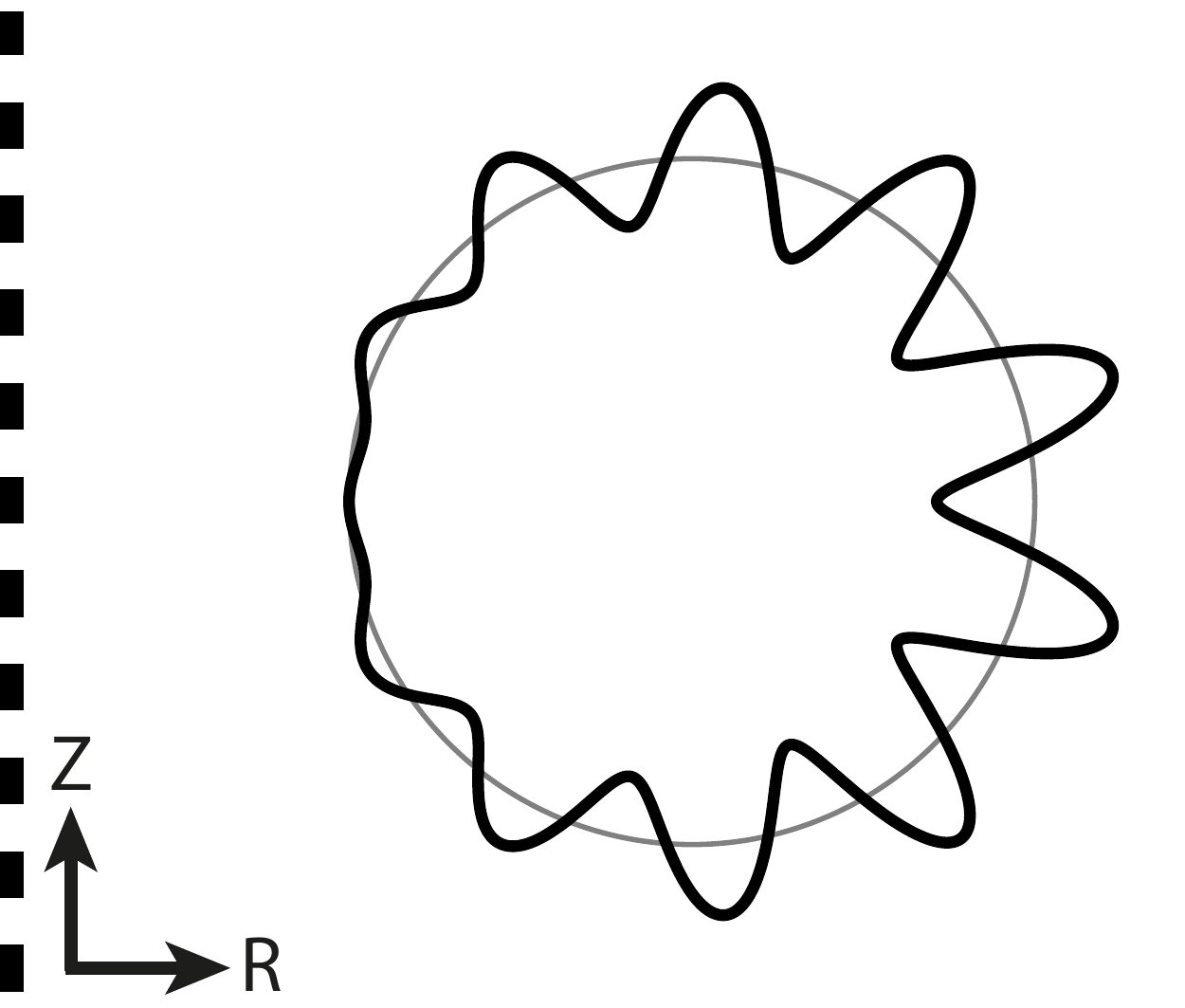}
		\includegraphics[width=0.3\textwidth]{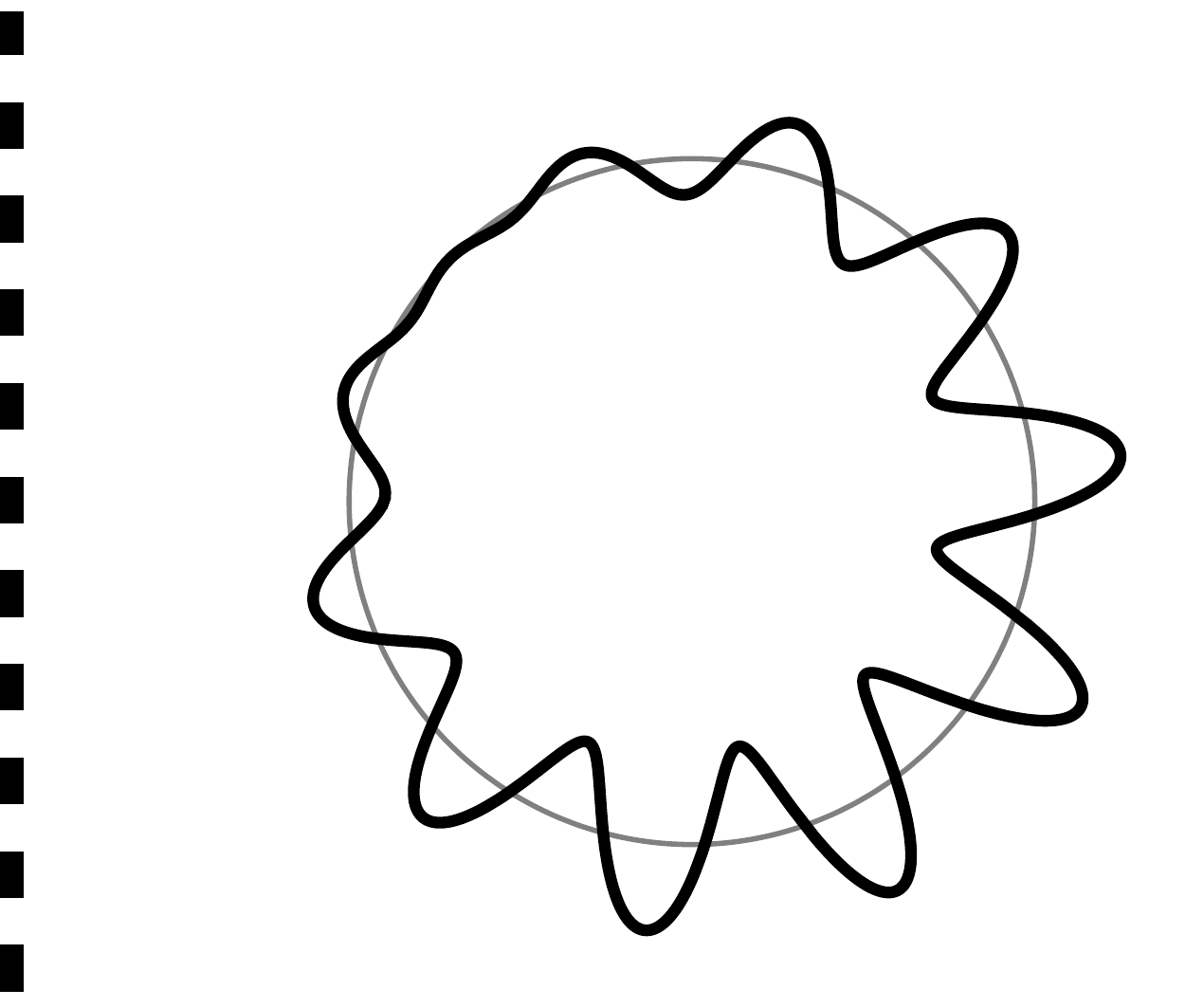}
		\includegraphics[width=0.3\textwidth]{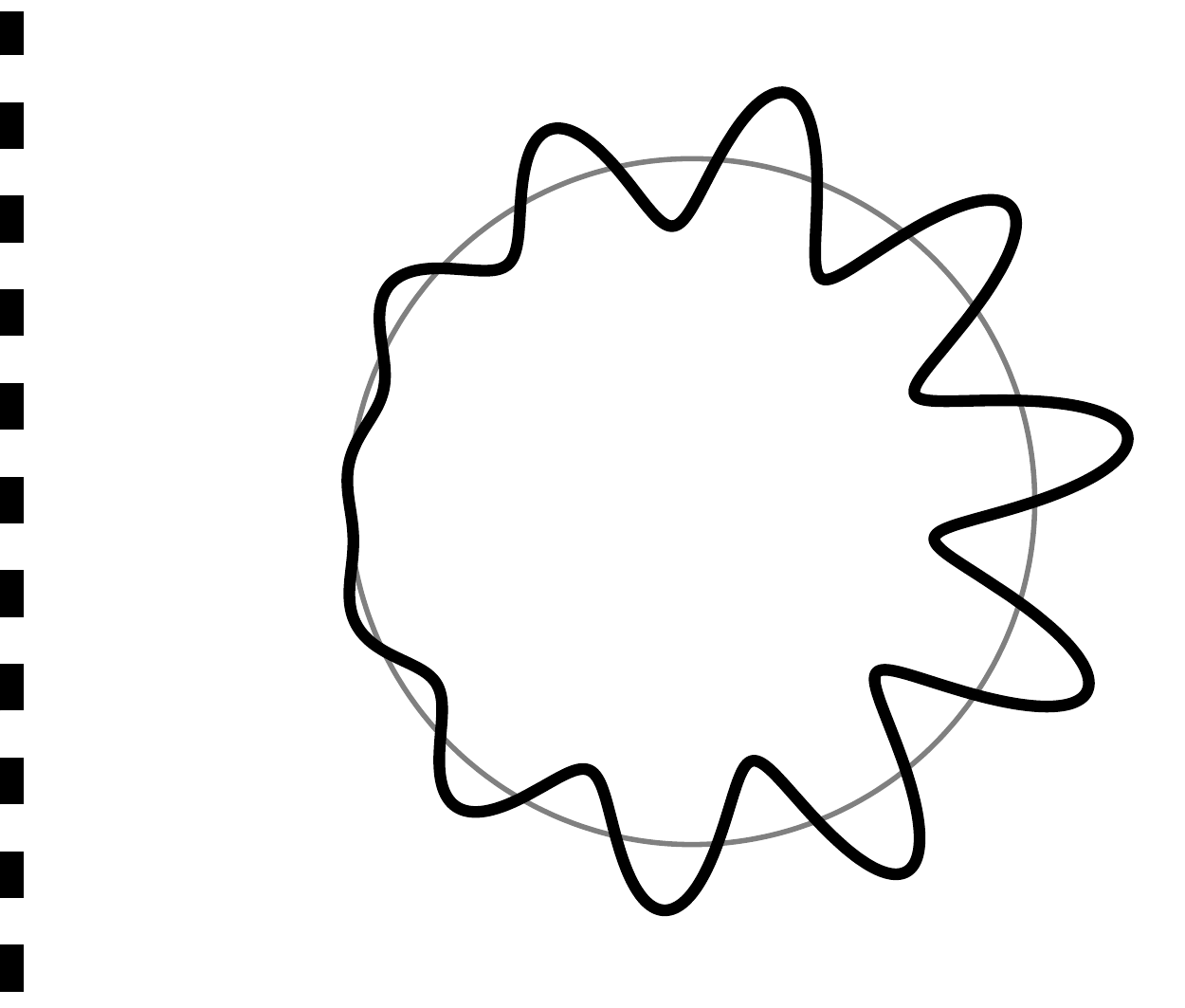}
	\end{center}
	%\hspace{0.15\textwidth} $\Pi_{\zeta s} = 0$ \hspace{0.12\textwidth} $\Pi_{\zeta s} \sim \Exp{- \beta \left( 1 \right)^{\gamma}}$ \hspace{0.08\textwidth} $\Pi_{\zeta s} \sim \Exp{- \beta \left( 10 \right)^{\gamma}}$
	\caption{Exaggerated flux surface geometries (black) with $m = 10$ and $m = 11$ Fourier shaping modes and (a) no tilt, (b) a $\pi/4$ tilt of the entire flux surface, and (c) a $\pi/4$ tilt of the fast poloidal variation, where circular flux surfaces (grey) are shown for comparison.}
	\label{fig:envelopeShaping}
\end{figure}

However, there is a subtlety concerning flux surfaces that possess slowly varying envelopes as well as fast variation. In figure \ref{fig:envelopeShaping}(a) we show a cartoon up-down symmetric flux surface specified by \refEq{eq:fluxSurfofInterestShapeWeak} with just two shaping modes, $m=10$ and $m=11$, which beat to create an $m=1$ envelope. We are free to consider all flux surface variation ``fast'' and tilt the entire flux surface. This produces figure \ref{fig:envelopeShaping}(b) which is a mirror symmetric flux surface. However, since we rotated the $m=1$ variation of the envelope, we have considered $m=1$ variation to be fast. This means that the difference in the fluxes produced by figures \ref{fig:envelopeShaping}(a) and \ref{fig:envelopeShaping}(b) is exponentially small in an expansion in $m_{c} = 1 \gg 1$, which is not particularly meaningful.

This example illustrates that if a flux surface is mirror symmetric it produces momentum transport that is exponentially small in the Fourier mode number of the mirror symmetric variation. However, the validity of this expansion may be questionable for certain cases such as geometries with low shaping effects or a slowly varying envelope created by the beating of several high mode number effects.

Conversely, we can choose to consider the $m=1$ envelope as slow and refrain from tilting it. To do so, we convert the flux surface specification into \refEq{eq:fluxSurfaceSpecScaleSep} and introduce the global tilt through the form of $z \left( \theta \right)$. This produces figure \ref{fig:envelopeShaping}(c), which is not mirror symmetric. However, it has a mirror symmetric fast shaping and up-down symmetric slow shaping (because the envelope remains unchanged). Hence, the difference in the fluxes produced by figures \ref{fig:envelopeShaping}(a) and \ref{fig:envelopeShaping}(c) is exponentially small in an expansion in $m_{c} = 10 \gg 1$.
	
This example shows that formally non-mirror symmetric configurations can still have exponentially small momentum flux if the slow variation is up-down symmetric and the fast variation is mirror symmetric. This becomes intuitive if we consider toroidicity as a second type of $m=1$ mode (in addition to the Shafranov shift). From this perspective up-down symmetry is just mirror symmetry with respect to the inherent, untilted mode from toroidicity. Hence we can add any slow shaping mode as long as it is aligned with the mode from toroidicity, keeping the slow shaping mirror symmetric.

Interpreted broadly, this symmetry demonstrates that variation on two spatial scales interact to generate momentum transport that is exponentially small in the scale separation. Hence, using neighbouring shaping modes to drive rotation is optimal.

%% file: Ch_08_GYRO_MomFluxScaling.tex
% !TEX root = /Users/Justin/Documents/Research/Writings/2016DoctoralThesis/DoctoralThesis.tex

\chapter{Envelopes: Scaling of momentum flux with shaping mode number}
\label{ch:GYRO_MomFluxScaling}

\begin{quote}
   \emph{Much of this chapter appears in reference \cite{BallMomFluxScaling2016}.}
\end{quote}

This chapter investigates slowly varying envelopes created by the beating of different fast flux surface shaping effects. We will prove that fast shaping effects can create slowly varying envelopes that generate momentum transport that is polynomially small in the fast shaping mode number. This contrasts with the exponential scaling found for mirror symmetric surfaces in chapter \ref{ch:GYRO_TiltingSymmetry} and suggests that up-down asymmetric envelopes created using low shaping effects could increase rotation.
%Broadly, these results indicate that the momentum flux generated by poloidal variation scales polynomially with the length scale of the variations.
In this chapter we restrict our attention to momentum transport generated in stationary plasmas (i.e. $\Omega_{\zeta} = d \Omega_{\zeta} / d r_{\psi} = 0$) and will assume the turbulence is electrostatic for simplicity.

In section \ref{sec:gyrokineticAnalysis}, we expand the gyrokinetic equation order-by-order in large shaping mode number to compare the momentum flux generated by different types of flux surface shaping. In section \ref{subsec:practicalNonMirrorSymShaping} we calculate how the momentum flux scales with the shaping effect mode number, given a specific set of simple geometries with two independently-tilted shaping modes. This is designed to give a concrete illustration of the more abstract and general scaling argument for geometries with up-down asymmetric envelopes presented in section \ref{subsec:genNonMirrorSymShaping}. Then, in section \ref{sec:numResults} we use nonlinear local gyrokinetic simulations to contrast the analytic results of section \ref{sec:gyrokineticAnalysis} with chapter \ref{ch:GYRO_TiltingSymmetry}. Lastly, section \ref{sec:interpretation} gives a broad interpretation of these analytic scalings.

%===================================================%
%===================================================%
\section{Analytic gyrokinetic analysis}
\label{sec:gyrokineticAnalysis}
%===================================================%
%===================================================%

First, we must calculate the local value of the eight geometric coefficients that appear in the stationary gyrokinetic equation from the local Miller equilibrium model in chapter \ref{ch:MHD_LocalEquil}. The full calculation is shown in appendix \ref{app:genGeoCoeff}, but here we will only calculate them to lowest order in large aspect ratio for use in section \ref{subsec:practicalNonMirrorSymShaping}. We will specify the flux surface geometry by the Expanded prescription (i.e. \refEq{eq:fluxSurfofInterestShapeWeak}, \refEq{eq:fluxSurfofInterestDerivWeak}, and \refEq{eq:gradShafranovLocalEqWeak} through \refEq{eq:geoAxialWeak}). These equations represent a completely general Fourier decomposition of the flux surface shape. Note that the change in the strength, $d \Delta_{m} / d r_{\psi}$, and tilt, $d \theta_{t m} / d r_{\psi}$, of each mode is determined by the global MHD equilibrium. In chapter \ref{ch:MHD_LocalEquil} we derive these quantities using a constant toroidal current profile in the limits of large aspect ratio and weak shaping. We will keep these parameters free and only use the constant current profile solutions to provide scalings for these quantities.

Additionally, the Miller model requires four scalar quantities, usually $I$ (the toroidal field flux function), $q$ (the safety factor), $dq/dr_{\psi}$ (the magnetic shear), and $d p / d r_{\psi}$ (the pressure gradient) of the flux surface of interest. However, for ease of notation, we will choose to replace $I$ by $B_{c 0} = I / R_{c 0}$ (the toroidal magnetic field at $R = R_{c 0}$ on the flux surface of interest) and $q$ by $d \psi/ d r_{\psi}$ (see \refEq{eq:dpsidrpsi}). Also, when we expand to lowest order in aspect ratio, we will find that we can replace both $dp/dr_{\psi}$ and $dq/dr_{\psi}$ (related to $d I / d \psi$ by \refEq{eq:dIdpsiFullForm}) with
\begin{align}
\hat{s}' &\equiv 2 + r_{\psi 0} \left( \frac{d \psi}{d r_{\psi}} \right)^{-1} \left( \mu_{0} R_{c 0}^{2} \left( \frac{d \psi}{d r_{\psi}} \right)^{-1} \frac{d p}{d r_{\psi}} + R_{c 0} B_{c 0} \frac{d I}{d \psi} \right) \label{eq:shiftedShatDef} \\
&= 2 - r_{\psi 0} \left( \frac{d \psi}{d r_{\psi}} \right)^{-1} \mu_{0} j_{\zeta} R_{c 0} . \nonumber
\end{align}
We can make this replacement because the toroidal current, which appears on the right side of the Grad-Shafranov equation, is a flux function to lowest order in aspect ratio (see \refEq{eq:gradShafEq}). We note that if the flux surfaces are exactly circular and $d p / d r_{\psi} = 0$, then $\hat{s}' = \hat{s} \equiv \left( r_{\psi 0} / q \right) dq/dr_{\psi}$.

To lowest order in aspect ratio we immediately see that $B \rightarrow B_{c 0}$ and $a_{|| s} \rightarrow 0$, so we can focus on the other six (i.e. $\hat{b} \cdot \Nabla \theta$, $v_{d s \psi}$, $v_{d s \alpha}$, $\left| \Nabla \psi \right|^{2}$, $\Nabla \psi \cdot \Nabla \alpha$, and $\left| \Nabla \alpha \right|^{2}$). In this limit the momentum flux, given by \refEq{eq:momFlux}, becomes
\begin{align}
  \Pi_{\zeta s} &= \frac{2 \pi i R_{c 0} B_{c 0}}{\oint d\theta \left( \hat{b} \cdot \Nabla \theta \right)^{-1}} \sum_{k_{\psi}, k_{\alpha}} k_{\alpha} \oint d\theta \left( \hat{b} \cdot \Nabla \theta \right)^{-1} \label{eq:momFluxSimple} \\
  &\times \int dw_{||} d\mu ~ w_{||} J_{0} \left( k_{\perp} \rho_{s} \right) \phi \left( k_{\psi}, k_{\alpha} \right) h_{s} \left( - k_{\psi}, - k_{\alpha} \right) . \nonumber
\end{align}
In order to calculate the poloidal field we must use
\begin{align}
   \vec{B}_{p} = \vec{\nabla} \zeta \times \vec{\nabla} r_{\psi} \frac{d \psi}{d r_{\psi}} , \label{eq:poloidalFieldDef}
\end{align}
our geometry specification (given by \refEq{eq:fluxSurfofInterestShapeWeak}, \refEq{eq:fluxSurfofInterestDerivWeak}, and \refEq{eq:gradShafranovLocalEqWeak} through \refEq{eq:geoAxialWeak}), and the vector identity
\begin{align}
   \vec{\nabla} u_{1} &= \frac{\partial \vec{r} / \partial u_{2} \times \partial \vec{r} / \partial u_{3}}{\partial \vec{r} / \partial u_{1} \cdot \left( \partial \vec{r} / \partial u_{2} \times \partial \vec{r} / \partial u_{3} \right)} \label{eq:gradIdentity}
\end{align}
for $\left( u_{1}, u_{2}, u_{3} \right)$, a cyclic permutation of $\left( r_{\psi}, \theta, \zeta \right)$. Having calculated the poloidal field we can find $\hat{b} \cdot \Nabla \theta$,
\begin{align}
   v_{d s \psi} &= \frac{m_{s}}{Z_{s} e} \hat{b} \cdot \Nabla \theta \frac{\partial R}{\partial \theta}  , \label{eq:driftVelPsiAspect}
\end{align}
and $\left| \Nabla \psi \right|^{2} = R^{2} B_{p}^{2}$ to lowest order in aspect ratio.

However, $\Nabla \alpha$ contains second-order radial derivatives, which are not specified as input. The Miller model determines them by ensuring that the Grad-Shafranov equation is satisfied. With considerable work (shown in appendix \ref{app:genGeoCoeff}), we can use the Grad-Shafranov equation to calculate that
\begin{align}
   \Nabla \alpha &= \frac{\partial \alpha}{\partial \psi} \Nabla \psi + \frac{\partial \alpha}{\partial \theta} \Nabla \theta , \label{eq:gradAlphaAspect}
\end{align}
where
\begin{align}
   \frac{\partial \alpha}{\partial \psi} &= - \left. \int_{\theta_{\alpha}}^{\theta} \right|_{\psi} d \theta' \frac{\partial A_{\alpha}}{\partial \psi} + A_{\alpha} \left( \psi, \theta_{\alpha} \right) \frac{d \theta_{\alpha}}{d \psi} \label{eq:gradAlphaPsiAspectUnsimplified} \\
   &= - \left. \int_{\theta_{\alpha}}^{\theta} \right|_{\psi} d \theta' \left( \frac{\partial A_{\alpha}}{\partial \psi} \right)_{\text{orthog}} + \left[ \frac{B_{c 0}}{R_{c 0}^{3} B_{p}^{3}} \dlpdthetaPrime \Nabla \psi \cdot \Nabla \theta' \right]_{\theta' = \theta_{\alpha}}^{\theta' = \theta} \label{eq:gradAlphaPsiAspect} \\
   &+ \left( \frac{B_{c 0}}{R_{c 0} B_{p}} \dlpdtheta \right)_{\theta = \theta_{\alpha}} \frac{d \theta_{\alpha}}{d \psi} . \nonumber
\end{align}
and
\begin{align}
   \frac{\partial \alpha}{\partial \theta} &= - A_{\alpha} \left( \psi, \theta \right) = - \frac{B_{c 0}}{R_{c 0} B_{p}} \dlpdtheta
\end{align}
to lowest order in aspect ratio. Here
\begin{align}
   \left( \frac{\partial A_{\alpha}}{\partial \psi} \right)_{\text{orthog}} = \frac{B_{c 0}}{R_{c 0}^{2} B_{p}^{2}} \dlpdthetaPrime \left( \frac{d \psi}{d r_{\psi}} \frac{\hat{s}' - 2}{r_{\psi 0} R_{c 0} B_{p}} + 2 \kappa_{p} \right) \label{eq:gradAlphaAspectIntegrand}
\end{align}
is the part of $\partial A_{\alpha} / \partial \psi$ that remains if the $\left( r_{\psi}, \theta, \zeta \right)$ coordinate system is orthogonal, $\kappa_{p}$ is the poloidal magnetic field curvature given by \refEq{eq:poloidalCurv}, and $l_{p}$ is the poloidal arc length (defined such that \refEq{eq:dLpdtheta} is true). The form of \refEq{eq:gradAlphaAspectIntegrand} is useful because it does not contain any radial derivatives (except $d \psi/ d r_{\psi}$ which is an input to the calculation) and distinguishes the important term: the poloidal curvature. This allows us to find $\Nabla \psi \cdot \Nabla \alpha$, $\left| \Nabla \alpha \right|^{2}$, and
\begin{align}
   v_{d s \alpha} &= \frac{1}{\Omega_{s}} \left( \frac{B_{c 0}}{R_{c 0}} \frac{\partial R}{\partial \psi} + \frac{\partial R}{\partial \theta} \frac{\partial \alpha}{\partial \psi} \Nabla \psi \cdot \left( \Nabla \theta \times \Nabla \zeta \right) \right) \label{eq:driftVelAlphaAspect}
\end{align}
to lowest order in aspect ratio.

%===================================================%
\subsection{Asymptotic expansion ordering}
\label{subsec:asymptoticExpansionScaling}
%===================================================%

We know from section \ref{sec:upDownSymArg} that, unless the up-down symmetry of the geometric coefficients is broken, the time-averaged momentum flux will always be zero to lowest order in $\rho_{\ast} \equiv \rho_{i} / a \ll 1$. We will investigate the consequences of breaking the up-down symmetry using different shaping effects. To do this we will expand \refEq{eq:gyrokineticEq}, \refEq{eq:quasineut}, and \refEq{eq:momFluxSimple} in large Fourier shaping mode number, i.e. $m_{c} \gg 1$ where $m_{c}$ is a characteristic mode number of the fast shaping effects. We will expand
\begin{align}
  h_{s} &= h_{s 0} + h_{s 1} + h_{s 2} + h_{s 3} + \ldots \\
  \phi &= \phi_{0} + \phi_{1} + \phi_{2} + \phi_{3} + \ldots ,
\end{align}
where the subscript indicates the order of the quantity in $m_{c}^{-1} \ll 1$. This expansion separates the long spatial scale coordinate $\theta$, from a short spatial scale coordinate, $z$, defined by \refEq{eq:zDef}. Distinguishing the variation on each scale, e.g. $h_{s} \left( \theta, z \right)$ and $\phi \left( \theta, z \right)$, means that we can rewrite poloidal derivatives according to \refEq{eq:poloidalDerivTransform}. Ultimately we will only be interested in large scale phenomena, so we will need to average quantities in $z$ using \refEq{eq:zAvg}, but we must still manipulate the $z$-dependent portion, given by
\begin{align}
  \widetilde{\left( \ldots \right)} \equiv \left( \ldots \right) - \overline{\left( \ldots \right)} . \label{eq:zDep}
\end{align}

%===================================================%
\subsection{Two mode shaping in the large aspect ratio gyrokinetic model}
\label{subsec:practicalNonMirrorSymShaping}
%===================================================%

In this section we will prescribe a specific geometry and expand the large aspect ratio gyrokinetic, quasineutrality, and momentum flux equations order-by-order to determine the scaling of the momentum flux with $m_{c} \gg 1$. Hence formally we require that $\epsilon \ll 1$ for the aspect ratio expansion and also that $\epsilon \ll m_{c}^{-1} \ll 1$ for the subsidiary expansion in shaping mode number. We perform a more general calculation for $\epsilon \sim 1$ in section \ref{subsec:genNonMirrorSymShaping}.

We completely specify the shape of the flux surface of interest (and how it changes with radius) by \refEq{eq:fluxSurfofInterestShapeWeak}, \refEq{eq:fluxSurfofInterestDerivWeak}, and \refEq{eq:gradShafranovLocalEqWeak} through \refEq{eq:geoAxialWeak}. We will choose the ordering
\begin{align}
  \Delta_{m} - 1 \sim m_{c}^{-2} \label{eq:weakShapingOrdering}
\end{align}
for the magnitude of the large mode number shaping effects to be consistent with \refEq{eq:reasonableShaping}. We can determine that
\begin{align}
\frac{d \Delta_{m}}{d r_{\psi}} \sim \frac{m_{c} \left( \Delta_{m} - 1 \right)}{r_{\psi}} . \label{eq:changeShapeOrdering}
\end{align}
from \refEq{eq:shapingRadialDeriv_rPsi} and $d \theta_{t m} / d r_{\psi} = 0$ from \refEq{eq:tiltAngleDeriv}, which were derived for a constant current profile.

In this calculation, we will use flux surfaces with simple fast shaping that beats together to form slowly varying envelopes. To create these flux surfaces, we include only two high-order shaping effects, $m$ and $n$, in \refEq{eq:fluxSurfofInterestShapeWeak} and \refEq{eq:fluxSurfofInterestDerivWeak}. We will order both $m \sim m_{c}$ and $n \sim m_{c}$, which are free to have different strengths, $\Delta_{m}$ and $\Delta_{n}$, and tilt angles, $\theta_{t m}$ and $\theta_{t n}$. However, we order $n - m \sim 1$, $\Delta_{n} - 1 \sim \Delta_{m} - 1 \sim m_{c}^{-2}$, and $d \Delta_{n} / d r_{\psi} \sim d \Delta_{m} / d r_{\psi} \sim m_{c} \left( \Delta_{m} - 1 \right) / r_{\psi}$. Given these orderings \refEq{eq:fluxSurfofInterestShapeWeak} and \refEq{eq:fluxSurfofInterestDerivWeak} become 
\begin{align}
   r_{0} \left( \theta \right) &= r_{\psi 0} \left( 1 - \frac{\Delta_{m} - 1}{2} \Cos{z_{m s}} - \frac{\Delta_{n} - 1}{2} \Cos{z_{n s}} \right) + O \left( m_{c}^{-4} r_{\psi} \right) \label{eq:fluxSurfaceSpecSimplified} \\
   \left. \frac{\partial r}{\partial r_{\psi}} \right|_{\psi_{0}} &= 1 - \frac{r_{\psi 0}}{2} \frac{d \Delta_{m}}{d r_{\psi}} \Cos{z_{m s}} - \frac{r_{\psi 0}}{2} \frac{d \Delta_{n}}{d r_{\psi}} \Cos{z_{n s}} + O \left( m_{c}^{-2} \right) , \label{eq:fluxSurfaceChangeSpecSimplified}
\end{align}
where
\begin{align}
  z_{m s} &\equiv m \left( \theta + \theta_{t m} \right) \label{eq:zmsDef} \\
  z_{n s} &\equiv n \left( \theta + \theta_{t n} \right) . \label{eq:znsDef}
\end{align}
Note that we are not using the form of \refEq{eq:fluxSurfaceSpecScaleSep}, instead we have the sum of two arbitrary fast modes. By doing so we will explicitly see how the mode numbers and tilt angles combine, sometimes generating a slowly varying envelope.

%===================================================%
\subsubsection{Geometric coefficients}
\label{subsubsec:geoCoeffs}
%===================================================%

To lowest order, $O \left( 1 \right)$, the geometric coefficients are those of a circular tokamak and are entirely independent of the short spatial scale coordinate, $z$. To next order the coefficients depend on $z$, but are algebraically intensive to find. The full expressions for all six coefficients (and several intermediate quantities that are useful in the derivation) are given in appendix \ref{app:nonMirrorGeoCoeffs}, but here we will only derive $v_{d s \alpha}$ to serve as an illustrative example. This coefficient signifies the magnetic drifts in the direction perpendicular to the magnetic field, but still within the flux surface. We will start with \refEq{eq:gradShafranovLocalEqWeak} through \refEq{eq:geoAxialWeak}, \refEq{eq:fluxSurfaceSpecSimplified}, and \refEq{eq:fluxSurfaceChangeSpecSimplified} and use them to construct all of the quantities appearing in \refEq{eq:driftVelAlphaAspect}, the expression for $v_{d s \alpha}$ to lowest order in aspect ratio.

It will be sufficient to calculate all quantities to $O \left( m_{c}^{-1} \right)$ with the exception of $\partial R / \partial \theta$ and $\partial Z / \partial \theta$, because they appear in the poloidal curvature with an extra poloidal derivative (see \refEq{eq:poloidalCurv}). This extra derivative creates an additional factor of $m_{c}$, which boosts $O \left( m_{c}^{-2} \right)$ effects to $O \left( m_{c}^{-1} \right)$. Directly differentiating \refEq{eq:geoMajorRadiusWeak} and \refEq{eq:geoAxialWeak} we find
\begin{align}
   \frac{\partial R}{\partial \theta} &= \frac{d r_{0}}{d \theta} \Cos{\theta} - r_{0} \Sin{\theta} \label{eq:partialRpartialTheta} \\
   \frac{\partial Z}{\partial \theta} &= \frac{d r_{0}}{d \theta} \Sin{\theta} + r_{0} \Cos{\theta} \label{eq:partialZpartialTheta} \\
      \frac{\partial^{2} R}{\partial \theta^{2}} &= \frac{d^{2} r_{0}}{d \theta^{2}} \Cos{\theta} - 2 \frac{d r_{0}}{d \theta} \Sin{\theta} - r_{0} \Cos{\theta} \label{eq:partialSqRpartialThetaSq} \\
   \frac{\partial^{2} Z}{\partial \theta^{2}} &= \frac{d^{2} r_{0}}{d \theta^{2}} \Sin{\theta} + 2 \frac{d r_{0}}{d \theta} \Cos{\theta} - r_{0} \Sin{\theta} , \label{eq:partialSqZpartialThetaSq}
\end{align}
where
\begin{align}
   \frac{d r_{0}}{d \theta} &= \frac{r_{\psi 0}}{2} \Big( m \left( \Delta_{m} - 1 \right) \Sin{z_{m s}} + n \left( \Delta_{n} - 1 \right) \Sin{z_{n s}} \Big) + O \left( m_{c}^{-3} r_{\psi} \right) \\
   \frac{d^{2} r_{0}}{d \theta^{2}} &= \frac{r_{\psi 0}}{2} \Big( m^{2} \left( \Delta_{m} - 1 \right) \Cos{z_{m s}} + n^{2} \left( \Delta_{n} - 1 \right) \Cos{z_{n s}} \Big) + O \left( m_{c}^{-2} r_{\psi} \right) .
\end{align}
From this point forward we will only need quantities to $O \left( m_{c}^{-1} \right)$ to accurately capture the up-down symmetry breaking. Substituting \refEq{eq:partialRpartialTheta} and \refEq{eq:partialZpartialTheta} into \refEq{eq:dLpdtheta} gives
\begin{align}
   \dlpdtheta &= r_{\psi 0} + O \left( m_{c}^{-2} r_{\psi} \right) . \label{eq:nonMirrordLpdTheta}
\end{align}
We can now substitute \refEq{eq:partialRpartialTheta} through \refEq{eq:nonMirrordLpdTheta} into \refEq{eq:poloidalCurv} to find
\begin{align}
   \kappa_{p} &= \frac{1}{r_{\psi 0}} \left( 1 - \frac{1}{r_{\psi 0}} \frac{d^{2} r_{0}}{d \theta^{2}} \right) + O \left( \frac{m_{c}^{-2}}{r_{\psi}} \right) \label{eq:nonMirrorPolCurv} \\
   &= \frac{1}{r_{\psi 0}} \left( 1 - \frac{1}{2} \left[ m^{2} \left( \Delta_{m} - 1 \right) \Cos{z_{m s}} + n^{2} \left( \Delta_{n} - 1 \right) \Cos{z_{n s}} \right] \right) + O \left( \frac{m_{c}^{-2}}{r_{\psi}} \right) .
\end{align}

Next we will calculate
\begin{align}
   \frac{\partial R}{\partial r_{\psi}} &= \left. \frac{\partial r}{\partial r_{\psi}} \right|_{\psi_{0}} \Cos{\theta} \label{eq:partialRpartialrPsi} \\
   \frac{\partial Z}{\partial r_{\psi}} &= \left. \frac{\partial r}{\partial r_{\psi}} \right|_{\psi_{0}} \Sin{\theta} \label{eq:partialZpartialrPsi}
\end{align}
straightforwardly from \refEq{eq:geoMajorRadiusWeak} and \refEq{eq:geoAxialWeak}. We can determine $\Nabla r_{\psi}$ through \refEq{eq:gradIdentity} and immediately find
\begin{align}
   \Nabla \theta &= \frac{\hat{e}_{\theta}}{r_{0} \left( \theta \right)} \label{eq:gradTheta} \\
   \Nabla \zeta &= \frac{\hat{e}_{\zeta}}{R} = \frac{\hat{e}_{\zeta}}{R_{c 0}} + O \left( \frac{\epsilon}{R_{0}} \right) , \label{eq:gradZeta}
\end{align}
where $\hat{e}_{\theta}$ and $\hat{e}_{\zeta}$ are the poloidal and toroidal angle unit vectors respectively. From this we know that the coordinate scalar triple product is
\begin{align}
   \Nabla \psi \cdot \left( \Nabla \theta \times \Nabla \zeta \right) = \frac{1}{J} = \frac{1}{r_{\psi 0} R_{c 0}} \frac{d \psi}{d r_{\psi}} \left( \left. \frac{\partial r}{\partial r_{\psi}} \right|_{\psi_{0}} \right)^{-1} + O \left( m_{c}^{-2} \frac{B_{0}}{R_{0}} \right) , \label{eq:tripleProd}
\end{align}
which is needed to calculate the second term of \refEq{eq:driftVelAlphaAspect}. Since we are using $d \psi / d r_{\psi}$ as an input instead of $q$, it is simple to find $\partial R / \partial \psi$ from \refEq{eq:partialRpartialrPsi} in order to calculate the first term of \refEq{eq:driftVelAlphaAspect}.

At this point we see that we have calculated all of the quantities appearing in \refEq{eq:driftVelAlphaAspect}, except for $\partial \alpha / \partial \psi$. This is specified by \refEq{eq:gradAlphaPsiAspect} and is made up of three terms. All of the terms require that we know
\begin{align}
   B_{p} = \frac{1}{J} \frac{\partial l_{p}}{\partial \theta} = \frac{1}{R_{c 0}} \frac{d \psi}{d r_{\psi}} \left( \left. \frac{\partial r}{\partial r_{\psi}} \right|_{\psi_{0}} \right)^{-1} + O \left( m_{c}^{-2} B_{p} \right) , \label{eq:BpAspect}
\end{align}
which is found using \refEq{eq:dLpdtheta}, \refEq{eq:poloidalFieldDef}, \refEq{eq:gradIdentity}, \refEq{eq:gradZeta}, and \refEq{eq:tripleProd}. Using \refEq{eq:nonMirrordLpdTheta}, \refEq{eq:nonMirrorPolCurv}, and \refEq{eq:BpAspect}, we can determine the integrand (i.e. \refEq{eq:gradAlphaAspectIntegrand}) that appears in the first term to be
\begin{align}
   \left( \frac{\partial A_{\alpha}}{\partial \psi} \right)_{\text{orthog}} &= B_{c 0} \left( \frac{d \psi}{d r_{\psi}} \right)^{-2} \left( \left. \frac{\partial r}{\partial r_{\psi}} \right|_{\psi_{0}} \right)^{2} \label{eq:nonMirrorAlphaIntegrand} \\
   &\times \left[ \left( \hat{s}' - 2 \right) \left. \frac{\partial r}{\partial r_{\psi}} \right|_{\psi_{0}} + 2 \left( 1 - \frac{1}{r_{\psi 0}} \frac{d^{2} r_{0}}{d \theta'^{2}} \right) \right] + O \left( \frac{m_{c}^{-2}}{r_{\psi}^{2} B_{0}} \right) \nonumber
\end{align}
to lowest order in aspect ratio. Calculating the indefinite integral of \refEq{eq:nonMirrorAlphaIntegrand} is straightforward and is explicitly given by \refEq{eq:dAalphadpsiIntegral}. The second term of \refEq{eq:gradAlphaPsiAspect} is found to be
\begin{align}
   \frac{B_{c 0}}{R_{c 0}^{3} B_{p}^{3}} \dlpdthetaPrime \Nabla \psi \cdot \Nabla \theta' = - \frac{B_{c 0}}{r_{\psi 0}} \left( \frac{d \psi}{d r_{\psi}} \right)^{-2} \left( \left. \frac{\partial r}{\partial r_{\psi}} \right|_{\psi_{0}} \right)^{2} \frac{d r_{0}}{d \theta'} + O \left( \frac{m_{c}^{-2}}{r_{\psi}^{2} B_{0}} \right) .
\end{align}
by substituting \refEq{eq:gradIdentity}, \refEq{eq:nonMirrordLpdTheta}, \refEq{eq:gradTheta}, and \refEq{eq:BpAspect}. At this point, by specifying the free parameter
\begin{align}
   \frac{d \theta_{\alpha}}{d \psi} &= \left( \frac{B_{c 0}}{R_{c 0} B_{p}} \dlpdtheta \right)_{\theta = \theta_{\alpha}}^{-1} \label{eq:nonMirrorThetaAlphaDerivCond} \\
   &\times \left[ - \left. \int_{\theta_{0}}^{\theta_{\alpha}} \right|_{\psi} d \theta' \left( \frac{\partial A_{\alpha}}{\partial \psi} \right)_{\text{orthog}} + \left( \frac{B_{c 0}}{R_{c 0}^{3} B_{p}^{3}} \dlpdthetaPrime \Nabla \psi \cdot \Nabla \theta' \right)_{\theta' = \theta_{\alpha}} \right] \nonumber
\end{align}
(not to be confused with \refEq{eq:thetaAlphaDerivTilted}, the value of $d \theta_{\alpha} / d \psi$ used in the previous chapter), we can use the third term of \refEq{eq:gradAlphaPsiAspect} to eliminate all of the terms in $\partial \alpha / \partial \psi$ that do not depend on $\theta$. Here $\theta_{0}$ is defined such that $\left. \int_{\theta_{0}}^{\theta_{\alpha}} \right|_{\psi} d \theta' \left( \partial A_{\alpha} / \partial \psi \right)_{\text{orthog}}$ does not have a term that is constant in poloidal angle. Additionally, we choose $\theta_{\alpha} \left( \psi_{0} \right) = 0$ for simplicity. Given this choice, \refEq{eq:gradAlphaPsiAspect} becomes
\begin{align}
   \frac{\partial \alpha}{\partial \psi} &= - \left. \int_{\theta_{0}}^{\theta} \right|_{\psi} d \theta' \left( \frac{\partial A_{\alpha}}{\partial \psi} \right)_{\text{orthog}} - \frac{B_{c 0}}{r_{\psi 0}} \left( \frac{d \psi}{d r_{\psi}} \right)^{-2} \left( \left. \frac{\partial r}{\partial r_{\psi}} \right|_{\psi_{0}} \right)^{2} \frac{d r_{0}}{d \theta} + O \left( \frac{m_{c}^{-2}}{r_{\psi}^{2} B_{0}} \right) . \label{eq:nonMirrorAlphaIntegral}
\end{align}
Substituting \refEq{eq:partialRpartialTheta}, \refEq{eq:partialRpartialrPsi}, \refEq{eq:tripleProd}, and \refEq{eq:nonMirrorAlphaIntegral} into \refEq{eq:driftVelAlphaAspect} gives
\begin{align}
   v_{d s \alpha} &= \frac{B_{c 0}}{R_{c 0} \Omega_{s}} \left( \frac{d \psi}{d r_{\psi}} \right)^{-1} \Bigg[ \frac{d r_{0}}{d r_{\psi}} \Cos{\theta} + \frac{1}{r_{\psi 0}} \frac{d r_{0}}{d r_{\psi}} \frac{d r_{0}}{d \theta} \Sin{\theta} \\
   &+ \frac{1}{B_{c 0}} \left( \frac{d \psi}{d r_{\psi}} \right)^{2} \left( \Sin{\theta} - \frac{1}{r_{\psi 0}} \frac{\partial r_{0}}{\partial \theta} \Cos{\theta} \right) \left( \frac{d r_{0}}{d r_{\psi}} \right)^{-1} \left. \int_{\theta_{0}}^{\theta} \right|_{\psi} d \theta' \left( \frac{\partial A_{\alpha}}{\partial \psi} \right)_{\text{orthog}} \Bigg] \nonumber \\
   &+ O \left( \frac{m_{c}^{-2}}{r_{\psi} R_{0} \Omega_{s}} \right) . \nonumber
\end{align}
To lowest order, this is the usual result for circular flux surfaces,
\begin{align}
   v_{d s \alpha 0} &= \frac{B_{c 0}}{R_{c 0} \Omega_{s}} \left( \frac{d \psi}{d r_{\psi}} \right)^{-1} \left( \Cos{\theta} + \hat{s}' \theta \Sin{\theta} \right) . \nonumber
\end{align}
To next order this is a complicated expression with the form of
\begin{align}
   v_{d s \alpha 1} &= D_{1} \theta \Sin{\theta} + \left( D_{2} \Sin{\theta} + D_{3} \theta \Cos{\theta} \right) \left( D_{4} \Sin{z_{m s}} + D_{5} \Sin{z_{n s}} \right) \nonumber \\
   &+ \left( D_{6} \Cos{\theta} + D_{7} \theta \Sin{\theta} \right) \left( D_{8} \Cos{z_{m s}} + D_{9} \Cos{z_{n s}} \right) + D_{10} \Sin{\theta} \label{eq:nonMirrorAlphaDrift} \\
   &\times \left[ \Sin{\left( n - m \right) \theta} \Cos{n \theta_{t n} - m \theta_{t m}} + \Cos{\left( n - m \right) \theta} \Sin{n \theta_{t n} - m \theta_{t m}} \right] . \nonumber
\end{align}
where $D_{i}$ are constants (the full expression is given in appendix \ref{app:nonMirrorGeoCoeffs}). Even after averaging over $z$ the last term remains, which has a coefficient of
\begin{align}
   D_{10} = \frac{r_{\psi 0}}{\left( n - m \right)} \left( m^{2} \left( \Delta_{m} - 1 \right) \frac{d \Delta_{n}}{d r_{\psi}} + n^{2} \left( \Delta_{n} - 1 \right) \frac{d \Delta_{m}}{d r_{\psi}} \right) . \label{eq:symBreakCoeff}
\end{align}
As we will show shortly, this term, which does not disappear after averaging over $z$, in general breaks the up-down symmetry of the gyrokinetic equations to $O \left( m_{c}^{-1} \right)$. We see that if we set $\Delta_{m} = 1$, $\Delta_{n} = 1$, or $n = m$ we produce mirror symmetric flux surfaces without an envelope and this symmetry-breaking term cancels. Also, we note that if $n \theta_{t n} = m \theta_{t m}$ the symmetry-breaking term also vanishes. This condition is only met when the envelope created by the beating of the two high-order shaping modes is up-down symmetric. Lastly, if we break our ordering of $n - m \sim 1$ and set $n=2m$, not only does the symmetry breaking term drop by an order (due to the $n-m$ in the denominator), but it varies on the fast spatial scale and will average to zero. This corresponds to flux surface shapes that lack an envelope like an ellipse, which only has Fourier components that are multiples of two.

Appendix \ref{app:nonMirrorGeoCoeffs} gives the full explicit expressions for all six geometric coefficients to lowest order in aspect ratio. We find those that do not depend on $\Nabla \alpha$ (i.e. $v_{d s \psi}$ and $\left| \Nabla \psi \right|^{2}$) are up-down symmetric in $\theta$ to $O \left( m_{c}^{-1} \right)$ after averaging over the fast spatial scale. However, the other three coefficients (i.e. $v_{d s \alpha}$, $\Nabla \psi \cdot \Nabla \alpha$, and $\left| \Nabla \alpha \right|^{2}$) lose their large-scale symmetry at $O \left( m_{c}^{-1} \right)$. The symmetry breaking terms arise from the interaction between $\kappa_{p}$ and $B_{p}^{-2}$ in \refEq{eq:gradAlphaAspectIntegrand}. Since $m_{c} \gg 1$ the second order derivatives in $\kappa_{p}$ (see \refEq{eq:poloidalCurv}) brings the effect of shaping from $O \left( m_{c}^{-2} \right)$ to $O \left( 1 \right)$. This shaping can then beat with the $O \left( m_{c}^{-1} \right)$ shaping in $B_{p}^{-2}$ and break the symmetry of the geometric coefficients to $O \left( m_{c}^{-1} \right)$. We note that $\kappa_{p}$ is ``normal'' curvature (i.e. perpendicular to the flux surface), as opposed to ``geodesic'' curvature (i.e. within the flux surface) \cite{RafiqNormalGeodesicCurv2005}. The importance of $\kappa_{p}$ is surprising because it arises from the {\it poloidal} field, not the {\it toroidal} field. Usually the focus is on the ``normal'' curvature of the {\it toroidal} field because it generates the largest contribution to the total field line curvature that appears in the magnetic drifts.

Ultimately, this beating between $\kappa_{p}$ and $B_{p}^{-2}$ is the dominant mechanism that breaks the up-down symmetry of the geometric coefficients to lowest order in aspect ratio. It is a subtle effect because the beating takes place in the integral in $\partial \alpha / \partial \psi$ (see \refEq{eq:gradAlphaPsiAspect} and \refEq{eq:gradAlphaAspectIntegrand}), which is contained in $\Nabla \alpha$ (see \refEq{eq:gradAlphaAspect}). However, it does not enter into the magnetic drift velocity itself. From studying these equations we can see that the beating between $\kappa_{p}$ and $B_{p}^{-2}$ alters the local magnetic shear (but without modifying the total magnetic shear). Therefore, in the perfect $m_{c} \gg 1$ limit, adding a small amount of the fast shaping modifies local field line pitch from one flux surface to the next (without changing the field line spacing). This perturbs the local cross-sectional shape (i.e. the shape in the plane perpendicular to the field line) of the turbulent eddies as they wrap around the torus. Specifically, it tilts the eddy cross-sectional shape a small amount one way or the other, depending on the location along the field line. This perturbation to the eddy has a slowly varying envelope, which is then acted on by the unperturbed up-down symmetric magnetic drifts.

The interaction of $\kappa_{p}$ and $B_{p}^{-2}$ certainly can create an envelope that breaks the up-down symmetry of the geometric coefficients and generates momentum flux, but it is still unclear at what order. By expanding the gyrokinetic and quasineutrality equations order-by-order in $m_{c}^{-1} \ll 1$ we will connect the symmetry-breaking of the geometric coefficients to symmetry-breaking of the distribution function and non-zero momentum flux.

\subsubsection{$O \left( m_{c} \right)$ gyrokinetic equation}

Expanding \refEq{eq:gyrokineticEq} to lowest order in $m_{c} \gg 1$ gives
\begin{align}
  w_{||} \left( \hat{b} \cdot \Nabla \theta \right)_{0} m \left. \frac{\partial \widetilde{h}_{s 0}}{\partial z} \right|_{\theta, w_{||}, \mu} = 0 . \label{eq:gyrokinEqOminus1}
\end{align}
We see from \refEq{eq:gradparO0} that $\left( \hat{b} \cdot \Nabla \theta \right)_{0}$ is a constant, so integrating over $z$ gives
\begin{align}
  \widetilde{h}_{s 0} &= 0 \label{eq:hs0tilde} \\
  \overline{h}_{s 0} &= h_{s 0} . \label{eq:hs0avg}
\end{align}

\subsubsection{$O \left( 1 \right)$ quasineutrality equation}

Expanding \refEq{eq:quasineut} to lowest order in $m_{c} \gg 1$ gives
\begin{align}
   \phi_{0} &= \left( \sum_{s} \frac{Z_{s}^{2} e^{2} n_{s}}{T_{s}} \right)^{-1} \sum_{s} \frac{2 \pi Z_{s} e B_{c 0}}{m_{s}} \int dw_{||} d \mu \left( J_{0} \left( k_{\perp} \rho_{s} \right) \right)_{0} h_{s 0} . \label{eq:phi0}
\end{align}
Using \refEq{eq:hs0avg} and \refEq{eq:FLRO0} we see that
\begin{align}
  \widetilde{\phi}_{0} &= 0 \label{eq:phi0tilde} \\
  \overline{\phi}_{0} &= \phi_{0} = \left( \sum_{s} \frac{Z_{s}^{2} e^{2} n_{s}}{T_{s}} \right)^{-1} \sum_{s} \frac{2 \pi Z_{s} e B_{c 0}}{m_{s}} \int dw_{||} d \mu \overline{\left( J_{0} \left( k_{\perp} \rho_{s} \right) \right)}_{0} \overline{h}_{s 0} . \label{eq:phi0avg}
\end{align}

\subsubsection{$O \left( 1 \right)$ gyrokinetic equation}

Expanding \refEq{eq:gyrokineticEq} to $O \left( 1 \right)$ gives
\begin{align}
  \frac{\partial h_{s 0}}{\partial t} &+ w_{||} \left( \hat{b} \cdot \Nabla \theta \right)_{0} \left( \left. \frac{\partial h_{s 0}}{\partial \theta} \right|_{z, w_{||}, \mu} + m \left. \frac{\partial \widetilde{h}_{s 1}}{\partial z} \right|_{\theta, w_{||}, \mu} \right) \nonumber \\
  &+ i \left( w^{2}_{||} + \frac{B_{c 0}}{m_{s}} \mu \right) \left( k_{\psi} v_{d s \psi 0} + k_{\alpha} v_{d s \alpha 0} \right) h_{s 0} \label{eq:gyrokinEqO0} \\
  &+ \left\{ \left( J_{0} \left( k_{\perp} \rho_{s} \right) \right)_{0} \phi_{0}, h_{s 0} \right\} - \frac{Z_{s} e F_{M s}}{T_{s}} \frac{\partial}{\partial t} \Big( \left( J_{0} \left( k_{\perp} \rho_{s} \right) \right)_{0} \phi_{0} \Big) \nonumber \\
  &+ i k_{\alpha} \left( J_{0} \left( k_{\perp} \rho_{s} \right) \right)_{0} \phi_{0}  F_{M s} \left[ \frac{1}{n_{s}} \frac{d n_{s}}{d \psi} + \left( \frac{m_{s} w^{2}}{2 T_{s}} - \frac{3}{2} \right) \frac{1}{T_{s}} \frac{d T_{s}}{d \psi} \right] = 0 . \nonumber
\end{align}
Averaging over $z$ after using \refEq{eq:hs0avg}, \refEq{eq:phi0avg}, and \refEq{eq:gradparO0} through \refEq{eq:FLRO0} gives
\begin{align}
  \frac{\partial \overline{h}_{s 0}}{\partial t} &+ w_{||} \overline{\left( \hat{b} \cdot \Nabla \theta \right)}_{0} \left. \frac{\partial \overline{h}_{s 0}}{\partial \theta} \right|_{z, w_{||}, \mu} + i \left( w^{2}_{||} + \frac{B_{c 0}}{m_{s}} \mu \right) \left( k_{\psi} \overline{v}_{d s \psi 0} + k_{\alpha} \overline{v}_{d s \alpha 0} \right) \overline{h}_{s 0} \nonumber \\
 &+ \left\{ \overline{\left( J_{0} \left( k_{\perp} \rho_{s} \right) \right)}_{0} \overline{\phi}_{0}, \overline{h}_{s 0} \right\} - \frac{Z_{s} e F_{M s}}{T_{s}} \frac{\partial}{\partial t} \left( \overline{\left( J_{0} \left( k_{\perp} \rho_{s} \right) \right)}_{0} \overline{\phi}_{0} \right) \label{eq:gyrokinEqO0avg} \\
 &+ i k_{\alpha} \overline{\left( J_{0} \left( k_{\perp} \rho_{s} \right) \right)}_{0} \overline{\phi}_{0}  F_{M s} \left[ \frac{1}{n_{s}} \frac{d n_{s}}{d \psi} + \left( \frac{m_{s} w^{2}}{2 T_{s}} - \frac{3}{2} \right) \frac{1}{T_{s}} \frac{d T_{s}}{d \psi} \right] = 0 , \nonumber
\end{align}
which does not depend on $z$. From \refEq{eq:gradparO0} through \refEq{eq:FLRO0} we see that \refEq{eq:phi0avg} and \refEq{eq:gyrokinEqO0avg} are unchanged by the $\left( k_{\psi}, k_{\alpha}, \theta, w_{||}, \mu, t \right) \rightarrow \left( - k_{\psi}, k_{\alpha}, - \theta, - w_{||}, \mu, t \right)$ coordinate system transformation when $\overline{h}_{s 0} \rightarrow - \overline{h}_{s 0}$ and $\overline{\phi}_{0} \rightarrow - \overline{\phi}_{0}$. This symmetry of the $O \left( 1 \right)$ gyrokinetic equations is important because, as discussed in section \ref{sec:upDownSymArg}, it can be used to demonstrate that the momentum flux must be zero.

Subtracting \refEq{eq:gyrokinEqO0avg} from \refEq{eq:gyrokinEqO0} we find
\begin{align}
  m w_{||} \overline{\left( \hat{b} \cdot \Nabla \theta \right)}_{0} \left. \frac{\partial \widetilde{h}_{s 1}}{\partial z} \right|_{\theta, w_{||}, \mu} = 0 .
\end{align}
Therefore, we know that
\begin{align}
  \widetilde{h}_{s 1} &= 0 \label{eq:hs1tilde} \\
  \overline{h}_{s 1} &= h_{s 1} . \label{eq:hs1avg}
\end{align}

\subsubsection{$O \left( 1 \right)$ momentum transport}

Expanding \refEq{eq:momFluxSimple} to lowest order gives
\begin{align}
  \Pi_{\zeta s 0} &= \frac{i R_{c 0} B_{c 0}}{\oint d\theta \left( \hat{b} \cdot \Nabla \theta \right)_{0}^{-1}} \sum_{k_{\psi}, k_{\alpha}} k_{\alpha} \oint d\theta \left( \hat{b} \cdot \Nabla \theta \right)_{0}^{-1} \label{eq:pi0} \\
  &\times \int dw_{||} d\mu w_{||} \left( J_{0} \left( k_{\perp} \rho_{s} \right) \right)_{0} \phi_{0} \left( k_{\psi}, k_{\alpha} \right) h_{s 0} \left( - k_{\psi}, - k_{\alpha} \right) . \nonumber
\end{align}
Using \refEq{eq:hs0avg}, \refEq{eq:phi0avg}, \refEq{eq:gradparO0}, and \refEq{eq:FLRO0} we find that
\begin{align}
    \Pi_{\zeta s 0} &= \frac{i R_{c 0} B_{c 0}}{2 \pi} \sum_{k_{\psi}, k_{\alpha}} k_{\alpha} \oint d\theta \int dw_{||} d\mu w_{||} \overline{\left( J_{0} \left( k_{\perp} \rho_{s} \right) \right)}_{0} \overline{\phi}_{0} \overline{h}_{s 0} . \label{eq:pi0avg}
\end{align}
Therefore by the $\left( k_{\psi}, k_{\alpha}, \theta, w_{||}, \mu, t \right) \rightarrow \left( - k_{\psi}, k_{\alpha}, - \theta, - w_{||}, \mu, t \right)$ symmetry outlined in section \ref{sec:upDownSymArg} we know that $\Pi_{\zeta s 0} = 0$ when averaged over a turbulent decorrelation time.

\subsubsection{$O \left( m_{c}^{-1} \right)$ quasineutrality equation}

Equation \refEq{eq:quasineut}, expanded to $O \left( m_{c}^{-1} \right)$, is
\begin{align}
   \phi_{1} &= \left( \sum_{s} \frac{Z_{s}^{2} e^{2} n_{s}}{T_{s}} \right)^{-1} \sum_{s} \frac{2 \pi Z_{s} e B_{c 0}}{m_{s}} \int dw_{||} d \mu \Big( \left( J_{0} \left( k_{\perp} \rho_{s} \right) \right)_{1} h_{s 0} + \left( J_{0} \left( k_{\perp} \rho_{s} \right) \right)_{0} h_{s 1} \Big) . \label{eq:phi1}
\end{align}
Using \refEq{eq:hs0avg}, \refEq{eq:hs1avg}, and  \refEq{eq:FLRO0}, then averaging over $z$ gives
\begin{align}
   \overline{\phi}_{1} &= \left( \sum_{s} \frac{Z_{s}^{2} e^{2} n_{s}}{T_{s}} \right)^{-1} \sum_{s} \frac{2 \pi Z_{s} e B_{c 0}}{m_{s}} \int dw_{||} d \mu \left( \overline{\left( J_{0} \left( k_{\perp} \rho_{s} \right) \right)}_{1} \overline{h}_{s 0} + \overline{\left( J_{0} \left( k_{\perp} \rho_{s} \right) \right)}_{0} \overline{h}_{s 1} \right) . \label{eq:phi1avg}
\end{align}
Note that $\widetilde{\phi}_{1} \neq 0$.

\subsubsection{$O \left( m_{c}^{-1} \right)$ gyrokinetic equation}

Expanding \refEq{eq:gyrokineticEq} to $O \left( m_{c}^{-1} \right)$, using \refEq{eq:hs0avg}, \refEq{eq:phi0avg}, \refEq{eq:hs1avg}, \refEq{eq:hs1tilde}, and \refEq{eq:gradparO0} through \refEq{eq:FLRO0}, gives
\begin{align}
  \frac{\partial \overline{h}_{s 1}}{\partial t} &+ w_{||} \left( \hat{b} \cdot \Nabla \theta \right)_{0} \left( \left. \frac{\partial \overline{h}_{s 1}}{\partial \theta} \right|_{z, w_{||}, \mu} + m \left. \frac{\partial \widetilde{h}_{s 2}}{\partial z} \right|_{\theta, w_{||}, \mu} \right) \nonumber \\
  &+ i \left( w^{2}_{||} + \frac{B_{c 0}}{m_{s}} \mu \right) \left( k_{\psi} \overline{v}_{d s \psi 0} + k_{\alpha} \overline{v}_{d s \alpha 0} \right) \overline{h}_{s 1} + \left\{ \overline{\left( J_{0} \left( k_{\perp} \rho_{s} \right) \right)}_{0} \overline{\phi}_{0}, \overline{h}_{s 1} \right\} \nonumber \\
 &+ \left\{ \overline{\left( J_{0} \left( k_{\perp} \rho_{s} \right) \right)}_{0} \phi_{1}, \overline{h}_{s 0} \right\} - \frac{Z_{s} e F_{M s}}{T_{s}} \frac{\partial}{\partial t} \left( \overline{\left( J_{0} \left( k_{\perp} \rho_{s} \right) \right)}_{0} \phi_{1} \right) \nonumber \\
 &+ i k_{\alpha} \overline{\left( J_{0} \left( k_{\perp} \rho_{s} \right) \right)}_{0} \phi_{1}  F_{M s} \left[ \frac{1}{n_{s}} \frac{d n_{s}}{d \psi} + \left( \frac{m_{s} w^{2}}{2 T_{s}} - \frac{3}{2} \right) \frac{1}{T_{s}} \frac{d T_{s}}{d \psi} \right] \label{eq:gyrokinEqO1simple} \\
 &= w_{||} \left( \hat{b} \cdot \Nabla \theta \right)_{1} \left. \frac{\partial \overline{h}_{s 0}}{\partial \theta} \right|_{z, w_{||}, \mu} - i \left( w^{2}_{||} + \frac{B_{c 0}}{m_{s}} \mu \right) \left( k_{\psi} v_{d s \psi 1} + k_{\alpha} v_{d s \alpha 1} \right) \overline{h}_{s 0} \nonumber \\
 &- \left\{ \left( J_{0} \left( k_{\perp} \rho_{s} \right) \right)_{1} \overline{\phi}_{0}, \overline{h}_{s 0} \right\} + \frac{Z_{s} e F_{M s}}{T_{s}} \frac{\partial}{\partial t} \Big( \left( J_{0} \left( k_{\perp} \rho_{s} \right) \right)_{1} \overline{\phi}_{0} \Big) \nonumber \\
 &- i k_{\alpha} \left( J_{0} \left( k_{\perp} \rho_{s} \right) \right)_{1} \overline{\phi}_{0}  F_{M s} \left[ \frac{1}{n_{s}} \frac{d n_{s}}{d \psi} + \left( \frac{m_{s} w^{2}}{2 T_{s}} - \frac{3}{2} \right) \frac{1}{T_{s}} \frac{d T_{s}}{d \psi} \right] . \nonumber
\end{align}
Averaging over $z$ we find that
\begin{align}
  \frac{\partial \overline{h}_{s 1}}{\partial t} &+ w_{||} \hat{b} \cdot \Nabla \theta \left. \frac{\partial \overline{h}_{s 1}}{\partial \theta} \right|_{z, w_{||}, \mu} + i \left( w^{2}_{||} + \frac{B_{c 0}}{m_{s}} \mu \right) \left( k_{\psi} \overline{v}_{d s \psi 0} + k_{\alpha} \overline{v}_{d s \alpha 0} \right) \overline{h}_{s 1} \nonumber \\
 &+ \left\{ \overline{\left( J_{0} \left( k_{\perp} \rho_{s} \right) \right)}_{0} \overline{\phi}_{0}, \overline{h}_{s 1} \right\} + \left\{ \overline{\left( J_{0} \left( k_{\perp} \rho_{s} \right) \right)}_{0} \overline{\phi}_{1}, \overline{h}_{s 0} \right\} - \frac{Z_{s} e F_{M s}}{T_{s}} \frac{\partial}{\partial t} \left( \overline{\left( J_{0} \left( k_{\perp} \rho_{s} \right) \right)}_{0} \overline{\phi}_{1} \right) \nonumber \\
 &+ i k_{\alpha} \overline{\left( J_{0} \left( k_{\perp} \rho_{s} \right) \right)}_{0} \overline{\phi}_{1}  F_{M s} \left[ \frac{1}{n_{s}} \frac{d n_{s}}{d \psi} + \left( \frac{m_{s} w^{2}}{2 T_{s}} - \frac{3}{2} \right) \frac{1}{T_{s}} \frac{d T_{s}}{d \psi} \right] \nonumber \\
 &= w_{||} \overline{\left( \hat{b} \cdot \Nabla \theta \right)}_{1} \left. \frac{\partial \overline{h}_{s 0}}{\partial \theta} \right|_{z, w_{||}, \mu} - i \left( w^{2}_{||} + \frac{B_{c 0}}{m_{s}} \mu \right) \left( k_{\psi} \overline{v}_{d s \psi 1} + k_{\alpha} \overline{v}_{d s \alpha 1} \right) \overline{h}_{s 0} \label{eq:gyrokinEqO1avg} \\
 &- \left\{ \overline{\left( J_{0} \left( k_{\perp} \rho_{s} \right) \right)}_{1} \overline{\phi}_{0}, \overline{h}_{s 0} \right\} + \frac{Z_{s} e F_{M s}}{T_{s}} \frac{\partial}{\partial t} \left( \overline{\left( J_{0} \left( k_{\perp} \rho_{s} \right) \right)}_{1} \overline{\phi}_{0} \right) \nonumber \\
 &- i k_{\alpha} \overline{\left( J_{0} \left( k_{\perp} \rho_{s} \right) \right)}_{1} \overline{\phi}_{0}  F_{M s} \left[ \frac{1}{n_{s}} \frac{d n_{s}}{d \psi} + \left( \frac{m_{s} w^{2}}{2 T_{s}} - \frac{3}{2} \right) \frac{1}{T_{s}} \frac{d T_{s}}{d \psi} \right] . \nonumber
\end{align}
From \refEq{eq:alphaDriftO1} and \refEq{eq:gradPsiDotGradAlphaO1} through \refEq{eq:FLRO1def} we see that \refEq{eq:phi1avg} and \refEq{eq:gyrokinEqO1avg} are \textit{not} symmetric in $\left( k_{\psi}, k_{\alpha}, \theta, w_{||}, \mu, t \right) \rightarrow \left( - k_{\psi}, k_{\alpha}, - \theta, - w_{||}, \mu, t \right)$ when $\overline{h}_{s 0} \rightarrow - \overline{h}_{s 0}$, $\overline{\phi}_{0} \rightarrow - \overline{\phi}_{0}$, $\overline{h}_{s 1} \rightarrow - \overline{h}_{s 1}$, and $\overline{\phi}_{1} \rightarrow - \overline{\phi}_{1}$. This is due to both the drift term $\overline{v}_{d s \alpha 1}$ as well as $\left( \Nabla \psi \cdot \Nabla \alpha \right)_{1}$ and $\left| \Nabla \alpha \right|^{2}_{1}$ in $\overline{\left( J_{0} \left( k_{\perp} \rho_{s} \right) \right)}_{1}$ (which accounts for finite gyroradius effects).

\subsubsection{$O \left( m_{c}^{-1} \right)$ momentum transport}

Expanding \refEq{eq:momFluxSimple} to $O \left( m_{c}^{-1} \right)$ and using \refEq{eq:gradparO0} and \refEq{eq:gradparO1}  gives
\begin{align}
  \Pi_{\zeta s 1} &= \frac{i R_{c 0} B_{c 0}}{2 \pi} \sum_{k_{\psi}, k_{\alpha}} k_{\alpha} \oint d\theta \int dw_{||} d\mu w_{||} \Bigg[ - \overline{\left( \hat{b} \cdot \Nabla \theta \right)}_{0}^{-1} \left( \hat{b} \cdot \Nabla \theta \right)_{1} \left( J_{0} \left( k_{\perp} \rho_{s} \right) \right)_{0} \phi_{0} h_{s 0} \nonumber \\
  &+ \left( J_{0} \left( k_{\perp} \rho_{s} \right) \right)_{1} \phi_{0} h_{s 0} + \left( J_{0} \left( k_{\perp} \rho_{s} \right) \right)_{0} \phi_{1} h_{s 0} + \left( J_{0} \left( k_{\perp} \rho_{s} \right) \right)_{0} \phi_{0} h_{s 1}  \Bigg] . \label{eq:pi1}
\end{align}
After applying \refEq{eq:FLRO0}, \refEq{eq:gradparO1}, \refEq{eq:hs0avg}, \refEq{eq:phi0avg}, and \refEq{eq:hs1avg} we find
\begin{align}
    \Pi_{\zeta s 1} &= i R_{c 0} B_{c 0} \sum_{k_{\psi}, k_{\alpha}} k_{\alpha} \oint d\theta \int dw_{||} d\mu w_{||} \left[ \overline{\left( J_{0} \left( k_{\perp} \rho_{s} \right) \right)}_{1} \overline{\phi}_{0} \overline{h}_{s 0} \right. \label{eq:pi1avg} \\
    &+ \left. \overline{\left( J_{0} \left( k_{\perp} \rho_{s} \right) \right)}_{0} \overline{\phi}_{1} \overline{h}_{s 0} + \overline{\left( J_{0} \left( k_{\perp} \rho_{s} \right) \right)}_{0} \overline{\phi}_{0} \overline{h}_{s 1} \right] . \nonumber
\end{align}
Since neither $\overline{\left( J_{0} \left( k_{\perp} \rho_{s} \right) \right)}_{1}$, $\overline{\phi}_{1}$, nor $\overline{h}_{s 1}$ have a definite parity in $\left( k_{\psi}, k_{\alpha}, \theta, w_{||}, \mu, t \right) \rightarrow \left( - k_{\psi}, k_{\alpha}, - \theta, - w_{||}, \mu, t \right)$, we cannot constrain $\Pi_{\zeta s 1}$ to be zero. This means that we expect the momentum flux to scale as $\Pi_{\zeta s} \sim m_{c}^{-1} \Pi_{gB}$. Since the energy flux $Q_{s}$ is non-zero to lowest order in $m_{c}$ (i.e. circular flux surfaces still have a non-zero energy flux), we can also say that $\left( v_{th i} / R_{c 0} \right) \Pi_{\zeta s} / Q_{s} \sim m_{c}^{-1}$.

%===================================================%
\subsection{General shaping in the gyrokinetic model}
\label{subsec:genNonMirrorSymShaping}
%===================================================%

Section \ref{subsec:practicalNonMirrorSymShaping} showed that the momentum flux scales as $O \left( m_{c}^{-1} \right)$, given a specific geometry (circular with two high-order cylindrical harmonic shaping effects) and a specific shaping ordering ($\Delta_{m} - 1 \sim m_{c}^{-2}$). However, this is a concrete, analytically tractable example of a more general argument. Here we will bound the symmetry breaking of the geometric coefficients by systematically ordering all of the quantities that compose them. We will make no presumptions about the low mode number shaping (other than to assume up-down symmetry) nor will we order the size of the fast mode number shaping (other than to assume $\Delta_{m} - 1 \ll 1$). We note that the analysis of this section does not use an expansion in aspect ratio.

\begin{table}
  \centering
  \caption{Scalings of the strength of fast plasma shaping effects for various geometric quantities, where $Q_{\text{low}}$ is the geometric quantity in the absence of any large mode number shaping (i.e. $\Delta_{m} = 1$) and all quantities are evaluated at $r_{\psi} = r_{\psi 0}$.}
  \begin{tabular}{ l c c c }
    $Q$ & Reference & $\widetilde{Q} / Q_{\text{low}}$ & $ \left( \overline{Q} - Q_{\text{low}} \right) / Q_{\text{low}}$ \\
    \hline
    $r$ & \refEq{eq:fluxSurfofInterestShapeWeak} & $\Delta_{m} - 1$ & $0$ \\
    $\partial r / \partial r_{\psi}$ & \refEq{eq:fluxSurfofInterestDerivWeak}, \refEq{eq:changeShapeOrdering} & $m_{c} \left( \Delta_{m} - 1 \right)$ & $0$ \\
    $R$ & \refEq{eq:geoMajorRadiusWeak} & $\Delta_{m} - 1$ & $0$ \\
    $Z$ & \refEq{eq:geoAxialWeak} & $\Delta_{m} - 1$ & $0$ \\
    $\partial R / \partial r_{\psi}$ & \refEq{eq:geoMajorRadiusWeak} & $m_{c} \left( \Delta_{m} - 1 \right)$ & $0$ \\
    $\partial Z / \partial r_{\psi}$ & \refEq{eq:geoAxialWeak} & $m_{c} \left( \Delta_{m} - 1 \right)$ & $0$ \\
    $\partial R / \partial \theta$ & \refEq{eq:geoMajorRadiusWeak} & $m_{c} \left( \Delta_{m} - 1 \right)$ & $0$ \\
    $\partial Z / \partial \theta$ & \refEq{eq:geoAxialWeak} & $m_{c} \left( \Delta_{m} - 1 \right)$ & $0$ \\
    $\partial^{2} R / \partial \theta^{2}$ & \refEq{eq:geoMajorRadiusWeak} & $m_{c}^{2} \left( \Delta_{m} - 1 \right)$ & $0$ \\
    $\partial^{2} Z / \partial \theta^{2}$ & \refEq{eq:geoAxialWeak} & $m_{c}^{2} \left( \Delta_{m} - 1 \right)$ & $0$ \\
    $\Nabla r_{\psi}$ & \refEq{eq:gradIdentity} & $m_{c} \left( \Delta_{m} - 1 \right)$ & $m_{c}^{2} \left( \Delta_{m} - 1 \right)^{2}$ \\
    $\Nabla \theta$ & \refEq{eq:fluxSurfofInterestShapeWeak} & $\Delta_{m} - 1$ & $\left( \Delta_{m} - 1 \right)^{2}$ \\
    $\Nabla \zeta$ & \refEq{eq:geoMajorRadiusWeak} & $\Delta_{m} - 1$ & $\left( \Delta_{m} - 1 \right)^{2}$ \\
    $B_{\zeta}$ & \refEq{eq:millerTorField} & $\Delta_{m} - 1$ & $\left( \Delta_{m} - 1 \right)^{2}$ \\
    $\Nabla \psi$ & \refEq{eq:gradIdentity} & $m_{c} \left( \Delta_{m} - 1 \right)$ & $m_{c}^{2} \left( \Delta_{m} - 1 \right)^{2}$ \\ \cline{4-4}
    $\left| \Nabla \psi \right|^{2}$ & \refEq{eq:gradIdentity} & $m_{c} \left( \Delta_{m} - 1 \right)$ & \multicolumn{1}{|c|}{$m_{c}^{2} \left( \Delta_{m} - 1 \right)^{2}$} \\ \cline{4-4}
    $B_{p}$ & \refEq{eq:poloidalFieldDef} & $m_{c} \left( \Delta_{m} - 1 \right)$ & $m_{c}^{2} \left( \Delta_{m} - 1 \right)^{2}$ \\ \cline{4-4}
    $B$ & \refEq{eq:poloidalFieldDef}, \refEq{eq:millerTorField} & $m_{c} \left( \Delta_{m} - 1 \right)$ & \multicolumn{1}{|c|}{$m_{c}^{2} \left( \Delta_{m} - 1 \right)^{2}$} \\ \cline{4-4}
    $\hat{b} \cdot \Nabla \theta$ & \refEq{eq:fluxSurfofInterestShapeWeak}, \refEq{eq:poloidalFieldDef} & $m_{c} \left( \Delta_{m} - 1 \right)$ & $m_{c}^{2} \left( \Delta_{m} - 1 \right)^{2}$ \\
    $\partial B / \partial \theta$ & \refEq{eq:poloidalFieldDef}, \refEq{eq:millerTorField} & $m_{c}^{2} \left( \Delta_{m} - 1 \right)$ & $m_{c}^{2} \left( \Delta_{m} - 1 \right)^{2}$ \\ \cline{4-4}
    $v_{d s \psi}$ & \refEq{eq:radialGCdriftvelocity} & $m_{c}^{2} \left( \Delta_{m} - 1 \right)$ & \multicolumn{1}{|c|}{$m_{c}^{3} \left( \Delta_{m} - 1 \right)^{2}$} \\ \cline{4-4}
    $a_{s ||}$ & \refEq{eq:parallelAcceleration} & $m_{c}^{2} \left( \Delta_{m} - 1 \right)$ & \multicolumn{1}{|c|}{$m_{c}^{3} \left( \Delta_{m} - 1 \right)^{2}$} \\ \cline{4-4}
    $d l_{p} / d \theta$ & \refEq{eq:dLpdtheta} & $m_{c} \left( \Delta_{m} - 1 \right)$ & $m_{c}^{2} \left( \Delta_{m} - 1 \right)^{2}$ \\
    $\kappa_{p}$ & \refEq{eq:poloidalCurv} & $m_{c}^{2} \left( \Delta_{m} - 1 \right)$ & $m_{c}^{2} \left( \Delta_{m} - 1 \right)^{2}$ \\
    $A_{\alpha}$ & \refEq{eq:IalphaDef} & $m_{c} \left( \Delta_{m} - 1 \right)$ & $m_{c}^{2} \left( \Delta_{m} - 1 \right)^{2}$ \\
    $\partial A_{\alpha} / \partial \psi$ & \refEq{eq:dIntegranddpsiRearrange} & $m_{c}^{2} \left( \Delta_{m} - 1 \right)$ & $m_{c}^{3} \left( \Delta_{m} - 1 \right)^{2}$ \\
    $\int d \theta ~ \partial A_{\alpha} / \partial \psi$ & \refEq{eq:gradAlphaInitial} & $m_{c} \left( \Delta_{m} - 1 \right)$ & $m_{c}^{3} \left( \Delta_{m} - 1 \right)^{2}$ \\
    $\Nabla \alpha$ & \refEq{eq:gradAlphaInitial} & $m_{c} \left( \Delta_{m} - 1 \right)$ & $m_{c}^{3} \left( \Delta_{m} - 1 \right)^{2}$ \\
    $\partial B / \partial r_{\psi}$ & \refEq{eq:dBpdpsi}, \refEq{eq:dBtordpsi} & $m_{c}^{2} \left( \Delta_{m} - 1 \right)$ & $m_{c}^{3} \left( \Delta_{m} - 1 \right)^{2}$ \\ \cline{4-4}
    $v_{d s \alpha}$ & \refEq{eq:alphaGCdriftvelocity} & $m_{c}^{2} \left( \Delta_{m} - 1 \right)$ & \multicolumn{1}{|c|}{$m_{c}^{3} \left( \Delta_{m} - 1 \right)^{2}$} \\ \cline{4-4}
    $\Nabla \psi \cdot \Nabla \alpha$ & \refEq{eq:gradIdentity}, \refEq{eq:gradAlphaInitial} & $m_{c} \left( \Delta_{m} - 1 \right)$ & \multicolumn{1}{|c|}{$m_{c}^{3} \left( \Delta_{m} - 1 \right)^{2}$} \\ \cline{4-4}
    $\left| \Nabla \alpha \right|^{2}$ & \refEq{eq:gradAlphaInitial} & $m_{c} \left( \Delta_{m} - 1 \right)$ & \multicolumn{1}{|c|}{$m_{c}^{3} \left( \Delta_{m} - 1 \right)^{2}$} \\ \cline{4-4}
  \end{tabular}
  \label{tab:shapingStrength}
\end{table}

Table \ref{tab:shapingStrength} gives a step-by-step summary of the results of the calculation. To begin, we must make some choices concerning the nature of the flux surface shape. The first two rows define the assumptions concerning the high-order flux surface shaping. We require that the high-order shaping must be periodic, that $\widetilde{r} \left( \theta, z \right) \sim O \left( \left( \Delta_{m} - 1 \right) r_{\psi 0} \right)$ on the flux surface of interest, and that $\widetilde{ \partial r / \partial r_{\psi}} \sim O \left( m_{c} \left( \Delta_{m} - 1 \right) \right)$ (which we discussed previously in arriving at \refEq{eq:changeShapeOrdering}). This is all consistent with \refEq{eq:fluxSurfofInterestShapeWeak} and \refEq{eq:fluxSurfofInterestDerivWeak}, which were used in the calculation of section \ref{subsec:practicalNonMirrorSymShaping}.

Now, we can derive the orderings for increasingly complex quantities and eventually find the geometric coefficients. For example, we can use \refEq{eq:geoMajorRadiusWeak} and \refEq{eq:geoAxialWeak} to derive the order that shaping enters into $R$ and $Z$. We also know that when we take a poloidal derivative of $\overline{Q}$, a $z$-independent quantity, it remains of the same order. However, when we take a poloidal derivative of $\widetilde{Q}$, the $z$-dependent part of a quantity, it gains an additional factor of $m$. Therefore, the orderings of the $z$-dependent parts of $\partial R / \partial \theta$ and $\partial Z / \partial \theta$ are larger than the orderings of $\widetilde{R}$ and $\widetilde{Z}$ by a factor of $m_{c}$. Also, when we calculate quantities such as $\Nabla r_{\psi}$ (see \refEq{eq:gradIdentity}) we get beating between the different high-order shaping effects. Therefore, when we Taylor expand in $m_{c} \gg 1$ and $\Delta_{m} - 1 \ll 1$, the shaping in the numerator and denominators of $\Nabla r_{\psi}$ can interact to produce terms that vary on the slow scale. This means that, when we use \refEq{eq:zAvg} to average over $z$, these slow terms remain and can break the up-down symmetry. The quantities $\Nabla \theta$ and $\Nabla \zeta$ can be expressed as $\hat{e}_{\theta} / r$ and $\hat{e}_{\zeta} / R$ respectively, so their scalings can be found by directly Taylor expanding \refEq{eq:fluxSurfofInterestShapeWeak} and \refEq{eq:geoMajorRadiusWeak}.

As discussed at the end of section \ref{subsubsec:geoCoeffs}, the poloidal curvature, $\kappa_{p}$, turns out to produce the most important symmetry-breaking term. In \refEq{eq:poloidalCurv} we see the two poloidal derivatives that bring the effect of shaping up to $O \left( m_{c}^{2} \left( \Delta_{m} - 1 \right) \right)$. However because of the relationship between $R \left( r_{\psi 0}, \theta \right)$ and $Z \left( r_{\psi 0}, \theta \right)$ (given by \refEq{eq:geoMajorRadiusWeak} and \refEq{eq:geoAxialWeak}), the beating between $\partial^{2} R / \partial \theta^{2}$ and $\partial Z / \partial \theta$ as well as $\partial^{2} Z / \partial \theta^{2}$ and $\partial R / \partial \theta$ cancels to $O\left( m_{c}^{3} \left( \Delta_{m} - 1 \right)^{2} \right)$ as shown in \refEq{eq:nonMirrorPolCurv}. Nevertheless, the poloidal curvature can still beat against the $O \left( m_{c} \left( \Delta_{m} - 1 \right) \right)$ shaping of $B_{p}^{-2}$ in \refEq{eq:gradAlphaAspectIntegrand}. This means that $\partial A_{\alpha} / \partial \psi$ (i.e. the integrand in $\Nabla \alpha$) contains $O \left( m_{c}^{3} \left( \Delta_{m} - 1 \right)^{2} \right)$ terms from the fast shaping that are independent of $z$ and break the up-down symmetry. When we take the integral to calculate $\Nabla \alpha$ the $z$-dependent terms lose a factor of $m_{c}$, but the $O \left( m_{c}^{3} \left( \Delta_{m} - 1 \right)^{2} \right)$ $z$-independent terms are not altered. Hence, the symmetry of the three geometric coefficients that contain $\Nabla \alpha$ is broken to $O \left( m_{c}^{3} \left( \Delta_{m} - 1 \right)^{2} \right)$.

We note that table \ref{tab:shapingStrength} only establishes an upper bound on the scaling of geometric quantities. When considering a specific geometry, it is always possible for terms to vanish or become small, giving zero to the expected order. For example, unless the flux surfaces have low order shaping, the $z$-dependent portion of $\partial l_{p} / \partial \theta$ will scale as $\Delta_{m} - 1$, rather than $m_{c} \left( \Delta_{m} - 1 \right)$. Similarly if the tokamak has a large aspect ratio or if the flux surfaces lack low order shaping the symmetry-breaking in $v_{d s \psi}$ and $a_{s ||}$ turns out to be $O \left( m_{c}^{2} \left( \Delta_{m} - 1 \right) \right)$, not $O \left( m_{c}^{3} \left( \Delta_{m} - 1 \right) \right)$. Another example can be seen from the simple geometry discussed in section \ref{subsec:practicalNonMirrorSymShaping}. If we create mirror symmetric flux surfaces without an envelope by setting $\Delta_{m} = 1$, $\Delta_{n} = 1$, or $n = m$ all of the symmetry-breaking terms in the geometric coefficients cancel (see \refEq{eq:nonMirrorAlphaDrift} and \refEq{eq:symBreakCoeff}) and the momentum transport is zero at all orders in the expansion. Additionally, from \refEq{eq:nonMirrorAlphaDrift} we see that if $n \theta_{t n} = m \theta_{t m}$ the symmetry-breaking terms also cancel. This condition is only met when the envelope created by the beating of the two high-order shaping modes is up-down symmetric.

We have just shown that, in general, the up-down symmetry breaking in the geometric coefficients can be no larger than $O \left( m_{c}^{3} \left( \Delta_{m} - 1 \right)^{2} \right)$. If we give $\Delta_{m} - 1$ a definite ordering in $m$, then we can expand the gyrokinetic equations (see \refEq{eq:gyrokineticEq}, \refEq{eq:quasineut}, and \refEq{eq:momFlux}) as we did in the previous section. Keeping all terms of $O \left( m_{c}^{4} \left( \Delta_{m} - 1 \right)^{2} \right)$ or larger leaves us with a completely up-down symmetric system of equations. From the expansion in section \ref{subsec:practicalNonMirrorSymShaping} we know that these up-down symmetric equations determine the momentum flux to $O \left( m_{c}^{4} \left( \Delta_{m} - 1 \right)^{2} \right)$. Hence, we know that $\Pi_{\zeta s}$ can scale no stronger than $m_{c}^{3} \left( \Delta_{m} - 1 \right)^{2}$.

However, there is one case that requires special treatment. Thus far we have only assumed that $\Delta_{m} - 1 \ll 1$, which means we are free to use the ordering $\Delta_{m} - 1 \sim m_{c}^{-1}$. This ordering requires convex regions in the flux surface shape (see section \ref{subsec:practicalNonMirrorSymShaping}), but it does not necessarily introduce x-points into the plasma (see appendix \ref{app:maxShaping}). When we adopt this ordering we see that the symmetry of the geometric coefficients is broken to $O \left( m_{c} \right)$, which causes problems when we try to repeat the order-by-order expansion performed in section \ref{subsec:practicalNonMirrorSymShaping}. Naively, as $\Nabla \psi \cdot \Nabla \alpha$ and $\left| \Nabla \alpha \right|^{2}$ become very large, we would expect the nonlinear and drive terms of the gyrokinetic equation to vanish (because $J_{0} \left( k_{\perp} \rho_{s} \right) \rightarrow 0$), meaning unstable solutions appear impossible. A more careful, sophisticated treatment of the Bessel functions (and the gyrokinetic equation as a whole) is beyond the scope of this thesis. Regardless, we have established that the momentum flux must scale as $O \left( 1 \right)$ at the very least, because we know the that the symmetry of the $O \left( 1 \right)$ gyrokinetic equation is broken. The same argument applies for $\Delta_{m} - 1 \gtrsim m_{c}^{-3/2}$.

In summary, we expect that high mode number flux surface shaping will beat together to create slowly varying envelopes that generate intrinsic momentum transport that scales as
\begin{align}
    \frac{v_{th i}}{R_{c 0}} \frac{\Pi_{\zeta s}}{Q_{s}} \sim m_{c}^{3} \left( \Delta_{m} - 1 \right)^{2} \label{eq:genMomFluxScaling}
\end{align}
when $\Delta_{m} - 1 \lesssim m_{c}^{-3/2}$. We note that dividing by the energy flux to get the momentum transport figure of merit derived in section \ref{sec:momTransportEst} does not change the scalings because the $O \left( 1 \right)$ energy flux (i.e. that of flux surfaces without high mode number shaping) is non-zero. Equation \refEq{eq:genMomFluxScaling} is consistent with section \ref{subsec:practicalNonMirrorSymShaping}, where we used a $\Delta_{m} - 1 \sim m_{c}^{-2}$ ordering with a particular geometry specification to derive that $\left( v_{th i} / R_{c 0} \right) \Pi_{\zeta s} / Q_{s} \sim m_{c}^{-1}$.

%===================================================%
\section{Numerical results}
\label{sec:numResults}
%===================================================%

\begin{figure}
 \centering
% (a) \hspace{0.95\textwidth}

 \includegraphics[width=0.15\textwidth]{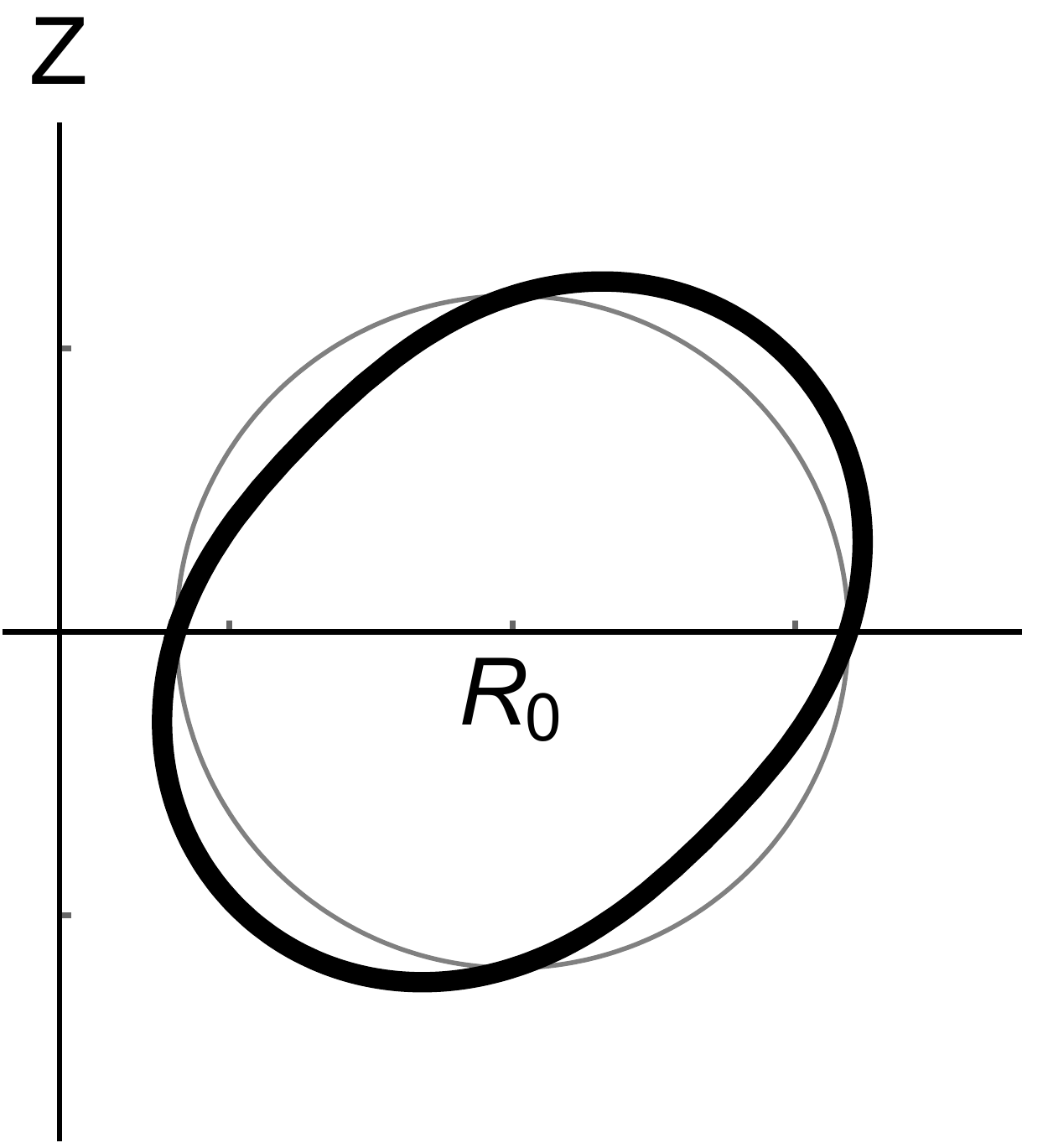}
 \includegraphics[width=0.15\textwidth]{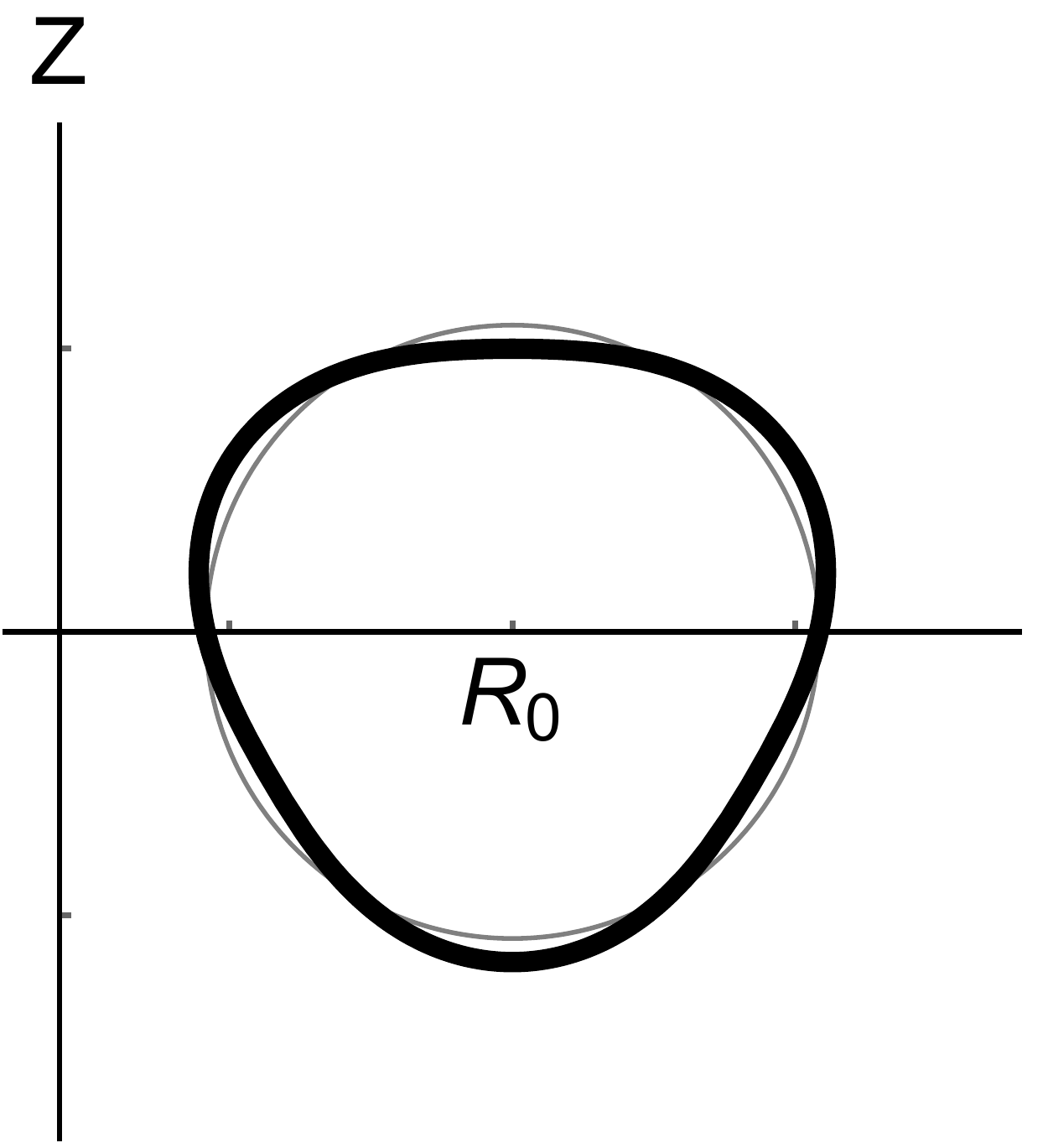}
 \includegraphics[width=0.15\textwidth]{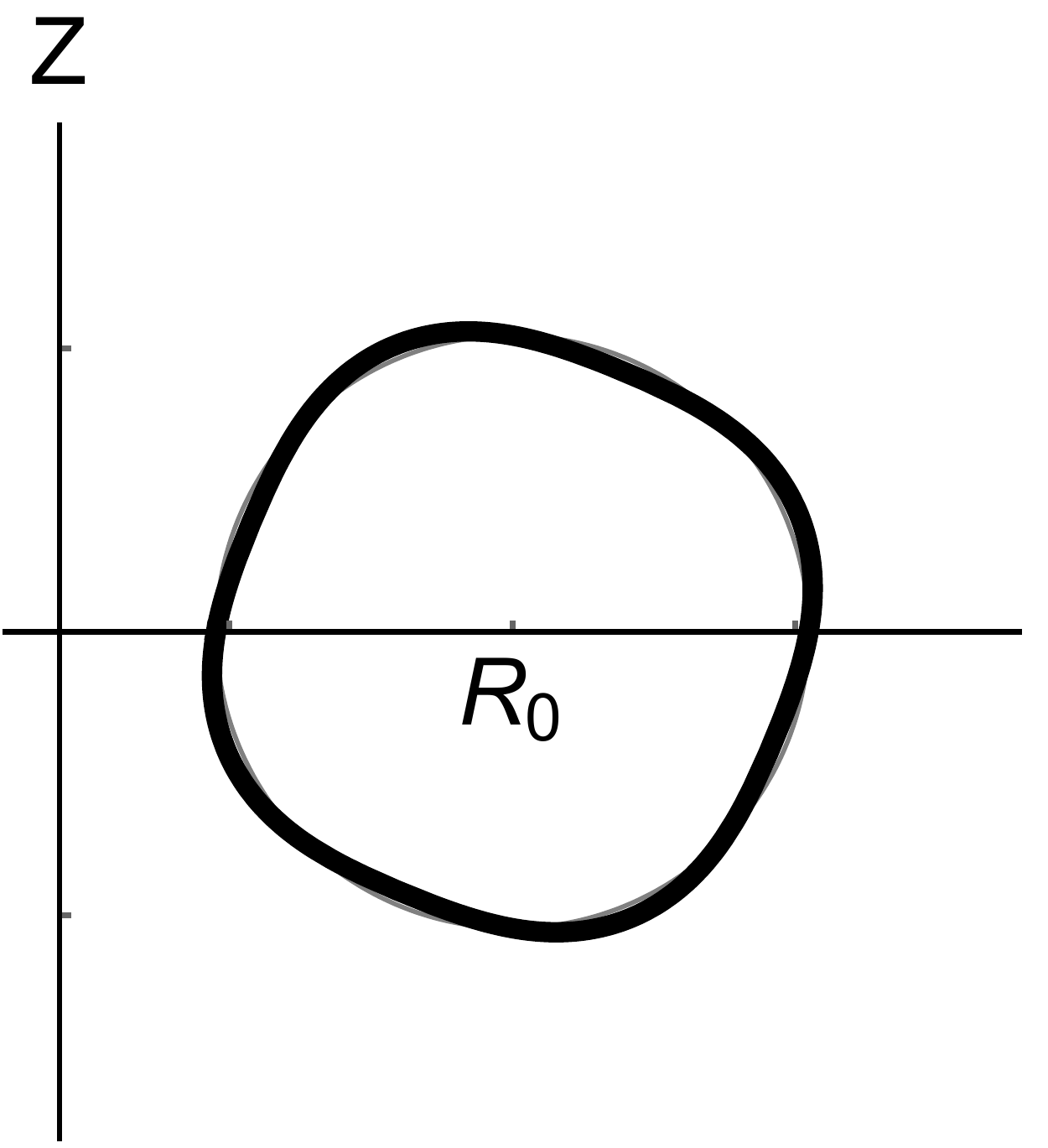}
 \includegraphics[width=0.15\textwidth]{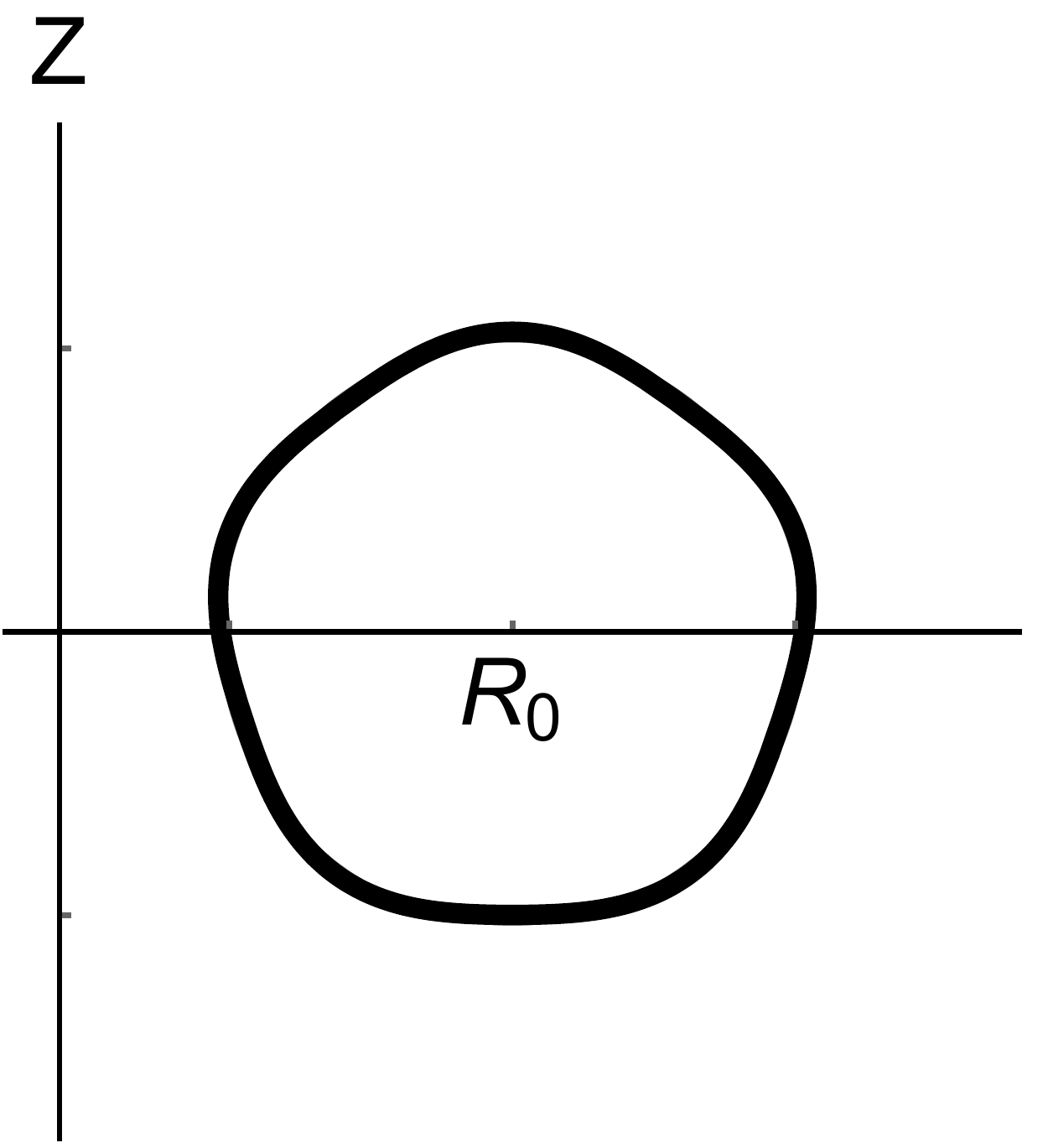}
 \includegraphics[width=0.15\textwidth]{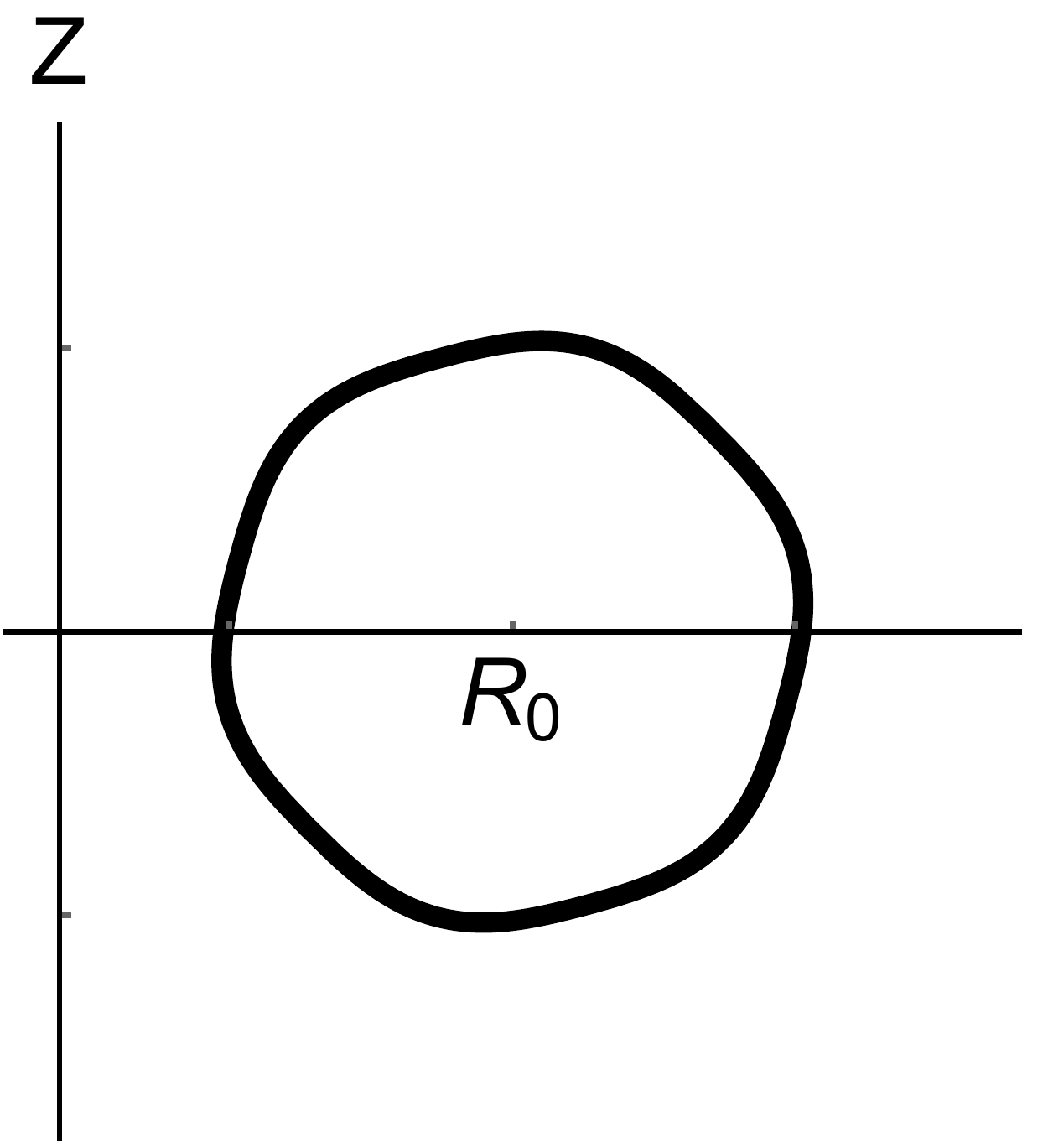}
 \includegraphics[width=0.0093\textwidth]{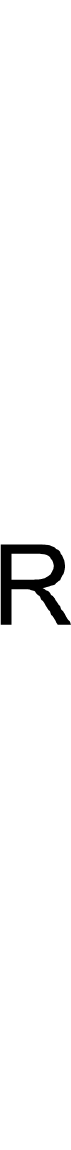}
 
 % (b) \hspace{0.95\textwidth}
 
 \includegraphics[width=0.15\textwidth]{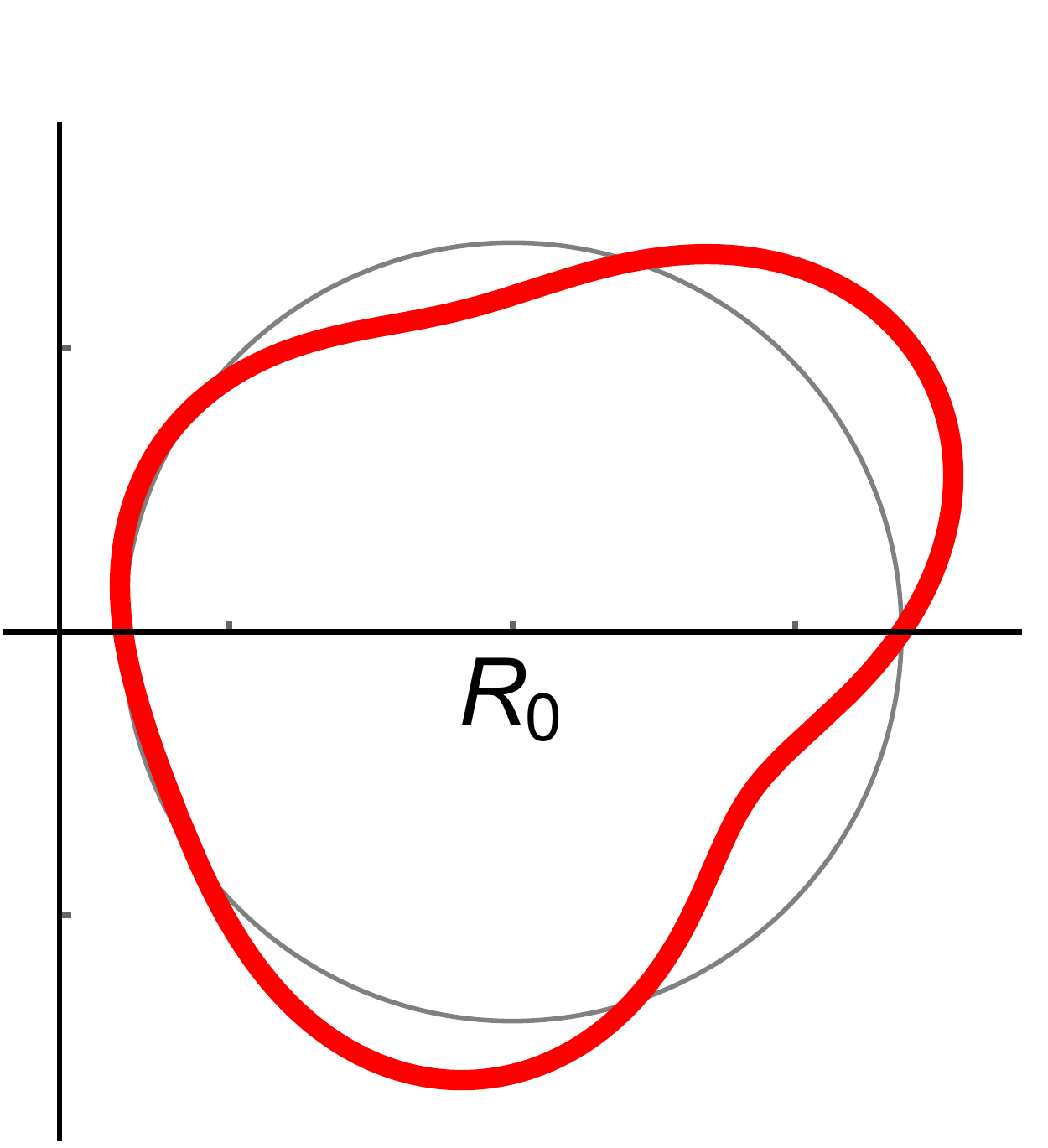}
 \includegraphics[width=0.15\textwidth]{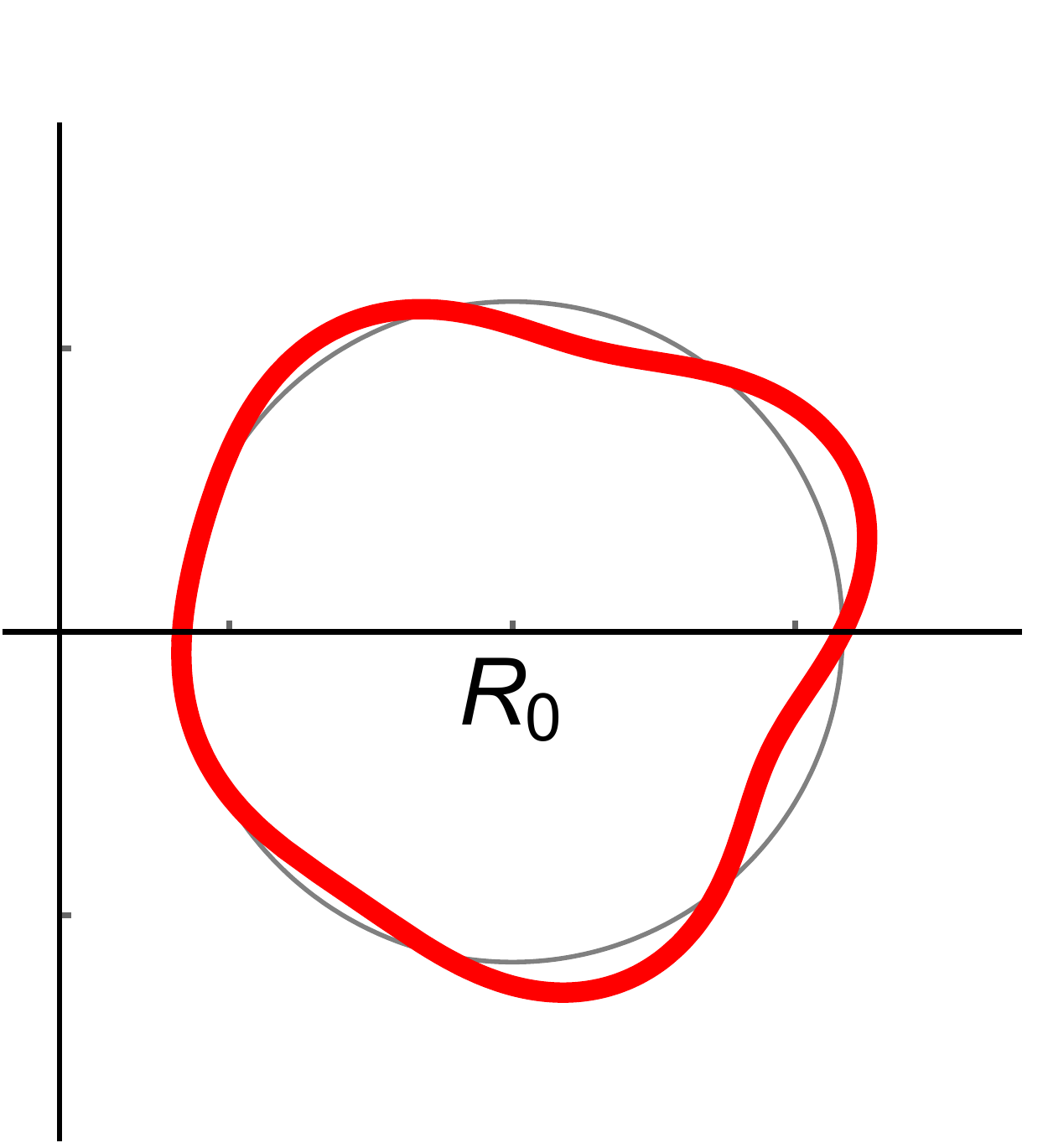}
 \includegraphics[width=0.15\textwidth]{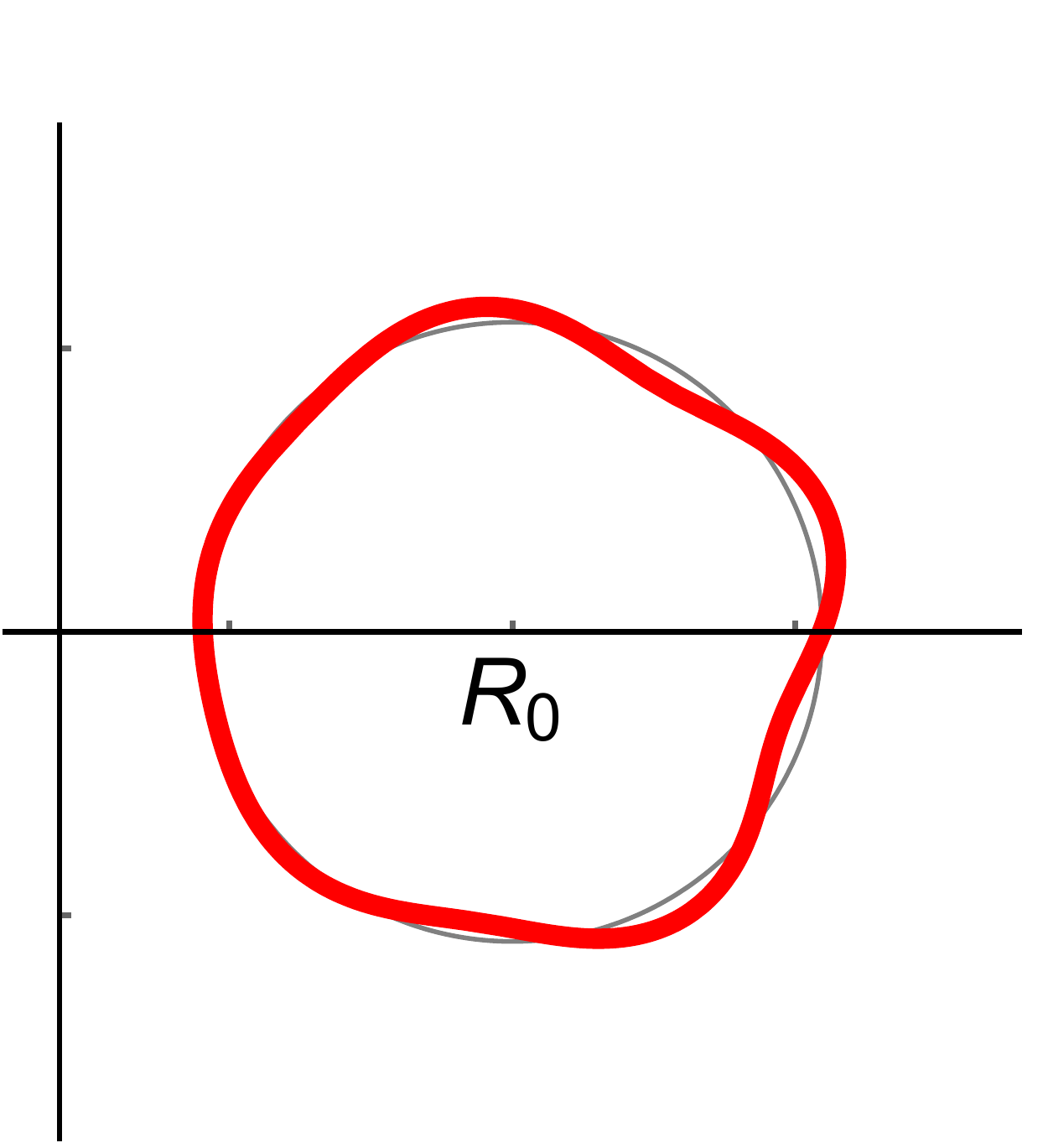}
 \includegraphics[width=0.15\textwidth]{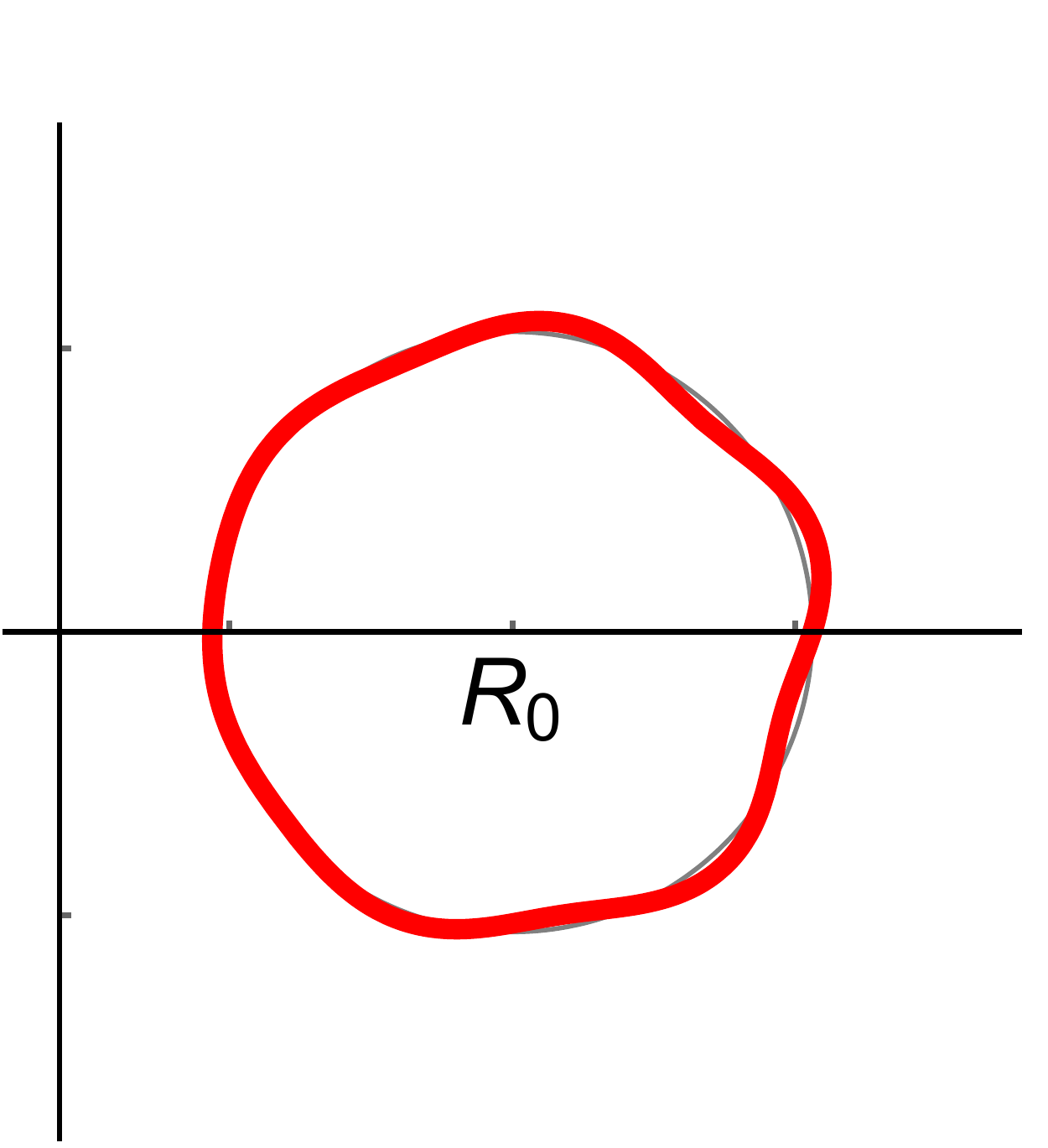}
 \includegraphics[width=0.15\textwidth]{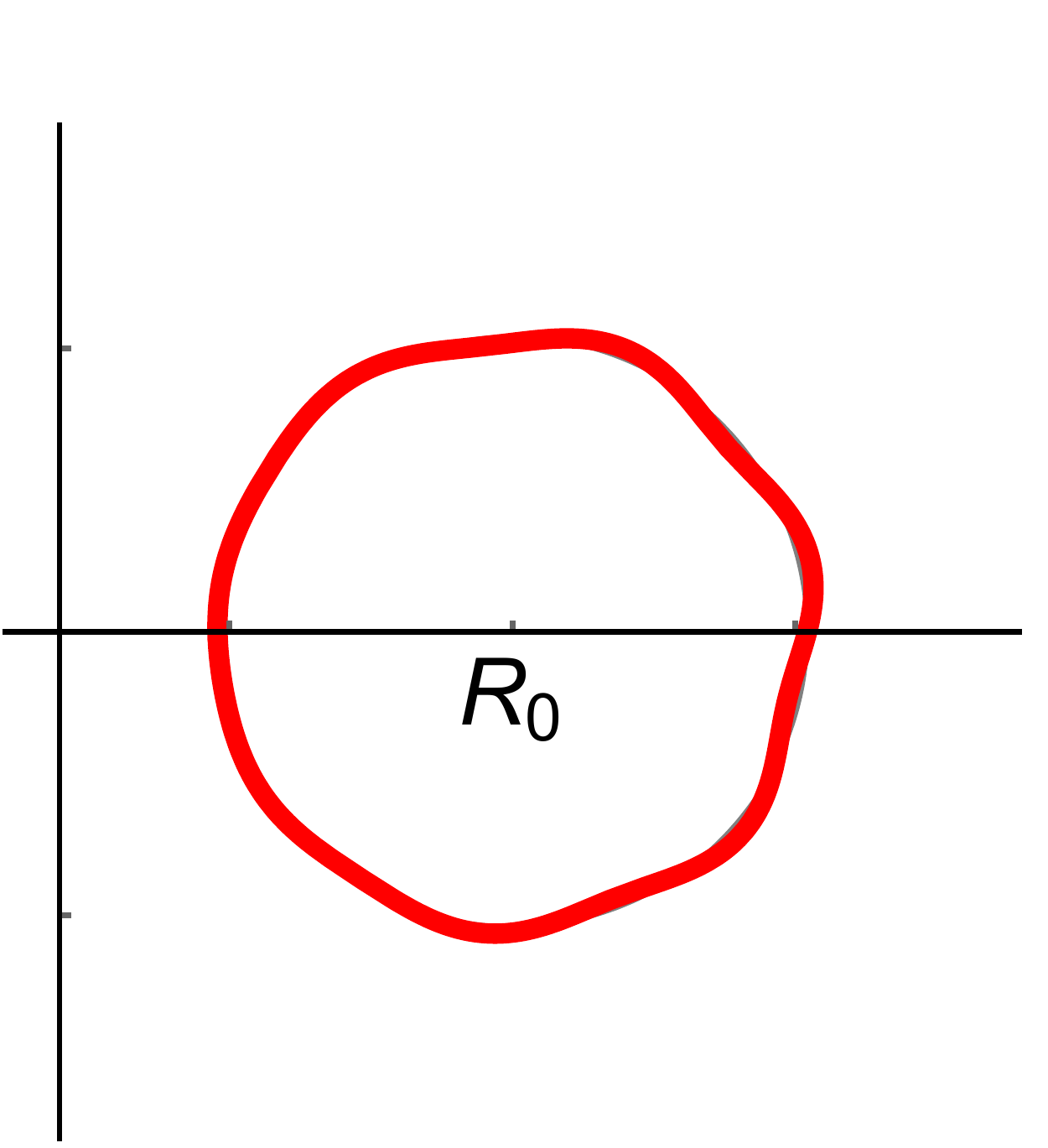}
 \includegraphics[width=0.0093\textwidth]{figs/Ch_08/xAxisLabelR.pdf}
 
 \includegraphics[width=0.15\textwidth]{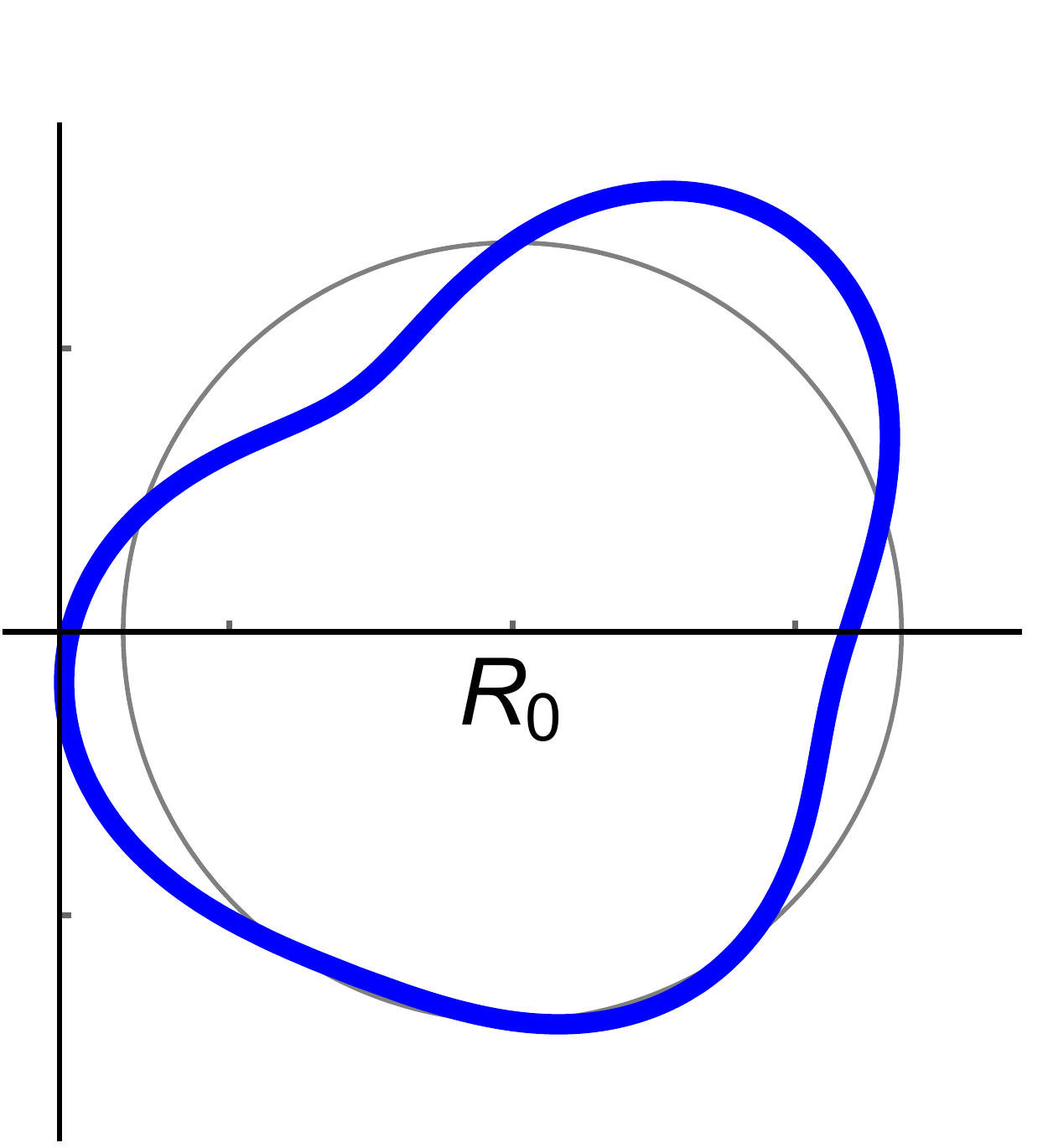}
 \includegraphics[width=0.15\textwidth]{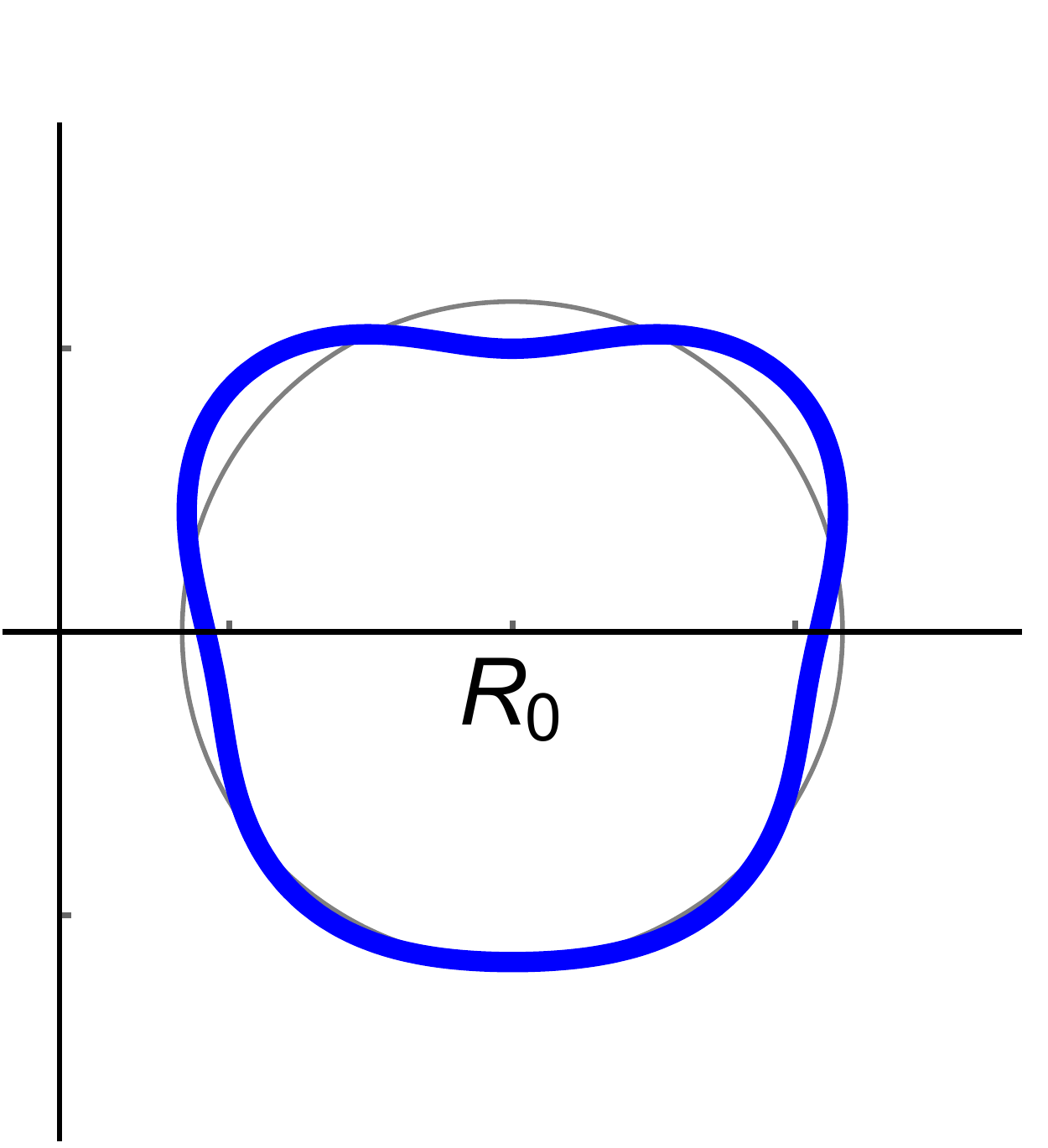}
 \includegraphics[width=0.15\textwidth]{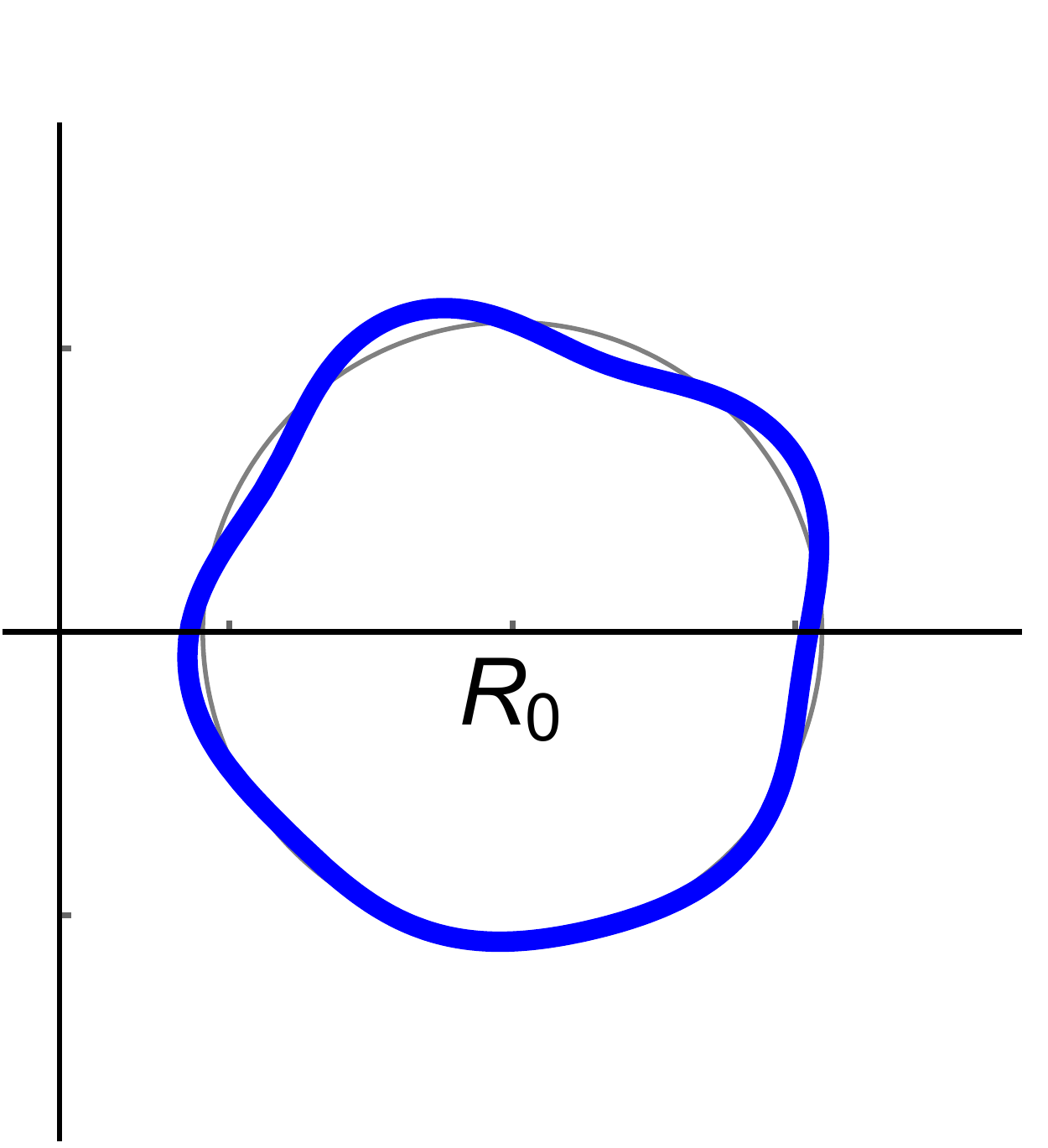}
 \includegraphics[width=0.15\textwidth]{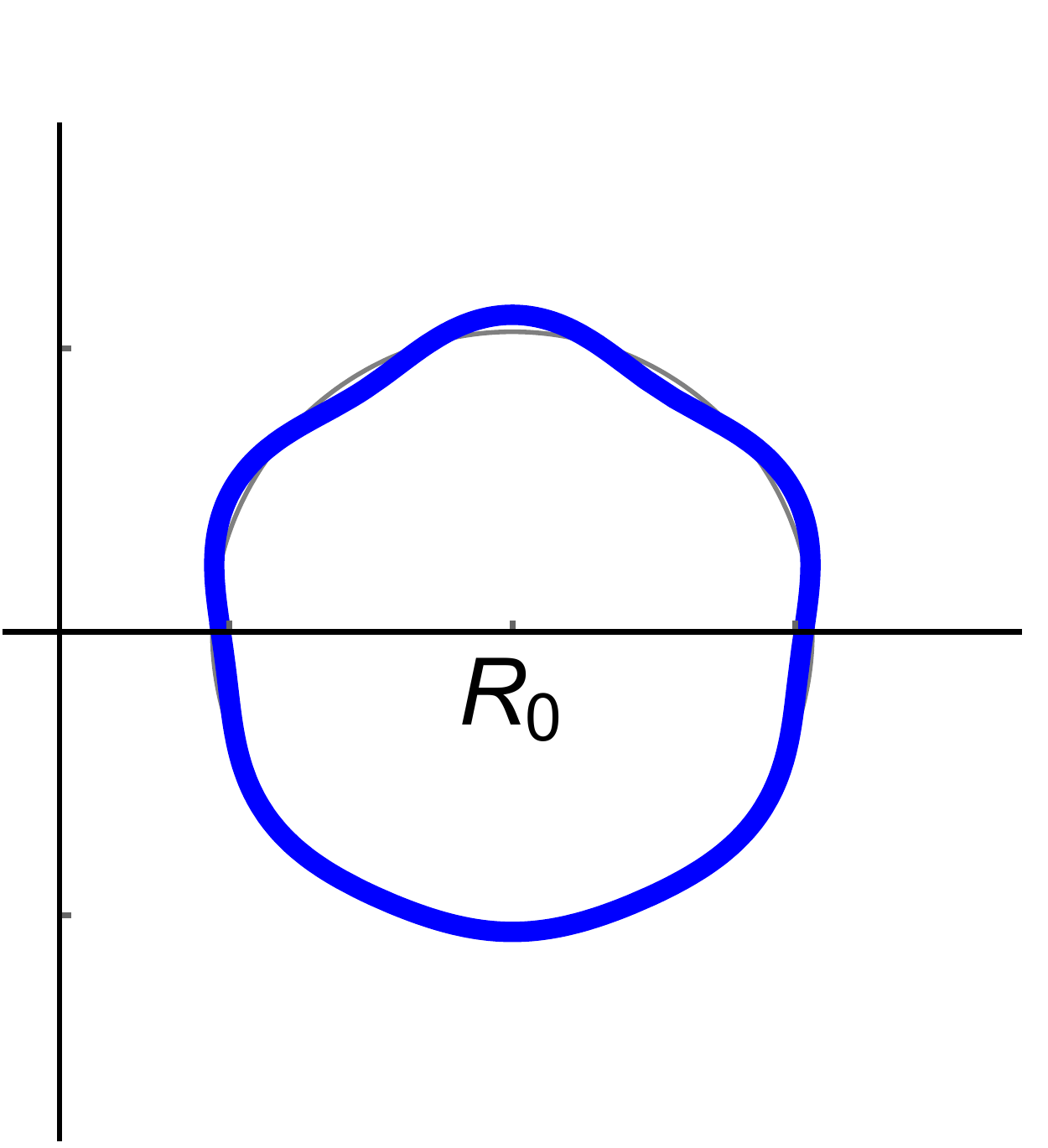}
 \includegraphics[width=0.15\textwidth]{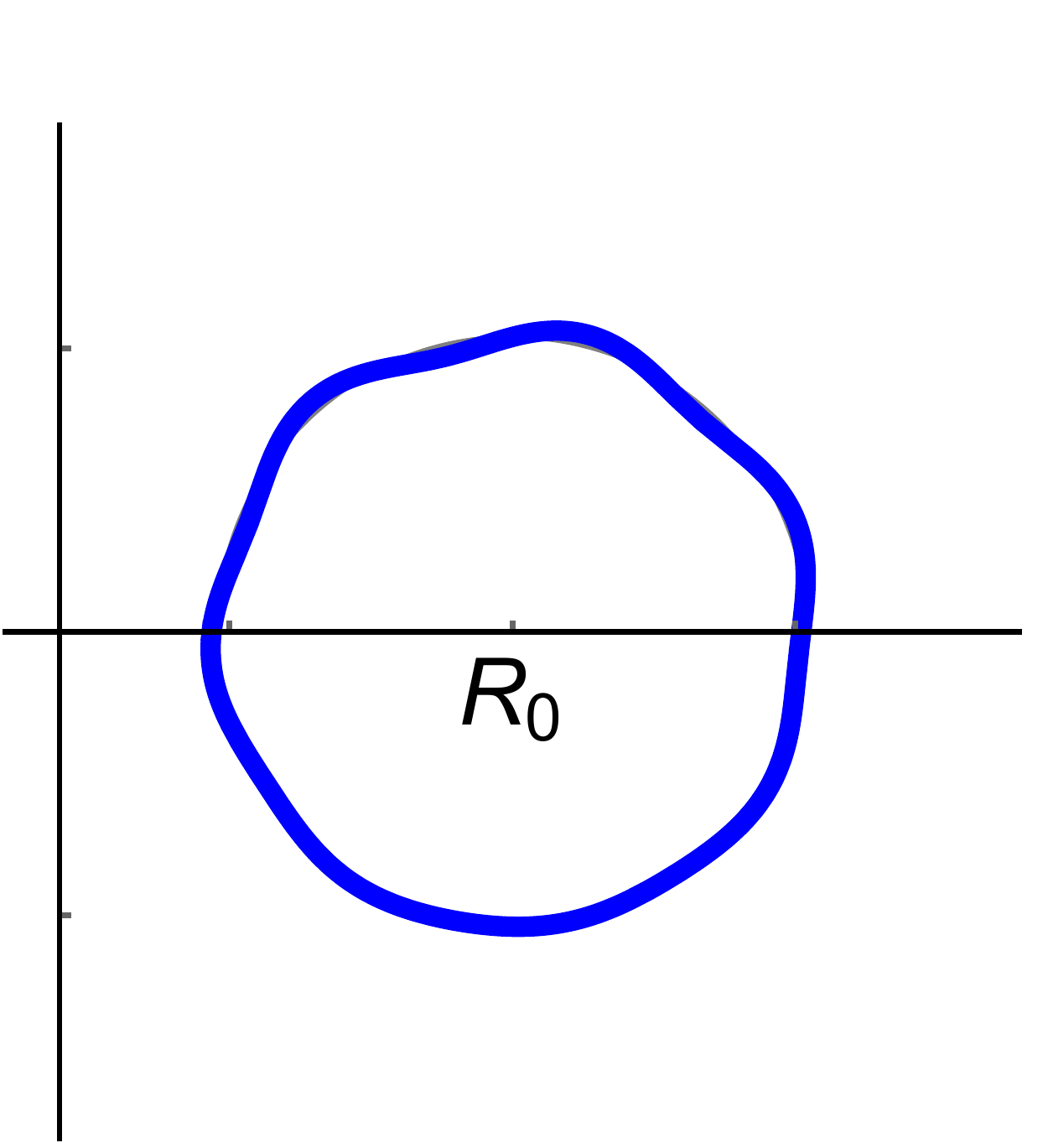}
 \includegraphics[width=0.0093\textwidth]{figs/Ch_08/xAxisLabelR.pdf}

 \caption{The $m_{c}=2$ through $m_{c}=6$ flux surface geometries in the mirror symmetric (top row), non-mirror symmetric with up-down symmetric envelope (middle row) and non-mirror symmetric with up-down asymmetric envelope (bottom row) scans, where circular flux surfaces are shown for comparison (grey).}
 \label{fig:simGeo}
\end{figure}

In this section we will present numerical results to test the analytic conclusions of sections \ref{subsec:practicalNonMirrorSymShaping} and \ref{subsec:genNonMirrorSymShaping} and compare with the results of chapter \ref{ch:GYRO_TiltingSymmetry}. We use GS2 to calculate the nonlinear turbulent fluxes generated by a given geometry and investigate the influence of the shape of the flux surface of interest by scanning $m_{c}$, a characteristic poloidal shaping mode number. We will compare the results of these numerical scans to the analytic scalings with $m_{c} \gg 1$ for mirror symmetric geometries (see section \ref{subsec:mirrorSymShaping}), non-mirror symmetric geometries with an up-down symmetric envelope, and non-mirror symmetric geometries with an up-down asymmetric envelope (see sections \ref{subsec:practicalNonMirrorSymShaping} and \ref{subsec:genNonMirrorSymShaping}).

All simulations are electrostatic and collisionless with deuterium ions and kinetic electrons. Unless specified, all parameters are fixed at Cyclone base case values (see \refEq{eq:cycloneBaseCase}). Since the non-mirror symmetric geometries have strong flux surface shaping these simulations needed to be run using $a / L_{T s} = 3.0$ to ensure that the turbulence was driven unstable. To estimate the impact of this on our results, a single mirror symmetric case was run at $a / L_{T s} = 3.0$, in addition to the run with $a / L_{T s} = 2.3$. This change in the temperature gradient was found to alter the ratio of the momentum to energy flux by less than a 5\%. All simulations used at least $48$ poloidal grid points, $127$ radial wavenumber grid points, $22$ poloidal wavenumber grid points, $12$ energy grid points, and $10$ untrapped pitch angle grid points (i.e. values of $\lambda \equiv w_{\perp}^{2} / \left( w^{2} B \right)$).

The geometry for the three scans is shown in figure \ref{fig:simGeo} and is specified by \refEq{eq:fluxSurfofInterestShapeWeak}, \refEq{eq:fluxSurfofInterestDerivWeak}, \refEq{eq:shapingRadialDeriv_rPsi} and \refEq{eq:tiltAngleDeriv} from the constant current Miller local equilibrium. The mirror symmetric scan has only one mode, $m = m_{c}$, while the two non-mirror symmetric scans have a second mode at $n = m_{c} + 1$. In section \ref{subsec:practicalNonMirrorSymShaping} we ordered $\Delta_{m} - 1 \sim m_{c}^{-2}$, so we will set the strength of the shaping such that $m_{c}^{2} \left( \Delta_{m} - 1 \right) = 1.5$ and $m_{c}^{2} \left( \Delta_{n} - 1 \right) = 1.5$ (if needed) is constant in the scan. For the mirror symmetric simulations we chose the tilt angle to be $\theta_{t m} = \pi / \left( 2 m_{c} \right)$, the angle halfway between the neighbouring up-down symmetric configurations (at $\theta_{t m} = 0$ and $\theta_{t m} = \pi/m_{c}$). For the non-mirror symmetric cases we must also set $\theta_{t n} = \left(m_{c} / n \right) \theta_{t m}$ (such that the envelope is up-down symmetric) or $\theta_{t n} = 0$ (such that the envelope is up-down asymmetric and halfway between neighbouring configurations with up-down symmetric envelopes). According to the symmetry breaking term in \refEq{eq:nonMirrorAlphaDrift}, a flux surface with only two shaping modes has an up-down symmetric envelope if and only if
\begin{align}
   \theta_{t n} = \frac{m_{c}}{n} \left( \theta_{t m} + Y_{1} \frac{\pi}{m_{c}} \right) \label{eq:twoModeEnvelopeCond}
\end{align}
for some integer $Y_{1}$.

In general, from the argument in section \ref{subsec:genNonMirrorSymShaping}, we would predict the momentum flux in an up-down asymmetric geometry to scale as $\left( v_{th i} / R_{c 0} \right) \Pi_{\zeta s} / Q_{s} \sim m_{c}^{-1}$. Indeed, we expect this to be the case for the non-mirror symmetric scan with an up-down asymmetric envelope, as we confirmed in section \ref{subsec:practicalNonMirrorSymShaping}. However, section \ref{subsec:mirrorSymShaping} shows the mirror symmetric scan is a special case where the momentum flux almost entirely cancels, giving the scaling $\left( v_{th i} / R_{c 0} \right) \Pi_{\zeta s} / Q_{s} \sim \Exp{- \beta m_{c}^{\gamma}}$ for constant $\beta$ and $\gamma$. Similarly, sections \ref{subsec:practicalNonMirrorSymShaping} and \ref{subsec:genNonMirrorSymShaping} show that configurations with an up-down symmetric envelope, even if they are non-mirror symmetric, see the same cancellation. Hence, they also have exponentially small momentum transport. We note that reference \cite{BallMomFluxScaling2016} contrasted a mirror symmetric scan with a scan in non-mirror symmetric geometries, but did not consider the effect of the up-down symmetry of the envelope. All of the non-mirror symmetric simulations that were performed had an up-down symmetric envelope, so the scaling should be identical to the mirror symmetric scan (i.e. exponential). Here we add a third scan of non-mirror symmetric geometries with an up-down asymmetric envelope, which we expect to produce momentum transport that decays much more slowly with $m_{c} \gg 1$ (i.e. polynomially).

As with the momentum flux, we expect that the energy flux in non-mirror symmetric configurations with an up-down asymmetric envelope to converge to that of circular flux surfaces like $m_{c}^{-1}$. However, in the other two sets configurations we expect the energy flux to have the same $m_{c}^{-1}$ scaling, as opposed to the exponential scaling expected for the momentum flux. This is because there are up-down symmetric fast shaping terms in the geometric coefficients (e.g. the first term in \refEq{eq:alphaDriftO1}, \refEq{eq:gradPsiDotGradAlphaO1}, and \refEq{eq:gradAlphaSqO1}) that cause a change in energy transport, whereas they do not cause momentum transport.

\begin{figure}
 \centering
 \includegraphics[width=0.7\textwidth]{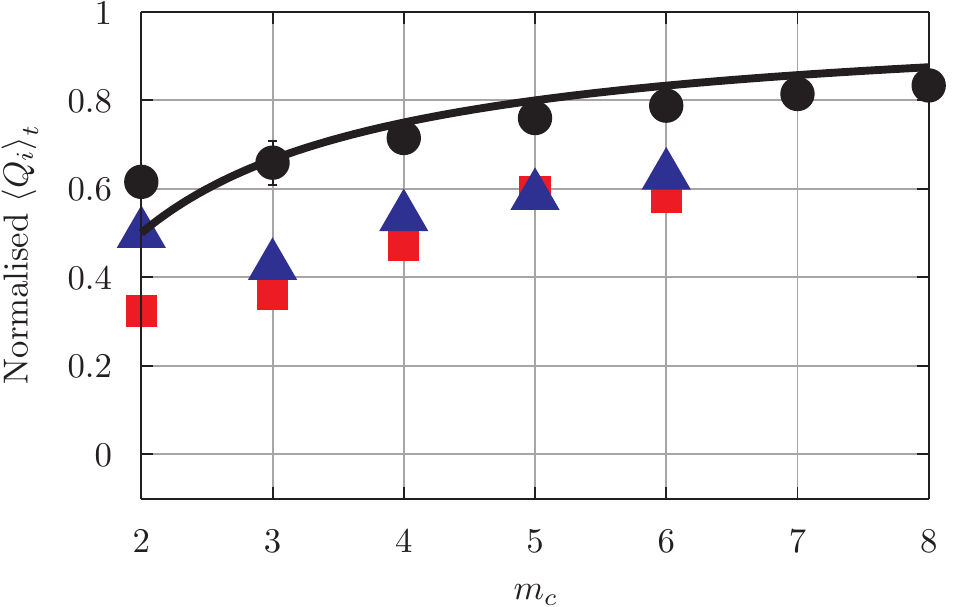}
 \caption{The radial ion energy flux from mirror symmetric flux surfaces (black, circles), non-mirror symmetric flux surfaces with an up-down symmetric envelope (red, squares), and non-mirror symmetric flux surfaces with an up-down asymmetric envelope (blue, triangles) normalised to the energy flux of a circular flux surface. Also shown is the $m_{c}^{-1}$ scaling (black, solid) expected for all three geometry scans.}
 \label{fig:energyFlux}
\end{figure}

\begin{figure}
 \centering
 \includegraphics[width=0.7\textwidth]{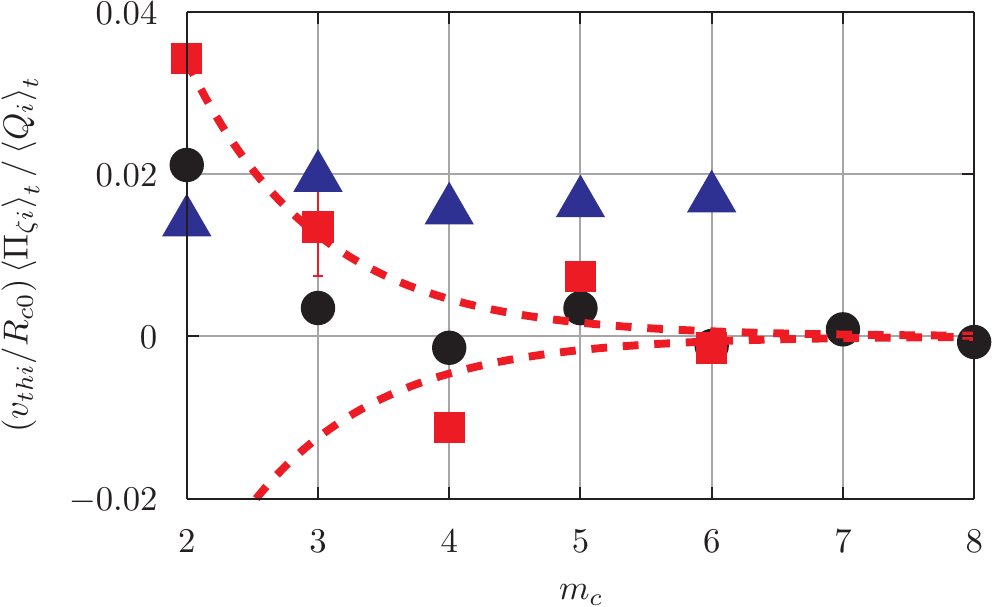}
 \caption{The momentum transport from mirror symmetric flux surfaces (black, circles), non-mirror symmetric flux surfaces with an up-down symmetric envelope (red, squares) and non-mirror symmetric flux surfaces with an up-down asymmetric envelope (blue, triangles). Also shown is an example exponential scaling (red, dotted) appropriate for the mirror symmetric and non-mirror symmetric with up-down asymmetric envelope scans.}
 \label{fig:momHeatFluxRatioAbs}
\end{figure}

Figure \ref{fig:energyFlux} shows the time-averaged ion energy flux calculated by GS2 for the three scans, which are all consistent with our theoretical expectations. In figure \ref{fig:momHeatFluxRatioAbs}, we see the time-averaged ratio of the ion momentum and energy fluxes from the GS2 simulations. Again, we see behaviour that is consistent with our expectations from analytic theory. Clearly the non-mirror symmetric configurations with an up-down asymmetric envelope decay more slowly than the other two scans with $m_{c} \gg 1$. We note that section \ref{subsec:mirrorSymShaping} only demonstrates that the momentum flux from mirror symmetric configurations cannot scale polynomially. It does not predict the scaling must be $\Exp{-m_{c}}$, as opposed to $\Exp{-m_{c}/2}$ or $m_{c} \Exp{-m_{c}^{2}}$ for example. However, the trendline shown in figure \ref{fig:momHeatFluxRatioAbs}, $\Exp{-m_{c}}$, seems to fit the data fairly well. Additionally, figure \ref{fig:momHeatFluxRatioAbs} shows that the non-mirror symmetric configurations with an up-down symmetric envelope produce more rotation than the configurations with an up-down asymmetric envelope at $m_{c} = 2$. It seems reasonable to attribute this to a failure to fully satisfy the assumption that $m_{c} \gg 1$.

%===================================================%
%===================================================%
\section{Interpretation}
\label{sec:interpretation}
%===================================================%
%===================================================%

This chapter determined that the intrinsic momentum flux generated by fast shaping with a slowly varying up-down asymmetric envelope is polynomially small in the scale of the fast shaping. Chapter \ref{ch:GYRO_TiltingSymmetry} concluded that the momentum flux in mirror symmetric equilibria is exponentially small in the scale of the shaping. In order to interpret the results of these analytic arguments we will distinguish between ``geometric'' effects and ``shaping'' effects. Geometric effects are those that give a poloidal dependence to the geometric coefficients, apart from the linear dependence built into $\alpha$ due to magnetic shear. Shaping effects are the subset of the geometric effects that are specified in the flux surface shape or its radial derivative (i.e. the $m \neq 0$ terms in \refEq{eq:fluxSurfofInterestShapeWeak} and \refEq{eq:fluxSurfofInterestDerivWeak}). Using this terminology we can view the tokamak as having an inherent $m=1$ geometric effect due to toroidicity, as discussed in section \ref{subsec:mirrorSymShaping}. However, we can see it is distinct from the $m=1$ shaping effect (i.e. the Shafranov shift) by looking at the geometric coefficients to lowest order in aspect ratio. The $m=1$ toroidal geometric effect only appears in the two magnetic drift coefficients (see \refEq{eq:psiDriftO0} and \refEq{eq:alphaDriftO0}), while the $m=1$ shaping effect affects all six (see \refEq{eq:gradparO1} through \refEq{eq:gradAlphaSqO1}). This toroidal geometric effect is not present in non-toroidal magnetic geometries such as the screw pinch.

Section \ref{sec:upDownSymArg} shows that if the magnetic geometry does not include at least two geometric effects with different tilt angles the momentum flux must be small in $\rho_{\ast} \ll 1$. Hence, in a screw pinch (without toroidicity to define up-down symmetry) the only option to drive intrinsic rotation is non-mirror symmetric shaping. Conversely, in a tokamak we can use either non-mirror symmetric shaping or an up-down asymmetric shaping effect with the toroidal geometric effect. The possibility of ignoring toroidicity entirely and using two modes with different tilt angles to drive momentum transport provides motivation to explore non-mirror symmetric configurations. This would be advantageous if, for some reason, toroidicity turns out to be very ineffective at driving rotation. %However, we will see from simulations in chapter \ref{ch:GYRO_NonMirrorSym} that toroidicity appears to dominate the momentum transport.

Sections \ref{subsec:practicalNonMirrorSymShaping} and \ref{subsec:genNonMirrorSymShaping} reveal another way to generate rotation. Namely, to beat two shaping effects together to make an slowly varying envelope that breaks up-down symmetry. This generates momentum flux that is polynomially small (for purely concave flux surfaces) in either mode number of the original shaping effects. Therefore, using a combination of low order shaping effects (e.g. elongation, triangularity) to create an up-down asymmetric envelope to drive rotation appears optimal. This reinforces the conclusions of chapter \ref{ch:MHD_ShapingIntuition}, which found that low order shaping modes were advantageous from MHD considerations. Specifically, we found that low order modes better penetrate to the magnetic axis from the plasma edge. Hence, they can make the inner flux surfaces of a device more asymmetric. Lastly, from the tilting symmetry presented in chapter \ref{ch:GYRO_TiltingSymmetry} we learned that the momentum flux generated by two geometrical effects is exponentially small in the difference in the spatial scales of the poloidal variation they create.
%This both motivates using shaping effects with neighbouring mode numbers and distinguishes mirror symmetric configurations from non-mirror symmetric configurations. In mirror symmetric tokamaks, the coupling between the toroidal geometric effect and shaping effects is the only mechanism that generates rotation. Non-mirror symmetric tokamaks have this same mechanism, but also allow two shaping effects to beat together and interact with toroidicity via an envelope.

Hence, the analytic arguments of chapters \ref{ch:MHD_ShapingIntuition}, \ref{ch:GYRO_TiltingSymmetry}, and \ref{ch:GYRO_MomFluxScaling} indicate that using low order, neighbouring shaping effects (e.g. elongation, triangularity) to break up-down asymmetry is best for creating rotation \cite{BallMomUpDownAsym2014, BallMastersThesis2013}. Chapter \ref{ch:GYRO_TiltingSymmetry} and the transport properties of screw pinches establish a distinction between mirror and non-mirror symmetric configurations and suggest that non-mirror symmetric configurations may be able to generate higher levels of rotation. We will explore these strategies at length in chapters \ref{ch:GYRO_ShafranovShift} and \ref{ch:GYRO_NonMirrorSym}.

%% file: Ch_09_GYRO_ShafranovShift.tex
% !TEX root = /Users/Justin/Documents/Research/Writings/2016DoctoralThesis/DoctoralThesis.tex

\chapter{Non-mirror symmetry: Shafranov shift and tilted elongation}
\label{ch:GYRO_ShafranovShift}

\begin{quote}
   \emph{Much of this chapter appears in reference \cite{BallShafranovShift2016}.}
\end{quote}

Chapters \ref{ch:MHD_ShapingIntuition}, \ref{ch:GYRO_TiltingSymmetry}, and \ref{ch:GYRO_MomFluxScaling} indicate that a non-mirror symmetric geometry with an up-down asymmetric envelope created by low order shaping effects is optimal for maximising intrinsic rotation. In this context, there are two options. The first is to introduce up-down asymmetric elongation using external poloidal field coils and then rely on the Shafranov shift (i.e. the shift in the magnetic axis due to toroidicity) to break mirror symmetry. This appears optimal because it makes use of the lowest possible shaping modes (i.e. $m = 1$ and $m = 2$). However, this strategy has the drawbacks that the effect of the Shafranov shift is formally small in aspect ratio and the direction/magnitude of the shift is a consequence of the plasma $\beta$ profile and the global MHD equilibrium. Hence it is not independently controlled by external coils. Additionally, including effects that are small in aspect ratio also introduces $\beta'$, which controls the effect of the pressure gradient on the magnetic equilibrium. We will see that $\beta'$ tends to strongly decrease the momentum transport. The second option is to use external coils to introduce both elongation and triangularity (i.e. $m = 2$ and $m = 3$ shaping) into the flux surface shape and directly break mirror symmetry. Both modes are lowest order in aspect ratio and can be directly controlled by external shaping magnets, but this relies on higher order shaping modes than the first option.

Practically speaking, these two strategies are intertwined as the divertor geometry nearly always introduces some triangularity into the flux surfaces and the Shafranov shift exists regardless of the shape of flux surfaces. Nevertheless, for simplicity it is useful to distinguish them and examine each option independently. In this chapter we will explore the former: the influence of the Shafranov shift and the effect of the $\beta$ profile on the turbulent momentum flux in the core of tokamaks. In chapter \ref{ch:GYRO_NonMirrorSym} we will explore the latter: the effect breaking flux surface mirror symmetry by independently varying the tilt angles of elongation and triangularity.

In this chapter we will perform nonlinear gyrokinetic simulations of tilted elliptical flux surfaces with a Shafranov shift. Section \ref{subsec:inputParameters} starts by summarising the input parameters used in the gyrokinetic simulations. Section \ref{subsec:results} details the results of several numerical scans aimed at illuminating the effect of the Shafranov shift and the $\beta$ profile on momentum transport. In section \ref{sec:betaPrime} we discuss the sensitivity of the momentum transport to changes in the magnetic equilibrium caused by altering the local gradient of $\beta$. Furthermore, in section \ref{sec:betaProfile} we consider the impact of changing the shape of the radial profile of $\beta$.

%===================================================%
%===================================================%
\section{Input Parameters}
\label{subsec:inputParameters}
%===================================================%
%===================================================%

As in chapters \ref{ch:GYRO_TiltingSymmetry} and \ref{ch:GYRO_MomFluxScaling}, we will use GS2 to calculate the turbulent transport in shaped variants of the Cyclone base case (specified by the parameters of \refEq{eq:cycloneBaseCase}). Most of our simulations will model tilted elliptical flux surfaces, all of which have an elongation of $\kappa = 2$. The turbulent fluxes calculated by GS2 will be normalised to gyroBohm values given by \refEq{eq:momFluxGyroBohm} and \refEq{eq:energyFluxGyroBohm}.

We will use a modified version of GS2 to simulate local Miller equilibria with flux surfaces prescribed with Exact specification, given by \refEq{eq:fluxSurfofInterestShapeExact}, \refEq{eq:fluxSurfofInterestDerivExact}, and \refEq{eq:gradShafranovLocalEqExact} through \refEq{eq:geoAxialExact}. In order to create a realistic Shafranov shift, we will take the results from section \ref{sec:nextOrderMillerEquil}, which uses the location of the magnetic axis shown for the constant current case in figure \ref{fig:ShafShiftLoc}(b). 
%Unfortunately, in the range $\theta_{\kappa b} \in \left[ 0, \pi/8 \right]$ these tilt angles nearly exactly coincide with the condition for an up-down symmetric envelope, given by \refEq{eq:twoModeEnvelopeCond}. This indicates that the configuration should generate an exponentially small momentum flux in $m_{c} \gg 1$, the characteristic Fourier mode number of the shaping effects. However, as we will see from the results, this does not appear to be significant for the $m = 1$ Shafranov shift.
Additionally, since the size of the Shafranov shift is closely connected to the plasma pressure, we included the effect of $\beta'$ on the magnetic equilibrium. To capture this GS2 requires a local value of
\begin{align}
   \beta' \equiv \frac{2 \mu_{0} a}{B_{0}^{2}} \frac{d p}{d a_{\psi}} \label{eq:betaPrimeDef}
\end{align}
because it constructs the poloidal magnetic field to be consistent with the Grad-Shafranov equation. We will find that the momentum transport is quite sensitive to $\beta'$, so it is an important parameter. In keeping with rough projections for ITER \cite{AymarITERSummary2001}, we use a pressure profile that is linear in $a_{\psi}$. This allows us to estimate that
\begin{align}
   \beta' \approx - \frac{2 \mu_{0} p_{\text{axis}}}{B_{0}^{2}} \approx - 0.06 , \label{eq:betaPrimeEstimate}
\end{align}
using an ITER-like value for $p_{\text{axis}}$. Since we are running electrostatic simulations the value of $\beta$ itself has no effect.

We note that assuming a constant $\beta'$ (i.e. $d p / d a_{\psi}$) profile is formally inconsistent with the constant $d p / d \psi$ profile used in the MHD calculation of the Shafranov shift. Hence, using the results shown in figure \ref{fig:ShafShiftLoc} together with \refEq{eq:betaPrimeEstimate} is not formally valid. However, figure \ref{fig:ShafShiftLocPressureProf} shows that the magnitude and direction of the Shafranov shift is insensitive to large changes in the shape of the pressure profile at constant $R_{0 b}$, $a$, $\kappa_{b}$, $I_{p}$, and $p_{\text{axis}} / \psi_{0 b}$. This suggests that, since we have kept the proper parameters fixed, the pressure profile mismatch will not have much effect.

%===================================================%
%===================================================%
\section{Parameter scan results}
\label{subsec:results}
%===================================================%
%===================================================%

A total of four scans in $\theta_{\kappa}$, the tilt angle of the flux surface of interest, were performed at
\begin{enumerate}
   \item[(1)] $\beta' = 0$ with no Shafranov shift, \\
   \item[(2)] $\beta' = 0$ with a modest Shafranov shift (approximately half the ITER-like Shafranov shift), \\
   \item[(3)] $\beta' = 0$ with an ITER-like Shafranov shift, and \\
   \item[(4)] an ITER-like $\beta' = -0.06$ with an ITER-like Shafranov shift.
\end{enumerate}
These were done to directly determine the influence of the Shafranov shift and $\beta'$. The magnitude and direction of the local ITER-like Shafranov shift was kept consistent with \refEq{eq:radialShafShiftLocal} and \refEq{eq:axialShafShiftLocal}. Additionally, a single simulation was performed with $\beta' = -0.06$ and no Shafranov shift in order to isolate the effect of $\beta'$.

Four scans in $\rho_{0}$, the minor radial coordinate of the flux surface of interest, were performed at
\begin{enumerate}
   \item[(1)] $\beta' = 0$ with no Shafranov shift, \\
   \item[(2)] $\beta' = 0$ with an ITER-like Shafranov shift, \\
   \item[(3)] an ITER-like $\beta' = -0.06$ with no Shafranov shift, and \\
   \item[(4)] an ITER-like $\beta' = -0.06$ with an ITER-like Shafranov shift.
\end{enumerate}
All simulations had elliptical flux surfaces with $\theta_{\kappa} = \pi / 8$. These scans were done in order to investigate the balance between the Shafranov shift, which we expect to enhance the momentum transport, and $\beta'$, which our GS2 simulations will reveal to reduce the momentum transport. For these scans we kept $\beta'$ constant to be consistent with ITER (according to \refEq{eq:betaPrimeEstimate}) and again calculated the local Shafranov shift at each minor radius according to \refEq{eq:radialShafShiftLocal} and \refEq{eq:axialShafShiftLocal}.

Lastly, a small scan was performed with circular flux surfaces in which $\theta_{\text{axis}}$, the direction of the Shafranov shift, was varied. This is unphysical, but it was done to explicitly isolate the effect of a pure flux surface Shafranov shift.

%===================================================%
%===================================================%
\subsection{Elliptical boundary tilt scans}
\label{subsubsec:tiltedEllipticalScans}
%===================================================%
%===================================================%

\begin{figure}
 \centering
 \includegraphics[width=0.7\textwidth]{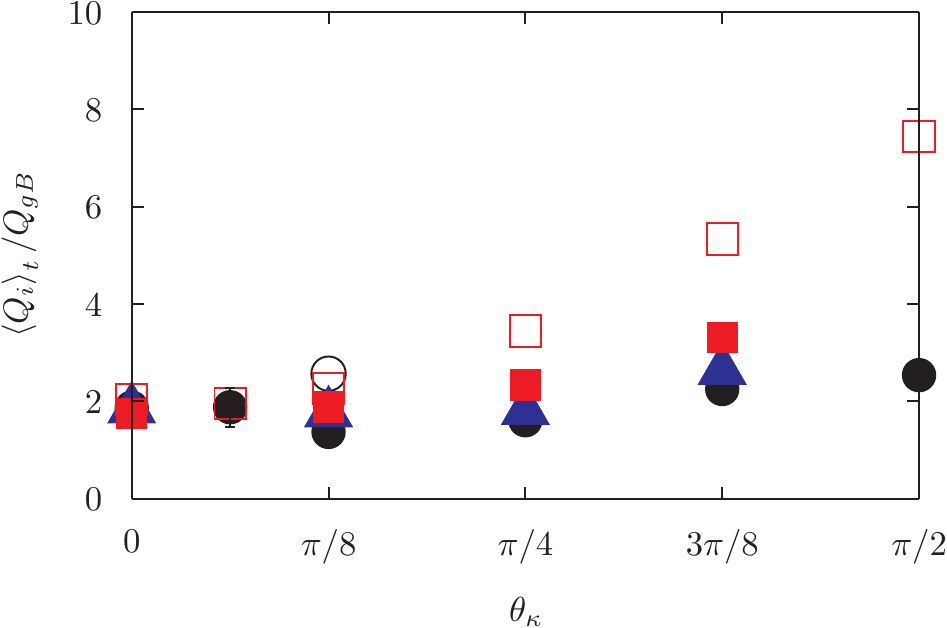}
 \caption{The ion energy flux for flux surfaces with no shift (black, circles), a modest shift (blue, triangles), and an ITER-like shift (red, squares) for $\beta' = 0$ (filled) and an ITER-like $\beta'$ (empty).}
 \label{fig:heatFlux}
\end{figure}

For the tilted elliptical scans, the ion energy flux calculated by GS2 is shown in figure \ref{fig:heatFlux}.  In all simulations the electron energy flux was consistently smaller than the ion energy flux, typically by a factor of four. We see that the energy flux is fairly insensitive to the effects of both the Shafranov shift and $\beta'$ in the domain of $\theta_{\kappa} \in \left[ 0, \pi / 8 \right]$. At more extreme tilt angles we see that $\beta'$ dramatically increases the energy flux, as does the shift (albeit to a lesser extent).

\begin{figure}
 \centering
 \includegraphics[width=0.7\textwidth]{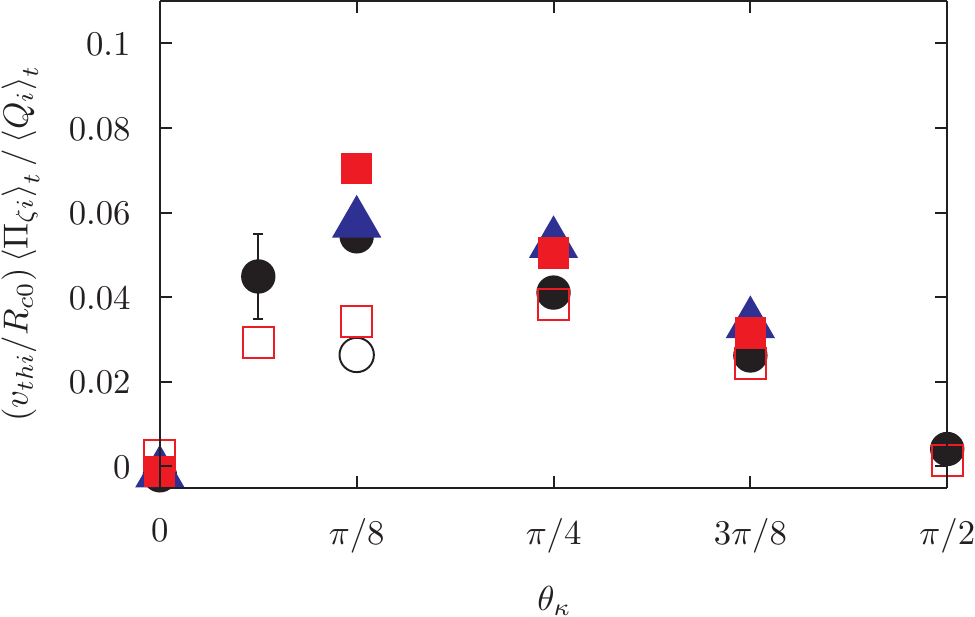}
 \caption{The momentum transport for flux surfaces with no shift (black, circles), a modest shift (blue, triangles), and an ITER-like shift (red, squares) for $\beta' = 0$ (filled) and an ITER-like $\beta'$ (empty).}
 \label{fig:momHeatFluxRatio}
\end{figure}

The ratio of the momentum flux to the energy flux, which is an estimate of intrinsic rotation through \refEq{eq:rotationGradEst}, is shown in figure \ref{fig:momHeatFluxRatio}. As expected, we see that the presence of an ITER-like Shafranov shift increases the momentum transport by approximately $30 \%$. However, a non-zero $\beta'$ significantly reduces the momentum transport. These two effects counteract one another and for ITER-like values at $\theta_{\kappa} = \pi / 8$ and $\rho_{0} = 0.54$ the shift is overshadowed by $\beta'$, leading to a net reduction in the momentum transport of about $30 \%$.

%Additionally, from figure \ref{fig:ShafShiftLoc} and \refEq{eq:twoModeEnvelopeCond} we see that the envelope created by the beating of elongation and the Shafranov shift in the $\theta_{\kappa b} = \pi / 8$ configuration is almost exactly up-down symmetric. This suggests that the up-down symmetry of the envelope is not important at these low mode numbers.

%===================================================%
%===================================================%
\subsection{Minor radial scans}
\label{subsubsec:minorRadialScans}
%===================================================%
%===================================================%

%\begin{figure}
% \centering
% \includegraphics[width=0.7\textwidth]{figs/Ch_09/nonlinearAvgIonHeatFluxWithMinorRadius.pdf}
% \caption{The radial dependence of the energy flux for flux surfaces with no shift (black, circles) and a strong shift (red, squares) varied according to \refEq{eq:radialShafShiftLocal} and \refEq{eq:axialShafShiftLocal}, for $\beta' = 0$ (filled) and an ITER-like $\beta'$ (empty).}
% \label{fig:heatFluxWithMinorRadius}
%\end{figure}
%
%[HEAT FLUX PLOT ANALYSIS]

These scans keep $\beta'$, $d \Ln{T_{s}} / d \rho$, $d \Ln{n_{s}} / d \rho$, $q$, and $\hat{s}$ constant with minor radius. We chose to keep $\beta'$ constant to be consistent with ITER (according to \refEq{eq:betaPrimeEstimate}). The others were kept fixed in order to avoid varying too many parameters in the scan. However, constant values for $d \Ln{T_{s}} / d \rho$ and $d \Ln{n_{s}} / d \rho$ are not an unreasonable approximation to many experiments, especially in the core of tokamaks \cite{BarnesTrinity2010}.
%The local shift is calculated at each minor radius to be consistent with \refEq{eq:radialShafShiftLocal} and \refEq{eq:axialShafShiftLocal}, which results from the global MHD calculation.

\begin{figure}
 \centering
 \includegraphics[width=0.7\textwidth]{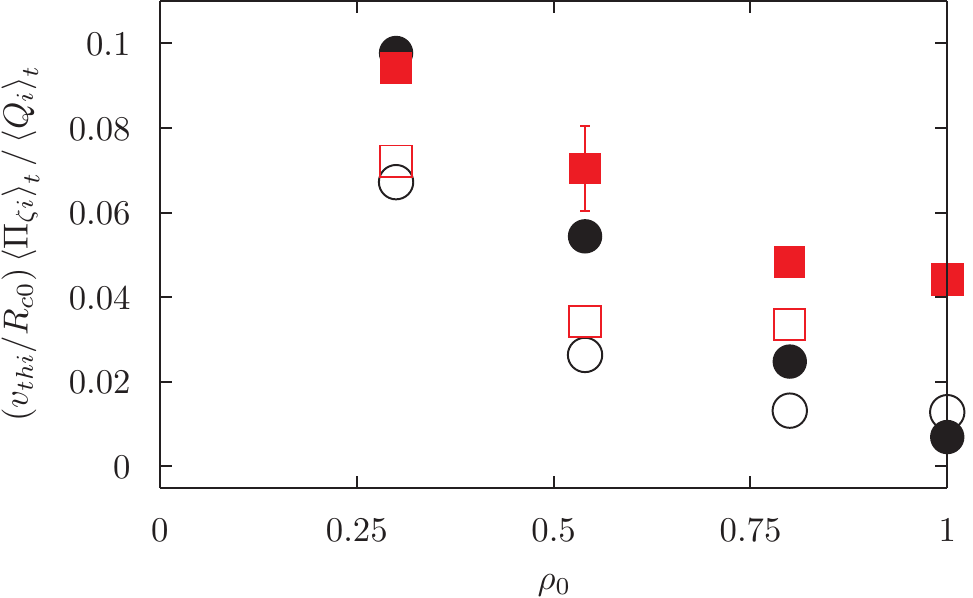}
 \caption{The radial dependence of the momentum transport for flux surfaces with no shift (black, circles) and a strong shift (red, squares) varied according to \refEq{eq:radialShafShiftLocal} and \refEq{eq:axialShafShiftLocal}, for $\beta' = 0$ (filled) and an ITER-like $\beta'$ (empty).}
 \label{fig:momHeatFluxRatioWithMinorRadius}
\end{figure}

The minor radial dependence of the momentum flux, shown in figure \ref{fig:momHeatFluxRatioWithMinorRadius}, is consistent with the previous scans. Comparing the two scans with $\beta' = 0$, we see that the difference in the momentum transport of the two scans increases with minor radius. This makes sense because the magnitude of the local shift increases linearly with minor radius according to \refEq{eq:radialShafShiftLocal} and \refEq{eq:axialShafShiftLocal}.

Additionally, throughout this scan the magnitude of the Shafranov shift changes, while $\beta'$ remains constant. Hence, we can compare the no shift, $\beta' = 0$ case to the ITER-like shift, ITER-like $\beta'$ case to demonstrate the counteracting effects of the shift and $\beta'$ on the momentum transport. Because the shift is weak at small values of $\rho_{0}$, the net effect of the shift and $\beta'$ is to lower the momentum transport. However, at large values of $\rho_{0}$ the shift is stronger, but $\beta'$ remains the same. Here the net effect of the shift and $\beta'$ is to enhance the momentum transport. Note that at $\rho_{0} = 1$ the momentum transport in the shifted configurations with and without $\beta'$ are indistinguishable.

Lastly, the dominant trend in figure \ref{fig:momHeatFluxRatioWithMinorRadius} is the roughly linear decrease of the momentum transport with minor radius. This is not currently understood as nearly all simulations of intrinsic rotation from up-down asymmetry were performed using $a_{\psi 0} / R_{c 0} \approx 1 / 6$. However, the decreasing trend with increasing $a_{\psi 0} / R_{c 0}$ is consistent with the results of several simulations performed at $a_{\psi 0} / R_{c 0} \approx 1 / 12$ and $a_{\psi 0} / R_{c 0} \approx 1 / 3$ in reference \cite{BallMomUpDownAsym2014}. These results suggest that in this parameter range the momentum transport increases with the aspect ratio.
%which initially seems counter intuitive. Thinking in terms of single-particle orbits, there is no equivalent to the concept of up-down asymmetry in a screw pinch. However, an infinite aspect ratio tokamak still has magnetic drifts that define up-down symmetry and lead to momentum transport.

%===================================================%
%===================================================%
\subsection{Circular flux surface scan}
\label{subsubsec:circularScan}
%===================================================%
%===================================================%

\begin{figure}
 (a) \hspace{0.27\textwidth} (b) \hspace{0.28\textwidth} (c)
 \begin{center}
  \includegraphics[height=0.24\textwidth]{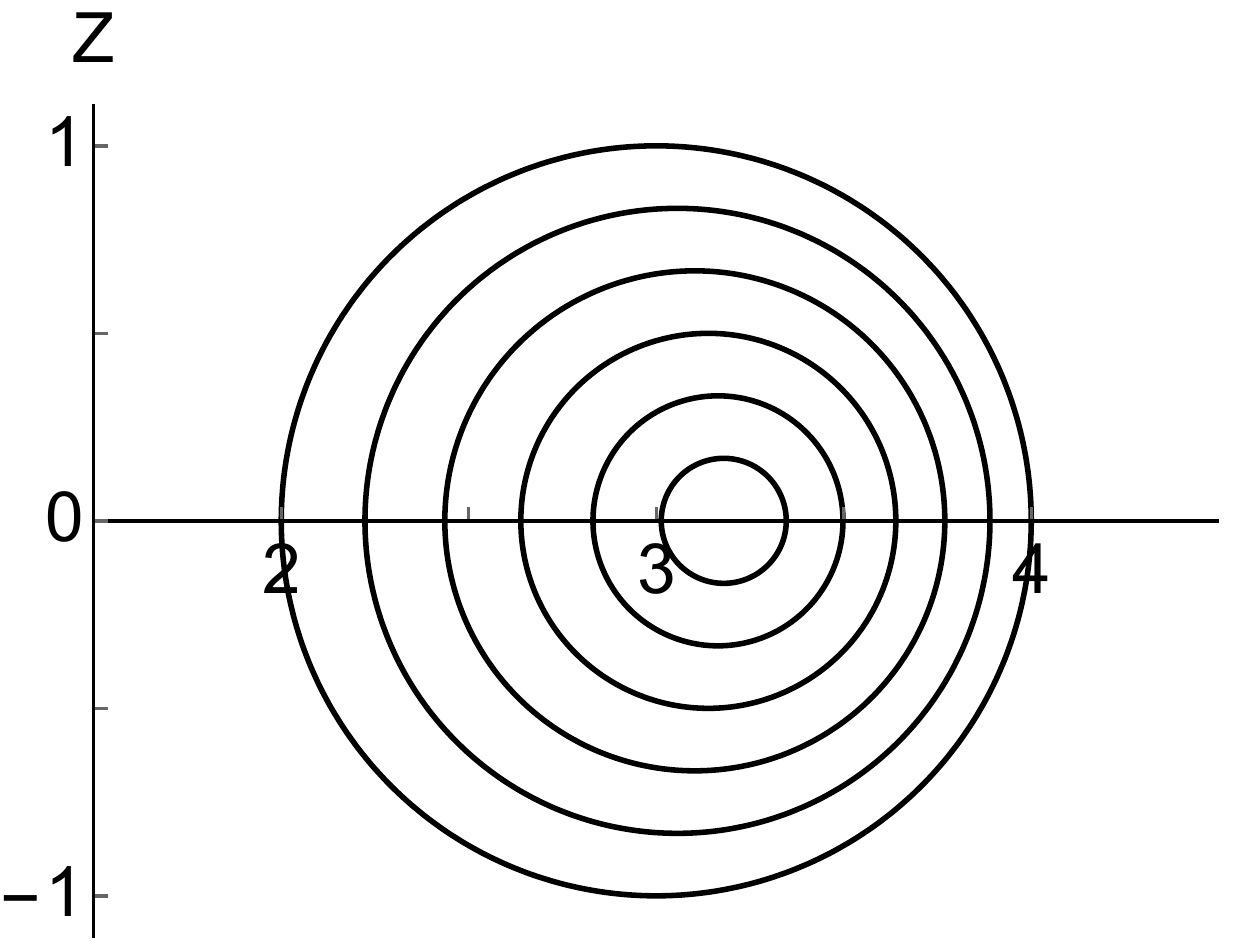}
  \includegraphics[height=0.24\textwidth]{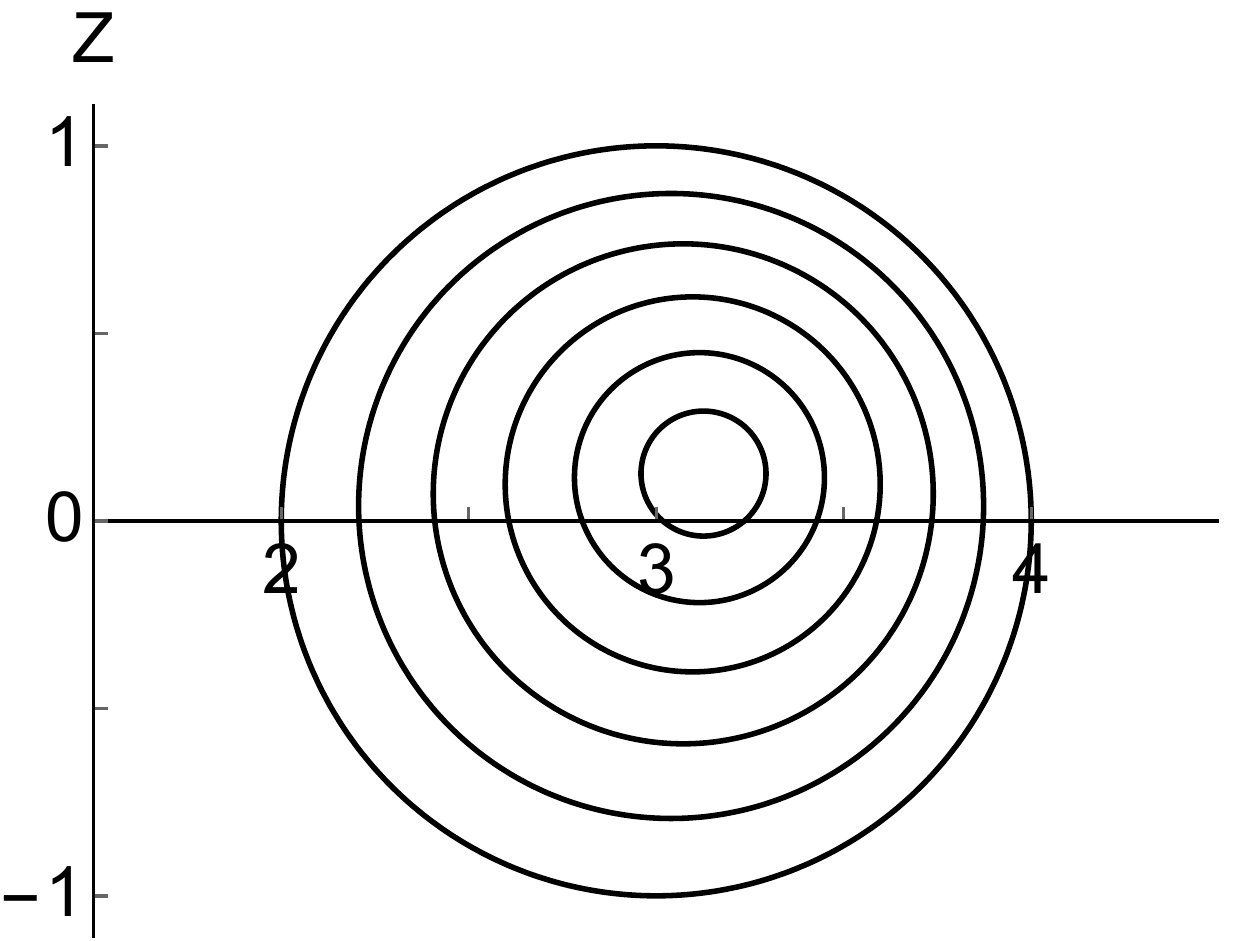}
  \includegraphics[height=0.24\textwidth]{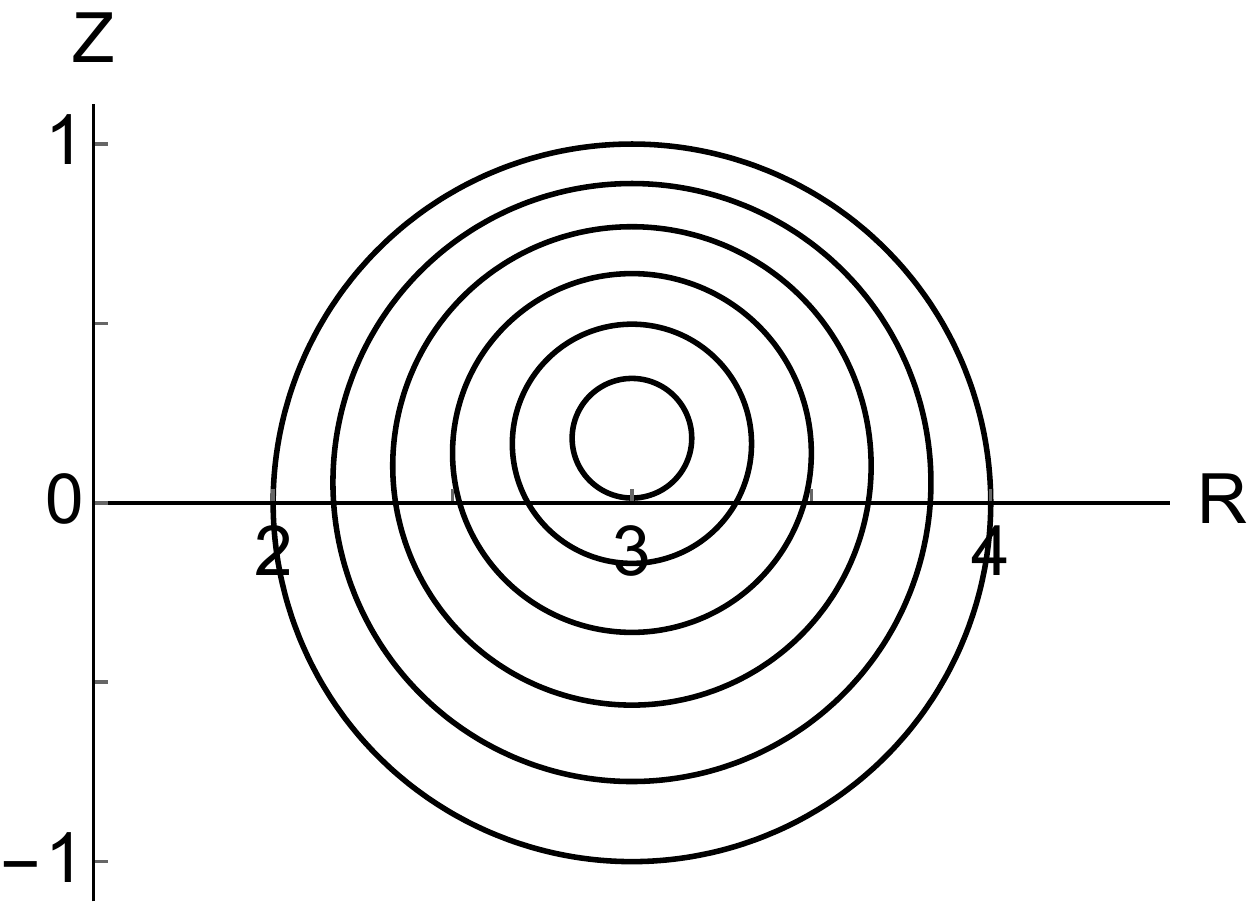}
 \end{center}
 \caption{The magnetic geometry for circular flux surfaces with an ITER-like (a) horizontal shift, (b) diagonal shift, or (c) vertical shift.}
 \label{fig:circScanGeo}
\end{figure}

To completely isolate the effect of the Shafranov shift on momentum transport we also ran simulations with shifted circular flux surfaces as shown in figure \ref{fig:circScanGeo}. To create up-down asymmetry and drive momentum transport we varied the direction of the tilt by changing the parameter $\theta_{\text{axis}}$ with the magnitude of the shift fixed at a value $\sim 30 \%$ larger than an ITER-like machine. Scanning $\theta_{\text{axis}}$ is unphysical because circular flux surfaces can only ever have a shift in the outboard radial direction, which corresponds to $\theta_{\text{axis}} = 0$. Though unphysical, this scan will help clarify the influence of the Shafranov shift.

\begin{figure}
 \centering
 \includegraphics[width=0.7\textwidth]{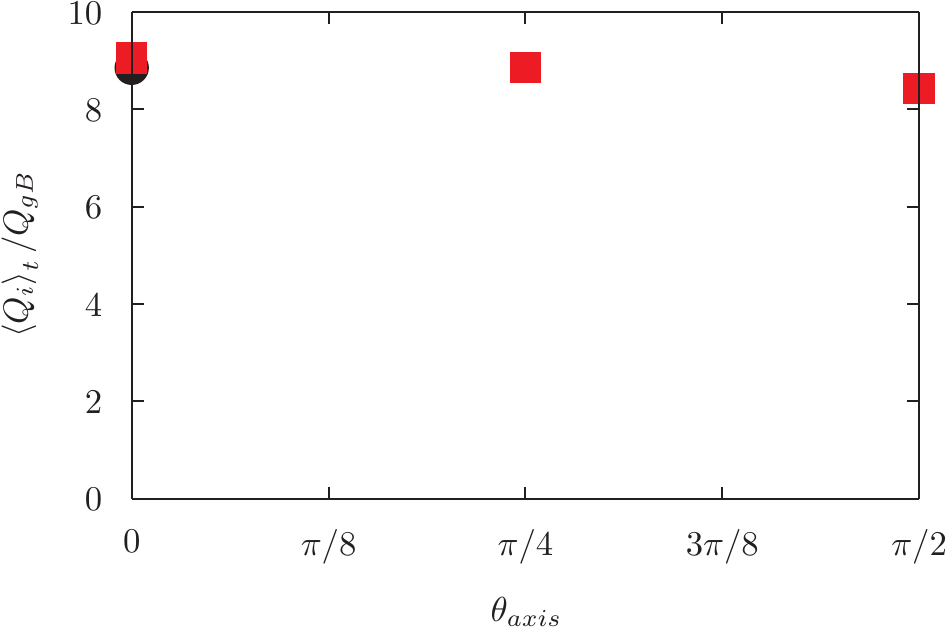}
 \caption{The energy flux for circular flux surfaces with no shift (black, circles) and an ITER-like shift (red, squares) as a function of the direction of the Shafranov shift, where all simulations have $\beta' = 0$.}
 \label{fig:heatFluxCirc}
\end{figure}

Figure \ref{fig:heatFluxCirc} shows that the presence and direction of the Shafranov shift has little effect on the energy flux from circular flux surfaces. This is akin to the tilted elliptical results in the range of $\theta_{\kappa} \in \left[ 0, \pi / 8 \right]$, but different from the tilted elliptical results in the range of $\theta_{\kappa} \in \left[ \pi / 8, \pi / 2 \right]$. A possible explanation for figure \ref{fig:heatFluxCirc} is that the magnitude of the shift in the circular equilibria is similar to that in the $\theta_{\kappa} = \pi / 16$ elliptical equilibria, but considerably less than the shift present in the elliptical equilibria with large tilt angles.

\begin{figure}
 \centering
 \includegraphics[width=0.7\textwidth]{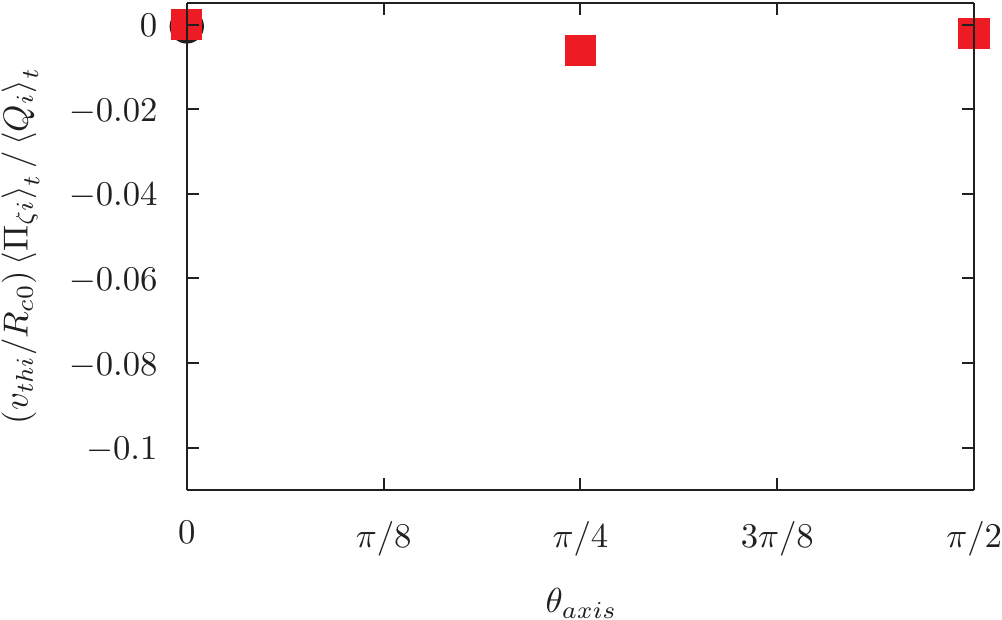}
 \caption{The momentum transport for circular flux surfaces with no shift (black, circles) and an ITER-like shift (red, squares) as a function of the direction of the Shafranov shift, where all simulations have $\beta' = 0$. Note that we have kept the range of the vertical axis the same as in figures \ref{fig:momHeatFluxRatio} and \ref{fig:momHeatFluxRatioWithMinorRadius} for ease of comparison.}
 \label{fig:momHeatFluxRatioCirc}
\end{figure}

\begin{figure}
 \centering
 \includegraphics[width=0.7\textwidth]{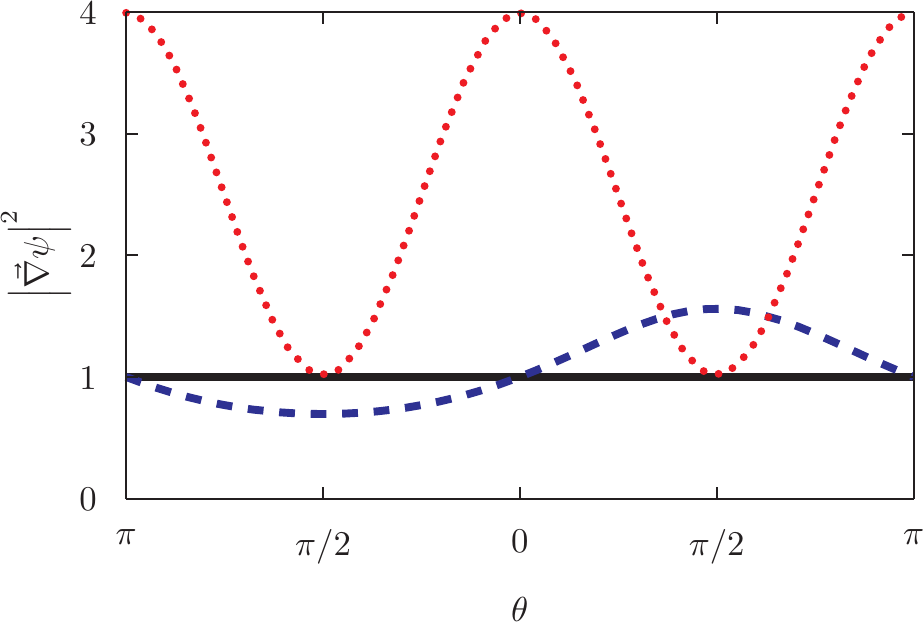}
 \caption{The geometric coefficient $\left| \Nabla \psi \right|^{2}$ for unshifted circular flux surfaces (black, solid), circular flux surfaces with a strong vertical shift (blue, dashed), and unshifted flux surfaces with a vertical elongation of $\kappa = 2$ (red, dotted) normalised to the unshifted circular value.}
 \label{fig:gds22}
\end{figure}

In figure \ref{fig:momHeatFluxRatioCirc}, we see the effect of a strong Shafranov shift on momentum transport. It appears that a pure shift in circular flux surfaces (even when it is diagonal or vertical) drives minimal rotation compared to that generated by elliptical flux surfaces. This is somewhat surprising since the shift is an $m = 1$ shaping effect and, from the analysis in chapter \ref{ch:GYRO_TiltingSymmetry}, we expect the momentum flux to scale as $\Exp{- \beta m^{\gamma}}$ in mirror symmetric configurations. However, there are two important caveats. Firstly, the exponential scaling is only true in the limit of $m \gg 1$, which is clearly not satisfied for $m = 1$. Secondly, the Shafranov shift has a relatively minor effect on the magnetic equilibrium compared with elongating the flux surfaces to $\kappa = 2$ (even when the shift is $30 \%$ stronger than that expected in ITER). This can be seen by looking at the geometric coefficients that appear in the gyrokinetic equation (see chapter \ref{ch:GYRO_Overview}). Since these coefficients are the only way the magnetic geometry enters the local gyrokinetic model, they must control the momentum transport. For example, in figure \ref{fig:gds22} we see that elongating an unshifted circular configuration changes the coefficient $\left| \Nabla \psi \right|^{2}$ by $300 \%$, while introducing the Shafranov shift only makes a $50 \%$ difference. Hence, we believe that the Shafranov shift and elongation are comparably effective at transporting momentum, but that practical limits on the maximum value of $\beta$ constrain the Shafranov shift to have a small effect on the magnetic equilibrium compared to externally applied plasma shaping.

%Lastly, the small level of momentum transport induced by a pure Shafranov shift suggests that there is a nonlinear interaction between elongation and the Shafranov shift. We see from figure \ref{fig:momHeatFluxRatioCirc} that the maximum value of rotation achieved in simulations of a purely shifted configurations is $\left( v_{th i} / R_{c 0} \right) \left\langle \Pi_{\zeta i} \right\rangle_{t} / \left\langle Q_{i} \right\rangle_{t} \approx 0.006$, while the introduction of a somewhat weaker shift into the tilted elliptical configurations increased the amount of rotation by as much as $0.017$. Hence, a simple linear superposition of the two rotation generation effects (i.e. shift and elongation) is insufficient to explain the enhancement seen in figure \ref{fig:momHeatFluxRatio}.

%Instead we suppose the primary reason for the enhancement is a nonlinear effect of superimposing the two shaping effects. Specifically, the introduction of a shift into the tilted elliptical configuration breaks the mirror symmetry of the equilibrium, which chapters \ref{ch:GYRO_TiltingSymmetry} and \ref{ch:GYRO_MomFluxScaling} suggest could significantly enhance the momentum transport.

%===================================================%
%===================================================%
\section{Effect of the value of $\beta'$}
\label{sec:betaPrime}
%===================================================%
%===================================================%

\begin{figure}
	\begin{center}
		(a) \hspace{0.42\textwidth} (b) \hspace{0.3\textwidth}
		
		\includegraphics[height=0.3\textwidth]{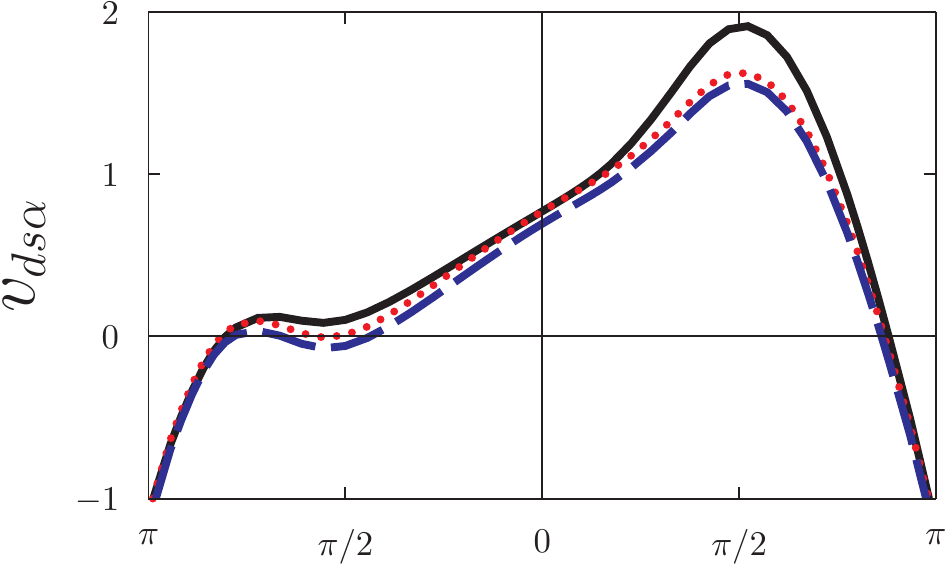}
		\includegraphics[height=0.3\textwidth]{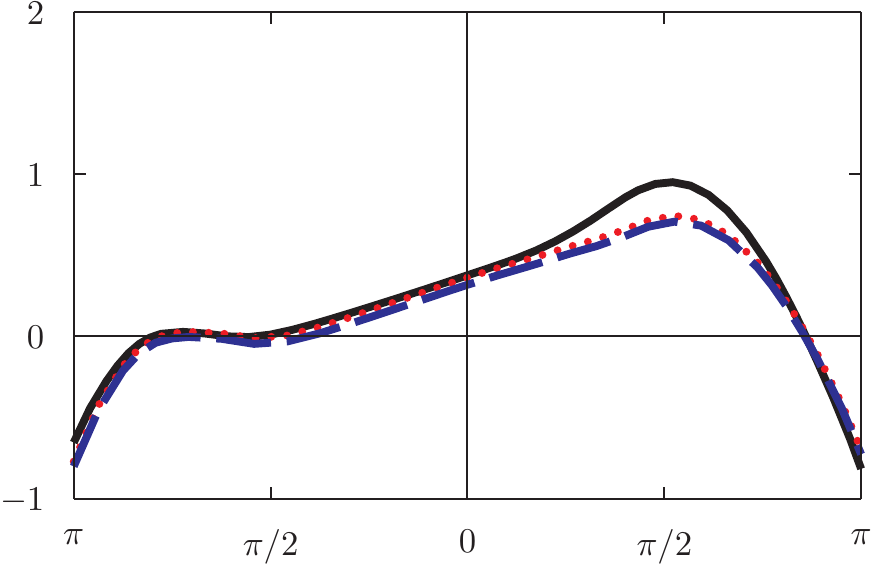}
		
		(c) \hspace{0.42\textwidth} (d) \hspace{0.3\textwidth}
		
		\includegraphics[height=0.35\textwidth]{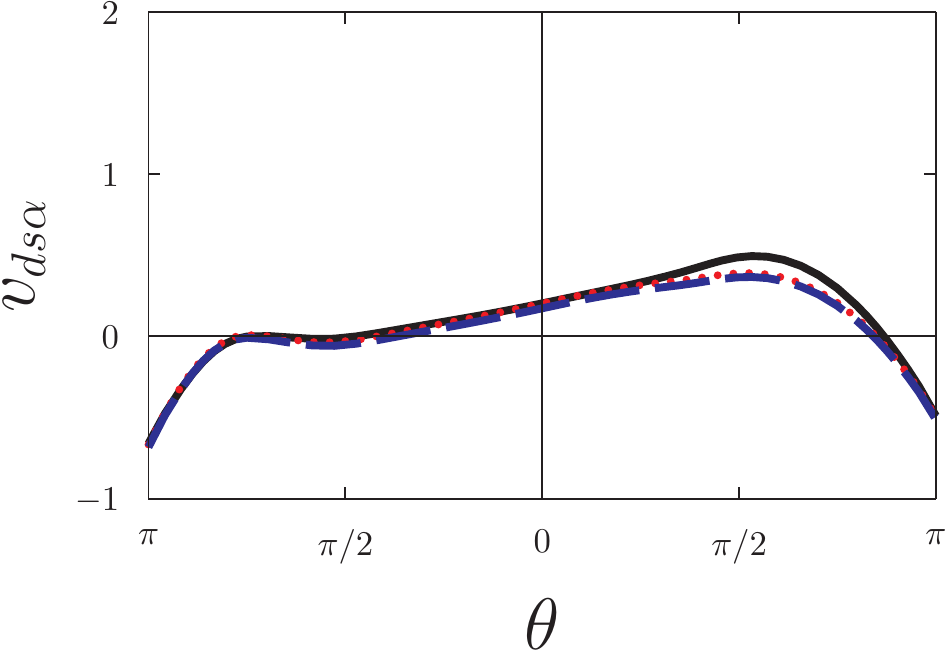}
		\includegraphics[height=0.35\textwidth]{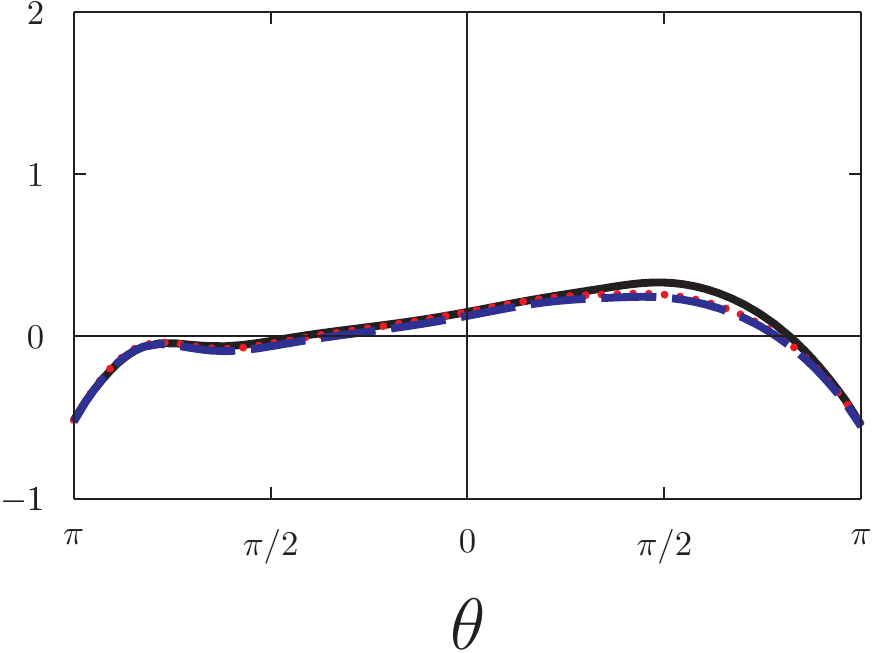}
	\end{center}
	\caption{The geometric coefficient $v_{d s \alpha}$ in units of $v_{th s}^{2} / \left( a^{2} \Omega_{s} \right)$ without Shafranov shift at (a) $\rho_{0} = 0.3$, (b) $\rho_{0} = 0.54$, (c) $\rho_{0} = 0.8$, and (d) $\rho_{0} = 1$ for no $\beta'$, $w_{||}^{2} = v_{th s}^{2}$, $w_{\perp}^{2} = 0$ (black, solid); an ITER-like $\beta'$, $w_{||}^{2} = v_{th s}^{2}$, $w_{\perp}^{2} = 0$ (red, dotted); and an ITER-like $\beta'$, $w_{||}^{2} = 0$, $w_{\perp}^{2} = 2 v_{th s}^{2}$ (blue, dashed).}
	\label{fig:driftCoeff}
\end{figure}

In section \ref{subsec:results} we included the effect of the Shafranov shift in nonlinear, local gyrokinetic simulations and found that it enhanced momentum transport as expected. Since the magnitude of the shift depends on the plasma pressure, we also included a non-zero $\beta'$. We found that $\beta'$ strongly reduced the momentum flux, often entirely cancelling the enhancement due to the Shafranov shift. Consequently, it is important to understand how $\beta'$ alters the geometric coefficients of gyrokinetics.

In appendix \ref{app:geoCoeff} we discuss how $\beta'$ enters into the analytic expressions for the geometric coefficients. We show that $\beta'$ vanishes in the large aspect ratio limit (for the orderings of \refEq{eq:gradShafOrderings}), like the Shafranov shift. This means that for large aspect ratio tokamaks $\beta'$ can be ignored and the results of reference \cite{BallMomUpDownAsym2014} (which ignores $\beta'$) apply. However, the Shafranov shift also vanishes in this limit, so it cannot be used to enhance the momentum transport.

Figure \ref{fig:driftCoeff} uses the simulations from the minor radial scan to show the quantitative effect of $\beta'$ on the geometric coefficient $v_{d s \alpha}$ (defined by \refEq{eq:alphaGCdriftvelocity}) with different values of $w_{||}$ and $w_{\perp}$. We chose $v_{d s \alpha}$ as chapter \ref{ch:GYRO_MomFluxScaling} indicates that it may be the most important geometric coefficient for understanding intrinsic rotation transport due to up-down asymmetry. We see that including a non-zero $\beta'$ tends to reduce the up-down asymmetry of $v_{d s \alpha}$, which is consistent with the observed reduction in momentum transport. % A more in-depth analysis is presented in \ref{app:geoCoeff}.

%===================================================%
%===================================================%
\section{Effect of the $\beta$ profile}
\label{sec:betaProfile}
%===================================================%
%===================================================%

\begin{figure}
	\begin{center}
		(a) \hspace{0.4\textwidth} (b) \hspace{0.35\textwidth}
		
		\includegraphics[width=0.43\textwidth]{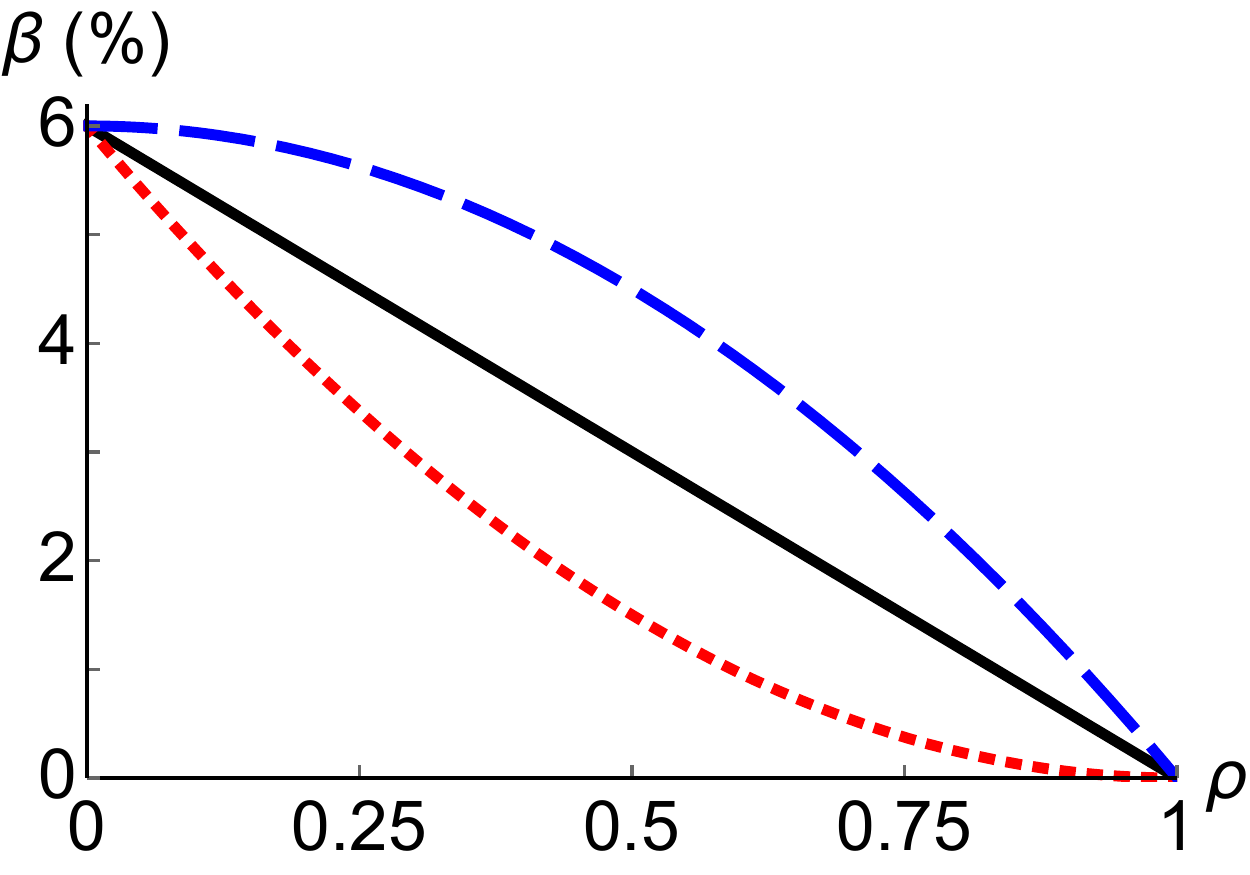}
		\includegraphics[width=0.47\textwidth]{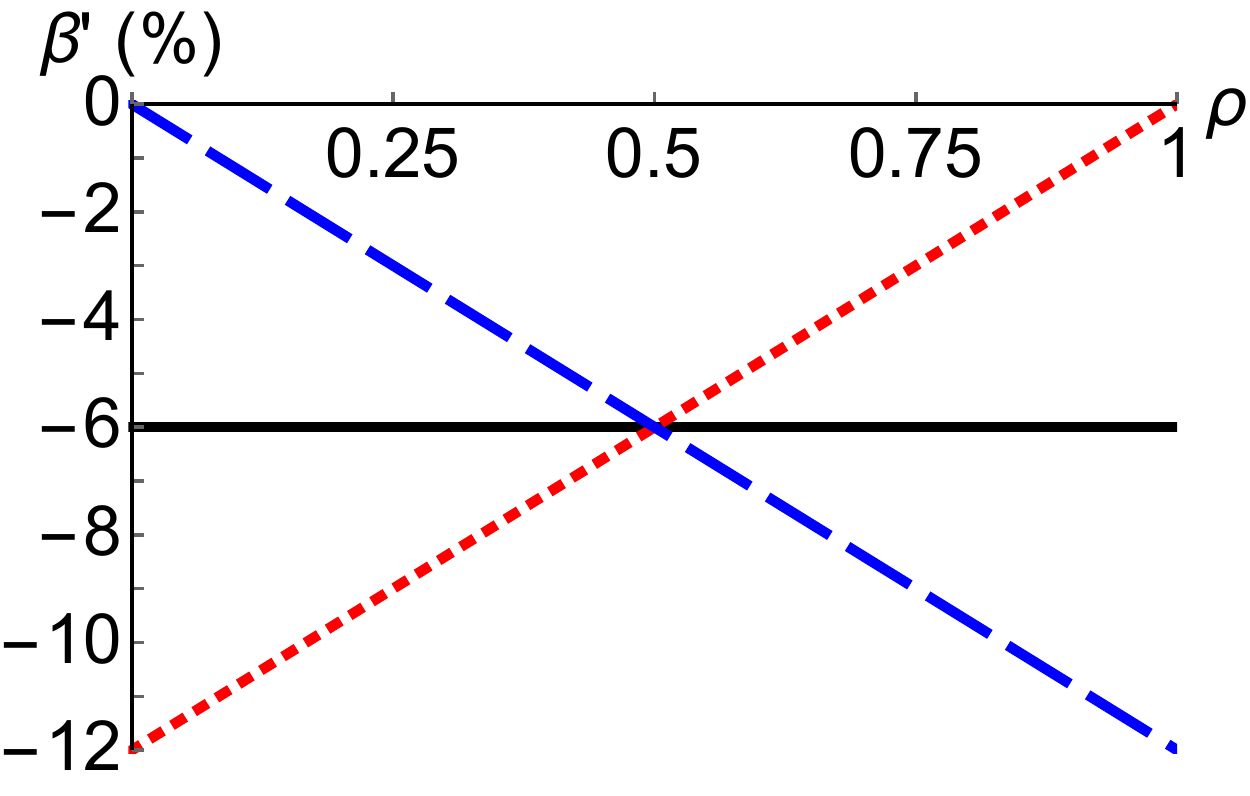}
		
		(c) \hspace{0.36\textwidth}
		
		\includegraphics[width=0.45\textwidth]{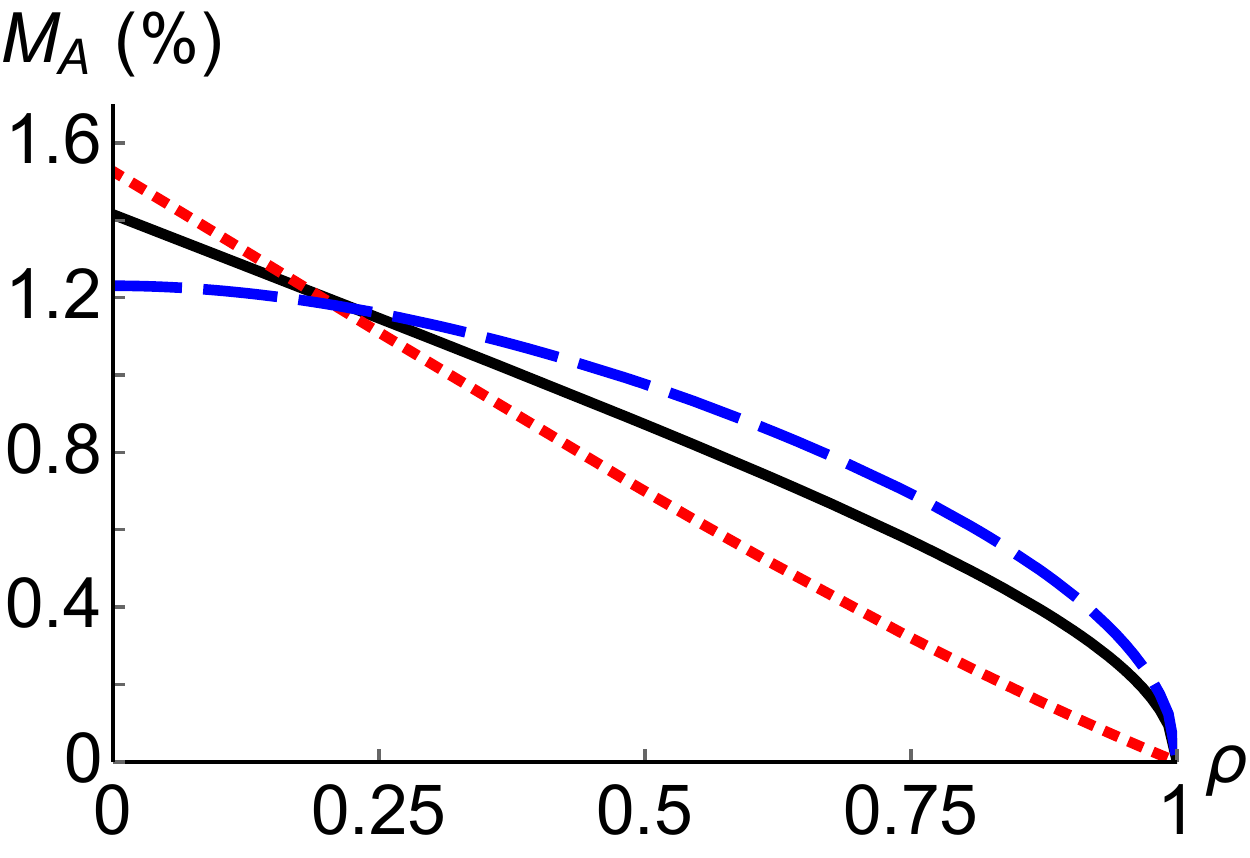}
	\end{center}
	\caption{Example (a) $\beta$ profiles, their corresponding (b) $\beta'$ profile, and their (c) Alfv\'{e}n Mach number profile, estimated using \refEq{eq:rotationGradEst} for a constant $\beta'$ (black, solid), linear peaked $\beta'$ (red, dotted), and linear hollow $\beta'$ (blue, dashed) profiles.}
	\label{fig:MachNumberProfile}
\end{figure}

In order to estimate a realistic value for $\beta'$, we used the on-axis value of $\beta$ predicted for ITER and assumed $\beta$ was linear with minor radius $a_{\psi}$. This gave a reasonable order of magnitude estimate. However, since the momentum transport is strongly and adversely affected by $\beta'$ it is worthwhile to discuss the implications of different radial profiles of $\beta'$. For example, we expect that in H-mode operation $\beta'$ would be larger at the plasma edge and smaller in the core compared to L-mode. Unfortunately, from the estimate of rotation given by \refEq{eq:rotationGradEst}, we see that $\beta'$ is necessary, even though including the effect of $\beta'$ in the geometric coefficients reduces the momentum flux. Physically the necessity of $\beta'$ is intuitive because the pressure gradient is the source of free energy that ultimately drives the momentum transport. Because of all these competing dependencies on $\beta$ and $\beta'$, both explicitly and through $\left( v_{th i} / R_{c 0} \right) \left\langle \Pi_{\zeta i} \right\rangle_{t} / \left\langle Q_{i} \right\rangle_{t}$, it is difficult to determine the $\beta$ profile that maximises rotation. In figure \ref{fig:MachNumberProfile} we use \refEq{eq:rotationGradEst} and the approximation
\begin{align}
   \frac{ v_{th i}}{R_{c}} \frac{\left\langle \Pi_{\zeta i} \right\rangle_{t}}{\left\langle Q_{i} \right\rangle_{t}} \approx c_{1} + c_{2} \rho + c_{3} \beta' + c_{4} \rho \beta' \label{eq:linearFit}
\end{align}
to estimate the rotation profile for different $\beta$ profiles. Here the coefficients $c_{1} = 0.11$, $c_{2} = -0.06$, $c_{3} = 0.34$, and $c_{4} = -0.28$ are determined by a fit to the data in figure \ref{fig:momHeatFluxRatioWithMinorRadius}. We include the dependence on $\beta'$ and $\rho$ in the fit because the momentum transport is sensitive to both. Figure \ref{fig:MachNumberProfile} shows that the shape of the $\beta$ profile can have a significant effect on the rotation profile. We see that a more peaked pressure profile seems to lead to a more peaked rotation profile. It is interesting to note that the rotation profiles with the lowest on-axis Mach number have the broadest rotation profiles. This means that the $\beta$ profile that maximises the {\it on-axis} Mach number is not necessarily optimal because broad rotation profiles are expected to be significantly more effective at stabilising MHD modes \cite{LiuITERrwmStabilization2004}.

%Equation \refEq{eq:rotationGradEst} also illuminates the results of reference \cite{BallMomUpDownAsym2014}. This work neglected the effect of a non-zero $\beta'$ on the geometric coefficients used in the gyrokinetic simulations, but relied on a non-zero $\beta'$ in order to calculate the rotation from the fluxes. According to \refEq{eq:rotationGradEst}, doing this can still be consistent, but only in the limit of $\sqrt{\beta} \to 0$. In this limit $\beta'$ can be small enough to have no effect on the gyrokinetic calculation of the fluxes, but still drive significant rotation.

%% file: Ch_10_GYRO_NonMirrorSym.tex
% !TEX root = /Users/Justin/Documents/Research/Writings/2016DoctoralThesis/DoctoralThesis.tex

\chapter{Non-mirror symmetry: Tilted elongation and tilted triangularity}
\label{ch:GYRO_NonMirrorSym}

From our analytic scaling studies in chapters \ref{ch:GYRO_TiltingSymmetry} and \ref{ch:GYRO_MomFluxScaling} we have identified two most promising types of non-mirror symmetry. In chapter \ref{ch:GYRO_ShafranovShift} we explored the first: using the Shafranov shift to break the mirror symmetry of tilted elliptical flux surfaces. In this chapter we will examine the second: using externally applied elongation and triangularity with independent tilt angles to directly make the flux surface shapes non-mirror symmetric. We will compare these two-mode geometries (i.e. flux surfaces with both elongation and triangularity) with single-mode geometries (i.e. flux surfaces with only elongation or triangularity) to see if the momentum transport can be significantly increased.

To do this, we will use GS2 to calculate the nonlinear electrostatic fluxes for a two dimensional scan in $\theta_{\kappa}$ and $\theta_{\delta}$, the tilt angles of elongation and triangularity respectively. In keeping with previous simulations, we will use a shaped variant of the Cyclone base case (specified by the parameters of \refEq{eq:cycloneBaseCase}) without collisions, except we must increase the temperature gradient of both species to $a / L_{T s} = 3.0$ to ensure that the turbulence is unstable. The strength of elongation and triangularity is set at $\Delta_{2} = 1.7$ and $\Delta_{3} = 1.3$ respectively, which were values estimated from the shape of the ITER plasma \cite{AymarITERSummary2001}. We note that this elongation is significantly lower than the value of $2$ used in chapter \ref{ch:GYRO_ShafranovShift}.

These unusual, highly-shaped flux surface shapes have quite detailed and sharp poloidal structure. Properly resolving this requires a fine poloidal grid, typically with 96 grid points. This means that these simulations are considerably more expensive than single-mode flux surfaces (which typically require between 32 and 64 grid points).

%===================================================%
%===================================================%
\section{Exact shaping geometry scan}
\label{sec:exactScan}
%===================================================%
%===================================================%

\begin{figure}
	\centering
	(a) %\hspace{1em}% \hspace{0.75\textwidth}
	\includegraphics[align=t,width=0.95\textwidth]{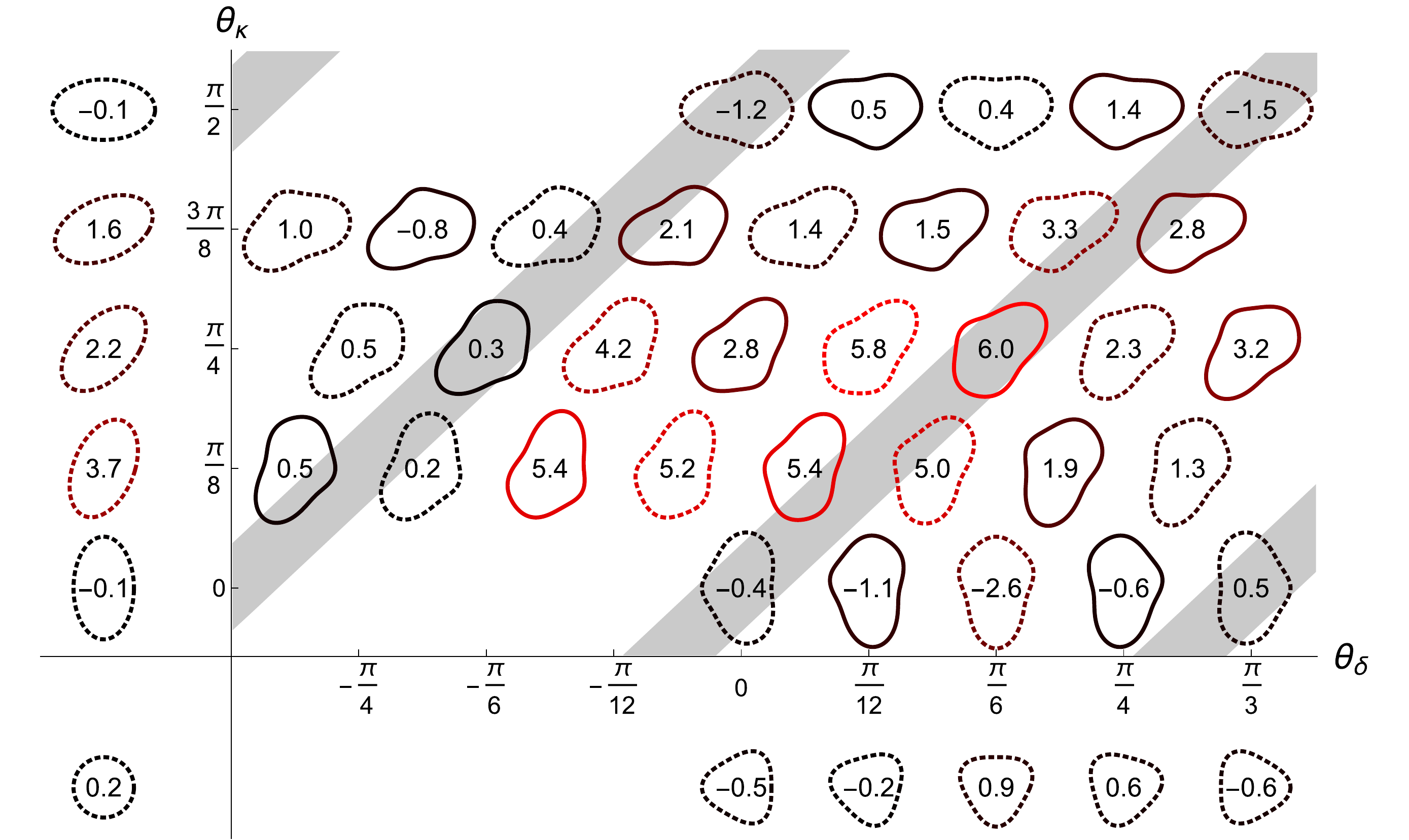}
	
%	\vspace{2em}
	
	(b) %\hspace{1em}% \hspace{0.75\textwidth}
	\includegraphics[align=t,width=0.95\textwidth]{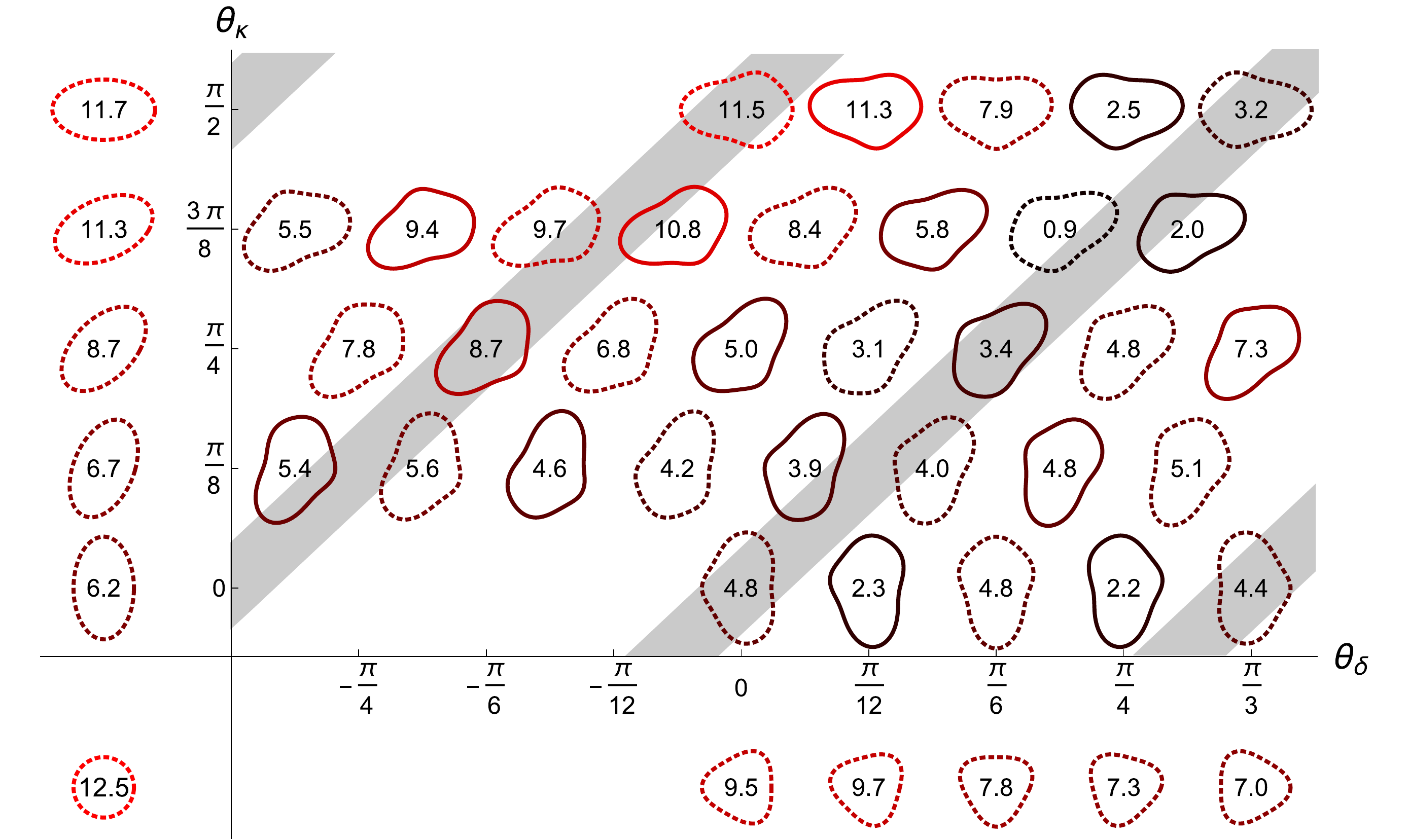}
	\caption{Values of (a) $100 \times \left( v_{th i} / R_{c 0} \right) \left\langle \Pi_{\zeta i} \right\rangle_{t} / \left\langle Q_{i} \right\rangle_{t}$ and (b) $\left\langle Q_{i} \right\rangle_{t} / Q_{gB}$ (with errors of $\pm 0.5$ and $\pm 0.3$ respectively) are indicated by the numbers/colours for various non-mirror symmetric (solid lines) and mirror symmetric (dotted lines) flux surfaces created with elongation and triangularity (using the Exact geometry specification). The thick grey bands indicate geometries with up-down symmetric envelopes according to \refEq{eq:twoModeEnvelopeCond}. Purely elongated flux surfaces are shown in quadrant \romanNum{2}, a circular flux surface is shown in quadrant \romanNum{3}, and purely triangular flux surfaces are shown in quadrant \romanNum{4} for comparison.}
	\label{fig:nonMirrorExact}
\end{figure}

%\begin{figure}
%	\centering
%	(a) \hspace{0.8\textwidth}
%	
%	\begin{subfigure}{\textwidth}
%		\includegraphics[width=1.0\textwidth]{figs/Ch_10/nonlinearAvgIonRatioMod.pdf}
%	\end{subfigure}
%	
%	\vspace{4em}
%	
%	(b) \hspace{0.8\textwidth}
%	
%	\begin{subfigure}{\textwidth}
%		\includegraphics[width=1.0\textwidth]{figs/Ch_10/nonlinearAvgIonMomMod.pdf}
%	\end{subfigure}
%\end{figure}
%\clearpage
%\begin{figure}
%	\ContinuedFloat % continue from previous page
%	\centering
%	(c) \hspace{0.8\textwidth}
%	
%	\begin{subfigure}{\textwidth}
%		\includegraphics[width=1.0\textwidth]{figs/Ch_10/nonlinearAvgIonHeatFlux.pdf}
%	\end{subfigure}
%	\caption{Values of (a) $100 \times \left( v_{th i} / R_{c 0} \right) \left\langle \Pi_{\zeta i} \right\rangle_{t} / \left\langle Q_{i} \right\rangle_{t}$, (b) $100 \times \left\langle \Pi_{\zeta i} \right\rangle_{t} / \Pi_{gB}$, and (c) $\left\langle Q_{i} \right\rangle_{t} / Q_{gB}$ (with errors of $\pm 0.5$, $\pm 10$, and $\pm 0.3$ respectively) are indicated by the numbers/colours for various non-mirror symmetric (solid lines) and mirror symmetric (dotted lines) flux surfaces created with elongation and triangularity (using the Exact geometry specification). The thick grey bands indicate geometries with up-down symmetric envelopes according to \refEq{eq:twoModeEnvelopeCond}. Purely elongated flux surfaces are shown in quadrant \romanNum{2}, a circular flux surface is shown in quadrant \romanNum{3}, and purely triangular flux surfaces are shown in quadrant \romanNum{4} for comparison.}
%	\label{fig:nonMirrorExact}
%\end{figure}

We performed the scan using the Exact geometry specified by \refEq{eq:fluxSurfofInterestShapeExact} and \refEq{eq:fluxSurfofInterestDerivExact} (with $\theta_{t 2} = \theta_{\kappa}$ and $\theta_{t 3} = \theta_{\delta}$) in order to produce figure \ref{fig:nonMirrorExact}. Note that the missing geometries are redundant as the up-down symmetry argument (see section \ref{sec:upDownSymArg}) implies that an up-down reflection of the flux surface shape changes the sign of the momentum flux and does not effect the energy flux.

Figure \ref{fig:nonMirrorExact}(a) shows how the intrinsic rotation generated by the two-mode geometries compares with that generated by flux surfaces with only elongation or triangularity. We see that the maximum intrinsic rotation from the two-mode geometry (i.e. that from the $\theta_{\kappa} = \pi / 4$, $\theta_{\delta} = \pi / 6$ case) is over $60\%$ larger than the maximum that can be generated with only elongation. Furthermore, the sum of the rotation generated by purely elongated flux surfaces and purely triangular flux surfaces falls short of the rotation generated by flux surfaces with both elongation and triangularity. This suggests that there is a nonlinear interaction between the different shaping effects.

One possible explanation is the breaking of flux surface mirror symmetry. This is the dominant mechanism in a screw pinch, where momentum transport is driven by the direct interaction of different shaping effects. However, this does not explain the size of the momentum flux in the $\theta_{\kappa} = \pi / 4$, $\theta_{\delta} = \pi / 12$ case, which is mirror symmetric. In fact, while the geometry with the largest momentum transport is non-mirror symmetric, on average the mirror and non-mirror symmetric configurations have roughly similar levels of turbulent transport. This indicates that the importance of non-mirror symmetry (e.g. the direct interaction of elongation and triangularity) for momentum transport is minimal in these low mode number configurations.

Another explanation is that the beating between elongation and triangularity creates an $m=1$ mode (i.e. an envelope), which drives the extra rotation. However, the configuration with the highest rotation has an up-down symmetric $m=1$ envelope, which we can see by substituting $\theta_{t 2} = \theta_{\kappa} =\pi / 4$ and $\theta_{t 3} = \theta_{\delta} = \pi / 6$ into \refEq{eq:twoModeEnvelopeCond}. This suggests that the tilting symmetry argument, which predicts that non-mirror symmetric configurations with an up-down symmetric envelope should have exponentially small momentum transport in $m_{c} \gg 1$, is not applicable to these configurations. This behaviour is consistent with figure \ref{fig:momHeatFluxRatioAbs}, which shows that configurations with an up-down asymmetric envelope have larger momentum transport than those with an up-down symmetric envelope at high $m$, but not at $m=2$. To understand why the tilting symmetry argument breaks down for elongation and triangularity we will study figure \ref{fig:FourierSpectrum}. We see that the poloidal distribution of momentum flux from circular flux surfaces tends to have a strong $m=2$ Fourier mode due to the inherent toroidicity of the tokamak. Furthermore, shaping the flux surface with a mode $m$ creates large Fourier components in the distribution of momentum flux at $m-1$ and $m-2$. These facts were used in chapter \ref{ch:GYRO_TiltingSymmetry} to explain why the agreement in figure \ref{fig:fullMomProfiles} breaks down for shaping modes with $m \leq 4$. Therefore, we expect triangularity to create an $m = 2$ Fourier mode in the momentum flux, which can beat against the inherent $m=2$ component from toroidicity to produce an $m=0$ component, which does not average to zero over the flux surface. This direct interaction between toroidicity and what we are considering ``fast'' shaping breaks the tilting symmetry. To put the issue simply, we relied on an expansion in $m_{c} \gg 1$ (or more precisely $m_{c} - 2 \gg 1$), which is not satisfied by elongation and triangularity (i.e. $m=2$ and $m=3$).

Instead, for these geometries it appears that the intrinsic rotation drive is dominated by the direct interaction of elongation and triangularity with toroidicity.

While it is true that in figure \ref{fig:nonMirrorExact}(a) the flux surfaces with two shaping modes generate significantly more rotation than the single-mode surfaces, it is not necessarily a fair comparison. The two-mode flux surfaces are more shaped, meaning they require more external coils/current and are less stable to axisymmetric modes. Looking at figure \ref{fig:momHeatFluxRatio}, we see that the maximum two-mode value (at $\Delta_{2} = 1.7$ and $\Delta_{3} = 1.3$) is only slightly above that produced by pure elongation with $\Delta_{2} = 2$ (simulated in chapter \ref{ch:GYRO_ShafranovShift}). However, the two-mode flux surfaces seem preferable for reasons apart from momentum transport. Figure \ref{fig:nonMirrorExact}(b) shows that the highest performing two-mode geometries directly stabilise turbulence and increase the confinement time. In fact the turbulence is completely stabilised at the standard Cyclone base case temperature gradient of $a / L_{T s} = 2.3$. Hence not only is the turbulence reduced, but the critical gradient is increased substantially (even compared to the elongated flux surfaces with $\Delta_{2} = 2$). We note that, as in the Shafranov shift scans, the electron energy flux was consistently smaller than the ion energy flux, typically by a factor of four.

%% file: Ch_11_Conclusions.tex
\chapter{Conclusions}
\label{ch:conclusions}

\begin{figure}
	\begin{center}
		\includegraphics[width=0.55\textwidth]{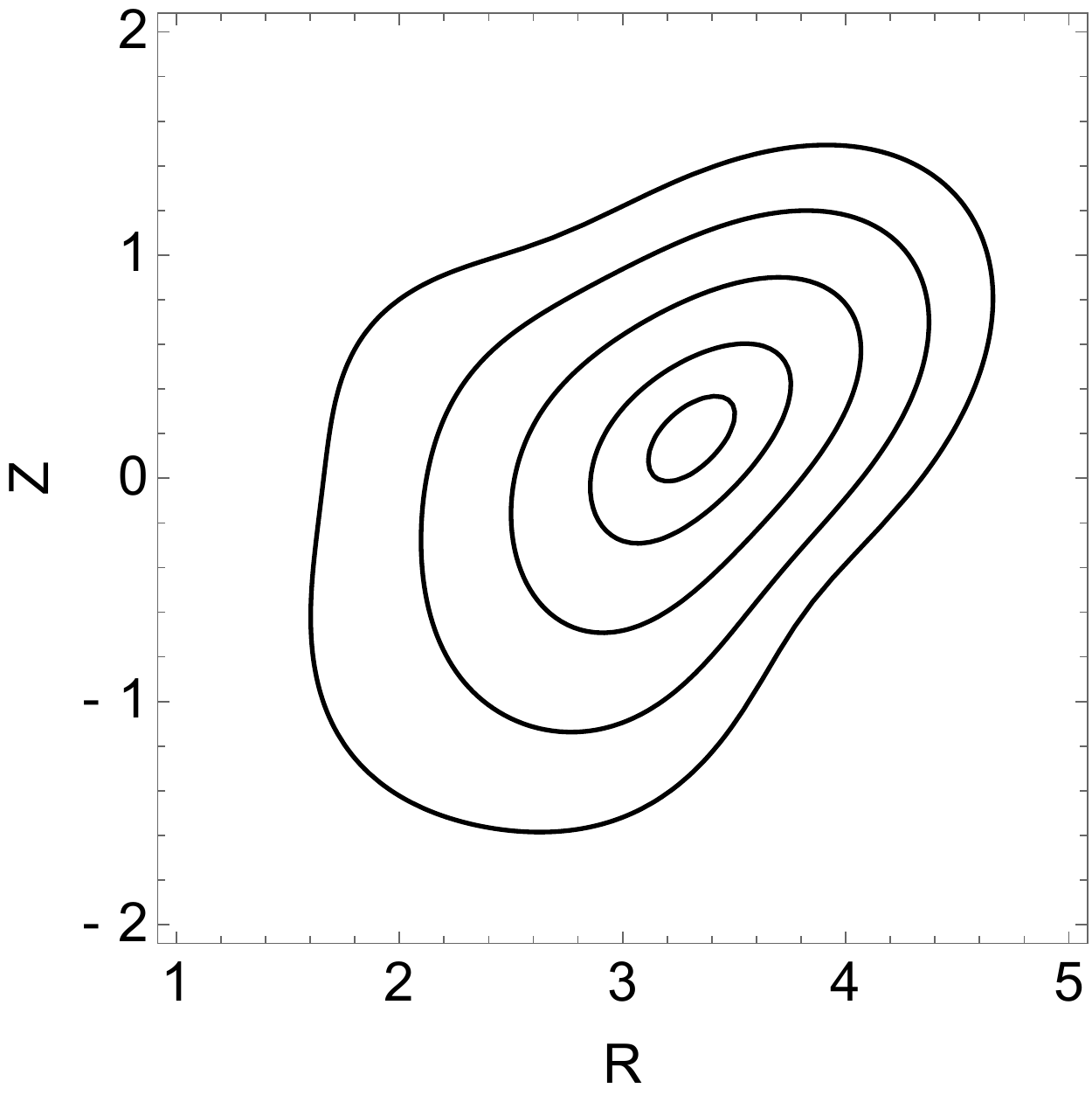}
	\end{center}
	\caption{The ``optimal'' magnetic geometry, i.e. the flux surface equilibrium that is expected to maximise intrinsic toroidal rotation generated by up-down asymmetry.}
	\label{fig:optimalGeo}
\end{figure}

The ``optimal'' flux surface geometry, as indicated by the analysis of this thesis, is shown in figure \ref{fig:optimalGeo}. It appears optimal because it generated the highest intrinsic rotation of any configuration simulated (see figure \ref{fig:nonMirrorExact}(a)) and has among the lowest energy flux (see figure \ref{fig:nonMirrorExact}(b)). These values could almost certainly be improved by going to stronger shaping, but we constrained our search to more practical geometries by using the magnitude of shaping from ITER. The boundary flux surface of the optimal geometry is specified by \refEq{eq:fluxSurfofInterestShapeExact}, \refEq{eq:geoMajorRadiusExact}, and \refEq{eq:geoAxialExact} with $a = a_{\psi 0} = 1$, $R_{0 b} = R_{c 0} = 3$, $\kappa = \Delta_{2} = 1.7$, $\delta \approx \Delta_{3} = 1.3$, $\theta_{\kappa} = \theta_{t 2} = \pi / 4$, and $\theta_{\delta} = \theta_{t 3} = \pi / 6$. The MHD equilibrium was calculated assuming a constant current profile according to \refEq{eq:gradShafLowestOrderSolsTiltConst} and \refEq{eq:psiNextOrderSolConst}. This is because a constant profile reasonably approximates experiments, although figure \ref{fig:exShaping} shows that a hollow current profile would increase the shaping of the inner flux surfaces.

The precise tilt angles of the optimal geometry indicate the mechanisms that transport momentum in flux surfaces with elongation and triangularity. The tilting symmetry presented in section \ref{sec:analyticArg} demonstrates that non-mirror symmetric geometries with an up-down symmetric envelope must have exponentially small momentum transport in the Fourier mode number of the fast shaping. This contrasts with the results of sections \ref{subsec:practicalNonMirrorSymShaping} and \ref{subsec:genNonMirrorSymShaping}, which show that a polynomial scaling holds when the envelope is up-down asymmetric. However, the optimal geometry turns out have an exactly up-down symmetric envelope. This suggests that the beating of elongation and triangularity to produce an $m=1$ envelope, which can then interact with toroidicity, is not particularly important. Additionally, while the optimal geometry is non-mirror symmetric, figure \ref{fig:nonMirrorExact}(a) shows several mirror symmetric configurations that generate fairly comparable momentum transport. This indicates that the momentum transported due to the direct interaction of elongation with triangularity (as would be dominant in a screw pinch) is fairly small in tokamaks. Hence, we believe that the momentum transport is dominated by the interactions of elongation and triangularity with toroidicity, but not through creating an $m=1$ envelope.

The nonlinear gyrokinetic simulations shown in figure \ref{fig:nonMirrorExact}(a) indicate that, without the effect of plasma pressure, the optimal geometry will generate momentum transport with $\left( v_{th i} / R_{c 0} \right) \left\langle \Pi_{\zeta i} \right\rangle_{t} / \left\langle Q_{i} \right\rangle_{t} \approx 0.06$. We note that the simulations from figure \ref{fig:nonMirrorExact}(a) were performed at a Cyclone base case value of $\rho_{0} = 0.54$, rather than at the edge. However, figure \ref{fig:momHeatFluxRatioWithMinorRadius} suggests that a mid-radius simulation without the effects of the Shafranov shift or $\beta'$ is not a bad estimate for the value at the edge including the Shafranov shift and $\beta'$. Therefore, even though including a constant ITER-like $d p / d a_{\psi}$ can enhance momentum transport through the Shafranov shift (see section \ref{subsec:results}) or reduce it through the way it modifies the magnetic geometry within the flux surface (see section \ref{sec:betaPrime}), we will assume these two effects roughly cancel with the effect of changing the value of $\rho_{0}$.

Using \refEq{eq:rotationGradEst}, we can estimate an on-axis Alfv\'{e}n Mach number of $1.5 \%$, assuming $\left( v_{th i} / R_{c 0} \right) \left\langle \Pi_{\zeta i} \right\rangle_{t} / \left\langle Q_{i} \right\rangle_{t} \approx 0.06$ is constant across the entire minor radius. This value of $1.5 \%$ is similar to figure \ref{fig:MachNumberProfile}(c), which uses a more sophisticated treatment of elongated configurations with a Shafranov shift. We note that both of these estimates ignores the momentum pinch, which may enhance the level of rotation by a much as a factor of three \cite{PeetersMomPinch2007}. Conversely, we know that, unlike elongation (see figure \ref{fig:exShaping}(b)), triangularity does not penetrate effectively to the core (see figure \ref{fig:exShaping}(e)). This causes the inner flux surfaces of figure \ref{fig:optimalGeo} to be less strongly shaped than the outer surfaces, so we might expect a reduction in the momentum transport. However, looking more closely at \refEq{eq:rotationGradEst}, we see that the on-axis rotation is the minor radial integral of $\left( v_{th i} / R_{c 0} \right) \left\langle \Pi_{\zeta i} \right\rangle_{t} / \left\langle Q_{i} \right\rangle_{t}$, weighted by the pressure gradient. Hence, a strategy to deal with the finite penetration of triangularity is to have a broad pressure profile with a steep gradient near the edge (i.e. H-mode). Not only will this make better use of the edge plasma shaping, but figure \ref{fig:MachNumberProfile} indicates it will lead to a broader rotation profile (which is better for stabilising MHD modes). \emph{Nevertheless, up-down asymmetry looks to be capable of generating rotation with an on-axis Alfv\'{e}n Mach number of $1.5 \%$ in large devices, which is approximately what is needed to stabilise certain MHD modes and allow for violation of the Troyon limit.}

%% file: App_A_MaxShaping.tex
% !TEX root = /Users/Justin/Documents/Research/Writings/2016DoctoralThesis/DoctoralThesis.tex

\chapter{Maximum achievable flux surface shaping}
\label{app:maxShaping}

\begin{quote}
	\emph{Much of this appendix appears in reference \cite{BallMomFluxScaling2016}.}
\end{quote}

If we try to create flux surfaces with extreme shaping, we will eventually introduce x-points into the plasma, opening the flux surfaces. Since open field lines cannot confine fusion plasmas, this provides a fundamental limit on the strength of plasma shaping. To quantify this we will take \refEq{eq:gradShafLowestOrderSolsTiltConst} with a single shaping mode $m$ from our analysis of the constant current profile and require that $\Nabla \psi_{N} = 0$. This gives us the condition that
\begin{align}
   N_{0, m} = \frac{\hat{j}_{0} b_{x}^{2-m}}{2 m} ,
\end{align}
where $b_{x}$ is the radial location of all $m$ of the x-points. Substituting this into \refEq{eq:deltaToFourier} and making use of \refEq{eq:deltaDef} gives 
\begin{align}
   \Delta_{x}^{-2} + \frac{2}{m} \Delta_{x}^{-m} = 1 - \frac{2}{m} , \label{eq:sepShapingCond}
\end{align}
where $\Delta_{x} \equiv b_{x} / a_{x}$ is the strongest flux surface shaping possible and $a_{x}$ is the minimum distance of the separatrix from the magnetic axis. This can be solved exactly using numerical methods or approximated analytically as
\begin{align}
   \Delta_{x} - 1 = \frac{1.2785}{m} + O \left( m^{-2} \right) , \label{eq:sepShapingSol}
\end{align}
in the limit that $m \gg 1$. The numerical constant in \refEq{eq:sepShapingSol} is the solution $x = 1.2785$ to
\begin{align}
   x - \Exp{-x} = 1 .
\end{align}
Hence, we can conclude that, given a constant current profile, $\Delta_{m} - 1 \sim m^{-1}$ is the strongest possible scaling. Any scaling stronger than this will necessarily introduce x-points into the plasma.

%% file: App_B_MagAxisLoc.tex
% !TEX root = /Users/Justin/Documents/Research/Writings/2016DoctoralThesis/DoctoralThesis.tex

\chapter{Location of the magnetic axis for a tilted elliptical boundary and constant current profile}
\label{app:exactMagAxisLoc}

\begin{quote}
	\emph{Much of this appendix appears in reference \cite{BallShafranovShift2016}.}
\end{quote}

In order to find the location of the magnetic axis for a constant toroidal current profile we will start with \refEq{eq:gradShafLowestOrderSolsTiltConst}. By requiring that $\psi_{0} \left( r_{b} \left( \theta \right), \theta \right) = \psi_{0 b}$ be constant on the tilted elliptical boundary parameterized by \refEq{eq:boundarySurf}, we find that
\begin{align}
\psi_{0 b} &= \frac{\hat{j}_{0}}{2} \frac{a^{2} \kappa_{b}^{2}}{\kappa_{b}^{2} + 1} \label{eq:boundaryFluxConst} \\
N_{0, 2} &= \frac{\hat{j}_{0}}{4} \frac{\kappa_{b}^{2} - 1}{\kappa_{b}^{2} + 1} \label{eq:fourierShapingConst}
\end{align}
as well as $\theta_{t 0 2} = \theta_{\kappa b}$ (according to \refEq{eq:tiltAngleSol}). All other lowest order Fourier coefficients are zero.

Calculating the next order Fourier coefficients from the boundary condition (i.e. requiring that $\psi_{1} \left( r_{b} \left( \theta \right), \theta \right) = \psi_{1 b}$ is constant) is algebraically intensive. We start with \refEq{eq:psiNextOrderSolConst}, the next order solution of the poloidal flux for a constant current profile. Note that while the current profile is assumed to be constant, we are allowing for a pressure gradient that is linear in $\psi$. First, we will postulate that the fifth, third, and first Fourier harmonics are the only ones required to match the boundary condition. All other next order Fourier coefficients are set to zero. Then we change poloidal angle to $\theta_{s} \equiv \theta + \theta_{\kappa b}$ in order to align the coordinate system with the minor and major axes of the elliptical boundary flux surface. Next we change from polar coordinates to Cartesian coordinates in the poloidal plane (i.e. $r = \sqrt{X^{2} + Y^{2}}$ and $\theta_{s} = \ArcTan{Y/X}$). This converts $\psi_{1} \left( r, \theta_{s} \right)$ into $\psi_{1} \left( X, Y \right)$, a fifth-order polynomial that contains products of $X$ and $Y$. For the boundary condition we use
\begin{align}
   \left( \frac{X}{a} \right)^{2} + \left( \frac{Y}{\kappa_{b} a} \right)^{2} = 1 , \label{eq:cartesianEllipse}
\end{align}
the traditional Cartesian formula for an ellipse, instead of \refEq{eq:boundarySurf}. Solving \refEq{eq:cartesianEllipse} for $Y \left( X \right)$ and substituting it into $\psi_{1} \left( X, Y \right)$ allows us to eliminate all appearances of $X^{2}$, $X^{4}$, $Y^{2}$, and $Y^{4}$. We are left with a fifth-order polynomial that only has six terms, one proportional to each of $X^{5}$, $Y^{5} \left( X \right)$, $X^{3}$, $Y^{3} \left( X \right)$, $X$, and $Y \left( X \right)$. Since we have already made use of the boundary condition, we know that the whole polynomial must be constant. Requiring that the coefficients of the six terms be zero gives
\begin{align}
C_{1, m} &= A_{C m} \Cos{m \theta_{\kappa b}} - A_{S m} \Sin{m \theta_{\kappa b}} \\
S_{1, m} &= - A_{S m} \Cos{m \theta_{\kappa b}} - A_{C m} \Sin{m \theta_{\kappa b}} ,
\end{align}
where
\begin{align}
   A_{C 5} &\equiv \left( \kappa_{b}^{2} - 1 \right) \frac{\hat{f}_{0 p} \hat{j}_{0 p}}{48 R_{0 b}} \frac{\left( \kappa_{b}^{2} - 1 \right) \hat{j}_{0} - \left( 7 \kappa_{b}^{2} + 5 \right) N_{0, 2}}{5 \kappa_{b}^{4} + 10 \kappa_{b}^{2} + 1} \Cos{\theta_{\kappa b}} \\
   A_{S 5} &\equiv - \left( - \kappa_{b}^{2} + 1 \right) \frac{\hat{f}_{0 p} \hat{j}_{0 p}}{48 R_{0 b}} \frac{\left( - \kappa_{b}^{2} + 1 \right) \hat{j}_{0} + \left( 5 \kappa_{b}^{2} + 7 \right) N_{0, 2}}{\kappa_{b}^{4} + 10 \kappa_{b}^{2} + 5} \Sin{\theta_{\kappa b}} \\
   A_{C 3} &\equiv \frac{1}{4 R_{0 b}} \frac{1}{3 \kappa_{b}^{2} + 1} \left( \left( \kappa_{b}^{2} - 1 \right) \left( \frac{\hat{j}_{0}}{4} + \hat{j}_{0 p} + N_{0, 2} \right) \right. \\
   &+ \left. \frac{1}{3} \frac{\left( - 5 \kappa_{b}^{4} + 2 \kappa_{b}^{2} + 3 \right) \hat{j}_{0} + 4 \left( 5 \kappa_{b}^{4} + 4 \kappa_{b}^{2} + 3 \right) N_{0, 2}}{5 \kappa_{b}^{4} + 10 \kappa_{b}^{2} + 1} \kappa_{b}^{2} a^{2} \hat{f}_{0 p} \hat{j}_{0 p} \right) \Cos{\theta_{\kappa b}} \nonumber \\
   A_{S 3} &\equiv \frac{1}{4 R_{0 b}} \frac{1}{\kappa_{b}^{2} + 3} \left( \left( - \kappa_{b}^{2} + 1 \right) \left( \frac{\hat{j}_{0}}{4} + \hat{j}_{0 p} - N_{0, 2} \right) \right. \\
   &+ \left. \frac{1}{3} \frac{\left( 3 \kappa_{b}^{4} + 2 \kappa_{b}^{2} - 5 \right) \hat{j}_{0} - 4 \left( 3 \kappa_{b}^{4} + 4 \kappa_{b}^{2} + 5 \right) N_{0, 2}}{\kappa_{b}^{4} + 10 \kappa_{b}^{2} + 5} \kappa_{b}^{2} a^{2} \hat{f}_{0 p} \hat{j}_{0 p} \right) \Sin{\theta_{\kappa b}} \nonumber \\
   A_{C 1} &\equiv - \frac{1}{4 R_{0 b}} \frac{\kappa_{b}^{2} a^{2}}{3 \kappa_{b}^{2} + 1} \left( \left( \hat{j}_{0} + 4 \hat{j}_{0 p} + 4 N_{0, 2} \right) \right. \\
   &- \left. \frac{4}{3} \frac{2 \left( \kappa_{b}^{2} + 1 \right) \hat{j}_{0} + \left( \kappa_{b}^{2} + 7 \right) N_{0, 2}}{5 \kappa_{b}^{4} + 10 \kappa_{b}^{2} + 1} \kappa_{b}^{2} a^{2} \hat{f}_{0 p} \hat{j}_{0 p} \right) \Cos{\theta_{\kappa b}} \nonumber \\
   A_{S 1} &\equiv \frac{1}{4 R_{0 b}} \frac{\kappa_{b}^{2} a^{2}}{\kappa_{b}^{2} + 3} \left( \left( \hat{j}_{0} + 4 \hat{j}_{0 p} - 4 N_{0, 2} \right) \right. \\
   &- \left. \frac{4}{3} \frac{2 \left( \kappa_{b}^{2} + 1 \right) \hat{j}_{0} - \left( 7 \kappa_{b}^{2} + 1 \right) N_{0, 2}}{\kappa_{b}^{4} + 10 \kappa_{b}^{2} + 5} \kappa_{b}^{2} a^{2} \hat{f}_{0 p} \hat{j}_{0 p} \right) \Sin{\theta_{\kappa b}} \nonumber
\end{align}
and $A_{C m} = A_{S m} = 0$ for all other $m$. These coefficients reduce to those found in section 2.1.2 of reference \cite{BallMastersThesis2013} when $\hat{f}_{0 p} = 0$ as expected.

The above equations give the full analytic solution to the Grad-Shafranov equation to lowest and next order in aspect ratio for a constant toroidal current profile, linear (in $psi$) pressure gradient, and tilted elliptical boundary. We want to substitute these solutions into \refEq{eq:magAxisCondition} and solve for $r_{\text{axis}}$ and $\theta_{\text{axis}}$, the minor radial and poloidal locations of the magnetic axis. The simplest approach is to first expand \refEq{eq:magAxisCondition} to lowest order in $\epsilon \ll 1$ and change to Cartesian coordinates to find
\begin{align}
   \left. \Nabla \psi_{0} \left( R, Z \right) \right|_{R = R_{\text{axis} 0}, Z = Z_{\text{axis} 0}} + \left. \Nabla \psi_{1} \left( R, Z \right) \right|_{R = R_{0 b}, Z = Z_{0 b}} = 0 , \label{eq:magAxisConditionExpansion}
\end{align}
where $R_{\text{axis} 0}$ and $Z_{\text{axis} 0}$ are the lowest order solutions for the major radial and axial locations of the magnetic axis respectively. The solution to this,
\begin{align}
   R_{\text{axis} 0} - R_{0 b} &= \frac{1}{2} \frac{S_{0, 2} S_{1, 1} - \left( \frac{\hat{j}_{0}}{4} - C_{0, 2} \right) C_{1, 1}}{\left( \frac{\hat{j}_{0}}{4} \right)^{2} - C_{0, 2}^{2} - S_{0, 2}^{2}} \label{eq:magAxisLocRadialLowest} \\
   Z_{\text{axis} 0} - Z_{0 b} &= \frac{1}{2} \frac{S_{0, 2} C_{1, 1} - \left( \frac{\hat{j}_{0}}{4} + C_{0, 2} \right) S_{1, 1}}{\left( \frac{\hat{j}_{0}}{4} \right)^{2} - C_{0, 2}^{2} - S_{0, 2}^{2}} , \label{eq:magAxisLocAxialLowest}
\end{align}
is easy to find and gives the location of the magnetic axis to first order in $\epsilon \ll 1$. However, this turns out to be a fairly poor approximation and does not produce close agreement with the numerical results from ECOM. However, if we solve \refEq{eq:magAxisCondition} exactly (even though $\psi$ is not calculated exactly) we get a much better approximation that matches ECOM. The crucial step to solving \refEq{eq:magAxisCondition} exactly is to assume that the lowest order solution for the location of the magnetic axis in \refEq{eq:magAxisLocRadialLowest} and \refEq{eq:magAxisLocAxialLowest} has the exactly correct tilt angle, i.e.
\begin{align}
   \theta_{\text{axis}} &= \theta_{\text{axis} 0} = \ArcTan{\frac{Z_{\text{axis} 0} - Z_{0 b}}{R_{\text{axis} 0} - R_{0 b}}} .  \label{eq:magAxisPoloidalLocExact}
\end{align}
We can see that this is indeed true by substituting \refEq{eq:magAxisPoloidalLocExact} into \refEq{eq:magAxisCondition}, which produces a quartic equation of the form
\begin{align}
   d_{4} r_{\text{axis}}^{4} + d_{2} r_{\text{axis}}^{2} + d_{1} r_{\text{axis}} + d_{0} = 0 \label{eq:magAxisRadialLocExactCondition}
\end{align}
with coefficients
\begin{align}
   d_{4} &\equiv - \frac{5 \hat{f}_{0 p} \hat{j}_{0}}{8 R_{0 b}} \left[ \frac{\hat{j}_{0} \Cos{\theta_{\text{axis} 0}}}{6} + \frac{C_{0, 2} \Cos{\theta_{\text{axis} 0}} + S_{0, 2}  \Sin{\theta_{\text{axis} 0}}}{3} \right. \\
   &+ \left. \frac{C_{0, 2} \Cos{3 \theta_{\text{axis} 0}} + S_{0, 2}  \Sin{\theta_{\text{axis} 0}}}{2} \right] + 5 \left( C_{1, 5} \Cos{5 \theta_{\text{axis} 0}} + S_{1, 5}  \Sin{5 \theta_{\text{axis} 0}} \right) \nonumber \\
   d_{2} &\equiv \frac{3}{4 R_{0 b}} \left[ \left( \frac{\hat{j}_{0} + 4 \hat{j}_{0 p}}{4} + C_{0, 2} \right) \Cos{\theta_{\text{axis} 0}} + S_{0, 2}  \Sin{\theta_{\text{axis} 0}} \right] \\
   &+ 3 \left( C_{1, 3} \Cos{3 \theta_{\text{axis} 0}} + S_{1, 3}  \Sin{3 \theta_{\text{axis} 0}} \right) \nonumber \\
   d_{1} &\equiv 2 \left( \frac{\hat{j}_{0}}{4} + C_{0, 2} \Cos{2 \theta_{\text{axis} 0}} + S_{0, 2}  \Sin{2 \theta_{\text{axis} 0}} \right) \\
   d_{0} &\equiv C_{11} \Cos{\theta_{\text{axis} 0}} + S_{1, 1}  \Sin{\theta_{\text{axis} 0}} .
\end{align}
The exact location of the magnetic axis is given by solution of this quartic and \refEq{eq:magAxisPoloidalLocExact}. Quartics have a very complicated analytic solution, so in practice it is simpler to solve computationally. However, for the special case of $\hat{f}_{0 p} = 0$ we see that $d_{4} = 0$ and the quartic reduces to a quadratic solved by
\begin{align}
   r_{\text{axis}} &= \frac{- d_{1} + \sqrt{d_{1}^{2} - 4 d_{2} d_{0}}}{2 d_{2}} . \label{eq:magAxisRadialLocExact}
\end{align}

%% file: App_C_TurbulentFluxes.tex
% !TEX root = /Users/Justin/Documents/Research/Writings/2016DoctoralThesis/DoctoralThesis.tex

\chapter{Electromagnetic turbulent fluxes and energy exchange}
\label{app:fluxes}

\begin{quote}
	\emph{Much of this appendix appears in reference \cite{BallMirrorSymArg2016}.}
\end{quote}

From references \cite{ParraUpDownSym2011, SugamaHighFlowGyro1998, AbelGyrokineticsDeriv2012} among others we see that the electromagnetic fluxes of particles, momentum, and energy as well as the energy exchange between species are the only turbulent quantities needed to evolve the transport equations for particles, momentum, and energy. Furthermore, it is convenient to calculate these fluxes in a frame rotating with the bulk plasma, using the velocity variable $\vec{w} \equiv \vec{v} - R \Omega_{\zeta} \hat{e}_{\zeta}$. To do so we will follow the procedure outlined in section II.D and appendix E of reference \cite{ParraUpDownSym2011}.

The complete electromagnetic turbulent flux of particles in a tokamak can be defined as
\begin{align}
   \Gamma_{s} &\equiv \left\langle \gamma_{s} \right\rangle_{\psi} \equiv - \left\langle R \left\langle \left\langle \int d^{3} w \underline{h}_{s} \hat{e}_{\zeta} \cdot \left( \delta \vec{E} + \vec{w} \times \delta \vec{B} \right) \right\rangle_{\Delta \psi} \right\rangle_{\Delta t} \right\rangle_{\psi} ,
\end{align}
where $\gamma_{s}$ is the poloidally-dependent particle flux (defined by \refEq{eq:polDistPartFlux} in the electrostatic limit), $\delta \vec{E} = - \vec{\nabla}_{\perp} \underline{\phi}$ is the turbulent electric field, $\delta \vec{B} = \underline{B}_{||} \hat{b} + \vec{\nabla} \underline{A}_{||} \times \hat{b}$ is the turbulent magnetic field, and the $\underline{\left( \ldots \right)}$ indicates that the quantity has not been Fourier analysed. After considerable manipulation we find the flux of particles to be
\begin{align}
   \Gamma_{s} &= \frac{4 \pi^{2} i}{m_{s} V'} \left\langle \sum_{k_{\psi}, k_{\alpha}} k_{\alpha} \oint d \theta J B \int dw_{||} d \mu ~ h_{s} \left( - k_{\psi}, - k_{\alpha} \right) \right. \nonumber \\
   & \times \Bigg[ \phi \left( k_{\psi}, k_{\alpha} \right) J_{0} \left( k_{\perp} \rho_{s} \right) \\
   &- A_{||} \left( k_{\psi}, k_{\alpha} \right) w_{||} J_{0} \left( k_{\perp} \rho_{s} \right) \nonumber \\
   &+ \left. B_{||} \left( k_{\psi}, k_{\alpha} \right) \frac{1}{\Omega_{s}} \frac{\mu B}{m_{s}} \frac{2 J_{1} \left( k_{\perp} \rho_{s} \right)}{k_{\perp} \rho_{s}} \Bigg] \right\rangle_{\Delta t} . \nonumber
\end{align}

The complete electromagnetic turbulent flux of toroidal angular momentum in a tokamak can be defined as
\begin{align}
   \Pi_{\zeta} &\equiv \sum_{s} \Pi_{\zeta s} + \Pi_{\zeta B} , \label{eq:totalMomFlux}
\end{align}
where
\begin{align}
   \Pi_{\zeta s} &\equiv \left\langle \pi_{\zeta s} \right\rangle_{\psi} \\
   &\equiv - \left\langle R \left\langle \left\langle \int d^{3} w \underline{h}_{s} m_{s} R \left( \vec{w} \cdot \hat{e}_{\zeta} + R \Omega_{\zeta} \right) \hat{e}_{\zeta} \cdot \left( \delta \vec{E} + \vec{w} \times \delta \vec{B} \right) \right\rangle_{\Delta \psi} \right\rangle_{\Delta t} \right\rangle_{\psi}
\end{align}
is the contribution from particles,
\begin{align}
   \Pi_{\zeta B} &\equiv - \left\langle R \left\langle \left\langle \hat{e}_{\zeta} \cdot \tensor{\sigma} \cdot \vec{\nabla} \psi \right\rangle_{\Delta \psi} \right\rangle_{\Delta t} \right\rangle_{\psi}
\end{align}
is the momentum transported by the electromagnetic fields, $\pi_{s}$ is the poloidally-dependent angular momentum flux (defined by \refEq{eq:polDistMomFlux} in the electrostatic limit),
\begin{align}
   \tensor{\sigma} \equiv \frac{1}{\mu_{0}} \vec{B} \vec{B} - \frac{1}{2 \mu_{0}} B^{2} \tensor{I}
\end{align}
is the Maxwell stress tensor, and $\tensor{I}$ is the identity matrix. After considerable manipulation we find the angular momentum transported by particles to be
\begin{align}
   \Pi_{\zeta s} &= \frac{4 \pi^{2} i}{V'} \left\langle \sum_{k_{\psi}, k_{\alpha}} k_{\alpha} \oint d \theta J B \int dw_{||} d \mu ~ h_{s} \left( - k_{\psi}, - k_{\alpha} \right) \right. \nonumber \\
   & \times \left\{ \phi \left( k_{\psi}, k_{\alpha} \right) \left[ \left( \frac{I}{B} w_{||} + R^{2} \Omega_{\zeta} \right) J_{0} \left( k_{\perp} \rho_{s} \right) + \frac{i}{\Omega_{s}} \frac{k^{\psi}}{B} \frac{\mu B}{m_{s}} \frac{2 J_{1} \left( k_{\perp} \rho_{s} \right)}{k_{\perp} \rho_{s}} \right] \right. \label{eq:speciesMomFlux} \\
   &- A_{||} \left( k_{\psi}, k_{\alpha} \right) \left[ \left( \frac{I}{B} w_{||} + R^{2} \Omega_{\zeta} \right) w_{||} J_{0} \left( k_{\perp} \rho_{s} \right) + \left( i \frac{w_{||}}{\Omega_{s}} \frac{k^{\psi}}{B} + \frac{I}{B} \right) \frac{\mu B}{m_{s}} \frac{2 J_{1} \left( k_{\perp} \rho_{s} \right)}{k_{\perp} \rho_{s}} \right] \nonumber \\
   &+ \left. \left. B_{||} \left( k_{\psi}, k_{\alpha} \right) \frac{1}{\Omega_{s}} \left[ \left( \frac{I}{B} w_{||} + R^{2} \Omega_{\zeta} \right) \frac{\mu B}{m_{s}} \frac{2 J_{1} \left( k_{\perp} \rho_{s} \right)}{k_{\perp} \rho_{s}} + \frac{i}{2 \Omega_{s}} \frac{k^{\psi}}{B} \frac{\mu^{2} B^{2}}{m_{s}^{2}} G \left( k_{\perp} \rho_{s} \right) \right] \right\} \right\rangle_{\Delta t} \nonumber
\end{align}
and the transport by the fluctuating fields to be
\begin{align}
   \Pi_{\zeta B} &= \frac{2 \pi i}{\mu_{0} V'} \left\langle \sum_{k_{\psi}, k_{\alpha}} k_{\alpha} \oint d \theta J A_{||} \left( k_{\psi}, k_{\alpha} \right) \right. \label{eq:elecMagMomFlux} \\
   &\times \left. \Big[ - i k^{\psi} A_{||} \left( - k_{\psi}, - k_{\alpha} \right) + I B_{||} \left( - k_{\psi}, - k_{\alpha} \right) \Big] \right\rangle_{\Delta t} , \nonumber
\end{align}
where $k^{\psi} \equiv \vec{k}_{\perp} \cdot \vec{\nabla} \psi = k_{\psi} \left| \vec{\nabla} \psi \right|^{2} + k_{\alpha} \vec{\nabla} \psi \cdot \vec{\nabla} \alpha$ and $G \left( x \right) \equiv 8 \left( 2 J_{1} \left( x \right) - x J_{0} \left( x \right) \right) / x^{3}$. Note that, when summing over all species, \refEq{eq:perpCur} can be used to show that the $B_{||}$ term in \refEq{eq:elecMagMomFlux} cancels the fourth $A_{||}$ term in \refEq{eq:speciesMomFlux}.

The complete electromagnetic turbulent flux of energy carried by particles can be defined as
\begin{align}
   Q_{s} &\equiv \left\langle q_{s} \right\rangle_{\psi} \equiv - \left\langle R \left\langle \left\langle \int d^{3} w \underline{h}_{s} \left( \frac{m_{s}}{2} w^{2} + Z_{s} e \Phi_{0} - \frac{m_{s}}{2} R^{2} \Omega_{\zeta}^{2} \right) \right. \right. \right. \\
   & \hat{e}_{\zeta} \cdot \left. \left. \left. \left( \delta \vec{E} + \vec{w} \times \delta \vec{B} \right) \right\rangle_{\Delta \psi} \right\rangle_{\Delta t} \right\rangle_{\psi} , \nonumber
\end{align}
where $q_{s}$ is the poloidally-dependent energy flux (defined by \refEq{eq:polDistHeatFlux} in the electrostatic limit). After considerable manipulation we find the energy transported by particles to be
\begin{align}
   Q_{s} &= \frac{4 \pi^{2} i}{V'} \left\langle \sum_{k_{\psi}, k_{\alpha}} k_{\alpha} \oint d \theta J B \int dw_{||} d \mu ~ h_{s} \left( - k_{\psi}, - k_{\alpha} \right) \left( \frac{w^{2}}{2} + \frac{Z_{s} e \Phi_{0}}{m_{s}} - \frac{m_{s}}{2} R^{2} \Omega_{\zeta}^{2} \right) \right. \nonumber \\
   & \times \left[ \phi \left( k_{\psi}, k_{\alpha} \right) \left( J_{0} \left( k_{\perp} \rho_{s} \right) \right) \right. \label{eq:speciesHeatFlux} \\
   &- A_{||} \left( k_{\psi}, k_{\alpha} \right) \left( w_{||} J_{0} \left( k_{\perp} \rho_{s} \right) \right) \nonumber \\
   &+ \left. \left. B_{||} \left( k_{\psi}, k_{\alpha} \right) \frac{1}{\Omega_{s}} \left( \frac{\mu B}{m_{s}} \frac{2 J_{1} \left( k_{\perp} \rho_{s} \right)}{k_{\perp} \rho_{s}} \right) \right] \right\rangle_{\Delta t} . \nonumber
\end{align}

The complete electromagnetic turbulent energy exchange between species can be written as
\begin{align}
   P_{Q s} &\equiv \left\langle p_{Q s} \right\rangle_{\psi} \equiv \left\langle \left\langle \left\langle \int d^{3} w Z_{s} e \underline{h}_{s} \frac{\partial \underline{\chi}}{\partial t} \right\rangle_{\Delta \psi} \right\rangle_{\Delta t} \right\rangle_{\psi} ,
\end{align}
where $\underline{\chi} \equiv \underline{\phi} - \vec{w} \cdot \vec{\underline{A}}$ is the generalised potential and $p_{Q s}$ is the poloidally-dependent turbulent energy exchange (defined by \refEq{eq:polDistHeating} in the electrostatic limit). After considerable manipulation we find the energy exchange to be
\begin{align}
   P_{Q s} &= \frac{4 \pi^{2}}{V'} \left\langle \sum_{k_{\psi}, k_{\alpha}} \oint d \theta J \Omega_{s} \int dw_{||} d \mu ~ h_{s} \left( - k_{\psi}, - k_{\alpha} \right) \right. \nonumber \\
   & \times \Bigg[ \frac{\partial}{\partial t} \left( \phi \left( k_{\psi}, k_{\alpha} \right) \right) J_{0} \left( k_{\perp} \rho_{s} \right) \\
   &- \frac{\partial}{\partial t} \left( A_{||} \left( k_{\psi}, k_{\alpha} \right) \right) w_{||} J_{0} \left( k_{\perp} \rho_{s} \right) \nonumber \\
   &+ \left. \frac{\partial}{\partial t} \left( B_{||} \left( k_{\psi}, k_{\alpha} \right) \right) \frac{1}{\Omega_{s}} \frac{\mu B}{m_{s}} \frac{2 J_{1} \left( k_{\perp} \rho_{s} \right)}{k_{\perp} \rho_{s}} \Bigg] \right\rangle_{\Delta t} . \nonumber
\end{align}

%% file: App_D_GeneralCoeff.tex
% !TEX root = /Users/Justin/Documents/Research/Writings/2016DoctoralThesis/DoctoralThesis.tex

\chapter{Calculation of the geometric coefficients within Miller local equilibrium}
\label{app:genGeoCoeff}

\begin{quote}
	\emph{Much of this appendix appears in reference \cite{BallMomFluxScaling2016}.}
\end{quote}

In this appendix we will calculate the ten geometric coefficients (i.e. $\hat{b} \cdot \Nabla \theta$, $B$, $v_{d s \psi}$, $v_{d s \alpha}$, $a_{s ||}$, $\left| \Nabla \psi \right|^{2}$, $\Nabla \psi \cdot \Nabla \alpha$, $\left| \Nabla \alpha \right|^{2}$, $R$, and $\left. \partial R / \partial \psi \right|_{\theta}$) that appear in the electromagnetic gyrokinetic equations with rotation. Here we will use the normal cylindrical poloidal angle $\theta$, but the expressions are general to an arbitrary poloidal angle. In order to calculate these coefficients for the local equilibrium specification (given in chapter \ref{ch:MHD_LocalEquil}) we must work within the local Miller geometry model. This means that we begin knowing the shape of the flux surface of interest (i.e. $R \left( \theta \right)$ and $Z \left( \theta \right)$), how it changes with minor radius (i.e. $\left. \partial R / \partial r_{\psi} \right|_{\theta}$ and $\left. \partial Z / \partial r_{\psi} \right|_{\theta}$), and several flux functions (e.g. the toroidal magnetic field flux function, the safety factor, the magnetic shear, and the pseudo-density (see \refEq{eq:pseudoDensity}) evaluated on the flux surface of interest. With only this information we can calculate the toroidal and poloidal magnetic fields using
\begin{align}
   \vec{B}_{\zeta} &= \frac{I \left( \psi \right)}{R} \hat{e}_{\zeta} \label{eq:millerTorField} \\
   \vec{B}_{p} &= \Nabla \zeta \times \Nabla r_{\psi} \frac{d \psi}{d r_{\psi}} \label{eq:millerPolField} ,
\end{align}
where $d \psi / d r_{\psi}$ can be calculated to be
\begin{align}
   \frac{d \psi}{d r_{\psi}} = \frac{I \left( \psi \right)}{2 \pi q} \left. \oint_{0}^{2 \pi} \right|_{\psi} d \theta \left( R^{2} \Nabla r_{\psi} \cdot \left( \Nabla \theta \times \Nabla \zeta \right) \right)^{-1} \label{eq:dpsidrpsi}
\end{align}
from the definition of the safety factor (i.e. \refEq{eq:safetyFactorDef}) and the gradients of $r_{\psi}$, $\theta$, and $\zeta$ can be found from \refEq{eq:gradIdentity}. Using only this information we can calculate $\hat{b} \cdot \Nabla \theta$, $B$, $v_{d s \psi}$ (defined by \refEq{eq:radialGCdriftvelocity}), $a_{s ||}$ (defined by \refEq{eq:parallelAcceleration}), $\left| \Nabla \psi \right|^{2}$, $R$, and $\left. \partial R / \partial \psi \right|_{\theta}$.

However calculating $v_{d s \alpha}$ (defined by \refEq{eq:alphaGCdriftvelocity}), $\Nabla \psi \cdot \Nabla \alpha$, and $\left| \Nabla \alpha \right|^{2}$ requires considerably more work as we must know $\Nabla \alpha$, $\left. \partial B_{\zeta} / \partial \psi \right|_{\theta}$, and $\left. \partial B_{p} / \partial \psi \right|_{\theta}$. Starting with $\left. \partial B_{p} / \partial \psi \right|_{\theta}$, we see from \refEq{eq:millerPolField} that it will depend on second order radial derivatives, which are not inputs to the Miller local equilibrium. The Miller model deals with this by calculating them through the Grad-Shafranov equation (i.e. \refEq{eq:gradShafEq}). We can rearrange \refEq{eq:gradShafEq} to get
\begin{align}
   \frac{R^{2}}{J} \frac{\partial}{\partial \psi} \left( J B_{p}^{2} \right) + \frac{R^{2}}{J} \frac{\partial}{\partial \theta} \left( \frac{J}{R^{2}} \Nabla \psi \cdot \Nabla \theta \right) = - \mu_{0} R^{2} \left. \frac{\partial p}{\partial \psi} \right|_{R} - I \frac{d I}{d \psi} ,
\end{align}
where the Jacobian is given by
\begin{align}
   J &\equiv \left| \vec{\nabla} \psi \cdot \left( \vec{\nabla} \theta \times \vec{\nabla} \zeta \right) \right|^{-1} = \left( \vec{B}_{p} \cdot \vec{\nabla} \theta \right)^{-1} = \frac{1}{B_{p}} \left. \frac{\partial l_{p}}{\partial \theta} \right|_{\psi} \label{eq:psiJacobian}
\end{align}
and $l_{p}$ is the arc length defined such that \refEq{eq:dLpdtheta} holds. We note that when $\Omega_{\zeta} = 0$ the quantity $\left. \partial p / \partial \psi \right|_{R} = d p / d \psi$ is an input to the calculation, otherwise it is determined by \refEq{eq:pseudoDensity}, \refEq{eq:backgroundQuasi}, and \refEq{eq:pressureGradDeriv}. Simplifying further and using \refEq{eq:gradIdentity} we finally find that
\begin{align}
   \frac{\partial B_{p}}{\partial \psi} &= - \frac{\mu_{0}}{B_{p}} \left. \frac{\partial p}{\partial \psi} \right|_{R} - \frac{I}{R^{2} B_{p}} \frac{d I}{d \psi} - B_{p} \left( \dlpdtheta \right)^{-1} \frac{\partial}{\partial \psi} \left( \dlpdtheta \right) \label{eq:dBpdpsi} \\
   &+ \left( \dlpdtheta \right)^{-1} \frac{\partial}{\partial \theta} \left( B_{p} \left( \dlpdtheta \right)^{-1} \frac{\partial \vec{r}}{\partial \psi} \cdot \frac{\partial \vec{r}}{\partial \theta} \right) . \nonumber
\end{align}
At this point we have yet to determine $d I / d \psi$, but will do so below.

Next we directly differentiate \refEq{eq:alphaDef} to find
\begin{align}
   \Nabla \alpha &= \left( - \left. \int_{\theta_{\alpha}}^{\theta} \right|_{\psi} d \theta' \frac{\partial A_{\alpha}}{\partial \psi} + A_{\alpha} \left( \psi, \theta_{\alpha} \right) \frac{d \theta_{\alpha}}{d \psi} - \frac{d \Omega_{\zeta}}{d \psi} t \right) \Nabla \psi - A_{\alpha} \left( \psi, \theta \right) \Nabla \theta + \Nabla \zeta , \label{eq:gradAlphaInitial}
\end{align}
where
\begin{align}
A_{\alpha} \left( \psi, \theta \right) \equiv \frac{I \left( \psi \right)}{R^{2} \vec{B} \cdot \Nabla \theta} \label{eq:IalphaDef}
\end{align}
is the integrand in the definition of $\alpha$. All quantities in \refEq{eq:gradAlphaInitial} are known except for the radial derivative of $A_{\alpha}$. We can calculate it by using the product rule on \refEq{eq:IalphaDef} to find
\begin{align}
   \frac{\partial A_{\alpha}}{\partial \psi} &= A_{\alpha} \left[ \left( 1 + \frac{I^{2}}{R^{2} B_{p}^{2}} \right) \frac{1}{I} \frac{d I}{d \psi} + \frac{\mu_{0}}{B_{p}^{2}} \left. \frac{\partial p}{\partial \psi} \right|_{R} - \frac{2}{R} \frac{\partial R}{\partial \psi} \right. \label{eq:dIntegranddpsi} \\
   &- \left. \frac{1}{B_{p}} \left( \dlpdtheta \right)^{-1} \frac{\partial}{\partial \theta} \left( B_{p} \left( \dlpdtheta \right)^{-1} \frac{\partial \vec{r}}{\partial \psi} \cdot \frac{\partial \vec{r}}{\partial \theta} \right) + 2 \left( \dlpdtheta \right)^{-1} \frac{\partial}{\partial \psi} \left( \dlpdtheta \right) \right] , \nonumber
\end{align}
where we have made use of \refEq{eq:psiJacobian} and \refEq{eq:dBpdpsi}. While this form is acceptable for the purposes of this thesis, we will rearrange it into a form that is more physically illuminating. To do so we will write
\begin{align}
   \frac{\partial \vec{r}}{\partial \psi} = \frac{1}{R B_{p}} \left( \dlpdtheta \right)^{-1} \frac{\partial \vec{r}}{\partial \theta} \times \hat{e}_{\zeta} + \left( \dlpdtheta \right)^{-2} \left( \frac{\partial \vec{r}}{\partial \psi} \cdot \frac{\partial \vec{r}}{\partial \theta} \right) \frac{\partial \vec{r}}{\partial \theta} 
\end{align}
using only \refEq{eq:dLpdtheta}, \refEq{eq:psiJacobian}, and vector identities such as $\partial \vec{r} / \partial \psi \cdot \left( \partial \vec{r} / \partial \theta \times \partial \vec{r} / \partial \zeta \right) = \left( \Nabla \psi \cdot \left( \Nabla \theta \times \Nabla \zeta \right) \right)^{-1}$. This allows us to see that
\begin{align}
   - \frac{2}{R} \frac{\partial R}{\partial \psi} &= - \frac{2}{R} \frac{\partial \vec{r}}{\partial \psi} \cdot \Nabla R \\
   &= - \frac{2}{R^{2} B_{p}} \left( \dlpdtheta \right)^{-1} \frac{\partial Z}{\partial \theta} + R^{2} \frac{\partial}{\partial \theta} \left( \frac{1}{R^{2}} \right) \left( \dlpdtheta \right)^{-2} \frac{\partial \vec{r}}{\partial \psi} \cdot \frac{\partial \vec{r}}{\partial \theta} . \label{eq:dRdpsi}
\end{align}
Combining this result with the second-to-last term in \refEq{eq:dIntegranddpsi} and applying the product rule several times, we find
\begin{align}
   - \frac{2}{R} \frac{\partial R}{\partial \psi} &- \frac{1}{B_{p}} \left( \dlpdtheta \right)^{-1} \frac{\partial}{\partial \theta} \left( B_{p} \left( \dlpdtheta \right)^{-1} \frac{\partial \vec{r}}{\partial \psi} \cdot \frac{\partial \vec{r}}{\partial \theta} \right) \label{eq:dRdpsiRearrange} \\
   &= - \frac{2}{R^{2} B_{p}} \left( \dlpdtheta \right)^{-1} \frac{\partial Z}{\partial \theta} + R^{2} B_{p} \left( \dlpdtheta \right)^{-1} \frac{\partial}{\partial \theta} \left( \frac{1}{R^{2} B_{p}} \left( \dlpdtheta \right)^{-1} \frac{\partial \vec{r}}{\partial \psi} \cdot \frac{\partial \vec{r}}{\partial \theta} \right) \nonumber \\
   &- 2 \left( \dlpdtheta \right)^{-1} \frac{\partial}{\partial \theta} \left( \left( \dlpdtheta \right)^{-1} \frac{\partial \vec{r}}{\partial \theta} \right) \cdot \frac{\partial \vec{r}}{\partial \psi} - 2 \left( \dlpdtheta \right)^{-2} \frac{\partial}{\partial \theta} \left( \frac{\partial \vec{r}}{\partial \psi} \right) \cdot \frac{\partial \vec{r}}{\partial \theta} . \nonumber
\end{align}
Using \refEq{eq:dLpdtheta} we see that the last term of \refEq{eq:dRdpsiRearrange} exactly cancels the final term appearing in \refEq{eq:dIntegranddpsi}. This shows that we can rewrite \refEq{eq:dIntegranddpsi} as
\begin{align}
   \frac{\partial A_{\alpha}}{\partial \psi} &= A_{\alpha} \left[ \frac{1}{I} \frac{d I}{d \psi} + \frac{I}{R^{2} B_{p}^{2}} \frac{d I}{d \psi} + \frac{\mu_{0}}{B_{p}^{2}} \left. \frac{\partial p}{\partial \psi} \right|_{R} - \frac{2}{R^{2} B_{p}} \left( \dlpdtheta \right)^{-1} \frac{\partial Z}{\partial \theta} \right. \label{eq:dIntegranddpsiRearrange} \\
   &- \left. 2 \left( \dlpdtheta \right)^{-1} \frac{\partial}{\partial \theta} \left( \left( \dlpdtheta \right)^{-1} \frac{\partial \vec{r}}{\partial \theta} \right) \cdot \frac{\partial \vec{r}}{\partial \psi} \right] + \frac{\partial}{\partial \theta} \left( \frac{I}{R^{2} B_{p}} \left( \dlpdtheta \right)^{-1} \frac{\partial \vec{r}}{\partial \psi} \cdot \frac{\partial \vec{r}}{\partial \theta} \right) \nonumber .
\end{align}
Lastly, substituting this into \refEq{eq:gradAlphaInitial} and using \refEq{eq:poloidalCurv} produces
\begin{align}
   \Nabla \alpha &= \left( - \left. \int_{\theta_{\alpha}}^{\theta} \right|_{\psi} d \theta' A_{\alpha} \left( \psi, \theta' \right) \left[ \frac{1}{I} \frac{d I}{d \psi} + \frac{I}{R^{2} B_{p}^{2}} \frac{d I}{d \psi} + \frac{\mu_{0}}{B_{p}^{2}} \left. \frac{\partial p}{\partial \psi} \right|_{R} - \frac{2}{R^{2} B_{p}} \left( \dlpdthetaPrime \right)^{-1} \frac{\partial Z}{\partial \theta'} \right. \right. \nonumber \\
   &+ \left. \left. \frac{2 \kappa_{p}}{R B_{p}} \right] + \left[ \frac{A_{\alpha} \left( \psi, \theta' \right)}{R^{2} B_{p}^{2}} \Nabla \psi \cdot \Nabla \theta' \right]_{\theta' = \theta_{\alpha}}^{\theta' = \theta} + A_{\alpha} \left( \psi, \theta_{\alpha} \right) \frac{d \theta_{\alpha}}{d \psi} - \frac{d \Omega_{\zeta}}{d \psi} t \right) \Nabla \psi  \label{eq:gradAlphaFinal} \\
   &- A_{\alpha} \left( \psi, \theta \right) \Nabla \theta + \Nabla \zeta . \nonumber
\end{align}
The first term inside the integral represents the change in the field line pitch that results from changing the toroidal field flux function on neighbouring flux surfaces. The second term in the integral accounts for the modification to the flux surface equilibrium that results from a radial gradient in the toroidal flux function. As we will explore in appendix \ref{app:geoCoeff}, the third term expresses the effect the pressure gradient has on the equilibrium. Despite appearances, the fourth term corresponds to how the toroidal magnetic field weakens as the major radial location changes. The last term in the integral accounts for the flux expansion (and weakening of the poloidal magnetic field) that occurs at regions of large poloidal curvature (see \refEq{eq:gradShafranovSimple}). The term immediately following the integral accounts for the particulars of how $\theta$ is defined, but we note this term vanishes if contours of constant $\theta$ are perpendicular to the flux surface of interest. The last term in the coefficient of $\Nabla \psi$ is a consequence of changing which field line is labelled $\alpha = 0$ from flux surface to flux surface. The final two terms of \refEq{eq:gradAlphaFinal} reflect the nonuniform spacing of the field lines in the poloidal direction and the uniform spacing in the toroidal direction respectively.

Equation \refEq{eq:gradAlphaFinal} allows us to calculate $\Nabla \psi \cdot \Nabla \alpha$, and $\left| \Nabla \alpha \right|^{2}$, but we must remember that we still lack an expression for $d I / d \psi$. This can be calculated by taking the radial gradient of the safety factor (i.e. \refEq{eq:safetyFactorDef}) in order to get the magnetic shear,
\begin{align}
   \frac{d q}{d \psi} &= \frac{1}{2 \pi} \left. \oint_{0}^{2 \pi} \right|_{\psi} d \theta \frac{\partial A_{\alpha}}{\partial \psi} .
\end{align}
This turns out to be very closely related to $\Nabla \alpha$, so we can use \refEq{eq:dIntegranddpsiRearrange} to find
\begin{align}
   \frac{d I}{d \psi} &= I \left( q + \frac{1}{2 \pi} \left. \oint_{0}^{2 \pi} \right|_{\psi} d \theta A_{\alpha} \left( \psi, \theta \right) \frac{I^{2}}{R^{2} B_{p}^{2}} \right)^{-1} \label{eq:dIdpsiFullForm} \\
   &\times \left( \frac{d q}{d \psi} - \frac{1}{2 \pi} \left. \oint_{0}^{2 \pi} \right|_{\psi} d \theta A_{\alpha} \left( \psi, \theta \right) \left[ \frac{\mu_{0}}{B_{p}^{2}} \left. \frac{\partial p}{\partial \psi} \right|_{R} - \frac{2}{R^{2} B_{p}} \left( \dlpdtheta \right)^{-1} \frac{\partial Z}{\partial \theta} + \frac{2 \kappa_{p}}{R B_{p}} \right] \right) . \nonumber
\end{align}
Lastly we can directly differentiate \refEq{eq:millerTorField} to find
\begin{align}
   \frac{\partial B_{\zeta}}{\partial \psi} &= \frac{1}{R} \frac{d I}{d \psi} - \frac{I}{R^{2}} \left( \frac{d \psi}{d r_{\psi}} \right)^{-1} \frac{\partial R}{\partial r_{\psi}} , \label{eq:dBtordpsi}
\end{align}
where we remember that $\partial R / \partial r_{\psi}$ is an input to the Miller model. This fully determines $v_{d s \alpha}$, which is defined by \refEq{eq:alphaGCdriftvelocity}.

The expressions in this section allow us to directly calculate all of the the gyrokinetic geometric coefficients within the framework of the Miller local equilibrium model.

%% file: App_E_NonmirrorCoeff.tex
% !TEX root = /Users/Justin/Documents/Research/Writings/2016DoctoralThesis/DoctoralThesis.tex

\chapter{Non-mirror symmetric geometric coefficients}
\label{app:nonMirrorGeoCoeffs}

\begin{quote}
	\emph{Much of this appendix appears in reference \cite{BallMomFluxScaling2016}.}
\end{quote}

In this section we give the full gyrokinetic geometric coefficients to lowest and next order in $m_{c} \gg 1$ for the geometry investigated in section \ref{subsec:practicalNonMirrorSymShaping}. These coefficients are accurate to lowest order in aspect ratio, given the orderings of \refEq{eq:weakShapingOrdering} and \refEq{eq:changeShapeOrdering}. In deriving these coefficients the following quantities are useful as way points:
\begin{align}
   \frac{\partial R}{\partial \theta} &= - r_{\psi 0} \left[ \Sin{\theta} - \frac{1}{2} \Cos{\theta} \Big( m \left( \Delta_{m} - 1 \right) \Sin{z_{m s}} + n \left( \Delta_{n} - 1 \right) \Sin{z_{n s}} \Big) \right. \\
   &- \left. \frac{1}{2} \Sin{\theta} \Big( \left( \Delta_{m} - 1 \right) \Cos{z_{m s}} + \left( \Delta_{n} - 1 \right) \Cos{z_{n s}} \Big) \right] + O \left( m_{c}^{-3} r_{\psi} \right) \nonumber \\
   \frac{\partial Z}{\partial \theta} &= r_{\psi 0} \left[ \Cos{\theta} + \frac{1}{2} \Sin{\theta} \Big( m \left( \Delta_{m} - 1 \right) \Sin{z_{m s}} + n \left( \Delta_{n} - 1 \right) \Sin{z_{n s}} \Big) \right. \\
   &- \left. \frac{1}{2} \Cos{\theta} \Big( \left( \Delta_{m} - 1 \right) \Cos{z_{m s}} + \left( \Delta_{n} - 1 \right) \Cos{z_{n s}} \Big) \right] + O \left( m_{c}^{-3} r_{\psi} \right) \nonumber
\end{align}
\begin{align}
   \Nabla r_{\psi} &= \left[ \Cos{\theta} + \frac{r_{\psi 0}}{2} \Cos{\theta} \left( \frac{d \Delta_{m}}{d r_{\psi}} \Cos{z_{m s}} + \frac{d \Delta_{n}}{d r_{\psi}} \Cos{z_{n s}} \right) \right. \nonumber \\
   &+ \left. \frac{1}{2} \Sin{\theta} \Big( m \left( \Delta_{m} - 1 \right) \Sin{z_{m s}} + n \left( \Delta_{n} - 1 \right) \Sin{z_{n s}} \Big) \right] \hat{e}_{R} \\
   &+ \left[ \Sin{\theta} + \frac{r_{\psi 0}}{2} \Sin{\theta} \left( \frac{d \Delta_{m}}{d r_{\psi}} \Cos{z_{m s}} + \frac{d \Delta_{n}}{d r_{\psi}} \Cos{z_{n s}} \right) \right. \nonumber \\
   &- \left. \frac{1}{2} \Cos{\theta} \Big( m \left( \Delta_{m} - 1 \right) \Sin{z_{m s}} + n \left( \Delta_{n} - 1 \right) \Sin{z_{n s}} \Big) \right] \hat{e}_{Z} + O \left( m_{c}^{-2} \right) \nonumber \\
      \Nabla \theta &= \frac{1}{r_{\psi 0}} \Big( - \Sin{\theta} \hat{e}_{R} + \Cos{\theta} \hat{e}_{Z} \Big) + O \left( \frac{m_{c}^{-2}}{r_{\psi}} \right)
\end{align}
\begin{align}
   \frac{\partial R}{\partial r_{\psi}} &= \Cos{\theta} - \frac{r_{\psi 0}}{2} \Cos{\theta} \left( \frac{d \Delta_{m}}{d r_{\psi}} \Cos{z_{m s}} + \frac{d \Delta_{n}}{d r_{\psi}} \Cos{z_{n s}} \right) + O \left( m_{c}^{-2} \right) \\
   \frac{\partial Z}{\partial r_{\psi}} &= \Sin{\theta} - \frac{r_{\psi 0}}{2} \Sin{\theta} \left( \frac{d \Delta_{m}}{d r_{\psi}} \Cos{z_{m s}} + \frac{d \Delta_{n}}{d r_{\psi}} \Cos{z_{n s}} \right) + O \left( m_{c}^{-2} \right)
\end{align}
\begin{align}
   \left( \frac{\partial A_{\alpha}}{\partial \psi} \right)_{\text{orthog}} &= B_{c 0} \left( \frac{d \psi}{d r_{\psi}} \right)^{-2} \left[ \hat{s}' + \frac{r_{\psi 0}}{2} \left( m^{2} \left( \Delta_{m} - 1 \right) \frac{d \Delta_{m}}{d r_{\psi}} + n^{2} \left( \Delta_{n} - 1 \right) \frac{d \Delta_{n}}{d r_{\psi}} \right) \right. \nonumber \\
   &- \left( m^{2} \left( \Delta_{m} - 1 \right) + \left( 3 \hat{s}' - 2 \right) \frac{r_{\psi 0}}{2} \frac{d \Delta_{m}}{d r_{\psi}} \right) \Cos{z_{m s}} \nonumber \\
   &- \left( n^{2} \left( \Delta_{n} - 1 \right) + \left( 3 \hat{s}' - 2 \right) \frac{r_{\psi 0}}{2} \frac{d \Delta_{n}}{d r_{\psi}} \right) \Cos{z_{n s}} \\
   &+ \frac{r_{\psi 0}}{2} \left( m^{2} \left( \Delta_{m} - 1 \right) \frac{d \Delta_{m}}{d r_{\psi}} \Cos{2 z_{m s}} + n^{2} \left( \Delta_{n} - 1 \right) \frac{d \Delta_{n}}{d r_{\psi}} \Cos{2 z_{n s}} \right) \nonumber \\
   &+ \frac{r_{\psi 0}}{2} \left( m^{2} \left( \Delta_{m} - 1 \right) \frac{d \Delta_{n}}{d r_{\psi}} + n^{2} \left( \Delta_{n} - 1 \right) \frac{d \Delta_{m}}{d r_{\psi}} \right) \nonumber \\
   &\times \left. \Big( \Cos{z_{m s} + z_{n s}} + \Cos{z_{m s} - z_{n s}} \Big) \right] + O \left( \frac{m_{c}^{-2}}{r_{\psi}^{2} B_{0}} \right) \nonumber
\end{align}
\begin{align}
   \left. \int_{\theta_{0}}^{\theta} \right|_{\psi} d \theta' \left( \frac{\partial A_{\alpha}}{\partial \psi} \right)_{\text{orthog}} &= B_{c 0} \left( \frac{d \psi}{d r_{\psi}} \right)^{-2} \left[ \hat{s}' \theta \right. \nonumber \\
   +& \frac{r_{\psi 0}}{2} \left( m^{2} \left( \Delta_{m} - 1 \right) \frac{d \Delta_{m}}{d r_{\psi}} + n^{2} \left( \Delta_{n} - 1 \right) \frac{d \Delta_{n}}{d r_{\psi}} \right) \theta \nonumber \\
   -& \frac{1}{m} \left( m^{2} \left( \Delta_{m} - 1 \right) + \left( 3 \hat{s}' - 2 \right) \frac{r_{\psi 0}}{2} \frac{d \Delta_{m}}{d r_{\psi}} \right) \Sin{z_{m s}} \nonumber \\
   -& \frac{1}{n} \left( n^{2} \left( \Delta_{n} - 1 \right) + \left( 3 \hat{s}' - 2 \right) \frac{r_{\psi 0}}{2} \frac{d \Delta_{n}}{d r_{\psi}} \right) \Sin{z_{n s}} \label{eq:dAalphadpsiIntegral} \\
   +& \frac{r_{\psi 0}}{4} \left( m \left( \Delta_{m} - 1 \right) \frac{d \Delta_{m}}{d r_{\psi}} \Sin{2 z_{m s}} + n \left( \Delta_{n} - 1 \right) \frac{d \Delta_{n}}{d r_{\psi}} \Sin{2 z_{n s}} \right) \nonumber \\
   +& \frac{r_{\psi 0}}{2} \left( m^{2} \left( \Delta_{m} - 1 \right) \frac{d \Delta_{n}}{d r_{\psi}} + n^{2} \left( \Delta_{n} - 1 \right) \frac{d \Delta_{m}}{d r_{\psi}} \right) \nonumber \\
   \times& \left. \left( \frac{1}{m + n} \Sin{z_{m s} + z_{n s}} + \frac{1}{m - n} \Sin{z_{m s} - z_{n s}} \right) \right] \nonumber \\
   +& O \left( \frac{m_{c}^{-2}}{r_{\psi}^{2} B_{0}} \right) . \nonumber
\end{align}
Here $z_{m s}$ and $z_{n s}$ are defined by \refEq{eq:zmsDef} and \refEq{eq:znsDef}, while the constant $\theta_{0}$ is defined such that $\left. \int_{\theta_{0}}^{\theta_{\alpha}} \right|_{\psi} d \theta' \left( \partial A_{\alpha} / \partial \psi \right)_{\text{orthog}}$ does not have a term that is independent of the poloidal angle.
%\begin{align}
%   \frac{d \theta_{\alpha}}{d \psi} &= \frac{1}{2} \left( \frac{1}{R_{c 0} B_{p}} \dlpdtheta \right)_{\theta = \theta_{\alpha}}^{-1} \left( \frac{d \psi}{d r_{\psi}} \right)^{-2} \left( m \left( \Delta_{m} - 1 \right) \Sin{m \theta_{t m}} + n \left( \Delta_{n} - 1 \right) \Sin{n \theta_{t n}} \right. \\
%   &+ \left. \frac{r_{\psi 0}}{n - m} \left( m^{2} \left( \Delta_{m} - 1 \right) \frac{d \Delta_{n}}{d r_{\psi}} + n^{2} \left( \Delta_{n} - 1 \right) \frac{d \Delta_{m}}{d r_{\psi}} \right) \Sin{m \left( \theta_{t m} - \theta_{t n} \right)} \right) + O \left( m_{c}^{-2} \right) \nonumber
%\end{align}
Continuing the calculation we find
\begin{align}
   \frac{\partial \alpha}{\partial \psi} &= - B_{c 0} \left( \frac{d \psi}{d r_{\psi}} \right)^{-2} \Bigg[ \hat{s}' \theta + \frac{r_{\psi 0}}{2} \left( m^{2} \left( \Delta_{m} - 1 \right) \frac{d \Delta_{m}}{d r_{\psi}} + n^{2} \left( \Delta_{n} - 1 \right) \frac{d \Delta_{n}}{d r_{\psi}} \right) \theta \nonumber \\
   &- \frac{1}{2} \Big( m \left( \Delta_{m} - 1 \right) \Sin{z_{m s}} + n \left( \Delta_{n} - 1 \right) \Sin{z_{n s}} \Big) \\
   &+ \frac{r_{\psi 0}}{2 \left( n - m \right)} \left( m^{2} \left( \Delta_{m} - 1 \right) \frac{d \Delta_{n}}{d r_{\psi}} + n^{2} \left( \Delta_{n} - 1 \right) \frac{d \Delta_{m}}{d r_{\psi}} \right) \nonumber \\
   &\times \Sin{\left( n - m \right) \theta + n \theta_{t n} - m \theta_{t m}} \Bigg] + O \left( \frac{m_{c}^{-2}}{r_{\psi}^{2} B_{0}} \right) \nonumber
\end{align}
\begin{align}
   \Nabla \alpha &= - B_{c 0} \left( \frac{d \psi}{d r_{\psi}} \right)^{-1} \Bigg\{ \nonumber \\
   & \Bigg[ - \Sin{\theta} + \hat{s}' \theta \Cos{\theta} + \frac{r_{\psi 0}}{2} \left( m^{2} \left( \Delta_{m} - 1 \right) \frac{d \Delta_{m}}{d r_{\psi}} + n^{2} \left( \Delta_{n} - 1 \right) \frac{d \Delta_{n}}{d r_{\psi}} \right) \theta \Cos{\theta} \nonumber \\
   &- \frac{1}{2} \Big( \Cos{\theta} - \hat{s}' \theta \Sin{\theta} \Big) \Big( m \left( \Delta_{m} - 1 \right) \Sin{z_{m s}} + n \left( \Delta_{n} - 1 \right) \Sin{z_{n s}} \Big) \nonumber \\
   &+ \frac{r_{\psi 0}}{2} \Big( \Sin{\theta} + \hat{s}' \theta \Cos{\theta} \Big) \left( \frac{d \Delta_{m}}{d r_{\psi}} \Cos{z_{m s}} + \frac{d \Delta_{n}}{d r_{\psi}} \Cos{z_{n s}} \right) \nonumber \\
   &+ \frac{r_{\psi 0}}{2 \left( n - m \right)} \left( m^{2} \left( \Delta_{m} - 1 \right) \frac{d \Delta_{n}}{d r_{\psi}} + n^{2} \left( \Delta_{n} - 1 \right) \frac{d \Delta_{m}}{d r_{\psi}} \right) \Cos{\theta} \nonumber \\
   &\times \Sin{\left( n - m \right) \theta + n \theta_{t n} - m \theta_{t m}} \Bigg] \hat{e}_{R} \\
   &+ \Bigg[ \Cos{\theta} + \hat{s}' \theta \Sin{\theta} + \frac{r_{\psi 0}}{2} \left( m^{2} \left( \Delta_{m} - 1 \right) \frac{d \Delta_{m}}{d r_{\psi}} + n^{2} \left( \Delta_{n} - 1 \right) \frac{d \Delta_{n}}{d r_{\psi}} \right) \theta \Sin{\theta} \nonumber \\
   &- \frac{1}{2} \Big( \Sin{\theta} + \hat{s}' \theta \Cos{\theta} \Big) \Big( m \left( \Delta_{m} - 1 \right) \Sin{z_{m s}} + n \left( \Delta_{n} - 1 \right) \Sin{z_{n s}} \Big) \nonumber \\
   &- \frac{r_{\psi 0}}{2} \Big( \Cos{\theta} - \hat{s}' \theta \Sin{\theta} \Big) \left( \frac{d \Delta_{m}}{d r_{\psi}} \Cos{z_{m s}} + \frac{d \Delta_{n}}{d r_{\psi}} \Cos{z_{n s}} \right) \nonumber \\
   &+ \frac{r_{\psi 0}}{2 \left( n - m \right)} \left( m^{2} \left( \Delta_{m} - 1 \right) \frac{d \Delta_{n}}{d r_{\psi}} + n^{2} \left( \Delta_{n} - 1 \right) \frac{d \Delta_{m}}{d r_{\psi}} \right) \Sin{\theta} \nonumber \\
   &\times \Sin{\left( n - m \right) \theta + n \theta_{t n} - m \theta_{t m}} \Bigg] \hat{e}_{Z} \Bigg\} + O \left( \frac{m_{c}^{-2}}{r_{\psi}} \right) . \nonumber
\end{align}

The $O \left( 1 \right)$ geometric coefficients are simply those of a circular tokamak and are given by
\begin{align}
  \left( \hat{b} \cdot \Nabla \theta \right)_{0} &= \overline{\left( \hat{b} \cdot \Nabla \theta \right)}_{0} = \frac{1}{r_{\psi 0} R_{c 0} B_{c 0}} \frac{d \psi}{d r_{\psi}} \label{eq:gradparO0} \\
  v_{d s \psi 0} &= \overline{v}_{d s \psi 0} = - \frac{1}{R_{c 0} \Omega_{s}} \frac{d \psi}{d r_{\psi}} \Sin{\theta} \label{eq:psiDriftO0} \\
  v_{d s \alpha 0} &= \overline{v}_{d s \alpha 0} = \frac{B_{c 0}}{R_{c 0} \Omega_{s}} \left( \frac{d \psi}{d r_{\psi}} \right)^{-1} \left( \Cos{\theta} + \hat{s}' \theta \Sin{\theta} \right) \label{eq:alphaDriftO0} \\
  \left| \Nabla \psi \right|^{2}_{0} &= \overline{\left| \Nabla \psi \right|}^{2}_{0} = \left( \frac{d \psi}{d r_{\psi}} \right)^{2} \\
  \left( \Nabla \psi \cdot \Nabla \alpha \right)_{0} &= \overline{\left( \Nabla \psi \cdot \Nabla \alpha \right)}_{0} = - B_{c 0} \hat{s}' \theta \\
  \left| \Nabla \alpha \right|^{2}_{0} &= \overline{\left| \Nabla \alpha \right|^{2}}_{0} = B_{c 0}^{2} \left( \frac{d \psi}{d r_{\psi}} \right)^{-2} \left( 1 + \hat{s}'^{2} \theta^{2} \right) \label{eq:gradAlphaSqO0} \\
  \left( J_{0} \left( k_{\perp} \rho_{s} \right) \right)_{0} &= \overline{\left( J_{0} \left( k_{\perp} \rho_{s} \right) \right)}_{0} = J_{0} \left( k_{\perp 0} \rho_{s} \right) , \label{eq:FLRO0}
\end{align}
where $\hat{s}'$ is defined by \refEq{eq:shiftedShatDef} and
\begin{align}
  k_{\perp 0} \rho_{s} &\equiv \sqrt{\frac{2 m_{s} \mu}{Z_{s}^{2} e^{2} B_{c 0}}} \sqrt{k_{\psi}^{2} \left| \Nabla \psi \right|^{2}_{0} + 2 k_{\psi} k_{\alpha} \left( \Nabla \psi \cdot \Nabla \alpha \right)_{0} + k_{\alpha}^{2} \left| \Nabla \alpha \right|^{2}_{0}} . \label{eq:FLR0def}
\end{align}
Note that all of the coefficients are independent of the short spatial scale coordinate, $z$.

To $O \left( m_{c}^{-1} \right)$ the geometric coefficients are
\begin{align}
  \left( \hat{b} \cdot \Nabla \theta \right)_{1} &= \frac{1}{2 R_{c 0} B_{c 0}} \frac{d \psi}{d r_{\psi}} \left( \frac{d \Delta_{m}}{d r_{\psi}} \Cos{z_{m s}} + \frac{d \Delta_{n}}{d r_{\psi}} \Cos{z_{n s}} \right) \label{eq:gradparO1} \\
  v_{d s \psi 1} &= \frac{1}{2 R_{c 0} \Omega_{s}} \frac{d \psi}{d r_{\psi}} \Bigg[ \Cos{\theta} \Big( m \left( \Delta_{m} - 1 \right) \Sin{z_{m s}} + n \left( \Delta_{n} - 1 \right) \Sin{z_{n s}} \Big) \nonumber \\
  &- r_{\psi 0} \Sin{\theta} \left( \frac{d \Delta_{m}}{d r_{\psi}} \Cos{z_{m s}} + \frac{d \Delta_{n}}{d r_{\psi}} \Cos{z_{n s}} \right) \Bigg] \label{eq:psiDriftO1}
\end{align}
\begin{align}
  v_{d s \alpha 1} &= \frac{B_{c 0}}{2 R_{c 0} \Omega_{s}} \left( \frac{d \psi}{d r_{\psi}} \right)^{-1} \nonumber \\
  \times& \Bigg[ r_{\psi 0} \left( m^{2} \left( \Delta_{m} - 1 \right) \frac{d \Delta_{m}}{d r_{\psi}} + n^{2} \left( \Delta_{n} - 1 \right) \frac{d \Delta_{n}}{d r_{\psi}} \right) \theta \Sin{\theta} \nonumber \\
  -& \Big( \Sin{\theta} + \hat{s}' \theta \Cos{\theta} \Big) \Big( m \left( \Delta_{m} - 1 \right) \Sin{z_{m s}} + n \left( \Delta_{n} - 1 \right) \Sin{z_{n s}} \Big) \nonumber \\
  -& r_{\psi 0} \Big( \Cos{\theta} - \hat{s}' \theta \Sin{\theta} \Big) \left( \frac{d \Delta_{m}}{d r_{\psi}} \Cos{z_{m s}} + \frac{d \Delta_{n}}{d r_{\psi}} \Cos{z_{n s}} \right) \label{eq:alphaDriftO1} \\
  +& \frac{r_{\psi 0}}{\left( n - m \right)} \left( m^{2} \left( \Delta_{m} - 1 \right) \frac{d \Delta_{n}}{d r_{\psi}} + n^{2} \left( \Delta_{n} - 1 \right) \frac{d \Delta_{m}}{d r_{\psi}} \right) \Sin{\theta} \nonumber \\
  \times& \Big( \Sin{\left( n - m \right) \theta} \Cos{n \theta_{t n} - m \theta_{t m}} \nonumber \\
  &+ \Cos{\left( n - m \right) \theta} \Sin{n \theta_{t n} - m \theta_{t m}} \Big) \Bigg] \nonumber \\
  \left| \Nabla \psi \right|^{2}_{1} &= r_{\psi 0} \left( \frac{d \psi}{d r_{\psi}} \right)^{2} \left( \frac{d \Delta_{m}}{d r_{\psi}} \Cos{z_{m s}} + \frac{d \Delta_{n}}{d r_{\psi}} \Cos{z_{n s}} \right) \label{eq:gradPsiSqO1} \\
  \left( \Nabla \psi \cdot \Nabla \alpha \right)_{1} &= - B_{c 0} \Bigg[ \frac{r_{\psi 0}}{2} \left( m^{2} \left( \Delta_{m} - 1 \right) \frac{d \Delta_{m}}{d r_{\psi}} + n^{2} \left( \Delta_{n} - 1 \right) \frac{d \Delta_{n}}{d r_{\psi}} \right) \theta \nonumber \\
  -& \Big( m \left( \Delta_{m} - 1 \right) \Sin{z_{m s}} + n \left( \Delta_{n} - 1 \right) \Sin{z_{n s}} \Big) \nonumber \\
  +& r_{\psi 0} \hat{s}' \theta \left( \frac{d \Delta_{m}}{d r_{\psi}} \Cos{z_{m s}} + \frac{d \Delta_{n}}{d r_{\psi}} \Cos{z_{n s}} \right) \label{eq:gradPsiDotGradAlphaO1} \\
  +& \frac{r_{\psi 0}}{2 \left( n - m \right)} \left( m^{2} \left( \Delta_{m} - 1 \right) \frac{d \Delta_{n}}{d r_{\psi}} + n^{2} \left( \Delta_{n} - 1 \right) \frac{d \Delta_{m}}{d r_{\psi}} \right) \nonumber \\
  \times& \Big( \Sin{\left( n - m \right) \theta} \Cos{n \theta_{t n} - m \theta_{t m}} \nonumber \\
  &+ \Cos{\left( n - m \right) \theta} \Sin{n \theta_{t n} - m \theta_{t m}} \Big) \Bigg] \nonumber \\
  \left| \Nabla \alpha \right|^{2}_{1} &= B_{c 0}^{2} \left( \frac{d \psi}{d r_{\psi}} \right)^{-2} \Bigg[ r_{\psi 0} \left( m^{2} \left( \Delta_{m} - 1 \right) \frac{d \Delta_{m}}{d r_{\psi}} + n^{2} \left( \Delta_{n} - 1 \right) \frac{d \Delta_{n}}{d r_{\psi}} \right) \hat{s}' \theta^{2} \nonumber \\
  -& 2 \hat{s}' \theta \Big( m \left( \Delta_{m} - 1 \right) \Sin{z_{m s}} + n \left( \Delta_{n} - 1 \right) \Sin{z_{n s}} \Big) \nonumber \\
  -& r_{\psi 0} \left( 1 - \hat{s}'^{2} \theta^{2} \right) \left( \frac{d \Delta_{m}}{d r_{\psi}} \Cos{z_{m s}} + \frac{d \Delta_{n}}{d r_{\psi}} \Cos{z_{n s}} \right) \label{eq:gradAlphaSqO1} \\
  +& \frac{r_{\psi 0}}{n - m} \left( m^{2} \left( \Delta_{m} - 1 \right) \frac{d \Delta_{n}}{d r_{\psi}} + n^{2} \left( \Delta_{n} - 1 \right) \frac{d \Delta_{m}}{d r_{\psi}} \right) \hat{s}' \theta \nonumber \\
  \times& \Big( \Sin{\left( n - m \right) \theta} \Cos{n \theta_{t n} - m \theta_{t m}} \nonumber \\
  &+ \Cos{\left( n - m \right) \theta} \Sin{n \theta_{t n} - m \theta_{t m}} \Big) \Bigg] \nonumber
\end{align}
\begin{align}
  \left( J_{0} \left( k_{\perp} \rho_{s} \right) \right)_{1} &= - k_{\perp 1} \rho_{s} J_{1} \left( k_{\perp 0} \rho_{s} \right) , \label{eq:FLRO1avg}
\end{align}
where
\begin{align}
  k_{\perp 1} \rho_{s} &\equiv \frac{k_{\perp 0} \rho_{s}}{2} \frac{k_{\psi}^{2} \left| \Nabla \psi \right|_{1}^{2} + 2 k_{\psi} k_{\alpha} \left( \Nabla \psi \cdot \Nabla \alpha \right)_{1} + k_{\alpha}^{2} \left| \Nabla \alpha \right|_{1}^{2}}{k_{\psi}^{2} \left| \Nabla \psi \right|_{0}^{2} + 2 k_{\psi} k_{\alpha} \left( \Nabla \psi \cdot \Nabla \alpha \right)_{0} + k_{\alpha}^{2} \left| \Nabla \alpha \right|_{0}^{2}} . \label{eq:FLRO1def}
\end{align}
From the last terms in each of \refEq{eq:alphaDriftO1}, \refEq{eq:gradPsiDotGradAlphaO1}, and \refEq{eq:gradAlphaSqO1} we see that (even after averaging over $z$) $v_{d s \alpha 1}$, $\left( \Nabla \psi \cdot \Nabla \alpha \right)_{1}$, and $\left| \Nabla \alpha \right|^{2}_{1}$ are all up-down asymmetric.

%% file: App_F_BetaDependence.tex
% !TEX root = /Users/Justin/Documents/Research/Writings/2016DoctoralThesis/DoctoralThesis.tex

\chapter{Dependence of the geometric coefficients on $\beta'$}
\label{app:geoCoeff}

\begin{quote}
	\emph{Much of this appendix appears in reference \cite{BallShafranovShift2016}.}
\end{quote}

In this appendix, we will study the sensitivity of the momentum flux to $\beta'$ by investigating how the gyrokinetic equation changes with $\beta'$. The magnetic geometry only enters the local gyrokinetic model (in the absence of rotation) through eight geometric coefficients (i.e. $\hat{b} \cdot \Nabla \theta$, $B$, $v_{d s \psi}$, $v_{d s \alpha}$, $a_{s ||}$, $\left| \Nabla \psi \right|^{2}$, $\Nabla \psi \cdot \Nabla \alpha$, and $\left| \Nabla \alpha \right|^{2}$), which are defined in chapter \ref{ch:GYRO_Overview}.

The calculation of the geometric coefficients in GS2 is done in the context of the Miller local equilibrium (see chapter \ref{ch:MHD_LocalEquil} and appendix \ref{app:genGeoCoeff}). This must be done carefully as the Miller model takes the flux surface shape and its radial derivative as input, but all second order radial derivatives are calculated by ensuring the Grad-Shafranov equation is satisfied. It is through these second order radial derivatives (as well as the explicit dependence appearing in $v_{d s \alpha}$) that $\beta'$ enters the geometric coefficients. Additionally, we note that we keep the safety factor, the magnetic shear, the background gradients, and the geometry fixed as we change $\beta'$. Therefore, while the Shafranov shift directly enters the flux surface geometry and affects all of the geometric coefficients, the effect of $\beta'$ is limited to a few coefficients. The parameter $\beta'$, which is a normalised form of $d p / d a_{\psi}$ (see \refEq{eq:betaPrimeDef}), only enters into three coefficients: $v_{d s \alpha}$, $\Nabla \psi \cdot \Nabla \alpha$, and $\left| \Nabla \alpha \right|^{2}$. We will start with equations derived in appendix \ref{app:genGeoCoeff} and show precisely how $\beta'$ enters into various quantities. Eventually we will find the three geometric coefficients and see that $\beta'$ has an effect that is small in the inverse aspect ratio $\epsilon \ll 1$, when using the ohmically heated tokamak ordering (i.e. \refEq{eq:gradShafOrderings}).

First we combine \refEq{eq:IalphaDef} and \refEq{eq:dIdpsiFullForm} to get
\begin{align}
   \underbrace{I \frac{d I}{d \psi}}_{B_{0}} &= \bigg( \underbrace{\frac{2 \pi q}{I^{3}}}_{R_{0}^{-3} B_{0}^{-3}} + \underbrace{\left. \oint_{0}^{2 \pi} \right|_{\psi} d \theta' \frac{1}{R^{4} B_{p}^{2} \vec{B} \cdot \Nabla \theta'}}_{\epsilon^{-2} R_{0}^{-3} B_{0}^{-3}} \bigg)^{-1} \Bigg( \underbrace{\frac{2 \pi}{I} \frac{d q}{d \psi}}_{\epsilon^{-2} R_{0}^{-3} B_{0}^{-2}} \label{eq:IdIdpsi} \\
   &- \left. \oint_{0}^{2 \pi} \right|_{\psi} d \theta' \underbrace{\frac{1}{R^{2} \vec{B} \cdot \Nabla \theta'}}_{R_{0}^{-1} B_{0}^{-1}} \bigg[ \underbrace{\frac{\mu_{0}}{B_{p}^{2}} \frac{d p}{d \psi}}_{\epsilon^{-2} R_{0}^{-2} B_{0}^{-1}} - \underbrace{\frac{2}{R^{2} B_{p}} \left( \dlpdthetaPrime \right)^{-1} \frac{\partial Z}{\partial \theta'}}_{\epsilon^{-1} R_{0}^{-2} B_{0}^{-1}} + \underbrace{\frac{2 \kappa_{p}}{R B_{p}}}_{\epsilon^{-2} R_{0}^{-2} B_{0}^{-1}} \bigg] \Bigg) , \nonumber
\end{align}
where the curly braces below the different terms give their ordering in $\epsilon \ll 1$. We see that introducing $\beta'$ creates a lowest order modification to $I \left( d I / d \psi \right)$. Next, using \refEq{eq:toroidalCur}, we can find that the right-hand side of the Grad-Shafranov equation can be written as
\begin{align}
   \underbrace{\mu_{0} j_{\zeta} R}_{B_{0}} &= - \bigg( \underbrace{\frac{2 \pi q}{I^{3}}}_{R_{0}^{-3} B_{0}^{-3}} + \underbrace{\left. \oint_{0}^{2 \pi} \right|_{\psi} d \theta' \frac{1}{R^{4} B_{p}^{2} \vec{B} \cdot \Nabla \theta'}}_{\epsilon^{-2} R_{0}^{-3} B_{0}^{-3}} \bigg)^{-1} \label{eq:betaPrimeInhomoGradShaf} \\
   \times& \Bigg[ \underbrace{\mu_{0} R^{2} \frac{\partial p}{\partial \psi}}_{B_{0}} \Bigg( \underbrace{\frac{2 \pi q}{I^{3}}}_{R_{0}^{-3} B_{0}^{-3}} + \underbrace{\left. \oint_{0}^{2 \pi} \right|_{\psi} d \theta' \frac{1}{R^{4} B_{p}^{2} \vec{B} \cdot \Nabla \theta'}}_{\epsilon^{-2} R_{0}^{-3} B_{0}^{-3}} - \underbrace{\frac{R_{0}^{2}}{R^{2}} \left. \oint_{0}^{2 \pi} \right|_{\psi} d \theta' \frac{1}{R_{0}^{2} R^{2} B_{p}^{2} \vec{B} \cdot \Nabla \theta'}}_{\epsilon^{-2} R_{0}^{-3} B_{0}^{-3}} \Bigg) \nonumber \\
   +& \underbrace{\frac{2 \pi}{I} \frac{d q}{d \psi}}_{\epsilon^{-2} R_{0}^{-3} B_{0}^{-2}} - \left. \oint_{0}^{2 \pi} \right|_{\psi} d \theta' \underbrace{\frac{1}{R^{2} \vec{B} \cdot \Nabla \theta'}}_{R_{0}^{-1} B_{0}^{-1}} \bigg( \underbrace{\frac{2 \kappa_{p}}{R B_{p}}}_{\epsilon^{-2} R_{0}^{-2} B_{0}^{-1}} - \underbrace{\frac{2}{R^{2} B_{p}} \left( \dlpdthetaPrime \right)^{-1} \frac{\partial Z}{\partial \theta'}}_{\epsilon^{-1} R_{0}^{-2} B_{0}^{-1}} \bigg) \Bigg] , \nonumber
\end{align}
which explicitly includes a term proportional to the pressure gradient (i.e. $\beta'$). However, to lowest order in aspect ratio the coefficient of this term is zero as it is composed of a safety factor term that is small and two integral terms that cancel with each other (because $R = R_{0} + O \left( \epsilon R_{0} \right)$). All other quantities in \refEq{eq:betaPrimeInhomoGradShaf} do not contain the pressure gradient and can be calculated directly from the flux surface geometry provided to the Miller model. Therefore, $\beta'$ only introduces an $O \left( \epsilon B_{0} \right)$ modification to $\mu_{0} j_{\zeta} R$.

We will see that the toroidal current density from the Grad-Shafranov equation (i.e. $\mu_{0} j_{\zeta} R$) will appear in several places in the geometric coefficients. For example, rearranging \refEq{eq:dBpdpsi} gives the radial derivative of the poloidal field as
\begin{align}
   \underbrace{\frac{\partial B_{p}}{\partial \psi}}_{a^{-1} R_{0}^{-1}} &= \underbrace{\frac{\mu_{0} j_{\zeta} R}{R^{2} B_{p}}}_{a^{-1} R_{0}^{-1}} - \underbrace{B_{p} \left( \dlpdtheta \right)^{-1} \frac{\partial}{\partial \psi} \left( \dlpdtheta \right)}_{a^{-1} R_{0}^{-1}} \\
   &+ \underbrace{\left( \dlpdtheta \right)^{-1} \frac{\partial}{\partial \theta} \left( B_{p} \left( \dlpdtheta \right)^{-1} \frac{\partial \vec{r}}{\partial \psi} \cdot \frac{\partial \vec{r}}{\partial \theta} \right)}_{a^{-1} R_{0}^{-1}} . \nonumber
\end{align}
Although the toroidal current term appears as $O \left( a^{-1} R_{0}^{-1} \right)$, the effect of $\beta'$ on $\partial B_{p} / \partial \psi$ is small by an order (i.e. $O \left( \epsilon a^{-1} R_{0}^{-1} \right)$) because $\beta'$ does not enter $\mu_{0} j_{\zeta} R$ to lowest order. To calculate the derivative of the toroidal field we can directly differentiate $B_{\zeta} = I / R$ to get \refEq{eq:dBtordpsi}. Ordering both terms we see that the effect of $d I / d \psi$ is small, so the effect of $\beta'$ on $\partial B_{\zeta} / \partial \psi$ through \refEq{eq:IdIdpsi} is small by one order, entering at $O \left( \epsilon a^{-1} R_{0}^{-1} \right)$.

Using \refEq{eq:IalphaDef} and \refEq{eq:gradAlphaFinal} gives
\begin{align}
   \underbrace{\Nabla \alpha}_{a^{-1}} &= \Bigg( - \left. \int_{\theta_{\alpha}}^{\theta} \right|_{\psi} d \theta' \underbrace{\frac{I}{R^{2} \vec{B} \cdot \Nabla \theta'}}_{1} \bigg[ \underbrace{\frac{1}{I} \frac{d I}{d \psi}}_{\epsilon^{2} a^{-2} B_{0}^{-1}} - \underbrace{\frac{\mu_{0} j_{\zeta} R}{R^{2} B_{p}^{2}}}_{a^{-2} B_{0}^{-1}} - \underbrace{\frac{2}{R^{2} B_{p}} \left( \dlpdthetaPrime \right)^{-1} \frac{\partial Z}{\partial \theta'}}_{\epsilon a^{-2} B_{0}^{-1}} + \underbrace{\frac{2 \kappa_{p}}{R B_{p}}}_{a^{-2} B_{0}^{-1}} \bigg] \nonumber \\
   &+ \underbrace{\left[ \frac{I \Nabla \psi \cdot \Nabla \theta'}{R^{4} B_{p}^{2} \vec{B} \cdot \Nabla \theta'} \right]_{\theta' = \theta_{\alpha}}^{\theta' = \theta}}_{a^{-2} B_{0}^{-1}} + \underbrace{\left( \frac{I}{R^{2} \vec{B} \cdot \Nabla \theta'} \right)_{\theta' = \theta_{\alpha}} \frac{d \theta_{\alpha}}{d \psi}}_{a^{-2} B_{0}^{-1}} \Bigg) \underbrace{\Nabla \psi}_{a B_{0}} \\
   &- \underbrace{\frac{I}{R^{2} \vec{B} \cdot \Nabla \theta} \Nabla \theta}_{a^{-1}} + \underbrace{\Nabla \zeta}_{\epsilon a^{-1}} . \nonumber
\end{align}
By ordering the various terms we find that the $d I / d \psi$ term is small by two orders in $\epsilon \ll 1$. However, the $\mu_{0} j_{\zeta} R$ term enters to lowest order, therefore the effect of $\beta'$ on $\Nabla \alpha$ is only small by one order (i.e. $\Order{\epsilon a^{-1}}$). The dependence of the coefficients $\Nabla \psi \cdot \Nabla \alpha$ and $\left| \Nabla \alpha \right|^{2}$ on $\Nabla \alpha$ is apparent. Hence $\beta'$ does not enter $\Nabla \psi \cdot \Nabla \alpha$ and $\left| \Nabla \alpha \right|^{2}$ to lowest order in $\epsilon \ll 1$. Instead it enters to next order due to the quantity $\mu_{0} j_{\zeta} R$, which is given by \refEq{eq:betaPrimeInhomoGradShaf}. The geometric coefficient $v_{d s \alpha}$ is more complicated. Substituting \refEq{eq:dBtordpsi} into \refEq{eq:alphaGCdriftvelocity} gives
\begin{align}
   \underbrace{v_{d s \alpha}}_{a^{-1} R_{0}^{-1} v_{th s}^{2} \Omega_{s}^{-1}} &= - \underbrace{\frac{w_{||}^{2}}{\Omega_{s}}}_{v_{th s}^{2} \Omega_{s}^{-1}} \Bigg( - \underbrace{\frac{\mu_{0} j_{\zeta} R}{R^{2} B}}_{\epsilon a^{-1} R_{0}^{-1}} - \underbrace{\frac{I^{2}}{R^{3} B} \frac{\partial R}{\partial \psi}}_{a^{-1} R_{0}^{-1}} + \underbrace{\frac{B_{p}}{B} \frac{\partial B_{p}}{\partial \psi}}_{\epsilon a^{-1} R_{0}^{-1}} - \underbrace{\frac{\partial B}{\partial \theta} \frac{ \hat{b} \cdot \left( \Nabla \theta \times \Nabla \alpha \right)}{B}}_{a^{-1} R_{0}^{-1}} \Bigg) \\
   &- \underbrace{\frac{\mu B}{m_{s} \Omega_{s}}}_{v_{th s}^{2} \Omega_{s}^{-1}} \Bigg( \underbrace{\frac{I}{R^{2} B} \frac{d I}{d \psi}}_{\epsilon a^{-1} R_{0}^{-1}} - \underbrace{\frac{I^{2}}{R^{3} B} \frac{\partial R}{\partial \psi}}_{a^{-1} R_{0}^{-1}} + \underbrace{\frac{B_{p}}{B} \frac{\partial B_{p}}{\partial \psi}}_{\epsilon a^{-1} R_{0}^{-1}} - \underbrace{\frac{\partial B}{\partial \theta} \frac{ \hat{b} \cdot \left( \Nabla \theta \times \Nabla \alpha \right)}{B}}_{a^{-1} R_{0}^{-1}} \Bigg) \nonumber .
\end{align}
We see that $\beta'$ will enter into the $\mu_{0} j_{\zeta} R$ term as well as both $\partial B_{p} / \partial \psi$ terms, but ordering these three terms reveals that the effect of $\beta'$ is $\Order{\epsilon^{2} a^{-1} R_{0}^{-1} v_{th s}^{2} \Omega_{s}^{-1}}$. The parameter $\beta'$ has a much larger $O \left( \epsilon a^{-1} R_{0}^{-1} v_{th s}^{2} \Omega_{s}^{-1} \right)$ effect through the two $\Nabla \alpha$ terms as well as the $d I / d \psi$ term. Figure \ref{fig:driftCoeff} illustrates the relative magnitudes of these two effects for several flux surface geometries. The difference between the dotted red line and the dashed blue line indicates the effect of the $d I / d \psi$ term, while the difference between the solid black line and the dotted red line indicates the effect of $\mu_{0} j_{\zeta} R$ acting through $\Nabla \alpha$. We see that the effect of $\mu_{0} j_{\zeta} R$ seems to dominate.

In conclusion, $\beta'$ only enters into three of the geometric coefficients: $v_{d s \alpha}$, $\Nabla \psi \cdot \Nabla \alpha$, and $\left| \Nabla \alpha \right|^{2}$. Because $\beta'$ has such a large effect on the momentum flux, this reinforces the idea from chapter \ref{ch:GYRO_MomFluxScaling} that $v_{d s \alpha}$ is a particularly important geometric coefficient for understanding intrinsic rotation transport due to up-down asymmetry. We also learned that the dominant effect of $\beta'$ on $\Nabla \psi \cdot \Nabla \alpha$ and $\left| \Nabla \alpha \right|^{2}$ is through the quantity $\mu_{0} j_{\zeta} R$ and is small in $\epsilon \ll 1$. The drift coefficient $v_{d s \alpha}$ also depends on $\beta'$ to next order because of $\mu_{0} j_{\zeta} R$. However, it has another separate dependence through the quantity $d I / d \psi$ that is formally the same size in $\epsilon \ll 1$, but it appears to be a weak effect (at least in the geometries of chapter \ref{ch:GYRO_ShafranovShift}). These dependences are the only way that the gyrokinetic model knows about $\beta'$. Hence they must be responsible for the significant reduction in the momentum transport.